# Human-Computer Interaction with
# Adaptable & Adaptive
# Motion-based Games for Health

by

Jan David Smeddinck

Edited for independent publication and based on my PhD thesis submitted for the degree of:

Doctor of Engineering (Dr.-Ing.)

Faculty 3: Mathematics / Computer Science

University of Bremen

Bremen, Germany

Submission: 2016 November 07

Defense: 2016 December 19

Edited publication: 2020 December 06

1st Supervisor: Prof. Rainer Malaka (University of Bremen, Germany)

2nd Supervisor: Assoc. Prof. Regan Mandryk (University of Saskatchewan, Canada)

External Examiner: Prof. Patrick Olivier (Newcastle University, United Kingdom)

*„Die Wahrheit kann nur im Geiste der Wissenschaft als Betrachtung des Ganzen im subjektiv fortbestehenden und reflektierten Sinne, beinhaltend sowohl Gestalt als auch Form/Prinzip, verstanden werden."*

*– Georg Wilhelm Friedrich Hegel*
*(Phänomenologie des Geistes)*

In Erinnerung an Eberhard

# Abstract

Technological and medical advances are leading to great improvements in overall quality of life and life expectancy. However, these positive developments are accompanied by considerable challenges. The modern sedentary lifestyle and common afflictions that become more prevalent with age are contributing to considerable burdens on health care systems and on a great number of individuals. In addition to specific primary treatments, physical activity plays a major role both in prevention and in the treatment of such afflictions, for example though the application area of physiotherapy. *Games for health* (GFH) in general and *motion-based games for health* (MGH) in particular are being discussed in research and industry for their ability to play a supportive role in health, by offering (a) *motivation* to engage in treatments, (b) *objective insights* on the status and development of individuals or groups based on data collection and analysis, and (c) *guidance* regarding treatment activities, which is especially promising when health professionals are not available in person. However, applications in health need to be tailored to the individual needs and abilities of patients in order to facilitate the best possible outcomes. While most games can be adjusted to a general level of player abilities, this is typically achieved with a single difficulty setting with a limited number of discrete tiers, such as "easy", "medium", and "hard". For most serious application use cases in health, more fine-grained and far-reaching adjustments are required. This can quickly lead to a need for applying adjustments on complex sets of parameters, which can be overwhelming for patient-players and even trained professionals. Automatic adaptivity and efficient manual adaptability are thus major concerns for the design and development of GFH and MGH. Despite a growing amount of research on specific methods for adaptivity, general considerations on human-computer interaction with adaptable and adaptive MGH are rare and scattered across reports from specific developments. Based on a thorough consideration of the existing background and related work, this thesis therefore focuses on establishing and augmenting theory for adaptability and adaptivity in human-computer interaction in the context of MGH. The considerations are supported by a series of studies and practical developments. Working with older adults and people with Parkinson's disease as frequent target groups that can arguably benefit from tailored activities, explorations and comparative studies that investigate the design, acceptance, and effectiveness of MGH are presented. The outcomes encourage the application of adaptivity for MGH following iterative human-centered design that considers the respective interests of the complex collage of involved parties and stakeholders, provided that the users receive adequate information and are empowered to exert control over the automated system when desired or required, and if adaptivity is embedded in such a way that it does not interfere with the users' sense of competence or autonomy.

# Acknowledgments

Over the course of the past six years, I was lucky enough to be met with great advice, collaboration, encouragement, and engagement, thanks to a great number of wonderful individuals. Without them, this work would not have been possible. I would like to thank my advisors *Rainer Malaka*, *Regan Mandryk*, and *Patrick Olivier* for their great support. Furthermore, this work would not exist without the lessons provided by *Marc Herrlich*, *Markus Krause*, and *Robert Porzel*, each of whom provided different, yet invaluable pieces of wisdom, practical advice, and led by example in their own respective personal styles. Words cannot express my gratitude to *Jenny Cramer*, who has cheered me on and supported me more than anyone else, and who has also provided a constant stream of valuable input to all my scientific work over the past years. I also want to thank my loving family, *Martin, Sabine, York, Eberhard, Brigitte, Andreas*, and *Regina*, who have always been there for me, as well as my extended family in the USA, *Allen, Dea Ann*, and *Kelsey*, and my newly won Capoeira family in Bremen at the *Centro Cultural Cazuá*. Most of my work has been cooperative and has relied on the great people at the *TZI Digital Media Lab*. Thank you, *Benjamin Walther-Franks, Nina Runge, Dirk Wenig, Insa Warms-Cangalovic, Irmgard Laumann, Gerald Volkmann, Dmitry Alexandrovsky, Himangshu Sarma*, and all other members and former members of the lab and the graduate school *Advances in Digital Media*. I owe further thanks to wonderful collaborators around the globe, including all my co-authors, *Kathrin Gerling, Max Birk*, and many members of the *HCI Lab* at the University of Saskatchewan, as well as of the *Open Lab* at Newcastle University, and the *Sony Computer Science Laboratories* in Paris, who have been wonderful hosts for research visits. Many of the research projects discussed in this thesis would not have been possible without the hard work and dedication of wonderful former students, including *Sandra Siegel, Saranat Tiemkeo, Jens Voges, Jorge Hey, Max Roll, Xiaoyi Wang, Guangtao Zhang, Simon Hermsdorf, Sophie Janzen*, and many others. I also have to thank all other members of the master projects *WuppDi* and *sPortal*. A number of projects were supported by professional collaborators, colleagues, companies and institutions. Thank you, *Jürgen Weemeyer* and the *vacances GmbH, Jan Wolters, Christine Jacobsen*, and the staff at *Rehamed, abraxas Medien*, the *German Parkinson's association* in Bremen, and many others. This work was also supported and enabled in many ways through great discussions with – and personal support by – friends, including *Henning Schmidtke, Maria Eckstein, Susan Wache*, and *the IK tribe*.

Funding for the projects that contributed to this thesis was provided by the *Wirtschaftsförderung Bremen* and the *EU fund for regional development* (EFRE), by the *German Federal Ministry of Education and Research* (BMBF), by the *Canadian GRAND NCE*, the *University of Bremen*, as well as by the *Klaus Tschira Stiftung gGmbH* (KTS), which granted the essential stipend that funded the first three years of my PhD work. Thank you, *Klaus Tschira*; sadly, I was late to tell you in person.

# Contents

# PART I: Overview of the Topic and Contributions

## 1  Introduction

Humanity has made historic advancements allowing more and more people to live long lives that do not require carrying out backbreaking work. However, the modern sedentary lifestyle and demographic change also lead to challenges that manifest through considerable individual and societal burden in the form of lifestyle diseases and other afflictions that are often chronic and become more prevalent with increasing age. Next to individual suffering, these conditions put increasing pressure on healthcare systems. Approaches towards alleviating these challenges are thus much needed.

Physical activity has been shown to be a major contributing factor in preventing lifestyle disease and in postponing or easing age-related afflictions (Böhm et al., 2009). Interactive digital media applications have been proposed and investigated as tools for persuasion and behavior change towards increased physical activity (Noar & Harrington, 2012). Video games in particular have been proposed in this context for their notable potential to *motivate*, as well as to provide *real-time guidance* regarding the quality and safety of ongoing exercise execution, and to provide *objective analyses* of abilities and progress over time (see section 3.1). Accordingly, video games have been investigated and found to be potentially beneficial in use cases for general fitness and prevention, as well as for physiotherapy and rehabilitation (Granic et al., 2013).

However, numerous actors in the field have also pointed out that considerable challenges remain with the design and development of such *exergames* or *motion-based games for health* (MGH) (Baranowski et al., 2008; Ricciardi & De Paolis, 2014; Sinclair et al., 2007). A *central challenge* results from the fact that players from the target groups of *games for health* (GFH) typically display a broad range of *different abilities and needs* (Göbel et al., 2010; Hocine et al., 2014). In order to correspond to the strongly heterogeneous needs of patients who become GFH players (thus *patient-players*), games must be rather flexible, offering *manual adaptability* and potentially *automatic adaptivity* that can facilitate a level of personalization that clearly goes beyond simple monolithic difficulty tiers such as *"easy", "medium",* or *"hard"* that are often used in commercial (movement-based) games. This general need for adaptation has been noted in related work and a range of approaches, especially in the direction of automatically adaptive solutions, have already been implemented (Alankus et al., 2010; Geurts et al., 2011; Göbel et al., 2010; Hocine et al., 2014). However, most work in these directions is still explorative, lacking

medium- or long-term validations, and work regarding the acceptance and situated functioning of such systems, as well as regarding the presentation and interaction with difficulty adjustments, is scarce to non-existent. Hence, this thesis is dedicated to the following general research problem: *"How can adaptability and adaptivity in MGH be realized in an efficient, effective, and enjoyable manner?"*

This broad general research problem is addressed through multiple projects with more specific sub-questions and according hypotheses that are oriented along three guiding research questions that address the underlying research problem. These guiding research questions are:

Q1; Regarding the *design*, or "how to realize such systems?": *What are the requirements of adaptable and adaptive systems for MGH and how can such systems be designed and implemented to respect the requirements of their specific application use-cases? E.g. with games targeting the support of physiotherapy for people with Parkinson's disease, games for the support of people with chronic unspecific lower back afflictions, or the requirements of more general groups, such as older adults?*

Q2; regarding the *acceptance*, or the "how will it be perceived?": *Will manual adaptation options and automatic adaptivity in MGH systems be accepted, and can they offer adequate user experience for both patients and professionals?*

Q3; regarding the *effectiveness*, or the "how well will it work?": *Will adaptation and adaptivity in MGH work effectively and efficiently in practice with different specific use-cases, as well as both for patients and professionals?*

## 1.1   Thesis Structure

The following chapters of this thesis contain investigations in theory and practice towards all three of these broad research questions along following approach: all further work is based on an overview of the background works on the general need and approaches for (motion-based) games for health (see section 2.1), a discussion of the terminology and definitions around serious games / games for health (section 2.2) and adaptable and adaptive systems (section 2.3). This is further complemented by a summary of the development of sensor devices (section 2.4) and theory that is rooted in motivational psychology (section 2.5), the state-of-the-art of exergames (section 2.6) and (adaptive / motion-based) games for health (section 2.7), as well as questions around privacy and security in digital health systems (section 2.9).

A summary of the theoretical developments that scaffold the approaches taken in this thesis is provided in chapter 3, touching on the basic promises of MGH (section 3.1), the general challenges of - and approaches to - adaptability and adaptivity in the context of MGH (section 3.2), a modular architecture for adaptive MGH (section 3.3), a structured approach to the modeling dimensions of adaptivity and adaptability for MGH (section 3.4) based on prior work on automation (Parasuraman et al., 2000; Sheridan, 2001) and modeling dimensions for adaptive systems (Andersson et al., 2009), the interdependent roles of accessibility, playability, player experience, and effectiveness as guiding considerations for the design and development of MGH (section 3.5), an extended model for the applicability of leading motivational theories in the context of serious games (section 3.6) that is based on self-determination theory (Ryan et al., 2006) and (dual-)flow (Chen, 2007; Csikszentmihalyi, 1990; Sinclair et al., 2009), as well as considerations on the role of human-centered and embodied aspects in the context of this work (section 3.7), culminating in a combined process model for *needs and abilities based human-centered design for adaptive systems* (section 3.8).

In the following practical part, a number of system developments and studies with individual specific research questions that contribute to the broader questions guiding this thesis are summarized and subsequently discussed in relation to the underlying theory and the guiding research questions. The included works consider preconditions and enabling technologies through research on game difficulty and difficulty adjustments explored with surveys (section 4.3), as well as the impact of simple gamification elements in fitness information application interfaces on motivation and performance (section 4.4). These summaries are followed by a discussion of the iterative human-centered development process of the MGH project *"Spiel Dich fit und gesund"* (SDF) (English: "play for fitness and health") that serves as a case study (see section 4.5) for the structuring theoretical considerations discussed in chapter 3. A series of research questions that stem from the work on the SDF project are then discussed and structured along the lines of two topic areas: (1) *interacting with difficulty adjustments*, including studies regarding the impact of certain game design and framing choices on player experience and performance. The study topics cover the impact of different modalities for presenting game difficulty choices (section 4.6.1), as well as embodied movement capability configurations on mobile devices (section 4.6.2). The studies in this first topic area are directly concerned with human-computer interaction around manual and automatic difficulty adjustments, contributing primarily to guiding research question Q1. (2) *Communicating feedback for motion-based health applications*, including studies on the impact of manipulations in visual complexity (section 4.7.1) and of different exercise instruction modalities (section 4.7.2) on player experience, performance, and acceptance, contributing both to guiding research question Q1 and Q2.

Two further studies that are foundational to this work are discussed in the next sections. The first study focuses on the general acceptance and situated functioning of adaptivity in MGH with the application use case of Parkinson's disease (section 4.8), contributing primarily to guiding research question Q2. The second study takes a medium-term situated approach comparing the functional impact and the influence of the prolonged use of adaptive or manually adaptable MGH on motivation and performance to regular physical therapy for chronic lower-back afflictions (section 4.9), contributing primarily to guiding research question Q3. Lastly, a number of explorative research projects are introduced that cover topics that connect to the main work discussed in this thesis, but go beyond adaptability and heuristic adaptivity for MGH implemented in the style of causal games (4.10). The limitations of both the theory and the practical projects are then discussed together with according indications for future work and an outlook on promising ongoing and upcoming work that connects to the guiding research questions of this thesis.

## 1.2   Thesis Contributions

While some prior work regarding the guiding research questions exists, this thesis makes novel contributions to the three questions both in building additional theory that connects to – and extends – related work, as well as through a series of system developments, evaluations, and studies. The *main contributions* are: (1) a summary of related work around human-computer interaction with adaptable and adaptive motion-based games for health, (2) a model for the design, implementation, and evaluation of adaptable and adaptive motion-based games for health based on prior work on automation, user-centered iterative design, and motivational psychology, and (3) a series of studies investigating different specific research questions and hypotheses that fall within the three general research questions that were introduced above. The studies that are the most central contributions of this thesis are: (a) an early situated case study with three subjects on the effectiveness, acceptance, and player experience of an adaptive motion-based game for the application area of Parkinson's disease, indicating that adaptivity in MGH, implemented following the theoretical considerations discussed in this thesis, is accepted and can lead to positive performance developments and positive game experiences [publication A.1]; (b) a medium-term situated study comparing adaptable motion-based games for the support of physiotherapy for chronic back afflictions to the same games with additional adaptive features and to traditional therapy without the games. This study was carried out in a situated manner and considered therapeutic effectiveness, usability for therapists, as well as user experience and acceptance for therapists and patients. The results indicate that such flexible MGH designed for upper body movement activation can lead to increased functional reach and to improvements in perceived autonomy and presence compared to traditional therapy, while also showing decreased tension-pressure and effort-importance [A.2]; and (c) a study and repeat-study of the effects of different modalities for presenting game difficulty choices to players (varying the level of control offered to players as well as the level of intrusiveness of the modality to the remaining game experience) in which the resulting player experience and affect are observed. Comparing traditional difficulty selection menus to difficulty choices that are embedded in the gameplay, and to automatic difficulty adjustments, these studies indicate that autonomy can be increased if players have manual control over difficulty settings, while the remaining and overall game experience measures remained largely unaffected [A.3].

Regarding the guiding research questions, the contributions in this thesis indicate a number of design decisions that can either lead to improved game experience, performance, or effectiveness regarding the targeted serious outcome(s), or that were found not to have a notable impact on these aspects for the sample populations. The discussion in theory, about how MGH with adaptability and adaptivity can be designed and analyzed, is underlined with practical examples and studies that deliver further evidence that adaptivity and adaptability in MGH are accepted by players and professionals alike and that show how flexible MGH can be effective for a number of application scenarios, such as augmenting therapy exercise options for people with Parkinson's disease or chronic lower-back afflictions.

---

*This is the edited version of a cumulative thesis. Most central arguments, insights, and figures that are presented and discussed in the body of this work have been published in peer reviewed papers. The following list of publications provides an overview of these works. While the publications are discussed throughout the body of this thesis, using the regular reference and citation style, content summaries of selected publications will be provided in places where they further the understanding of the line of arguments. Each publication is explicitly mentioned at least once in the text as **[publication X.Y]** in bold to indicate that a consideration of the publication supports or furthers the matter of discussion at that point. Excerpts from summative publications will be used where adequate to augment this text. References within these sections are re-mapped to the reference list included with this thesis. Figures are also adjusted to match the list of figures in this thesis. The following list of publications contains works that have benefitted substantially from the participation of students, or from the involvement of the listed co-authors, who have contributed to – or at times led – the underlying projects and writing. Part II of this thesis contains brief summaries of the contribution of the publications with regard to the context of this work, as well as brief statements on the personal contribution that provide a general impression regarding the scope of the involvement of the author of this thesis in the respective works.*

## A. Foundational publications

*[A.1] Smeddinck, J., Siegel, S., & Herrlich, M. (2013).* **Adaptive Difficulty in Exergames for Parkinson's disease Patients**. *In Proceedings of Graphics Interface 2013 (pp. 141–148). Regina, SK, Canada: Canadian Human-Computer Communications Society.*

*[A.2] Smeddinck, J. D., Herrlich, M., & Malaka, R. (2015).* **Exergames for Physiotherapy and Rehabilitation: A Medium-term Situated Study of Motivational Aspects and Impact on Functional Reach**. *In Proc. of the 33rd Annual ACM Conf. on Human Factors in Computing Systems (CHI 15) (p. 4143–4146). New York, NY, USA: ACM.*

*[A.3] Smeddinck, J. D., Mandryk, R. L., Birk, M. V., Gerling, K. M., Barsilowski, D., & Malaka, R. (2016).* **How to Present Game Difficulty Choices? Exploring the Impact on Player Experience**. *In Proceedings of the 2016 CHI Conference on Human Factors in Computing Systems (S. 5595–5607). New York, NY, USA: ACM.*

## B. Summative and supportive publications

*[B.1] Gerling, K. M., Miller, M., Mandryk, R. L., Birk, M. V., & Smeddinck, J. D. (2014).* **Effects of Balancing for Physical Abilities on Player Performance, Experience and Self-esteem in Exergames**. *In Proc. of the 32nd Annual ACM Conf. on Human Factors in Computing Systems (pp. 2201–2210). New York, NY, USA: ACM.*

*[B.2] Malaka, R., Herrlich, M., & Smeddinck, J. (2017).* **Anticipation in Motion-Based Games for Health**. *In M. Nadin (Ed.), Anticipation and Medicine (pp. 351–363). Springer International Publishing, Berlin / Heidelberg.*

*[B.3] Smeddinck, J. D. (2016).* **Games for Health**. *In R. Dörner, S. Göbel, M. Kickmeier-Rust, M. Masuch, & K. Zweig (Eds.), LNCS Entertainment Computing and Serious Games (Vol. 9970, pp. 212–264). Springer International Publishing, Berlin / Heidelberg.*

*[B.4] Smeddinck, J. D., Gerling, K. M., & Malaka, R. (2014).* **Anpassbare Computerspiele für Senioren**. *Informatik-Spektrum, 37(6), pp. 575–579.*

*[B.5] Smeddinck, J., Gerling, K.M., & Tiemkeo, S. (2013).* **Visual Complexity, Player Experience, Performance and Physical Exertion in Motion-Based Games for Older Adults**. *In Proceedings of the 15th ACM SIGACCESS International Conference on Computers and Accessibility (ASSETS '13), Bellevue, WA, USA.*

*[B.6] Smeddinck, J., Herrlich, M., Krause, M., Gerling, K., & Malaka, R. (2012).* **Did They Really Like the Game? -Challenges in Evaluating Exergames with Older Adults**. *Proceedings of the CHI Game User Research Workshop. CHI 2012, Game User Research Workshop, Austin, Texas, USA.*

*[B.7] Smeddinck, J. D., Herrlich, M., Roll, M., & Malaka, R. (2014).* **Motivational Effects of a Gamified Training Analysis Interface**. *In A. Butz, M. Koch, & J. Schlichter (Hrsg.), Mensch & Computer 2014 - Workshopband (pp. 397–404). Berlin: De Gruyter Oldenbourg.*

*[B.8] Smeddinck, J. D., Herrlich, M., Wang, X., Zhang, G., & Malaka, R. (2018).* **Work Hard, Play Hard: How Linking Rewards in Games to Prior Exercise Performance Improves Motivation and Exercise Intensity**. *Entertainment Computing (Vol. 29, pp. 20–30). Elsevier, Amsterdam, Netherlands.*

*[B.9] Smeddinck, J. D., Hey, J., Runge, N., Herrlich, M., Jacobsen, C., Wolters, J., & Malaka, R. (2015).* **Movi-Touch: Mobile Movement Capability Configurations**. *In Proceedings of the 17th International ACM SIGACCESS Conference on Computers & Accessibility (ASSETS '15) (pp. 389–390). New York, NY, USA: ACM.*

*[B.10] Smeddinck, J. D., Voges, J., Herrlich, M., & Malaka, R. (2014).* **Comparing Modalities for Kinesiatric Exercise Instruction**. *In CHI '14 Extended Abstracts on Human Factors in Computing Systems (pp. 2377–2382). New York, NY, USA: ACM.*

*[B.11] Streicher, A., & Smeddinck, J. D. (2016).* **Personalized and Adaptive Serious Games**. *In R. Dörner, S. Göbel, M. Kickmeier-Rust, M. Masuch, & K. Zweig (Eds.), Entertainment Computing and Serious Games (Vol. 9970, pp. 332–377), LNCS. Cham: Springer International Publishing.*

## C. Additional related publications

*[C.1] Eggert, C., Herrlich, M., Smeddinck, J., & Malaka, R. (2015).* **Classification of Player Roles in the Team-Based Multi-player Game Dota 2**. *In K. Chorianopoulos, M. Divitini, J. B. Hauge, L. Jaccheri, & R. Malaka (Eds.), Entertainment Computing - ICEC 2015 (Vol. 9353, pp. 112–125), LNCS. Cham: Springer International Publishing.*

# 2   Background and Related Work

The following sections summarize the background on motion-based games for health, adaptivity, and adaptability, including related work that the theoretical considerations, as well as the studies and developments presented in the later chapters of this thesis, build upon. This includes an overview of the broader challenges in health and demographic change that provide the underlying motivation, as well as discussions of the terminology around motion-based games for health, adaptability, and adaptivity. It also includes summaries of the progress in sensor devices and applications that support the development of GFH, of related theories from motivational psychology, as well as notes of seminal work in the areas that are relevant to motion-based games for health, and a brief summary of the aspects of security, privacy, and ethics, which are important in the context of player – and especially patient – data processing.

## 2.1   Basic Considerations on Games for Health

### 2.1.1   The Need and Market for Games for Health

Advances in modern medicine, agriculture and nutrition, transport, jobs with lower immediate physical demand, and other related developments have spurred considerable improvements in life expectancy. The number of people aged 65 or older is projected to grow from an estimated 524 million in 2010 to almost 1.5 billion in 2050, marking a jump from around 8% to 16% percent of the world population (WHO / NIH, 2011). Yet, at the same time, modern lifestyles have changed and factors such as poor diets and pervasive sedentary activities are major challenges that aggravate the prevalence of chronic noncommunicable diseases and threaten the otherwise improved quality of life. This applies to all age groups and to older adults in particular (Bloom et al., 2012). Next to considerable individual burden, this produces notable societal cost. As an example from a developed nation, the estimated direct and indirect costs of sedentary lifestyle to chronic health conditions in the US alone have been reported to be in excess of \$150 billion, or approximately 15% of the total national healthcare budget per year (adjusted for USD in the year 2000) (Booth & Chakravarthy, 2002). For older US Americans, 95% of health care costs result from chronic disease and health care spending is expected to increase by 25% from 2011 to 2030, even without accounting for inflation or higher costs of new technologies (Centers for Disease Control and Prevention, US Dept of Health and Human Services, 2013). At the same time, a minimum of 2.5 hours of *low to medium intensity physical activity* per week has been associated with a 30% reduction in across a range of morbidity factors related to sedentary lifestyle, old age, and obesity (e.g. coronary artery disease, stroke, and type 2 diabetes) (Hu et al., 2000, 2001; Manson et al., 1999). *Regular physical activity* has also been linked to *positive influences on overall well-being and healthy aging* (Böhm et al., 2009). Still, less than 5% of US adults engage in 30 minutes of physical activity each day, and only ~33% achieve the recommended amount of physical activity per week (National Center for Health Statistics (U.S.), 2012; President's Council on Fitness, Sports & Nutrition, 2016).

Next to public health messaging through traditional media and health-related programs in public institutions, such as schools or day care centers, *digital interactive media* can play a key role in improving awareness, in providing specific information and guidance, as well as in supporting motivation and social exchange for sustained adjustments towards a healthier lifestyle (Wylie & Coulton, 2009). Digital interactive applications have also been shown to provide substantial benefits in the support of the treatment of noncommunicable diseases through telerehabilitation (Galiano-Castillo et al., 2016; Moffet et al., 2015). In addition to the general role of computing, data processing, and information systems in health, *video games* have more recently been identified as another promising candidate digital technology (Granic et al., 2013). This is due to their – in principle – *inherently motivating nature*, the ability to *provide real-time guidance* and feedback, as well as due to the potential to support *data aggregation and medium- to long-term analysis* (cf. section 3.1). While prominent voices in general public debate have long dismissed video games as a niche medium with little appeal to the broad populace, video games are now the largest entertainment medium in many countries (Ryan et al., 2006). Big titles from series such as *Call of Duty* or *GTA*[i] amass more release day revenues than any book, record album, or movie (Ryan et al., 2006), can achieve multi-billion-dollar accumulated revenues (Thier, 2014) from game sales alone, and drive the development of whole franchises. At the same time, video game players today defy the traditional stereotypes. In 2009, 67% of US households owned either a console or PC used for entertainment software. Furthermore, the average US gamer is already more than 34 years old, roughly 40% of players are female, and approximately 26% are aged 50 years or older (Primack et al., 2012). This helps explain the considerable value proposition for serious games in health in well-being with multi-billion-dollar market estimates (Sánchez et al., 2012). In summary, games are a successful medium with wide reach into modern households and while traditional sedentary games certainly contribute to the time spent on sedentary activities, research on games is now looking past the traditional controversy about game addiction, violence and other potential negative effects (Rigby & Ryan, 2011), focusing on serious purposes and benefits that can range from improved aspects of cognition (Gao & Mandryk, 2012) to improved physical health and well-being (Gekker, 2012) and beyond.

---

[i] https://www.callofduty.com/, last viewed 2016-11-07, https://www.rockstargames.com/grandtheftauto/, last viewed 2016-11-07

### 2.1.2   Use Cases in Physiotherapy, Rehabilitation & Prevention

The application of games or interactive tools with game elements for serious purposes in health is seeing a surge in attention both from research and industry. Use cases are found in different areas, ranging from information processing, disease self-management, distraction from discomfort, professional training, and clinician skills or public education, to the active support of *physiotherapy, rehabilitation, and prevention* (Gekker, 2012; Primack et al., 2012), here PRP. The PRP group of use cases draws largely on *motion-based games* and is closely related to *exergames*. Exergames can be seen as a more general class of games that require that players perform physical exercises in order to achieve goals in a game (Yim & Graham, 2007). They are an established form of games with a broad reach into the consumer games market. However, motion-based games have also been adopted for a wide variety of different target groups that are in the scope of conventional PRP, such as people recovering from *stroke* (Alankus et al., 2010; Burke et al., 2009), people living with *Parkinson's disease* (Assad et al., 2011; Natbony et al., 2013), children and adolescents with *cerebral palsy* (Deutsch et al., 2008; Hernandez et al., 2012; Holt et al., 2013), and broader groups such as *older adults* in general (Gerling et al., 2012; Gerling & Mandryk, 2014; Gerling, 2014b). Much of this more recent work continues work along the lines of earlier investigations that had a focus on *virtual reality* (VR) (Jack et al., 2001; Jung et al., 2006; Weiss et al., 2004). VR shares large commonalities in the technological basis and interactive nature with games and both can also fuse in the form of VR games.

Within the larger health market, PRP represent a considerable segment. Taking the example of Germany as a developed nation with a large and high-quality public healthcare system, the number of active physiotherapists has doubled between the year 2000 and 2011. In a similar trend, the expenses of public health insurance providers for the area of physiotherapy have doubled between 1993 and 1999 from approx. 0.9 to approx. 1.8 billion Euros (Gesundheitsberichterstattung des Bundes, 2014) and yet, the system operates at capacity limits. Approaches to saving costs are required and one central aim is to increase the number of activities that are performed by patients in a more self-dependent and pro-active manner. Such dynamic and "hands-off" exercises, where the patients – as opposed to therapists – assume a leading role, are playing an increasingly important role (Deemter & Wolters, 2012) and the clinical indications of such exercises compared to less dynamic applications are positive (Stenström et al., 1997; van Tulder et al., 1997; Waddell et al., 1997). This also entails that therapeutic exercises are often executed without immediate therapist supervision, potentially even at home, leading to sustained periods without supervision or feedback. Interestingly, the common baseline for therapeutic exercise instructions is (still) formed by simple booklets with a textual description and some guiding pictures or illustrations that are handed to patients, usually after a brief professional introduction to the exercises (Uzor et al., 2012).

While such booklets have been shown to significantly increase therapeutic exercise adherence compared to purely verbal instructions (Schneiders et al., 1998), the approach does suffer from a number of notable shortcomings, such as patients (often subconsciously) skipping exercises, performing more or less repetitions than advised, missing out on therapeutic benefits by performing movements too quickly, and more (Uzor & Baillie, 2013). Achieving a good quality for treatments in the area of PRP is a challenge not only due to patient fallibility in self-directed exercising, but also since therapists often follow individual schools, making it difficult to deduct conclusive objective judgments of exercise execution accuracy when considering the agreement between multiple therapists (Pomeroy et al., 2003). Furthermore, prescribing too many exercises can lead to significantly worse adherence (Henry et al., 1999) and these challenges in exercise program design, instruction, and compliance are amplified by behavioral and socio-economic factors. It has been shown that – in extreme cases, for example in in cardio-vascular rehabilitation (Ice, 1985) – even a life-threatening diagnosis does not necessarily notably increase adherence rates. Other correlating factors with non-compliance are perceived / factual barriers, lack of positive feedback, and feelings of hopelessness (Sluijs et al., 1993).

Approaches to improving adherence include (a) coordination of exercise programs with patients' schedules, (b) frequent encouragement, (c) stimulus control (e.g. wearing training clothes), (d) cognitive strategies drawing attention to thoughts and feelings of patients about exercising, (e) time-oriented goals that are adjusted from session to session, (f) behavioral contracting regarding longer-term goals, (g) involving family members, (h) increased physician presence, approachability, and interaction, (i) goal-oriented clubs, and (j) increased feedback regarding progress (Ice, 1985). With the partial exception of (h), since virtual presence does not equal real presence, all of these approaches can be embedded in a motion-based game for PRP system. Notably, if seen through the lens of a game designer, most physical therapy sessions could already be called "gamified" to a certain extent. Just like workers in repetitive jobs sometimes invent challenges to keep themselves interested (Sansone, Weir, Harpster, & Morgan, 1992), therapists and care-givers at times create small games or rely on playful or game-like metaphors (e.g. asking patients to imagine picking apples from a tree) in order to keep patients – and to a certain extent likely also themselves – motivated (cf. section 4.5).

In this light, the large number of research projects noted in the beginning of this chapter is not surprising and neither are the significant advantages of using playful digital tools for exer-cising at home compared to traditional instructions via exercise sheets or booklets (Uzor & Baillie, 2014). More recently, increasingly ambitious controlled medium- to long-term studies have appeared that show first clinical evidence for a range of application areas. This includes extremity-related motor functioning in subacute stroke using *EyeToy*[2] games (Yavuzer et al., 2008), slowing the progression of mild cognitive impairments (Anderson-Hanley et al., 2012), improving cognitive skills (Kempermann et al., 2010), such as multi-tasking and cognitive control that decline with age (Anguera et al., 2013), as well as accompanying examples from a growing amount of application areas outside of PRP, for example improved adherence to cancer medication and cancer-related self-efficacy in younger audiences (Kato et al., 2008).

In summary, the application area of PRP is of high societal relevance. However, major challenges exist, including high costs of manual one-to-one treatment, a lack in objective comparability due to differences in therapeutic schools, potential for improvement of the safety and efficiency of performing therapeutic exercises without human oversight, lack of communication due to spatio-temporal constraints, and achieving sustained motivation to perform repetitive exercises. This makes the area of PRP a very promising candidate for applying MGH with their potential to provide sustained motivation, positive feedback and guidance, clear goals and a sense of self-direction (through in-game choices or personalization), as well as measures for objective comparison and analysis. These strengths can additionally be combined with surrounding system architectures that feature social and patient-to-therapist communication channels.

### 2.1.3 The Dual-Task Nature of Motion-based Games for Health

In typical motion-based games, players *use their bodies* to control a game which also requires a *cognitive effort* for solving tasks, puzzles, or for engaging with other game mechanics. It is due to this interwoven nature of *cognitive* and *physical activity* in motion-based games that they are almost always intrinsically of a *dual-task* nature (Bruin et al., 2010; Bruin et al., 2011). Dual-tasks are valued in the area of PRP for their efficiency and have been shown to be more beneficial than either cognitive or motor-training alone (Pichierri et al., 2011). In many ways, the efficiency of dual-task interventions can be connected to insights from embodiment theory (cf. section 3.7.3), where it has been shown that body images are not only formed by experiential action, but also influence behavior (enabling feedback loops), and a "healthy body and a healthy mind" have been shown to be highly interconnected (Gallagher, 2006). The case of the interdependence of cognitive and motor challenges, as well as potential benefits from motion-based games, can also be supported with recent findings specifically regarding the cognitive benefits of motion-based game activity (Anguera et al., 2013; Gao & Mandryk, 2012).

---

[2] https://en.wikipedia.org/wiki/EyeToy, last viewed 2016-11-07

### 2.1.4  Uncommon Target Groups

As indicated above, motion-based games offer potential in a wide range of application use cases. Frequently, the target groups that could benefit most from such applications differ considerably from the stereotypical gamer, which is often still characterized as a white male adolescent with deficient social skills. While this stereotype has already been largely disproven as indicated by more recent average gamer statistics (see section 2.1.1), target groups such as older adults or people with Parkinson's disease, which are in the focus of much of the work presented in this thesis, appear so uncommon that their preferences, general stance, expectations, and requirements with regard to video games warrant further discussion.

Contrary to the widespread opinion that video games do not play a role in the everyday life of older adults, related work has shown that many older adults do play video games (Ijsselsteijn et al., 2007; Nap et al., 2009). The potential of video games to provide stimulation, to be used for enhancing thinking processes, exercises in hand and eye coordination, and – perhaps most importantly – to provide opportunities to enhance self-esteem by mastering new materials, has been discussed as early as 1983 (Weisman, 1983). Based on a study carried out in 1982, Weisman reported that existing arcade type games had to be adapted before it was possible for them to be used by older adults in a nursing home. He noted that games which can be programmed so that the participants can start at a level that can easily be mastered and which also provide swift feedback on progress as well as the chance to improve in small increments, were most accepted. This clearly predates later findings in a similar direction regarding the need for *positive feedback and encouragement* as a more favored motivation mechanic compared to *providing extremely challenging tasks* (Assad et al., 2011; Gerling et al., 2011; Smeddinck, Siegel et al., 2013). Weisman also highlighted further findings that were confirmed much later in related work; e.g. the need to employ large and well defined visual symbols, as well as distinct and clear audio cues (Gerling et al., 2012; Nap et al., 2009; Weisman, 1983), and a preference of a social setting with one active player and multiple onlookers (cf. section 4.5), or the potential of video games to be used as diagnostic tools (Smeddinck, Herrlich, & Malaka, 2015). More recent research has found that older adults prefer video game implementations of common card or board games that they already know, or games of a similar nature (Nap et al., 2009), and that they also prefer games to which they attribute a realistic potential to help improve their physical or mental well-being (Nap et al., 2009; Smeddinck, Siegel, et al., 2013). This encompasses typical "brain-training" games, such as quizzes, or mathematical puzzles, but also motion-based implementations of popular games such as bowling (Clark & Kraemer, 2009; Deutsch et al., 2008; Hsu et al., 2011; Marston, 2010; Wittelsberger et al., 2012; Wollersheim et al., 2010). Older adults have also been found to show strong negative emotional reactions to displays of violence in video games (Nap et al., 2009), even if they were of comparatively low visual fidelity and of low ethical concern. In some cases, actions as comparatively benign as having to smash virtual bugs were disliked (Assad et al., 2011).

Next to the fact that games for health for older adults and more specifically MGH for older adults have shown to allow for positive effects on aspects of cognition (Anderson-Hanley et al., 2012) and on physical condition (Kempermann et al., 2010), they have also shown to have the potential to improve social aspects (Maier et al., 2014; Theng et al., 2012) and to elicit positive effects on specific mortality risks, such as achieving a fall risk reduction (Campbell et al., 1997; Clark & Kraemer, 2009; Uzor & Baillie, 2014). Lastly, the target group of older adults for GFH in general is set to grow due to demographic change and due to the likely expectation of gamers who are not yet older adults to be able to continue playing not only for entertainment, but also for serious purposes, as they grow older.

### 2.1.5   Challenging Heterogeneities

Despite the abovementioned potential benefits of (motion-based) games for health, there are still considerable challenges involved with designing, implementing, and successfully employing such games. In order to work efficiently and to offer a good experience, games for health must match the abilities and needs of the target users. Failing to meet this challenge can result in undesirable game experiences, in a failure to achieve the serious goals, or even in injury or other damage to the well-being of the players. Following this argument clearly establishes the need to respond to the large variances in requirements towards GFH, or in other words, of responding to heterogeneities, as a central challenge. Heterogeneity, in this context, results from differences in abilities and needs between individuals (see Figure 1 and Figure 2 for example) due to variance in age-related afflictions, differences in the progression of chronic disease, or differences in permanent physical ability. Additional variances are introduced since GFH and MGH developments can target heterogeneous application areas, with heterogeneous target groups (where developments often target multiple groups at once), and heterogeneous interested parties next to the targeted player audience (such as physicians, caregivers, family and friends) that are often overlooked in the design and development of GFH and MGH. However, considering the requirements that result from these heterogeneities is critical to successful deployments. A detailed discussion of these aspects can be found in (Smeddinck, 2016) [**see publication B.3, section 2.5**].

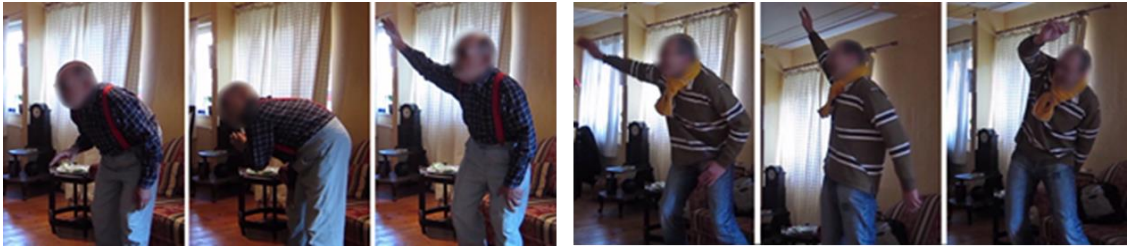

Figure 1: A series of images of participants in a study of motion-based games for the support of physiotherapy for Parkinson's disease patients (Smeddinck, Siegel, et al., 2013). One participant (left) shows a notably smaller range of motion, and less dynamic movement than the other (right).

Furthermore, social interaction around games and multiplayer situations between members of heterogeneous target groups require careful consideration and oftentimes additional flexibility, or even different approaches to adaptability and adaptivity altogether. This topic was approached, for example, in a study on dynamic balancing strategies that included dyads of players with heterogeneous abilities, as well as dyads of players with more homogeneous abilities (Gerling et al., 2015). Figure 2 shows an impression of these different dyads of players and the study is discussed in further detail in section 4.10.1.

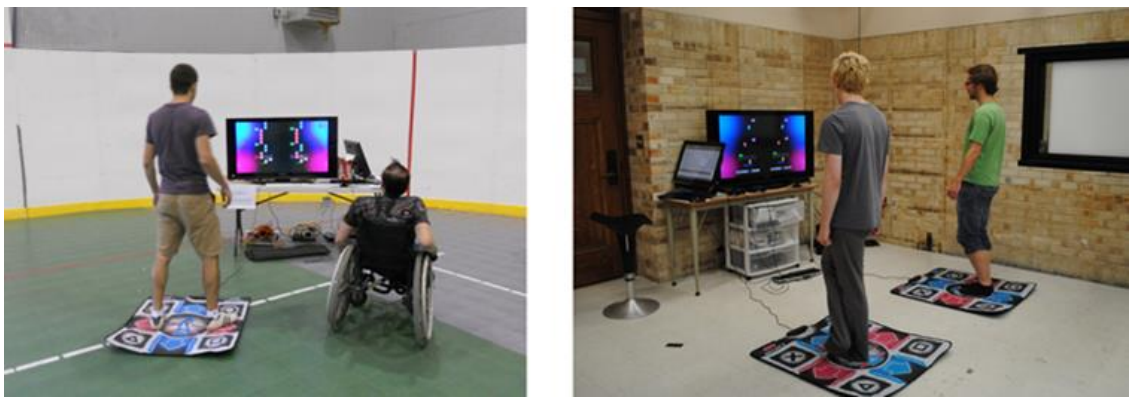

Figure 2: Two photos taken as part of a study of the impact of different levels of the notability of difficulty adjustments on player performance and experience. It showed different results in dyads of players with homogeneous abilities (right), compared to dyads of players with more heterogeneous abilities (left).

Drawn together, GFH – in many cases – need to cater to very specific needs and abilities of their players due to a broad range of heterogeneous target groups. The groups themselves are typically composed of people with considerable variance regarding abilities and needs in terms of both the interaction with the game and the interactions linked to the targeted serious outcome. They also differ with respect to their prior knowledge, and expectations of games.

### 2.1.6   The Need for Adaptability and Adaptivity

The aforementioned heterogeneous abilities and needs of potential players of GHF and MGH require a flexibility in adjustments which frequently surpasses manual difficulty selection offered in regular video games in terms of the required level and extent of control. This necessitates the development of adequate parameter sets for adjustments, including proper translation into game variables, as well as the development of interfaces for manual adaptation along these parameter sets that offer a good usability and user experience (cf. chapter 3). In many cases this also leads to a need for adaptivity, meaning a need to automate the adjustments in order to allow for a desirable frequency and level of detail of adjustments whilst not overburdening players or third parties, such as therapists, by requiring unreasonable manual effort in order to decide on – and enact – adjustments (cf. sections 3.7.6, 4.5, and 4.8).

> *Despite notable setbacks and disappointments since the early beginnings of adaptive software (Horvitz et al., 1998), the area has made great progress, and many commercial and mass-market applications nowadays have adaptive features. It is therefore not surprising that adaptive elements are increasingly being explored for usage within games, be it for personalization regarding a single player, or for balancing between multiple players (Gerling, Miller et al., 2014; Mueller et al., 2012). Adaptive techniques in the context of games have been shown to have the potential to improve game experiences (Andrade et al., 2006) and are being discussed intensively, especially in the context of motion-based games for health (Alankus et al., 2010; Rego et al., 2010), due to the complexity of body-based input (Hocine et al., 2014; Hoffmann et al., 2014; Smeddinck, Siegel, et al., 2013; Smeddinck, Herrlich, et al., 2015) and the related physiological health targets. While* **implicit** *game performance data and physiological data play a self-evident role in measurements to perform adaptation in GFH, psychological data and* **explicit** *user feedback can complement that information both as feedback data for "on-line" adaptive mechanisms (Liu et al., 2009) and for determining presets (Lindley & Sennersten, 2006). Methods that are based on user-centered feedback measurements also offer great potential for efficient generalizations and transfer from one GFH application to another, since the outcome measures are not relating to aspects of the game, but to aspects of the player (Hocine et al., 2011).*

> *Excerpt from: (Smeddinck, 2016)*

Adaptability and adaptivity in this sense must allow for games for health to be adjusted to meet the requirements of the target group (Geurts et al., 2011), e.g. considering typical age-related limitations that occur in the target group of older adults (Gerling et al., 2012), as well as to meet the requirements of the individual players who add further variance to the spectrum of needs and abilities (Aarhus et al., 2011). Even beyond these groups, design for adaptability and adaptivity should arguably consider the broader context of situated use. Through a generalizing lens, adaptability and adaptivity in this manner can be seen as an extension of the user-centered iterative design process (cf. sections 3.5 and 3.7). This process itself has been criticized as being too narrow as an approach to truly consider the stakeholders of an application and their concerns (Gajendar, 2012), for reasons such as failing to consider situated and user generated context (Dourish, 2001; Hartson & Pyla, 2012) after product release, taking user-centered iterative adjustments beyond the point of release to market with ongoing adjustments along a set of parameters with thresholds and adjustment options. This, of course, means that a considerate parameter selection and thresholding or discretization is of central importance, which motivates the discussion of the dimensions of adaptivity later in this thesis (section 3.3). It also means that, next to performance analysis and adjustment mechanisms that are commonly discussed as the basis for dynamic difficulty adjustments as a form of adaptivity in games (Adams, 2010), classic endpoints from user-centered iterative design, such as usabilitiy, playability, user experience, and player experience play an important role when considering outcomes of adjustments in adaptive and adaptable systems (cf. section 3.5). Notably, in the context of MGH, the typical focus on cognitive abilities / game skills and experience must be extended with conscious simultaneous considerations of the physical / motor abilities and experience of users. This notion has been partially discussed in related work under the term dual-flow (Sinclair et al., 2009) and it will be discussed in more detail in the theory chapter of this thesis. The need to consider physiological and psychological states and development means that the reliable and comparable assessment of the psychophysiological state of users plays an important role (Drachen et al., 2010; Mandryk et al., 2006). In addition to building the foundation for motion-based game controls, modern consumer-ready sensor devices and techniques in sensor fusion and data analysis play an important role in enabling games for health projects with regard to these assessment needs in an affordable manner (cf. section 2.4).

## 2.2   Definitions: Serious Games, Games for Health, and Motion-based Games

Broadly speaking, the concept of using games for serious purposes dates back to early non-digital games. Arguably as early as the 18th century, where dice games for musical learning were popular (Hedges, 1978), while – even more broadly speaking – most animals have of course relied on play for learning since ancient times (Crawford, 1984). With the occurrence of video games, the range of potential application scenarios has increased and the frequent need for the presence of co-players has been diminished. The term serious games has been established employing various definitions with a common and general summary being: *software that merges a non-entertaining (serious) purpose with a video game structure* (Djaouti et al., 2011). Serious games have been explored for a wide range of use cases, which can be separated into different classes based on *gameplay* (e.g. focus on ludic game or paidiaic play), *purpose* (messaging / information, training, etc.), and *scope* (by markets / target groups) (Djaouti et al., 2011). The main focus of this work is more specific, and lies on *motion-based games for health*, as a subclass of *games for health*, which are themselves a subclass of *serious games*.

> The **Games for Health Project** *defines* **games for health** *simply as "game technologies that improve health and the delivery of health care"* [3]. *GFH are sometimes referred to as* **serious games for health**. *Since the term* **serious** *is redundant, the addition can be abandoned. The term* **health games** *is used synonymously. Other less common terms that are used synonymously are* **eHealth games** *(Ciaramitaro, 2011),* **digital health games** *(Berkowitz & McCarthy, 2012; Brooks et al., 2014),* **healthy gaming** *(Brox et al., 2011), and more. In order to avoid fragmentation of the field, it appears advisable to consider whether the term* **games for health** *may be used in future discourse. Other terms such as* **exergames**, **fitness games**, **virtual rehabilitation**, **kinesiatric games**, **motion-based games for health**, **mental health games**, *or* **cognition games**, *are sometimes used synonymously, although they can arguably be seen to describe specific subclasses of GFH (here: different physical and mental health targets).*

> Excerpt from: (Smeddinck, 2016)

It is clear that the terminology on GFH is considerably fragmented. This is further complicated by concerns around the term *gamification*. The following excerpt offers a pragmatic treatment:

> The terms **gamified health**, **health gamification**, *or alternatively* **gHealth** *bring the aspect of* **gamification** *into the terminology and are not easily dismissed since they do arguably match the definition provided for GFH above. Since the debate on gamification and games is not the topic of this section, this text operates on the premise that both approaches reflect different angles of the same concept of using game technologies (Deterding et al., 2011) to improve health and the delivery of health care: While gamified health highlights the underlying serious purpose and application as the origin to which gamification elements can be added to a variable degree, GFH highlight the motivational potential of fully fledged games which can encompass a certain range of serious health purposes.*

> Excerpt from: (Smeddinck, 2016)

More recent related work further addresses this differentiation along the concept of *levels of gamification* and approaches that are either *casually gamified* or primarily *gameplay-driven* (Pfau et al., 2018). Another important aspect of GFH that requires clarifications on the terminology is the understanding, or underlying motivation / agenda behind the *serious purpose* that can differ in practical and ideological terms:

---

[3] https://gamesforhealth.org/about/, last viewed 2016-11-07

*While games for health can often also be **games for behavior change** (Fogg, 2009; Hekler et al., 2013) or **persuasive** (Oinas-Kukkonen & Harjumaa, 2009) **games**, those two classes encompass many other application areas. Moreover, GFH can also be non-persuasive and not tailored towards behavior change [as for example in many GFH that support the education or training of health care professionals (Baranowski et al., 2008)]. There is, however, considerable overlap between **education** or **learning games** that target professional or public education in health-related areas and **educational** or **learning games for health** that is difficult to avoid since the framing usually depends on the professional backgrounds of the researchers or game designers. It is therefore important to be aware of the various angles that different researchers, designers, developers, and GFH projects can take regarding the same core subject matter.*

*Excerpt from: (Smeddinck, 2016)*

Together, these discussions show how similar concepts are frequently explored from the perspective of different fields without much initial interaction. This is the case, for example, with motion-based games for health, which have been studied under this term but also under the term *virtual rehabilitation* (Jack et al., 2001; Kizony et al., 2006; Schüler et al., 2015). In this light it is evident that common taxonomies and classifications have to be established in order to allow for efficient scientific discourse across the involved fields. The underlying publication (Smeddinck, 2016) [**see publication B.3, section 2.6**] discusses – and suggests augmentations to – existing classification schemes for serious games for rehabilitation based on a bottom-up analysis of game characteristics (Rego et al., 2010), as well as more top-down approaches that generate a structure along (usage) fields and areas of application (cf. Table 1) (Sawyer & Smith, 2008; Gekker, 2012).

**Table 1: A taxonomy for GFH as suggested by Sawyer and Smith (2008).**

| Fields -> / Areas of Application | Personal | Professional Practice | Research/ Academia | PublicHealth |
|---|---|---|---|---|
| *Preventative* | Exergaming Stress | Patient Communication | Data Collection | Public Health Messaging |
| *Therapeutic* | Rehabilitainment Disease Management | Pain Distraction Cyber Physiology Disease Management | Virtual Humans | First Responders |
| *Assessment* | Self-Ranking | Measurement | Inducement | Interface / Visualization |
| *Educational* | First Aid Medical Information | Skills / Training | Recruitment | Management Simulations |
| *Informatics* | Personal Health Records | Electronic Health Records | Visualization | Epidemiology |

This structure is especially helpful for planning the GFH design or research, since it induces a conscious reflection about the potential interests and requirements of the involved parties based on their fields and helps with clearly defining the targeted areas of application. Akin to many other aspects around motion-based games for health and serious games in general, such classifications and taxonomies are just beginning to be established and are subject to change not only due to the relatively young research field, but arguably also due to the fact that studying serious games, like computer science (Newell et al., 1967), is a study of dynamic artifacts, a "study of the artificial", where new technological occurrences as phenomena are likely to drive a need for expanding / evolving classifications, taxonomy, as well as theories and models. Therefore, current models are relatively coarse, and the work at hand will, for example, demonstrate how the characteristic of *adaptivity* (Rego et al., 2010) alone has multiple sub-classes that can be connected to different (PRP) game experiences (cf. section 3.4).

## 2.3   Definitions: Adaptability, Adaptivity, Personalization, and Customization

In a similar manner to the sometimes ambiguous, sometimes inconsistent, and sometimes inept usage of different names for different types of games, a broad variety of terms are being used to describe software systems that can be adapted or that adapt automatically. While the section on seminal work on adaptability and adaptivity will cover a number of frequently used terms, the discussion in this thesis evolves largely around the two general term pairs of personalization and adaptivity, as well as customization and adaptability. These terms are often used differently in related work. Therefore, Table 2 contains descriptions of the understanding of the terms that is employed in this work.

**Table 2: Definitions of the various adaptation terms.**

| Concept | Definition |
| --- | --- |
| Adaptability | The fact that a system is not fixed, but can be changed (to the needs of users, to changing environmental contexts, etc.; changes are usually understood to be performed manually). |
| Customization | The act of changing a system to the needs of a user group or individual user (manually or automatically; may can be done by the group itself or by the user him- or herself, but may also be done by third parties; often related to the appearance or content of the given system). |
| Personalization | The act of changing a system to the needs of a specific individual user (often automatic but does not have to be, i.e., can be understood as a specific form of customization with a focus on individuality; personalization is also often related to appearance or content). |
| Adaptivity | The fact that a system is not fixed, but dynamically changes over time (to adjust to the needs of users or an individual user, or to adjust to changing environmental contexts, etc.; typically happens automatically; often related to settings and parameters present in the given system). |

The alternative term *tailoring* is often used as a substitute for either customization or personalization can encompass both manual and automatic adjustments. A more nuanced discussion of the terms is provided in (Streicher & Smeddinck, 2016) [**see publication B.11, section 3.2**]. Notably, these terms describe apparent properties of systems and do not define the techniques that are employed to achieve these ends. It is also important to point out that many systems will employ designs that mix elements of adaptivity and adaptability, so users may have a say in the adjustments considered by the adaptive system.

A final delimitation for clarification is that it is in the nature of digital interactive systems that they react to user input through action-reaction cycles in a manner that can be seen, or appear, as adjusting, or even adapting. *Adaptability* in the context of this work is not concerned with action-reaction cycles that form the core of the user interaction with the system, but it is concerned with facilitating a context in which action-reactions can happen in an optimized fashion, typically making adjustments that change – or configure – the system beyond the point of an atomic action-reaction. While adaptability and adaptive systems that focus on performance analysis by examining system-inherent interactions can be beneficial, much merit for improved adaptivity arguably lies in the processing of further contextual information regarding the user and the situated context of user-system interactions. Although such processes can, to a certain extent, rely on explicit user input, recording implicit cues requires sensor devices, which may – of course – also be required as controllers or input devices for motion-based games for health to begin with.

## 2.4   Background in Sensor Devices, Digital Health, and Quantified-Self

*Sensors* and *sensing devices* have been subject to considerable advances in recent years. This includes the range of sensing capabilities, sensitivity, accuracy, etc., but it also includes advancing miniaturization and magnitudinal cost reductions. Accelerometers, for example, were first built in the 1960s for the guidance systems of the *Titan 2* intercontinental missiles. Weighing about 3.4 kg and with the size of a small coffee maker, the guidance system had a cost per unit of ~100,000 US\$ (1966 value). Since then, accelerometers have shrunken to the size of a few cubic millimeters and cost cents (Ceruzzi, 2010). The data collection, analysis, and feedback that is enabled by such small-scale and highly affordable sensors and sensor-fusion devices has already changed the world, most evidently in recent times in the form of smartphones, arguably following early predictions from cybernetics (Wiener, 1961).

Technologically minded enthusiasts have seized the abilities that were brought into reach by these developments in hardware to start gathering information about themselves and their environment that would otherwise not have been available, at least on a comparable level of detail, in what is most frequently referred to as the *quantified self* (QS) movement. Important areas such as education (Rivera-Pelayo et al., 2012) and health (Swan, 2009) can benefit from new pathways to gaining insights on both individual users and multi-user systems. At the same time, digital applications such as games can also benefit from additional sensing abilities, since sensors can, in an abstract sense, always be employed as implicit or explicit control devices. It is therefore not surprising that recent developments in sensor devices, especially in the area of affordable, robust, reliable, and easy to use consumer devices, play a large role as facilitators of serious games of all kinds, and of motion-based games for health in particular.

MGH frequently rely on special sensing and tracking devices, be it for direct player input and control, for performance-, action execution quality-, or player health status-estimation, or for providing data to performance measures for adaptivity. Next to costs and maintenance, accuracy, and reliability can often be challenging to achieve and maintain, but the increasing availability of affordable and durable consumer (multi-)sensor devices, such as optical body trackers or wearables with motion-tracking capabilities, are easing some of these challenges. For example, the *Kinect*[4] depth-camera based body tracking has been shown to provide comparable performance to professional optical tracking systems (Chang et al., 2012; Galna et al., 2014), with some limitations due to the fixed tracking angle, yet also advantages in terms of ease of use and calibration drift. Further aspects of requirements for sensor devices in the complex situated use of GFH or MGH, of sensor-blindness, and the importance of actuation, are discussed in (Smeddinck, 2016) **[see publication B.3, section 5.4]**.

As indicated above, the MGH that this thesis focuses on aim at supporting ranges of different exercises, mostly from application use-cases in PRP, in a flexible manner, which means that a broad range of full-body movements need to be tracked. Modern sensor devices for affordable and consumer-ready full-body tracking, such as the Kinect v1 and v2 typically rely on infrared tracking with stereo cameras together with a regular high definition camera for color image representations. This technology is comparatively robust to changes in lighting and allows for the non-invasive detection of depth-information in a wide-angle window in front of the device. It also allows for a good approximation of the skeleton pose at any given moment and supports multiple users simultaneously. The devices also produce color image streams that can be used to provide non-distorted visual feedback or to support further analytic measures. As studies presented in the later parts of this thesis will underline, such tracking devices and the typical interaction schemes they support have been shown to be understood and accepted by older adults (Gerling et al., 2013). However, despite the increased tracking accuracy of the later generations of such devices, they still present challenges especially with regard to clinical use. For example, it is not always possible to gain reliable estimates of the exact configuration of all body parts, especially when the user leaves the tracking window, when body parts are occluded, or when users perform exercises that require additional props or that are performed whilst kneeling or lying on the ground. More recently, research and development projects on MGH, such as *Adaptify*[5], have begun to explore the integration of additional sensing and tracking devices, such as body-worn motion sensors, or pressure detection surfaces both of low (Gerling et al., 2010) and high spatial resolution (Sundholm et al., 2014; Sungkarat et al., 2011).

---

[4] https://en.wikipedia.org/wiki/Kinect, last viewed 2016-11-07

Developments in sensor devices have simultaneously been driven by – and facilitated – additional growth in in *digital health* and the *QS movement*. The latter can be seen as a movement that accompanies developments in *digital health* through a crowd (and enthusiasts) based movement that also aims at tracking data from – and gaining insights into – aspects that include and go beyond health.

### 2.4.1  Background in Digital Health and Quantified Self

The background in *digital health* (or *eHealth*) is discussed in further detail in the underlying related work (Smeddinck, 2016), highlighting the rapid growth, the emergence of aggregating platforms by the big players in technology, and the important role of smartphones as primary gateway devices that also offer gaming abilities [**see publication B.3, section 2.2**].

These developments in digital health and fitness tracking are closely related to the QS movement (Li et al., 2010), which, as mentioned above, is the label for an ongoing trend of using information systems to provide exact, objective and quantitative measurements of various personal data and performance. Receiving detailed feedback on personal performance through applications can be a strong motivating factor, for example in the domain of fitness training (Annesi, 1998). While games for health seek to add to this motivation with game elements, even comparatively simple gamification elements (Deterding et al., 2011) have been shown to provide notable motivation (see section 4.4) for exercise execution. Next to momentary motivation, personal sensors and feedback can also enhance human perception and allow insights into aspects that were previously imperceptible, which offers great opportunities for driving sustained behavior change, as will be discussed in more detail in section 2.5.

---

[5] https://www.adaptify.de, last viewed 2016-10-06

### *2.4.2  Multi-Modal Interaction and Natural User Interfaces*

If additional sensor devices are integrated into games for health, not only for gathering information, but in order to act directly as input or control devices, or to influence the system behavior in an indirect manner that means, in many cases, that these GFH will include *multi-modal interaction* (Moggridge, 2007), which arguably fosters intuitive *natural user interfaces* (NUIs) (Wigdor & Wixon, 2011). The Kinect devices, for example, feature microphone arrays, and audio input is already used in games for health, such as in voice therapy for people with Parkinson's disease (Krause et al., 2013). In recent years, with the exploration of a much broader array of health and exercise application use cases for serious games, a wide range of fitness devices that are equipped with their own sensors are starting to be integrated into these games, acting both as data sources and as game controllers. The types of devices range from mainstream devices, such as exercise machines, to more exotic devices for specific purposes, such as electrical muscle stimulation suits (Smeddinck, Herrlich, Roll, & Malaka, 2014). Explorations on alternative game control modalities, such as physical walking (Walther-Franks et al., 2013b), further augment these developments. The approach of natural user interfaces, despite its promises, such as being more accessible to people who are less experienced with standard dedicated computer input devices (making them potentially a good fit for target groups such as older adults), has been criticized for many shortcomings (Norman, 2010). Some points of critique transfer to current generation MGH, such as shortcomings regarding expected physical feedback. However, the lenses of multi-modal interaction and NUIs that have been discussed widely in related work on interaction design and HCI still offer important insights towards the design and implementation of games for health that employ comparable technology stacks. Furthermore, modern frameworks, such as *reality-based interaction* (Jacob et al., 2008) can arguably help designers to avoid some of the dissonance between promises, expectations, and system performances that has been criticized in the context of NUIs.

## 2.5  Motivational Psychology and Behavior Change

The potential of games to motivate players to perform actions, or to engage with content that might otherwise be perceived as tedious or boring, is likely the most common argument for employing games for health and for many other serious application areas. Since other research fields, particularly psychology, offer established generalized theories and models on prerequisites, triggers, and enabling factors for human motivation, it is important to consider these theories in GFH and MGH design. While this approach has arguably been largely overlooked in the early years of game user research, more recently, a substantial body of work from the field has focused on applying, evaluating, and further conceptualizing theoretical constructs in game studies (Nacke et al., 2010).

Prominent theories on motivation and games for health are summarized and discussed in the included publications [**primarily publication B.3, section 2.4**]. This includes the theory of *flow*, which is used to explain how games can motivate players to become so engaged that they are apparently completely drawn "out of their bodies" and "into an activity" (Csikszentmihalyi, 1990) and has been linked to digital games both by Csikszentmihalyi et al. (2014) and other authors (Chen, 2007). An important prerequisite for the emergence of a state of flow is a balance between the skills of a player and challenges presented (by the game), as illustrated in Figure 3.

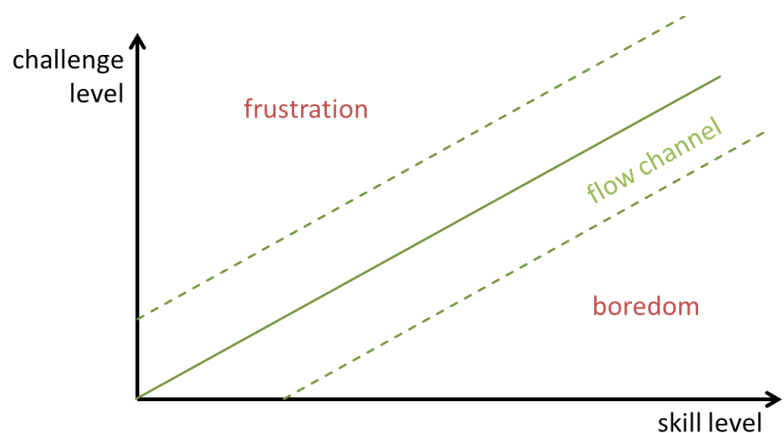

**Figure 3: An illustration of the flow channel between the right level of challenge and skill that allows for experiences to arise that are neither frustrating nor boring [green lines: after Csikszentmihalyi (1990), red "dangers": after Crawford (1984)].**

While this aspect shows a clear connection to the concerns of game design in general, as well as to concerns around adaptability and adaptivity in particular, it is important to point out that Csikszentmihalyi lists further prerequisites for flow that should also be considered in game design and in the design of systems for adjustments to game experiences. The concept of *dual-flow* (Sinclair et al., 2007) is also discussed in further detail in the aforementioned publication [B.3], describing how the balance between player skills and game challenges goes beyond the traditional focus on digital game design and mechanics when considering MGH. Due to their motion-based nature, it is also important to include the balance between physical abilities of a player and the challenges presented and this balance plays an equally important role (Sinclair et al., 2009). This argument can be linked to the *dual-task nature* (Pichierri et al., 2011) of MGH. As noted before, dual-tasks have been shown to be more beneficial than either cognitive or motor training alone (Bruin et al., 2010). Furthermore, the dual-flow approach to the balance of skills and challenges can be generalized to highlight the need to consider any physicality or otherwise expressed ability or skill that is directly linked to the *effectiveness* of an *intended outcome* (or *serious purpose*) of any GFH (i.e. not only motion-based) or serious game (see Figure 4). Lastly, this perspective can be also connected to the three central promises of GFH (cf. section 3.1), where *motivation* corresponds to *attractiveness*, and *guidance* and *analysis* are linked to *effectiveness* (momentary and prolonged).

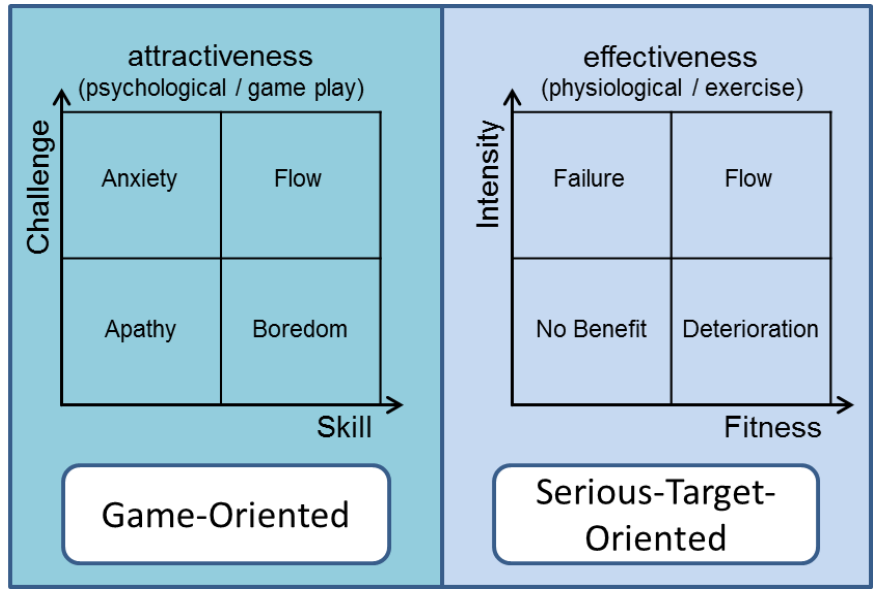

**Figure 4: The dual-flow model for exergames. A generalized adaptation after Sinclair et al. (2009).**

The same publication [**B.3, section 2.4**] also contains a discussion how *self-determination theory* (SDT) (Ryan & Deci, 2000) is increasingly being applied in the context of games, serious games, GFH, and MGH. As a *need satisfaction theory*, SDT states that three basic human needs must be fulfilled to support *intrinsic motivation* via a *self-determination motive* (cf. Figure 5). The three needs are *competence* (the need to feel competent at things we do), *autonomy* (the need to feel free in our decisions and goals), and *relatedness* (the need to feel related and socially connected). While Rigby and Ryan (Rigby & Ryan, 2011; Ryan et al., 2006) have convincingly established the application of SDT in games, SDT has also been successfully employed in the context of sports / exercising and general health behavior motivation. It is therefore a very interesting tool in the context of GFH and MGH and existing, as well as newly created psychometric instruments from SDT can support game user research, offering reliable insight into aspects of motivation (Birk & Mandryk, 2013; Birk et al., 2015; Smeddinck, Herrlich, et al., 2015).

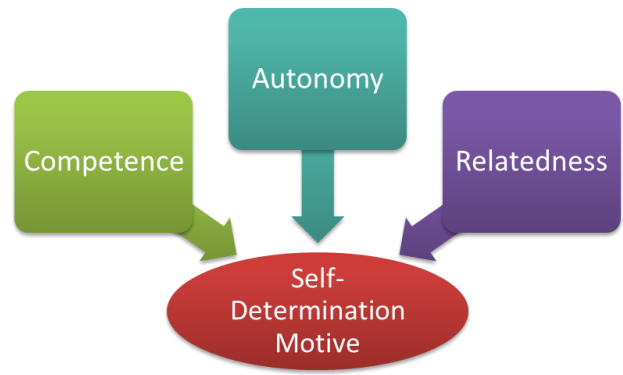

**Figure 5:** *Competence, autonomy,* and *relatedness* **needs satisfaction combined facilitate the** *self-determination motive.* **After Ryan & Deci (2000).**

Furthermore, the theory of *self-efficacy* (Bandura, 1982) has also been used in the context of games (Song, Peng, & Lee, 2011). It evolves around the extent of one's belief in one's own ability to complete tasks and reach goals, and it has been linked as a predictor to achieving health related tasks and goals. The aforementioned theories are frequently discussed in the context of *persuading* users into first starting a treatment or change and of triggering lasting *behavior change*. Accordingly, models and approaches from *persuasive applications design* (Orji et al., 2013) and *behavior change* (Fogg, 2009) have been used in the context of serious games (Baranowski et al., 2008; Thompson, 2012).

In brief, the *Fogg Behavior Model* constitutes that a certain threshold of a combination of *motivation, abilities,* and *triggers* must be exceeded in order to promote *behavior change* (Fogg, 2009). Notably, GFH can support all three elements. Arguably, the requirement of *triggers,* such as motivational messages (de Vries et al., 2016), hints at the larger context of MGH and indicates that a good balance of challenge and skills alone will not suffice, if the context of the potential player does not lead her of him to playing a game repeatedly (e.g. via reminders on a smartphone app). In this context, it should be noted that the term *persuasive systems* (or *persuasive technology*) is controversial due to the questionable element of unconsented coercion that may be present (Hekler et al., 2013). However, the respective related work (Bogost, 2010; Oinas-Kukkonen & Harjumaa, 2009) still holds insightful perspectives on behavior change. Relating back to *extrinsic motivation* and *intrinsic motivation* as discussed in the context of SDT, purely external rewards have been shown to work well with motivating simple motor activities (Ryan & Deci, 2000), so that certain levels of increased motivation with simple casual games are perhaps not surprising. Yet, such rewards can also undermine intrinsic motivation, if they are not *internalized* (Deci, 1971; Deci et al., 2001). This underlines calls for a careful and serious framing of GFH as meaningful parts of larger digital health and treatment systems.

Additional theories and frameworks, such as *perceived control* (Osnabrügge et al., 1985), *theory of planned behavior*, the *health belief model*, and the *transtheoretical model*, can also add to the discussion of serious games (Hekler et al., 2013), especially regarding their immediate and lasting motivational and behavioral impact. However, it is interesting to note that most theories explain, at best, 20 - 30% of the total variance in a given health behavior when tested in an intervention (Hekler et al., 2013). Thus, many effects and factors are currently not accounted for, motivating the development of more detailed domain-specific models.

## 2.6  Mass-Market and History of Motion-based Games

Today, controller devices for motion-based games are available for almost all platforms ranging from PCs to consoles and mobile devices. Fitness applications, such as *Wii Fit*, *Move Fitness*, *Kinect Sports*, or *Your Shape Fitness Evolved*[6] have big name manufacturer and publisher support and reach large audiences. Especially in the mobile area, applications that augment highly popular sports are breaking records in terms of user numbers. For running, exemplary applications, such as *"Zombies, Run!"*[7], can attract user numbers in the millions (Dredge, 2015). The developments in mass-market consumer motion-based games are grounded in multiple decades of history that started with early arcade machines, and currently continue to evolve with the growing markets for wearables (Swan, 2012), as well as room-scale virtual reality and augmented reality (Ma et al., 2014).

The history of consumer-oriented motion-based games has been discussed in related work (Herrlich, Wenig et al., 2014) [**see publication C.6**] and can be summarized along the lines of two surges that have been facilitated by advances in sensor and controller hardware, with a first wave being supported mainly by controller devices that are touched or moved by players, while a second, more recent, wave was driven by optical tracking technologies and integrated motion tracking devices (*EyeToy*, *WiiMote*, wearable fitness trackers, and the *Kinect*). This second wave has arguably facilitated the application of video games in an increasing number of health-related areas [**see publication B.3, section 4.2**] and is closely related to research from the field of virtual rehabilitation, such as hand function rehabilitation for people recovering from stroke (Jack et al., 2001). As with other GFH, personalization through adaptability and adaptivity remains an important challenge.

This need for adaptability and adaptivity, as well as a need to design games for the intended serious purpose from the ground up, result from a number of limitations that are present in typical mass-market motion-based games. While some successful examples show that mass-market motion-based games can – under certain circumstances – be employed for serious purposes (Clark & Kraemer, 2009; Deutsch et al., 2008; Hsu et al., 2011), the games are not designed to meet the heterogeneous abilities, needs, and preferences that result from the health-specific application areas, as indicated in section 2.1.5. In many cases, however, mass-market games do boast a production quality that is worth aspiring towards. Accordingly, more productions targeting health care use cases aim at utilizing this existing high production quality and the connected potential to motivate players. Approaches such as augmenting the games to facilitate motion-based input (Walther-Franks et al., 2013a), or linking rewards in sedentary games to physical exercises that are performed and tracked in an asynchronous manner (Smeddinck et al., 2018), are already being explored. These developments are discussed further in section 4.10.

---

[6] http://wiifit.com/, https://www.playstation.com/en-gb/games/move-fitness-ps3/, https://en.wikipedia.org/wiki/Kinect_Sports, https://en.wikipedia.org/wiki/Your_Shape:_Fitness_Evolved, last viewed 2016-11-07

[7] https://zombiesrungame.com/, last viewed 2016-11-07

## 2.7   Seminal Related Work

Adding to the general background and related work, the following sections focus on notable projects and studies in games for health, systems from digital health or quantified self that support guidance, analysis, or behavior change, as well as related work where methods from artificial intelligence, machine learning, and user modeling have been applied to these ends.

### 2.7.1   Games for Health

General surveys of GFH with strict formalized sampling and analysis methods have been presented in related work. Ricciardi et al. (2014) provide a survey of serious games in health professions, and Baranowski et al. (2008) provide an overview of games with a focus on health-related behavior change, whereas Kato (2010) reports on various examples of games in professional health care applications. Since such review articles already exist, a curated look at a selection of outstanding systems appears appropriate.

*In order to name some specific projects for further reading, this section highlights GFH work of seminal character. This means that each project has considerable novelty (usually due to the serious target), features a convincing production quality, and is in the best case accompanied by research to supply evidence in favor of the approach. Validated with clinical trials, the games **Re-Mission** one and two, where players enter "their own bodies" to take on the fight against malicious cells, help improve drug administration adherence and indicators of cancer-related self-efficacy and knowledge in adolescents and young adults diagnosed with different types of cancers (Kato et al., 2008). The playful VR experience **SnowWorld** (Hoffman et al., 2001, 2008), where patients visit a world full of shades of light blue and other colors associated with cold and imagery of snow and ice, has been shown to be a potentially viable adjunctive*

*nonpharmacologic analgesia both in the context of wound treatment for burn victims and in dental pain control. The game **Relive** (Semeraro et al., 2014) focuses on cardiopulmonary resuscitation (CPR) training and puts players into a compelling space station scenario with a high production quality. It features competitive multiplayer and has been released as a free game on the distribution platform Steam where it competes with many commercial titles. In **Project:EVO**, a game created to detect and track the development of Alzheimer's disease, players play a flying race type of game, in order to improve cognitive control via interference processing (Anguera et al., 2013). In the creature-care game **Monster Manor**, children with type 1 diabetes are encouraged to take responsibility of controlling their own blood sugar levels, earning them upgrades and support items for a monster they take care of (Kamel Boulos et al., 2015). **Meister Cody** (Kuhn et al., 2015) is a game to support children with dyscalculia (Kuhn et al., 2013), which features diagnostics, rich adaptability, adaptivity, as well as therapist and parental information and control; it is currently undergoing a large-scale trial. As an example from the large subgroup of motion-based games for health, **Valedo** is a gaming system where two motion-sensors are attached to a patient's body in order to enable full-body control of a suite of well-designed mini games for the support of back pain movement exercises (Jarske & Kolehmainen, 2013).*

*Excerpt from: (Smeddinck, 2016)*

### 2.7.2   Tools for Guidance, Analysis, and Behavior Change

As noted above, in the space of digital health and fitness tracking platforms, a number of notable tools for guidance, analysis, and behavior change already exist. This includes platforms for vendor specific hardware, such as the *Nike Plus* shoe inlets, applications for specific classes of hardware, typically wearables, such as *Fitbit*, applications for specific types of sports or activities, such as *Strava*, *Runkeeper*, etc[8]. Larger platforms by technological big players, such as *Microsoft*, *Google*, and *Apple Health*[9] also exist and aim to integrate multiple data sources, devices, and analytics from other applications and platforms. Furthermore, public health and governmental platforms are being built that aim to provide guidance, analysis, and behavior change, next to keeping personal health records, such as the public health service platform of Portugal[10] (Rodolfo, 2016). Such applications underline the importance of surrounding ecosystems for games for health. Not only are lessons in interaction design and data visualization sometimes transferrable to GFH, such as the importance of frequent clear reminders, visual summaries that are easy to parse, etc. (Ramanathan et al., 2013), as well as findings based on the introduction of some level of gamification to some of these systems; the existence of the platforms also suggests for GFH projects to consider aiming for interoperability and for supporting open standards for the exchange of health records, which includes strict treatment of privacy and security concerns (cf. section 2.9). The situated nature of (public) health records and the importance of an embodied lens on interaction design in this area is discussed prominently by Dourish (2001) and Detmer (2008), and the practical explorations and studies performed in the context of this thesis further underline these needs (see sections 4.5 and 4.9).

---

[8] https://en.wikipedia.org/wiki/Nike%2B, https://fitbit.com/, https://strava.com/, https://runkeeper.com/, last viewed 2016-11-07

[9] https://microsoft.com/microsoft-health, https://google.com/fit/, http://apple.com/ios/health/, last viewed 2016-11-07

[10] https://servicos.min-saude.pt/utente/, last viewed 2016-06-26

## 2.8   Adaptability and Adaptivity

As discussed in section 3.1, MGH have considerable potential to increase patient motivation to perform exercises, thus increasing adherence rates, and to support quantitative insights into the quality of exercise execution and into the overall treatment. Achieving sustained motivation, which depends on a good player experience, as well as achieving effective and reliable positive treatment results in the many indicated application areas, requires a large degree of comprehensible and efficient personalization. While the need for customization and personalization in software has been recognized even before the arrival of personal computers, early attempts at adaptive software, such as automation in *MS Word* (Horvitz et al., 1998) failed, arguably due to disparities between actual system abilities and user expectations, as well as due to premature automation as users were not ready to surrender control (Sheridan, 2001).

Approaches to adaptability and adaptivity in particular have been discussed in a large amount of related work both in relation to general interactive software systems, persuasive systems (under the terms *tailoring* and *personalization*) (Oinas-Kukkonen & Harjumaa, 2009), and in relation to more specific application scenarios, such as serious games, and (motion-based) games for health. Following the much contended early days of "smart" software which aimed to adapt to individual users typically targeting at more personalized and efficient work experiences (Horvitz et al., 1998), recent advances in this area have led to a current boom in work on personal assistants such as *Siri* by *Apple, Cortana* by *Microsoft, Google Assistant / Now*[11], and in work on multimedia enabled chatbot systems by larger messaging platforms, such as *Telegram, WhatsApp,* or *Facebook Messenger*[12].

---

[11] http://apple.com/ios/siri/, https://en.wikipedia.org/wiki/Cortana_%28software%29, https://assistant.google.com/, last viewed 2016-11-07

[12] https://telegram.org/, https://www.whatsapp.com/, https://www.messenger.com/, last viewed 2016-11-07

### 2.8.1  Adaptivity in Mainstream Games

It is thus not surprising that the technological as well as the software developments underlying these advances are also being applied to video games both for automatically adjusting games to individual player abilities and needs, as well as for balancing in multiplayer games (Herbrich et al., 2007; Mueller et al., 2012). In relation to the development of commercial titles it has been noted that the player experience can be improved (Andrade et al., 2006) through adaptive systems that support games in better corresponding to the fact that each player is different, that each player has a different preference for pace and style of gameplay, and that even players with similar overall levels of playing ability will often find separate aspects of a game to be more difficult to them individually (Charles et al., 2005). Renown mainstream titles that have employed adaptive techniques are, for example, the RPG *Fable*[13], in which the landscape changes in response to player actions and a player's evolving character, the first-person shooter *Max Payne*[14], which features so-called auto-dynamic difficulty technology that alters numerous aspects that affect difficulty, such as the number or strength of enemies in a room, or *Mario Kart 64*[15], where weaker players are supported with very powerful items. While adaptivity has long been applied in related areas, such as *intelligent tutoring systems* (ITS) (Beal et al., 2002; Brown et al., 1975) or intelligent game interfaces (Livingstone & Charles, 2004), adaptive techniques have only recently started to become more prevalent in games of all types, most commonly through *dynamic difficulty adjustments* (DDA). DDA typically describes the automatic adaptation of the difficulty to the current level of abilities of the user, based on predefined general parameter ranges, or according to a user model (cf. section 3.2.2).

A common manifestation of the DDA balancing technique is *rubber banding* (Pagulayan et al., 2012). With rubber banding, the possibilities (as expressed in game resources) of players are artificially boosted to increase their performance when the actual performance drops below a certain threshold. It is often used in racing games. In the case of *Mario Kart*, DDA has arguably helped making the game inviting for novice players. However, the visible boosts provided to trailing players and NPCs can potentially harm the experience of advanced players, since weaker players receive advantages that may be perceived as being unfair (Gerling et al., 2014). While some forms of DDA, such as rubber banding systems, can thus suffer from disbalance and exploitability (Missura & Gärtner, 2009), DDA can be employed to great success, as the mainstream titles mentioned above have shown. Notably, next to DDA there are also many other aspects of games that can be subject to adaptability and adaptivity. E.g. the content of the game, especially if procedural content generation is employed (Jennings-Teats et al., 2010), the audio-visual appearance, the larger framing of a game, or the availability of options for manual adjustments. This will be discussed further in section 3.4 of this thesis, together with a number of design decisions regarding adaptability and adaptivity that have to be made regarding questions such as what, when, and how to adjust.

---

[13] https://en.wikipedia.org/wiki/Fable_%28video_game%29, last viewed 2016-11-07

[14] https://en.wikipedia.org/wiki/Max_Payne, last viewed 2016-11-07

[15] https://en.wikipedia.org/wiki/Mario_Kart_64, last viewed 2016-11-07

### 2.8.2  Adaptivity and Adaptability in Games for Health

Adaptability and adaptivity have been proposed in the context of GFH, e.g. by (Göbel et al., 2010), and in the broader context of exergames (Rego et al., 2010). In a very general and generic sense, every adaptive system requires some form of *performance evaluation* together with an *adjustment mechanism* (Adams, 2010). A more detailed discussion of approaches to adaptivity in MGH can be found in sections 3.2.2 and following. For MGH, movement and physical performance account for important performance measures next to purely in-game performance measures that play a central role in sedentary games. However, psychometric and psychophysiological measures can also play a role in providing feedback for online adjustments (Liu et al., 2009), or for the selection of presets (Lindley & Sennersten, 2006), capturing facial expressions, for example, in order to determine the emotional status of the players (Grafsgaard et al., 2013). Interestingly, such measures that rely on human-centric, as opposed to game-centric feedback also bear potential for generalized use (Hocine et al., 2011).

Furthermore, players have been found to display a surprising preference of comparatively low difficulty and quick paths to achievements on average in the context of serious games based on objective gameplay logs (Lomas et al., 2013). This stands in some opposition to the frequently observed self-reported preference of challenging games (cf. section 4.3) and consistently low challenges can be detrimental to the targeted serious outcomes, and especially with MGH consistent effort is an important aspect. Adaptive systems for MGH thus have to balance between maximizing for player experience, and maximizing for effect with regard to the targeted outcome, as indicated above with the dual-flow model (Sinclair et al., 2009).

Regarding the adjustment mechanism, adaptive systems for SG can in general either be heuristic (often frequentist), or take a more open ended, typically probabilistic approach with methods from machine learning or artificial intelligence, or employ a mixture of both. The heuristic approach is usually based on following fixed responses to a predefined set of measures (Hocine et al., 2011), by employing mapping functions from performance measure outcomes to in-game variables with fixed lower and upper limits and thresholds for triggering adjustments (cf. section 3.2.2). While heuristic approaches allow for great certainty that the adjustments will lead to the desired effect, they require great effort and caution during development with exhaustive testing and are limited by the extent to which players during development actually represent players in real-world situated use. Furthermore, the parameter spaces of video games in general and MGH in particular can be very complex, thus requiring a large number of mappings that can quickly become challenging to establish and mange accurately. Methods from artificial intelligence and specifically machine learning can offer support with these challenges, but they also come with their own set of challenges and uncertainties, as discussed in the following section.

### 2.8.3   Artificial Intelligence, Machine Learning, and User Modeling

Since user skills are not a constant, both adaptable and adaptive systems face further challenges, as both users and the games undergo complex changes over time. Related work has explored the application of a broad range of techniques from artificial intelligence in general, and machine learning, as well as user modeling in particular, to video games, serious games, GFH, and MGH. Examples include the application of *Bayesian methods* for overcoming cold-start challenges (Zigoris & Zhang, 2006) or for player matching (Herbrich et al., 2007), employing *support vector machines* in order to classify the emotional status based on physiological player data such as electrodermal activity, sweating, and body temperature (Chanel et al., 2008), *reinforcement learning* for selecting activities that are sensitive to individual levels of challenge (Andrade et al., 2005), *genetic algorithms* for difficulty adjustments in platform games (Watcharasatharpornpong & Kotrajaras, 2009), *neural networks* for adaptive rehabilitation games (Barzilay & Wolf, 2013; Wong, 2008), *self-organizing maps* for game play behavior analysis during development with potential for adaptivity (Drachen et al., 2009), *decision trees* (Walonoski & Heffernan, 2006), or mixed *recommender systems* methods (Medler, 2009).

Hence, both supervised and unsupervised methods have been applied, with some examples in commercial games, for example with the use of *genetic algorithms* (GAs) in the games *Creatures* and *Black & White* (Schwab, 2004). While more open-ended approaches, such as GAs may function well in situations where there is a risk of getting stuck in local minima or maxima and can explore very complex dimensional spaces, they would arguably require too much training or online trial and error, even if spawn-pool optimization would be employed for different player types. Generally speaking, methods from optimization are potentially viable for application with MGH, since many aspects of human performance, such as physical abilities, usually follow gradual developments on increasing or decreasing trajectories. However, non-linear methods are still usually required, due to the non-linear progression of biophysiological phenomena (cf. section 3.7.1).

Due to the heterogeneity of users in terms of preferences, abilities, and needs together with individual patterns for learning, and medium- to long-term cognitive and physiological development, *user modeling* (UM) (Fischer, 2001; Kobsa, 2001) can be employed, especially if continued personalization is targeted over longer periods of regular use, since the aggregate and estimate changes in players over time (Chiou & Wong, 2008; Janssen et al., 2007). It is a central argument in this thesis that such personalization is a desirable goal in many MGH. Hence, it is not surprising that user models have been adopted both for parameter adjustment during game development (Langley & Hirsh, 1999), as well as for continued individualized adjustments during GFH use (Charles et al., 2005; Missura & Gärtner, 2009), for example with the use case of depression prevention (Janssen et al., 2007).

Naturally, next to individual user models, group modeling can also be applied, especially if a fast initial adaptivity is required. Missura and Gärtner, for example, have shown how observing manual difficulty selections by players can serve to form group clusters which can then be used for DDA (in their case using change over time instead of static cluster average difficulty) together with cluster prediction, outperforming *either sole solution*, as well as *constant difficulty* in terms of player experience (Missura & Gärtner, 2009). Notably the application of user models often blurs the lines between design time and use time (Fischer, 2001). Limitations for user or group modeling consist in "sensor blindness", since compared to information rich multimodal human communication, even advanced modern sensor devices are very limited (Fischer, 2001), as well as challenges in acquiring and managing dense usage data. Advances are being made with respect to both of these challenges, however, not only through a renewed interest in user tracking that goes hand in hand with investments in virtual and augmented reality technologies, but also through the creation of online (serious) gaming platforms with large user numbers that allow for data driven continued adaptivity, for example in the case of games for math learning (Lomas et al., 2013).

As the preceding sections have shown, a number of projects have explored the utilization of adaptive techniques in GFH and MGH. However, adoption in publicly available titles, or even the exploration over medium- to long-term duration still remains rare, which motivated a medium-term study that will be discussed in the later parts of this thesis (section 4.9). Furthermore, automation always presents the danger of interfering with user interests even if very accurate models of conservative methods are employed. The implications of the tradeoff between automation and user autonomy will be discussed in further detail in section 3.7.2.

## 2.9   Application Security, Data Protection, and Privacy

In modern game development, data collection and player tracking is prevalent and readily available, even as *software as a service* together with game analytics. This is arguably becoming even more important as many games are turning into software services that undergo continuous transformation beyond the initial release to market. In order to optimize game experiences, local play testing is augmented with real-world performance data, which must be captured and transferred for analysis. In a similar manner, while some adaptive systems can function on local machines, for example if they are of a purely heuristic nature, many ML and UM approaches require online data processing, e.g. in order to benefit from insights regarding the development of abilities and needs of similar players. If player information and performance or even sensor data are transferred online and processed remotely, the privacy of the players is at risk and complex local privacy and data security regulations must be considered.

The underlying related publication [**see publication B.3, section 5.8**] discusses the importance of considerations on *ethics*, *data privacy*, and *regulations* with regard to GFH, where especially sensitive data is processed and potentially stored that often can be uniquely identifying of a user, akin to a fingerprint, due to the physiological origins, and where potential undermining coercion due to persuasive techniques presents a danger (Hekler et al., 2013). In brief, while many users will often provide access to their data and allow remote processing, at least given *pseudonymization* or *anonymization*, considerations on privacy and security are important issues by law, as well as for ethical reasons, and from the perspective of many users. The use of personal data must be well justified and should make fundamental contributions that lead to personal benefits of the users (Daglish & Archer, 2009; Tang et al., 2006).

# 3   Structuring Theoretical Considerations

The following sections gather a number of structuring theoretical considerations that have - in part - resulted from a number of practical projects that will be discussed later in this thesis, while other parts have been constructed a priori, based on related work and considerations that were made to guide the practical work. Beginning with the basic promises of MGH, approaches to harness the benefits these promises entail with the help of techniques for adaptability and adaptivity, which are discussed from the perspectives of design, engineering, and research. Further sections add a perspective of motivational psychological constructs, a discussion of evaluation targets for development, and typical challenges that arise during the realization of adaptable and adaptive MGH projects. Lastly, a generalizing section presents a summary of an overarching process for *needs and abilities based human-centered design for adaptive systems* that has the potential to generalize to serious games, or even to the general conceptualization and implementation of adaptable and adaptive interactive digital media user experiences.

## 3.1   Three Central Promises of SG/MGH

As discussed in further detail in the included underlying publication (Smeddinck, 2016) [**see publication B.3, section 2.3**], the potential benefits of motion-based games for health for the support of physiotherapy, rehabilitation, and prevention, can be summarized as three central promises that also generalize to MGH, GFH, and even serious games (SG) with other application areas. The three promises are *motivation*, *guidance*, and *analysis* (Malaka et al., 2016; Smeddinck et al., 2014), and each aspect may be present in any given GFH to a different degree (cf. Figure 6). *Motivation* that arises from the employed game design and technologies promises to support players in performing activities that might otherwise be perceived as more repetitive, strenuous, or generally undesirable, be it physiotherapy exercises or schoolwork. Due to the interactive nature of SG, *guidance* can be provided in the form of real-time, or post session feedback regarding the quality of the performance of the activity, which is especially promising if feedback from human professionals is not available. *Analysis* of the objective development of a player can be potentially be provided, since game controller and performance data (and possibly further psychophysiological measurements) are processed to drive the game that can also augment the picture of a patient-player beyond what would usually be available for self-reflection, or for analysis in cooperation with professionals. This especially applies to adaptive SG, since the same measures that can support the adjustments frequently are of potential interest for the aforementioned purposes.

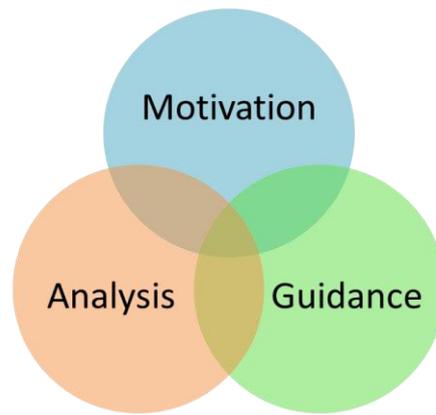

**Figure 6: Three central areas of potential benefits of GFH; via Malaka et al. (2016); Smeddinck et al. (2014).**

While these promises validate research and development efforts in GFH, it is important to underline that the potentials are frequently abused to exaggerate the practical benefits of GFH and even to challenge the relevance of human health professionals. As the sections regarding challenges of GFH in this thesis show, it is much more reasonable to approach research and development regarding GFH with an aim at augmenting the available palette of tools for supporting health and the delivery of health care. Health professionals will likely continue to play a central role in steering the general direction of treatments and in assuring the adequacy of using specific GFH with apt configurations.

As mentioned above, implementations differ regarding the extent to which each potential can be realized. However, they can also differ with regard to the *type or level* of each area of potential that they target. The potential for *motivation*, for example, can be helpful for casual single interaction episodes, as well as for long-term sustained interactions. *Guidance* in the context of MGH could mean small scale feedback on exercise execution quality, or it can relate to broader treatment planning. *Analysis* can, for example, facilitate reflections on a single session, but it can also mean full-fledged medical diagnostics, or even prediction. Therefore, these areas of potential serve to structure approaches in SG, and they present a reminder of the diverse opportunities in MGH beyond the aspect of motivation that is often presented without much consideration of the remaining potentials.

These broad classes can also be seen to encompass general design principles that were derived for persuasive systems, such as primary task support (motivation / guidance), self-monitoring (guidance / analysis), dialogue support (analysis), etc. (Oinas-Kukkonen & Harjumaa, 2009). Factors around personalization and customization are notably absent from this list of potentials, since adaptability and adaptivity can be argued to be of use as technical facilitators to each of the areas of potential, rather than being perceived as having an inherently valuable potential in and of themselves. Assuming the use of adaptivity and adaptability to achieve highly automated MGH experiences which support leveraging the potentials described above to a considerable extent, the case for digital games becomes evident. Compared to manual games, which can also fulfill self-determination needs very well, digital interactive games can be more flexible and scalable with regard to the level of personalization they offer for each individual player, since the effort for personalization is comparatively low, once a functioning mechanism has been established. If implemented in an efficient and reliable manner, the areas of guidance and analysis, in particular have a lot to offer in health applications where human judgment plays an important role, such as quality of motion execution in physiotherapy. Therapists have been found to disagree with each other on quality of movement (Pomeroy et al., 2003) and automated tools with digital measurements have the potential to support more objective and comparable analyses. While helpful automatic adjustments or the personalization performed by an adaptive MGH require considerate development and testing, manual adaptations performed by therapists also do not guarantee good game and treatment experiences.

## 3.2   Adaptability and Adaptivity

Summarizing the limitations of video games and exergames for use cases in health, Göbel et al. (2010) note that the most salient obstacles for existing systems include a lack of concepts for personalization, as well as a strong need for long-term motivation and sustained use. Arguably, adaptability and adaptivity can play an important role in overcoming these obstacles both through helping to present adequately flexible settings and challenges, and through providing adequately adjusted content. The terms adaptability and adaptivity are kept separate on most occasions in this work in order to underline that the required flexibility and achievement of adequate configurations can be attained both through manual adaptations, automatic adaptivity, or a mixture of techniques.

### 3.2.1   Approaches to Adaptability

The focus in the design and implementation of adaptable games that allow for manual adjustments usually lies on achieving a good usability and the user experience for the user interaction with the settings interfaces. For development goals such as efficiency, effectiveness, and a positive experience, related work from interaction design (Hartson & Pyla, 2012) provides important approaches, such as iterative user-centered (Allen, 1996) or participatory (Schuler & Namioka, 1993) design that play an increasingly important role in the development of settings interfaces and settings integration for serious games (Smeddinck, Herrlich, et al., 2015). Next to determining the mapping of settings parameters to in-game variables, which can take on different levels of *breadth* and *explicitness* (cf. section 3.4.2), designers and developers can decide whether they want to take the traditional route of *game-centric parameterization*, where settings are made per-game and the parameters typically represent aspects of the specific game, or whether they aim to implement *player-centric parameterization*, a concept which was inspired by developments in the context of this thesis (cf. section 4.5). In the latter case, settings are typically made on a per-user basis and the settings parameters relate to the psychophysiological abilities and needs of a user (e.g., "range of motion" in a motion-based game for health). Mixed methods, where, e.g., settings are made on player-centric parameters while still being made per game are also possible, as are group-based settings. The underlying publication (Streicher & Smeddinck, 2016) provided in Part II **[see publication B.II, section 3.3]** provides a more detailed overview of considerations around these aspects.

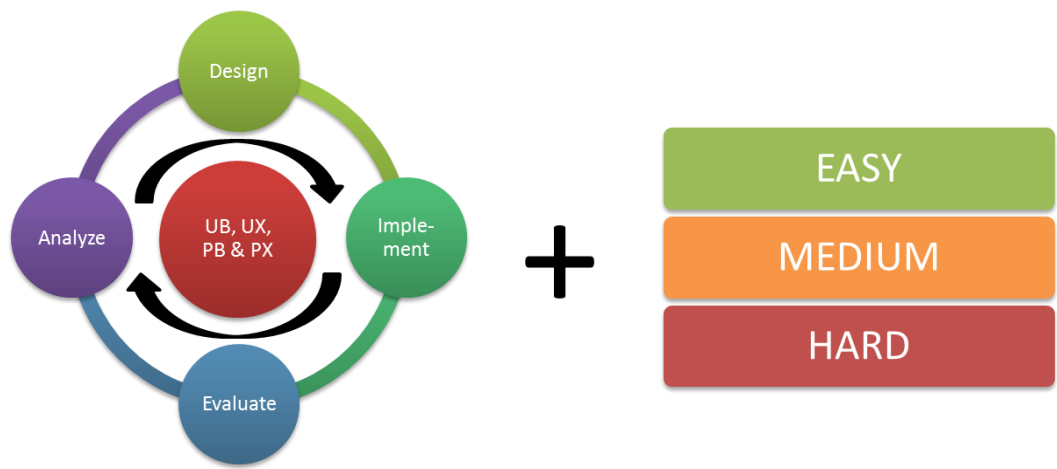

**Figure 7: A common approach to adaptability in games: Frequent manual adjustments during development and manipulation of monolithic variable bundles through menus in a game after release.**

The simple difficulty settings present in most video games to date do present a plain form of manual adaptability. Such settings are, however, limited in several ways:

a) They bundle a larger number of specific game variables under a few or most commonly just one overarching difficulty parameter(s). Such monolithic variable bundles facilitate coarse adjustments which result in a range of outcomes that may or may not correspond to the best settings with regard to player experience or with regard to an intended serious outcome (cf. discussion in sections 3.4.2, 3.7.6, and 4.5).

b) Interacting with manual settings can interrupt a play session, which may disturb immersion and be detrimental to the game experience and motivation (cf. section 4.6.1).

c) The existence of manual settings makes players acutely aware that the settings are being or have been adjusted, which can interact with the perception of the game performance. It should be noted that negative impacts (e.g. "I had to make it easier, because I was not good enough"), as well as positive impacts (e.g. "I was good enough to set the difficulty to 'hard'") appear plausible (cf. sections 3.4.2, 4.8, and 4.10.1).

d) The labels of difficulty levels often follow the convention of "easy", "medium", "hard", etc. but they are also often pertly labeled, e.g. "Hey, not too rough.", "Hurt me plenty.", etc. (examples from the original Doom game). This may appear pleasant and lead to desirable effects, but it may also appear unpleasant or disrespectful and can affect player motivation in ways that are difficult to predict (cf. section 4.6.1).

e) While game difficulty, for example, usually follows an increasing trajectory that is determined through game testing, relies on anticipating a certain development in player skills and abilities, the actual given player skills and abilities are subject to fine-grained temporal fluctuations (see section 3.7.1). Corresponding to these fluctuations manually could arguably appear strenuous and lead to bad player experiences.

These reasons that apply to general video games, as well as SG and MGH combine with additional reasons, such as the limited ability of players or patients to anticipate and self-regulate corresponding to the state and development of individual abilities and needs. This forms a clear case for the application of techniques for automating such adjustments, be it for difficulty adjustment or for the adjustment of other aspects (e.g. parameters for content generation).

### 3.2.2   Approaches to Adaptivity

When designing adaptive (serious) games, meaning games that automatically adjust to the player abilities and needs, the focus lies on a good playability and player experience, and it can be argued that getting these aspects right is even more important than it is in the case of purely adaptable games, since adaptivity is a form of automation and if automation does not adequately respect the intentions of the users, or even comes to hinder them, the most basic principles of usability and user experience can easily be violated (Sheridan, 2001). This aspect is discussed in further detail in the included publications [**see publication B.11, section 3.4**], together with the approach of partial manual involvement (cf. section 3.4) for avoiding maladaptation, and the basic building blocks of any adaptive SG, namely *performance evaluation* for measuring or estimating the impact of the current parameter settings on the player performance, as well as an *adjustment mechanism* that adjusts the parameter settings depending on the outcome of the performance evaluation (Adams, 2010). While the exact realization of such an automated optimization process can take many forms and the adjustments do not have to be limited to tuning existing parameters, one of the most common forms of adaptivity in games is *dynamic difficulty adjustment* (DDA). As illustrated in Figure 8, DDA can be described as heuristics-based adaptivity for dynamic game difficulty balancing, where difficulty is either increased or decreased, when the player performance crosses according thresholds.

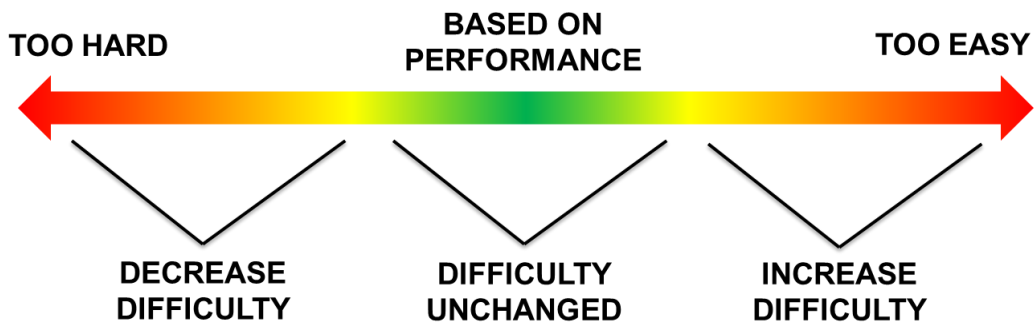

**Figure 8: Difficulty adjustment based on performance. The thick arrow indicates a possible range of performance from low to high. Three sections on the range of performance determine whether difficulty for the player should be increased, decreased, or left unchanged.**

Notably, adjustments can be performed on the basis of a single parameter, or with a direct influence on multiple variables. Parameter mappings and thresholds are usually determined and fine-tuned with iterative testing. Depending on the amount of adjustments and type of the parameters, the adjustments can be either *visible* or more or less *invisible* (cf. *saliency* in section 3.4.2). This is closely linked to one of the largest challenges with DDA: Since the goal of the adaptive system is to notably impact player performance or experience, the adjustments usually lead to notable changes in the game, which in turn lead to a resulting difficulty that deviates around the current theoretically optimal settings for any given player. This effect is (often) called *rubber banding* (Pagulayan et al., 2012) and it can be perceived as annoying, or even unfair. As mentioned above, a frequently referenced example for this rubber banding effect is the Mario Kart series. Despite the notable challenges, more subtle adjustments, such as slightly supporting the user with aiming (Vicencio-Moreira et al., 2014), or more readily available supply of resources, such as health packs (Hunicke, 2005), have been shown to be successful in improving the overall game experience or performance.

While some approaches exist that are purely based on heuristics and do not require any prior or accumulated information in order to determine the settings or adjustments for a specific player (cf. section 2.8), most adaptive SG or MGH aim to adjust as precisely as possible to the individual needs and abilities as they evolve and change during continued interaction with the system. In this context, gathering data about the performance of each individual player is a common prerequisite that entails a number of challenges.

### 3.2.3  Cold-Start Problem

The cold-start problem frequently arises with adaptive systems. It is known as a common challenge from machine learning and describes the problem of making predictions in the absence of a proper amount of data or information. The problem applies to adaptive SG, since they are usually designed to adapt to individual players, who, at one point in time, are all new users for whom no performance data have been recorded. Next to assuring careful user-centered iterative design and balancing and creating games that works comparatively well for an average population, the challenge can be approached with a number of measures, including calibration procedures, models that require manual settings before first play sessions, taking the performance and development of similar user groups into account, relying on established models of development based on the application use-case, or by refraining from any adaptivity until enough data has been captured [**see publication B.II, sec. 3.8**].

### 3.2.4  The Challenge of Co-Adaptation

The challenge of co-adaptation occurs, since player capabilities and needs are not only complex and multivariate, they also change (adapt) over time in a nonlinear fashion in any realistic prolonged use of any serious game, and in part, any user adapts to changes in the adaptive system itself. This can lead to situations where a system adapts settings in a specific way and the user adapts to these settings although they are not objectively optimal, and may even lead to harmful interactions. Approaches to controlling co-adaptation include prudent user-centered iterative design cycles, which assure that test groups are frequently changed before test users adjust to problems in the system, so that relevant problems can be isolated. Manual means to interfere with changes made by the adaptive system, or to correct them, can also avoid malicious co-adaptation. However, the challenge remains elusive, and controlling for problems with co-adaptation requires detailed observations of both the system and the user behavior [**see publication B.II, section 3.9**]. Despite these concerns, many adaptable and adaptive MGH have been successfully realized, at least in research and development.

## 3.3  Modular Development

The implementation of MGH is complex and can differ notably between projects. There are, however, common modules that may or may not be present in a specific MGH, but that tend to occur for consideration in most MGH projects and which can be described and discussed on a generalized level.

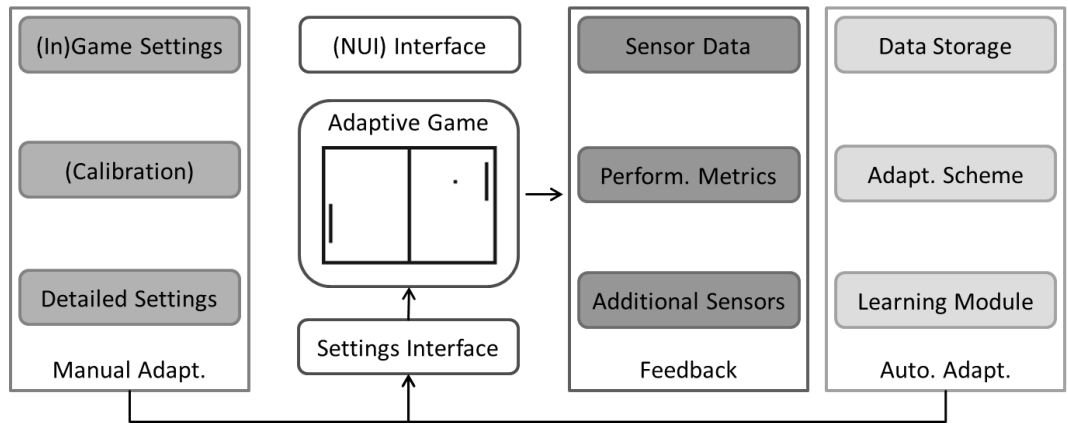

**Figure 9: A modular approach for engineering motion-based games for health featuring modules that compose the typical surrounding ecosystem that is developed for adaptable and / or adaptive MGH.**

At the core of any adaptable and / or adaptive MGH development stands a game. While the game design can vastly differ from one use case to the other, the surrounding technical infrastructure is often very similar (cf. Figure 9). Due to the motion-based nature of the game, the players interact through some form of natural user interface, such as a motion-tracking device. In order to allow for adjustments to the game, it has proven beneficial to consider a settings interface as a general abstraction layer that can be manipulated either through manual settings that are enacted via some form of (graphical) user interface, or automatically through some form of automatically adaptive system. In this manner, specific game variables can be mapped to one or multiple adjustment parameters in a clearly defined module.

If a manual settings layer is present, a selection of (usually comparatively coarse) settings that are presented directly in the game or in a game menu can be expressed as a module (here: "game settings"). In many use cases, however, more fine-grained settings have to be made available either to the players themselves, or to involved health professionals, such as therapists. These are frequently realized through separate (web) applications in order to facilitate remote settings, or to allow for making adjustments without interrupting the game client and are represented as the "detailed settings" module shown in Figure 9. Lastly, calibration procedures are often required to guarantee the correct functioning of sensor devices that are part of the natural user interface, or to support finding the most adequate settings for a given player. While calibration procedures can be partially or even largely automated, they typically require manual user involvement or must be triggered manually. The corresponding module "calibration" is thus included in the area for manual adaptation.

The feedback area of modules encapsulates functionalities that gather feedback about a system user through interactions that are captured and analyzed. Due to the motion-based nature of the games, some form of data is usually available from the natural user interface, for example skeleton data when using a Kinect device. In addition to driving the core interaction loop, information about player abilities can often be derived from the captured interaction data (such as the range of motion in the case of full-body posture inference). Such information is frequently augmented with game performance metrics that provide information about the players' progress and / or performance with respect to the game mechanics. Such information is important to guide the progression of the game. It is also an important target for adjustments, since good game experiences usually require players to be able to perform within a certain threshold range, relating to the balance in the flow channel of challenges and skills (cf. Figure 3). Additional sensors, such as wearable tracking devices, are often included to derive further information about the state of the player, such as heart rate, local accelerations, or sympathetic nervous system activation, even if they are not required to drive the core game experience (which would put them into the sensor data module in Figure 9). Even if this information is not used to drive an adaptive system, it can be used to provide objective analyses to players or interested third parties, facilitating the respective category of promises indicated in section 3.1.

The section for automatic adaptivity in Figure 9 contains three common modules. The data or information gathered through analysis is typically fully or partially persisted in some form of data storage (usually a database). This way, a learning module can, if present, employ that data or information to build and improve models for individual users or groups. Either way the persisted data can be used for performance aggregation and analysis. The adaptation scheme can then rely on the learning outcomes, on heuristics that do not depend on aggregated data, or frequently on a mixture of methods that combine learnt aspects with elements that are learning-free, in order to allow for automatic adjustments to game difficulty (dynamic difficulty adjustments), to game content, or to other dynamic aspects, such as generative elements. A large number of design dimensions typically determine the exact implementation of an adaptation scheme for an adaptive system. The following section provides a detailed discussion of these dimensions of adaptivity.

## 3.4 Dimensions of Adaptivity

### 3.4.1 Related Approaches and Models

While most practical benefits of computers rely on automation (Sheridan, 2001), adaptive systems usually automate adjustments to the appearance or functioning of the system itself. These adjustments aim to make the system respond more adequately to the needs, abilities, or goals of the users, to the usage context, or to the environment (Andersson et al., 2009). Due to this nature, the adjustments can interfere with, or even run contrary to the goals, understanding, accustomed usage schemata, or mental models of users. This can endanger not only the user experience, but also the efficient and effective system use. The design of adaptive systems should thus be carefully planned and tested. Sheridan et al. (2001) and Parasuraman et al. (2000) provide a general discussion and model for the tradeoff between the level of automation, and the levels of user involvement, as well as the saliency of changes (or user informedness) entailed in that automation. Figure 10 shows a simplified general scale of automation that ranges from systems that offer no assistance (no automation) to fully automatic execution that does neither take user input nor even inform them about adjustments.

Arguably, adaptivity is a specific case of automation, and a conscious consideration of separate dimensions that appear mixed in the scale of degrees of automation shown in Figure 10 can foster a much more deeply informed design. While Sheridan acknowledges that the scale folds together several relevant dimensions, mentioning (a) the degree of specificity required by a human for putting in requests, (b) the degree of specificity with which the machine communicates alternatives or recommendations, (c) the degree to which a human is responsible for initiating action execution or implementation, and (d) the timing and detail of feedback provided to the human after action execution, other authors have since discussed the dimensions of (self-)adaptive systems specifically and in a more fine-grained manner.

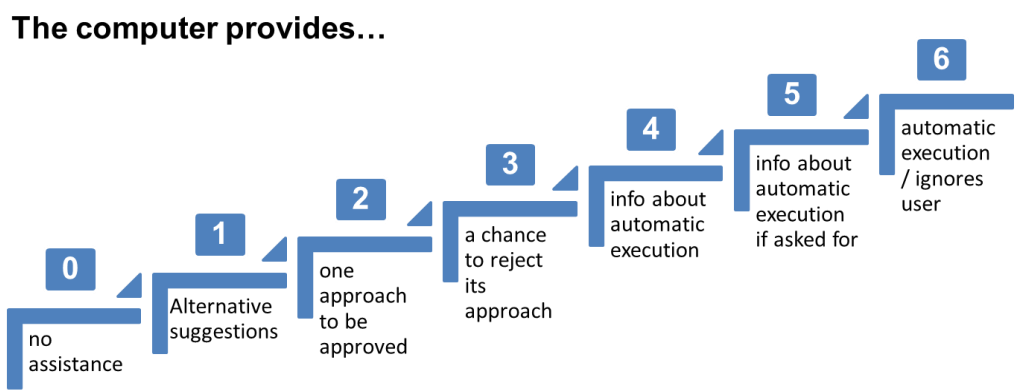

Figure 10: A simplified scale of degrees of automation following the model presented by Sheridan et al. (2001) meant to support design decisions around each of (a) acquiring information, (b) analysis and display, (c) deciding on actions, and (d) implementing actions.

A more detailed consideration with a direct relation to games is presented by Baldwin et al. (2013) in their framework of DDA in competitive multiplayer video games. They discuss a number of components found in games with DDA that were aggregated in a formal review after having recorded three key elements for each game: the trigger for activation (of DDA), the affected game rules, and the scope of the effect. The three elements were then broken into more specific classes – the components –, each with a number of associated attributes that are mutually exclusive per component. The components are:

1. **Determination:** In which game state or at what time is the decision to use DDA is made?

2. **Automation:** Is the decision to use DDA made by the player or the system?

3. **Recipient:** Which players are affected by the DDA?

4. **Skill Dependency:** Does it require skill to benefit from the dynamic adjustment?

5. **User Action:** Is a user interaction required to initiate DDA effects?

6. **Duration:** How long does a DDA instance last, or how often can an effect be used?

7. **Visibility / Awareness:** Is the DDA effect visible or not and is the player aware?

The attributes suggested for each component are summarized in Table 3 and examples for the components and attributes in the applied context of games are provided in the original publications (Baldwin et al., 2014; Baldwin et al., 2013).

**Table 3: Components and the respective attributes of DDA in video games (Baldwin et al., 2013).**

| Component | Attributes |
|---|---|
| 1) Determination | • Pre-gameplay<br>• Gameplay |
| 2) Automation | • Applied by system (automated)<br>• Applied by player(s) (manual) |
| 3) Recipient | • Individual<br>• Team |
| 4) Skill Dependency | • Skill dependent<br>• Skill independent |
| 5) User Action | • Action required<br>• Action not required |
| 6) Duration | • Single-use<br>• Multi-use<br>• Time-based |
| 7) Visibility | • Visible to beneficiary only<br>• Visible to non-beneficiaries only<br>• Visible to all players<br>• Not visible |

Although the scale by Sheridan aims to cover automation in general and the model by Baldwin aims at a very specific domain of automation, a closer comparison can show whether the more specific model can be folded into the more general approach. While there is some relation between the classes identified by Baldwin et al. and the dimensions that are folded into the scale of automation as noted by Sheridan (e.g. the degree to which a human is responsible for initiating action can be related to the component called "user action"), the match is only partial and the categories overlap (e.g. the dimensions of feedback timing and specificity mentioned by Sheridan overlap with the visibility component, but they also cover further aspects). The strength of the approach by Baldwin et al. lies in its explicit empirical foundation, being based on a formal survey of existing games. However, the decision to express the model classes as components with seemingly fixed levels of attributes can be questioned due to the actual freedom in (game) design to realize systems that present solutions that combine, or only partially realize the attributes. This aspect is arguably more aptly expressed with design dimensions as suggested by Sheridan. The framing of the component descriptions as questions, as chosen for the summary in this discussion, clearly hints at the potential of the approach for design in addition to evaluation and analysis. It seems apparent that this approach can be applied to other types of games beyond the class of competitive multiplayer games which Baldwin et al. targeted. Over the course of their work, Baldwin et al. have shifted from the term "visibility" to the term "awareness" for the same component. This brings the advantage of being independent of the modality, yet awareness can be difficult to assess (from an evaluation perspective) and very difficult to predict (from a design perspective). The term "saliency" could be considered as a more matching alternative, since it supports both perspectives and is independent of modality. Overall, it appears that Baldwin et al. provide a model that is more specific than the broad scale of automation presented by Sheridan and thus supports a more nuanced application in the conceptualization, design, evaluation, and study of adaptive and adaptable MGH. However, it does not cover all aspects that are folded into the scale by Sheridan (e.g. the timeliness of feedback provided to the user).

A more detailed approach to modeling dimensions of (self-)adaptive software systems has been presented by Andersson et al. (2009) who provide a table of dimensions that fall into the four categories of (system) *goals*, *change* (as in cause for adaptation), *mechanisms* (the system reaction towards change), and *effects* (impact of adaptation upon the system). These categories notably relate to the coarse classes of analysis employed by Baldwin et al. during their formal review: *change* relates to the *"trigger for activation"*, *mechanisms* relates to *"rules affected"*, and *effects* can be related to *"scope of effect"*. While the latter class appears to overlap with the category *"goals"*, the category of goals adds additional considerations that can be valuable in design and analysis (such as the number of goals, or whether they change within the lifetime of the system). The modeling dimensions suggested by Andersson et al. are listed in Table 4.

**Table 4: Modeling dimensions for (self-)adaptive software systems; after Andersson et al. (2009).**

| Dimensions | Degree | Definition |
|---|---|---|
| **Goals – goals are objectives the system under consideration should achieve** | | |
| evolution | static to dynamic | whether the goals can change within the lifetime of the system |
| flexibility | rigid, constrained, unconstrained | whether the goals are flexible in the way they are expressed |
| duration | temporary to persistent | validity of a goal through the system lifetime |
| multiplicity | single to multiple | how many goals there are? |
| dependency | independent to dependent (complementary to conflicting) | how the goals are related to each other |
| **Change – change is the cause for adaptation** | | |
| source | external (environmental), internal (application, middleware, infrastructure) | where is the source of change? |
| type | functional, non-functional, technological | what is the nature of change? |
| frequency | rare to frequent | how often a particular change occurs? |
| anticipation | foreseen, foreseeable, unforeseen | whether change can be predicted |
| **Mechanisms – what is the reaction of the system towards change** | | |
| type | parametric to structural | whether adaptation is related to the parameters of the system components or to the structure of the system |
| autonomy | autonomous to assisted (system or human) | what is the degree of outside intervention during adaptation |
| organization | centralized to decentralized | whether the adaptation is done by a single component or distributed amongst several components |
| scope | local to global | whether adaptation is localized or involves the entire system |
| duration | short, medium, long term | how long the adaptation lasts |
| timeliness | best effort to guaranteed | whether the time period for performing self-adaptation can be guaranteed |
| triggering | event-trigger to time-trigger | whether the change that triggers adaptation is associated with an event or a time slot |
| **Effects – what is the impact of adaptation upon the system** | | |
| criticality | harmless, mission-critical, safety-critical | impact upon the system in case the self-adaptation fails |
| predictability | non-deterministic to deterministic | whether the consequences of adaptation can be predictable |
| overhead | insignificant to failure | the impact of system adaptation upon the quality of services of the system |
| resilience | resilient to vulnerable | the persistence of service delivery that can justifiably be trusted, when facing changes |

This approach is very fine-grained, whilst also offering broader categories that facilitate a systematic approach on different levels of detail. The perspective, however, is almost exclusively system-centric (with the exception of the *autonomy* dimension), whereas Baldwin and Sheridan acknowledge the importance of the user perspective by highlighting aspects around the level and type of interaction with the automated system. While this core issue could be remedied by a fusion with the detailed approach by Andresson et al. by adding a further category of dimensions called, for example, "user involvement", the existing categories can also be extended to encompass user-centric considerations. This approach appears warranted given the significance of the role of the user in modern interactive digital systems and the fact that users themselves should be considered as complex (co-)adaptive entities (cf. section 3.2.4). This also applies to the context of (serious) games, GFH, and MGH, given the necessity of the involvement of the user perspective, especially with respect to facilitating the achievement of the targeted serious outcome.

### 3.4.2 Combined and Extended Approach

Thus, a combined approach to the modeling dimensions of adaptive and adaptable software systems can be composed as follows (following Andersson et al. unless noted otherwise):

**Table 5: A combined approach to modeling dimensions of adaptive and adaptable software systems.**

| Dimension | Degree | Definition |
|---|---|---|
| *Goals* – objectives that a system under consideration should achieve or allow a user to achieve | | |
| *evolution* | static < –– > dynamic | The degree to which the goals can change. |
| *flexibility* | rigid < – constrained – > unconstrained | The degree to which goals are flexible in the way they are expressed. |
| *persistence* | temporary < –– > persistent | The validity of a goal over time. Replaces duration. |
| *multiplicity* | single < –– > multiple | The number of goals that are explicitly considered. |
| *dependency* | independent < –– > dependent | The degree to which goals are related. |
| *ability* | ability-dependent < –– > ability-independent | The degree to which abilities are required to achieve goals. Via Baldwin et al. (skill dependency). |
| *Change* – the cause for adaptation as enacted on the system automatically or by a user | | |
| *source* | external < –– > internal | The source of change. |
| *nature* | functional < –– > technological | The nature of the change (replaces "type"). |
| *frequency* | rare < –– > frequent | How often a change occurs. |
| *anticipation* | foreseeable < –– > unforeseeable | The degree to which a change can be predicted. |
| *determination* | during < –– > between | The degree to which a change happens during or after interaction. Via Baldwin et al. |
| *explicitness* | explicit < –– > implicit | How directly feedback that causes change points towards that change. Own addition. |
| *Mechanisms* – reaction of the system towards change and given possible interactions by a user | | |
| *type* | parametric < –– > structural | Whether adjustments affect parameters or structures. |
| *autonomy* | autonomous < –– > assisted | The degree of outside intervention in triggering adjustments (after they have been determined). |
| *organization* | centralized < –– > decentralized | Whether adjustments are controlled from a single or multiple distributed components. |
| *scope* | local < –– > global | Whether the adaptations are locally contained or involve the entire system. |
| *duration* | short-term < –– > long-term | How long adjustments last. |
| *timeliness* | best-effort < –– > guaranteed | The degree to which the point in time for adjustments can be guaranteed. |
| *triggering* | event < –– > timer | Whether adjustments are triggered by events of after fixed periods of time. |

| *exploitativeness* | explorative < -- > exploitative | The degree to which adjustments are based on the next estimated optimum or whether further alternatives are explored. Own addition. |
|---|---|---|
| *stepping* | small < -- > large | The step size of the adjustments. Own addition. |
| *automation* | automated < -- > manual | The degree to which adjustments are determined by the system or the user. Following Sheridan / Baldwin et al. |
| ***Effect** – the impact of adaptation upon the system and the user* | | |
| *criticality* | none < – mission-critical – > safety-critical | Impact in case of failure to adjust. |
| *predictability* | non-deterministic < -- > deterministic | The degree to which consequences of the adjustments are predictable. |
| *overhead* | insignificant < -- > failure | The impact of adjustments on processing load. |
| *resilience* | resilient < -- > vulnerable | The level of persistence of functioning that can be expected given any changes. |
| *recipient* | single-party < -- > multiple-party | The number of entities affected by any adjustments. Via Baldwin et al. |
| *positioning* | control < -- > feedback | The degree to which adjustments affect control or feedback. |
| *breadth* | one-to-one < -- > many-to-many | Whether adjustments are mapped from one or multiple parameters to one or multiple variables. |
| *saliency* | indiscernible < -- > unmistakable | Whether adjustments are perceivable. Combines user notification by Sheridan and visibility / awareness by Baldwin et al. |
| *intrusiveness* | unobtrusive < -- > interrupting | Whether adjustments are perceived as being intrusive to current goals. Own addition. |

While also this model cannot reasonably claim to cover all potential aspects of adaptable and adaptive interactive systems it arguably allows for a comparatively detailed and informed structuring of approaches to the design, implementation, and evaluation of such systems. The total number of dimensions in the four categories underlines the complexity of the (31-dimensional) design space and provides a good intuition on the complexity of controlled developments and studies in this area, since all dimensions should theoretically be considered as design or experimental variables and the levels on each dimension are - in principle - continuous. Furthermore, common terms in flexible interactive (game) design, such as balancing, personalization, and customization, can be structurally anchored with this approach and modeled to encompass a certain n-dimensional sub-space in order to guide considerations on these aspects in design or analysis. Due to the level of detail, this extended model can also serve to augment classification schemes for GFH that have worked with rather broad classes and attributes such as "adaptability (yes/no)" so far (Rego et al., 2010). In order to allow for a swift approach to broader considerations on adaptability and adaptivity in interactive systems, a visual summary of the categories in the model can provide a bird's eye view.

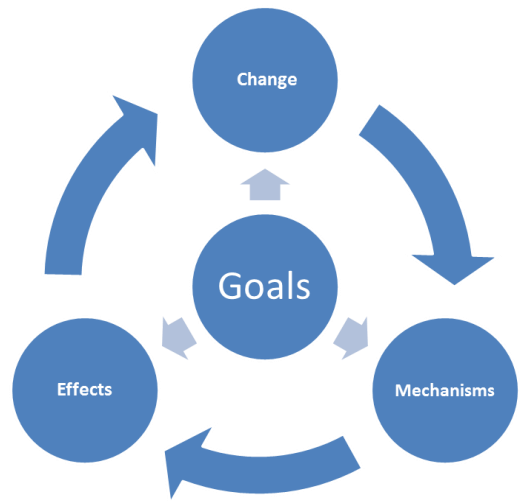

**Figure II: The process relation of the categories from the combined approach to modeling dimensions of adaptive and adaptable software systems.**

From this broad perspective, considerations on the goals of the entities involved in an adaptable or adaptive interactive system inform decisions on how to deal with changes, and how mechanisms respond to these changes. This, in turn leads to effects, which generate further changes in the interactive system as a whole. At this point, the close relation to the model for the design of adaptive games by Adams (Adams, 2010), which separates two aspects, the *performance evaluation* (which would fall under the category of considerations on *change* in this model) and the *adjustment mechanism* (which corresponds to the *mechanisms* in this model) becomes visible, although considerations on the overarching goals, as well as on the effects of the system operation are not explicitly highlighted in that approach. In a similar manner, Hunicke (2005) separates between *adjustment goals* (which correspond to the category of considerations on *goals* in this model) and *intervention strategies* (which can be interpreted to encompass the categories of considerations on *change* and *mechanisms* in this model, yet does not discuss the *effects* as explicitly as the model introduced above does).

Altogether the combined and extended model introduced above encompasses considerations presented by Sheridan et al., Baldwin et al., and can be related to the models by Adams and Hunicke, whilst also offering an arguably more complete broad picture composed of fine-grained dimensions. These terms and considerations can support the conceptualization, implementation, and analysis of adaptive and adaptable interactive systems.

## 3.5  Accessibility, Playability, Player Experience, and Effectiveness

While many specific aspects around potential experiences through interactions by users with (adaptive) GFH have been highlighted in the previous section, the *user* or *player experience* as an important recurring consideration and main endpoint for user-centered iterative (game) design (Fullerton et al., 2004) requires further discussion due to the complex nature of the goals and situated use of (motion-based) GFH. As this section will show, thinking merely along the lines of classic usability and user experience can endanger the successful realization and evaluation of MGH projects, since important aspects would not be explicitly considered.

*Most importantly, next to traditional game user experience outcomes, the specific health outcomes of a GFH matter. This duality of target outcomes suggests that both should be considered in design and testing. As noted before, a theoretical approach to this end is presented by Sinclair et al. (Sinclair et al., 2009) with the dual-flow framework. This work suggests a split of the general concept of flow into psychological flow and physiological flow, which is relevant for any motion-based playful application. The physical fitness and abilities of any given user present a physiological counterpart to the aspect of skill (which would likely contain elements of hand-eye coordination, fine-grained muscle control, and reflexes with any sedentary game as well) and have to be balanced with the level of physical challenges that are presented by the game. If the challenges exceed the physical fitness or abilities of a player, there is an increased risk of overstraining and injury. If the challenges notably undercut the physical fitness or abilities of the player, there is a risk of deteriorative effects due to a lack of training intensity, or at least of a lack of an impact regarding the targeted health outcomes. This has to be considered in game design, implementation, and testing, but it also has to be taken into account by adaptive systems for motion-based GFH. Next to the temporal fluctuations, the general interpersonal difference in abilities also entail that accessibility plays an important role in the design of (motion-based) games for health. The question whether the game system at hand contains barriers that could prevent users from the target group from reaping the potential benefits of the application is even pre-conditional to the questions of playability and player experience that are commonly discussed in game design literature [**see publication B.4**]. If the games are not accessible, any theoretical playability or player experience and any resulting motivation to be active or any resulting positive impact on health cannot come to fruition.*

*Given that aspects of accessibility are explicitly encompassed, user-centered design, repeated prototype testing and early pilot studies play an important role, if the complex challenges outlined in the preceding sections are to be tackled. GFH teams will benefit from following the classic user-centered design cycle (Hartson & Pyla, 2012) of analysis, design, implementation, and evaluation. However, it is important to point out that the iterations evolve around **multiple evaluative criteria** [see Figure 12].*

*Excerpt from: (Smeddinck, 2016)*

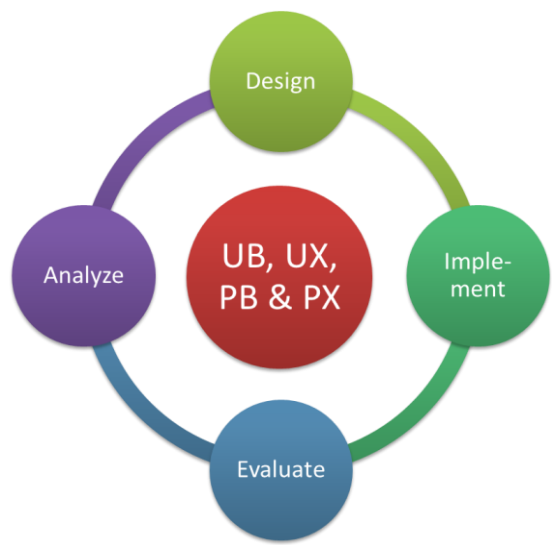

**Figure 12: The figure illustrates how user-centered iterative design in GFH revolves around the four cornerstones of usability (UB), user experience (UX), playability (PB), and player experience (PX).**

*Usability* matters in games, because they come with menus, settings, and controllers that are not necessarily part of the game design in terms of mechanics or aesthetics, yet are elementary for facilitating unhindered game play. It is thus useful to evaluate them with classic usability criteria such as efficiency, ease of use, performance, etc. Additionally, the **user experience** that is perceived by the players in the interactions with the application surrounding the core game is important. Unlike non-game applications, games, serious games, and GFH must be designed for a good player experience as well. **Player experience** results from the interaction with the core game, the game mechanics, the rewards, potential social interactions, etc. and is measured with different tools (see for example [section 2.5] in this chapter) than usability [e.g. SUS (Brooke, 1996)] or user experience [e.g. (Laugwitz et al., 2008)]. As a fourth evaluative focus, **playability** regarding the interaction with the core game itself should be considered separately from the usability of the immediately surrounding interactions, and it is important to note that, while inefficiencies and challenges in use will only be tolerated to a certain level, they do not always exert a clearly negative impact as they would be from the perspective of usability, since challenge and inefficient behaviors are important elements of many games (Smeddinck et al., 2016). Lastly, it is also helpful to recall the different parties of interest introduced in the preceding sections, which should be considered separately with regard to how important usability (UB), user experience (UX), playability (PB), and player experience (PX) in a specific GFH are, or are likely to be, to them. This overview clearly illustrates the complexity of the iterative user-centered design approach with GFH. Given an awareness of these challenges, however, these considerations can help prioritize evaluations and design targets in GFH where both patients and therapists are important target groups. While playability and player experi-

*ence are usually most important to the patients, the usability and user experience of the inter-actions surrounding the core game play are of usually most important to the therapists or other professionals.*

*Excerpt from: (Smeddinck, 2016)*

Furthermore, it is important to notice that classic criteria from usability often do play a role in games, for example in interfaces, but also concerning the game controls. However, maximum efficiency and effectiveness are usually not the focus of the core game mechanics. To the contrary, since playful and/or game mechanics often evolve around purposefully challenging the user, they can oppose the judgment by such usability criteria (cf. section 4.6.1). This presents a crux for many GFH, since the serious purpose often results in designers having to find compromises between providing enjoyable game experiences and supporting the serious goals [**see publication B.3, section 3.3**].

The limitations of usability testing within game testing due to a mismatch in its standard metrics have also been noted in related work (Pagulayan et al., 2003). Nacke et al. (2010) aggregate this aspect with a reference to the necessity of further extensions in order to include serious aspects (Gee, 2003; Prensky, 2001) and a discussion of the terminology around playability and player experience as well as pointers towards specific empirical research methods (L. Nacke et al., 2010). Arguably, the need to consider further requirements with serious games applies also to the context of adaptable and adaptive serious games. In this case, the temporal component in personalization or customization presents an additional challenge to the application of many established testing methods. Alankus et al. (2010) make an argument for the importance of further considerations beyond usability in the context of adaptable and adaptive MGH by highlighting not only playability, but also therapeutic value and fun as important goals for customizable games to support patients in stroke rehabilitation.

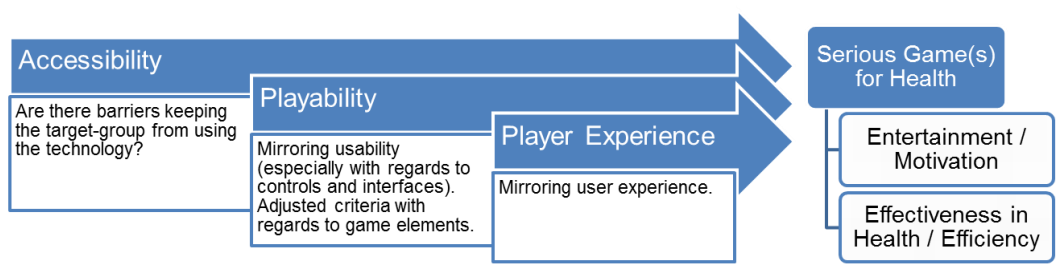

**Figure 13: A schematic summary showing how accessibility and playability are preconditions to player experience. When all layers are given, positive outcomes can arise both in terms of entertainment and in effectiveness in achieving the health target (Smeddinck et al., 2014).**

In summary, serious games in general and motion-based games for health in particular must be accessible to the target group, and reports in related work indicate that entertainment technologies will be used even in difficult application scenarios such as residential homes for older adults, if they are presented in an accessible manner [e.g. simple physical controls might prove beneficial as compared to a complex tablet-device for self-control tasks (Ziat et al., 2016)]. Furthermore, usability and playability need to be taken into account to ensure a low-effort and efficient interaction with game or gamified interfaces (cf. section 4.4) and other game components. Only if a solid foundation of accessibility and playability is present, actual user experience or player experience can arise and reasonably be taken into account, facilitating great game experiences and results regarding the desired serious outcome (cf. Figure 13).

## 3.6   Adaptability, Adaptivity, and the Motivational Pull of Video Games

As discussed above (cf. section 3.1), the potential to provide motivation is one of the core promises of serious games such as MGH. A common goal of adaptability and adaptivity in video games is to optimize motivation through a well-adjusted game experience. It is thus important to build approaches to adaptability and adaptivity on a solid understanding of the psychological foundations of motivation.

*Self-Determination Theory* has been used with considerable success for investigating and explaining motivation in video games and the effects of game play on well-being, stating that, in essence the psychological "pull" of games results from their capacity to engender feelings of *autonomy* (due to afforded volition), *competence* (due to affording effectiveness), and *relatedness* (due to affording social connection). These factors not only motivate further play but can also enhance aspects of psychological well-being (such as self-esteem, positive affect, subjective vitality, etc.) (Ryan et al., 2006). The factors competence and autonomy are closely related to game enjoyment and well-being after a playing session. Competence and autonomy perceptions relate to the intuitive nature of game controls and to presence / immersion in play. Ryan and Rigby explicitly connect their work to that of Bartle (1996) and Yee (2006) on player types in online multiplayer games, arguing that these approaches tend to reflect structure and content of games rather than underlying motives and satisfaction of players. With the components of autonomy, relatedness, and competence, SDT offers fruitful perspectives on both GFH and games for the target group older adults. As one main target of MGH in therapy is to eventually improve exercising in the absence of professionals, this can potentially alter the perception of autonomy [e.g. as captured via the PENS questionnaire (Rigby & Ryan, 2007)], which has been shown to be increased when using MGH as opposed to traditional physiotherapy without games (Smeddinck, Herrlich, et al., 2015). In addition to the regular established potential of video games to offer need satisfaction as described in SDT, GFH can arguably fulfill both extrinsic and intrinsic needs with regard to health issues. A theoretical framework to capture elements of intrinsic motivation (such as the level of "feeling in control of one's condition") is thus required. Older adults are usually burdened by increasing limitations in cognitive and motor function that can have a negative impact on the SDT factors. Perceived competence can be limited as an immediate effect of functional decline. Perceived autonomy can follow as well as relatedness, if the ability to participate in social interaction in the ways one is accustomed to becomes impacted. The potential of video games to improve all three factors of needs satisfaction (at least temporarily) is especially promising when they are unusually low before intervention. The three factors can also be employed to reason about the potential impact of adaptability and adaptivity, as well as about the specific dimensions that are discussed in section 3.4. For example, with an increasing level of *automation*, the satisfaction of autonomy needs can be expected to be negatively impacted, while competence may be improved due to a more adequate level of challenge, which can be presented without endangering the current level of immersion due to interacting with menus for manual adaptations. Evidence for these hypotheses has been gathered in a study on the impact of different modalities for difficulty adjustments that is discussed in section 4.6.1.

At the same time, a large amount of work on motivation in video games builds on flow theory following Csikszentmihalyi (1990), as noted in section 2.5, which also yields convincing insights and can be linked to considerations around adaptability and adaptivity, especially regarding the balance of challenges and skills. This relation is expressed in Figure 14 where the hypothetical development in skills of a given player (blue arrow-line) is sketched onto the model of the component that treats the required balance between skills and challenge as a facilitating factor for flow experiences.

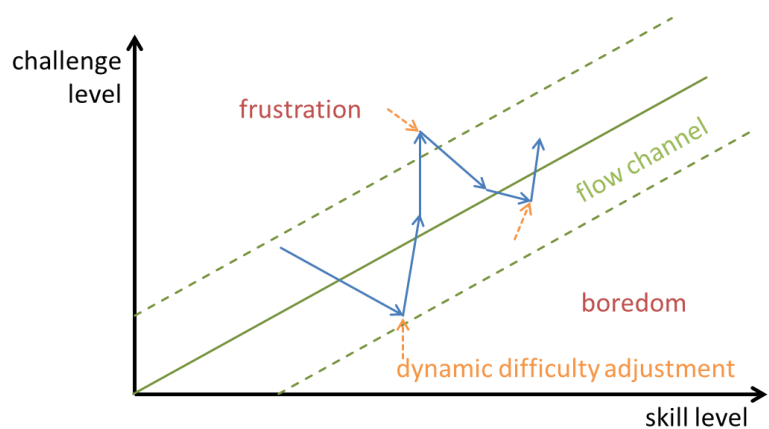

**Figure 14: Dynamic (or manual) difficulty adjustments (orange arrows) can be used to adapt the level of challenge presented to an individual player over time, as both the player abilities (blue arrow-line) and the basic level of challenge of the game progress [after (Streicher & Smeddinck, 2016)].**

Figure 14 indicates how the approach of DDA in games often directly aims at addressing the prerequisite of presenting an adequate balance of challenges relative to the skills of an individual user. This connection has been discussed by Chen (2007), who also remarks on the potential challenges that result when control is taken away from the player in fully automated systems. Chen (2007) suggests a system where manual game difficulty choices are available, but integrated as credible choices in the game world, in order to avoid "breaking the magic circle" of a game (Smeddinck et al., 2016). Such player oriented difficulty adjustments (or embedded difficulty choices) have been shown to increase users perceived autonomy compared to fully automated DDA (cf. section 4.6.1).

It can be argued that flow theory has a narrow focus compared to SDT, and that flow experiences are not the definite goal for every situation and in any game. With motion-based games for health, for example, flow will likely be appreciated during exercise execution, where players focus on performing movements and only small-bandwidth feedback regarding the execution quality can be processed without drawing players out of the immediate gameplay. After a period of motion-execution there is, however, an opportunity to break with the "lulling" (Horkheimer & Adorno, 2002) of immersed flow and to encourage conscious reflection of the self-perceived and objectively achieved performance. In these moments, a comparatively larger-bandwidth and complexity of feedback can reasonably be expected to be conveyable, although this entails the willful breaking of an ongoing flow experience as a desirable element of the overall game experience. While each theory offers room for arguments, both flow and SDT aim to explain states of high motivation that afford task executions and it thus reasonable to consider their potential overlap and compatibility.

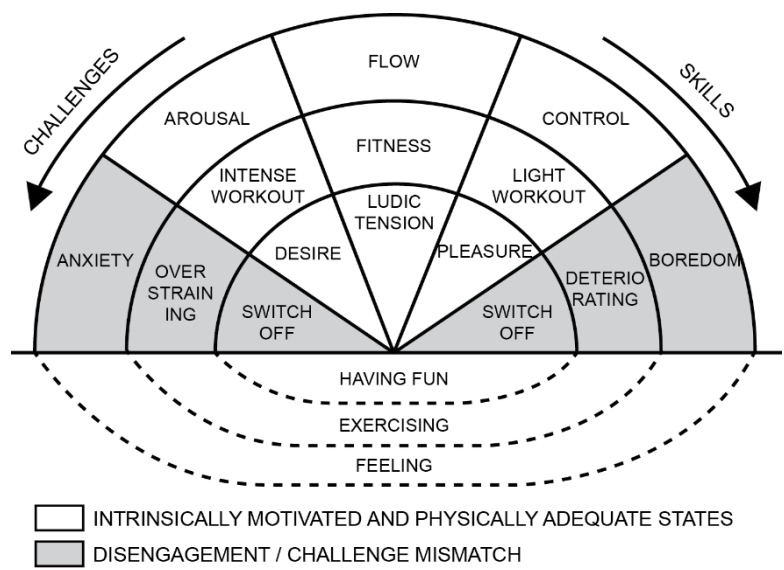

**Figure 15: Intrinsic motivation as a balance between challenges and skills. Adapted from learning games to exergames, following the original model by Denis and Jouvelot (2005).**

Accordingly, flow theory and SDT do recognize each other and research can benefit from the use of both theories. It can be argued that SDT focuses mainly on the basic premises that shape a self-determination motive that enables intrinsic motivation in the first place, while flow theory is mostly concerned with questions around how a momentary situation of very strong intrinsic motivation arises and how it can be maintained (Smeddinck et al., 2016). Denis and Jouvelot (2005) have suggested a model that relies on SDT, however, they establish motivation in the context of games as a balance of challenge and skills in a manner akin to the application of flow theory in the context of games (Chen, 2007). Figure 15 shows an appropriation of that model, which was originally constructed for the application area of games for learning, for the application area of MGH, where aspects around exercising replace aspects of learning. Drawing on the SDT framework of *intrinsic motivation*, which relates to the push to act freely and on one's own volition, *extrinsic motivation*, which relates to factors external to the activity itself, and *amotivation*, which describes the absence of motivation (Deci, 1971), the model that suggests a central corridor of intrinsically motivated states which only arise in a state of balance between challenges and skills. This model can arguably be further augmented to represent dual-flow and to evolve around a generic serious target outcome in the case of other GFH or serious games.

The aforementioned balance between challenge and skills that is central to flow theory is also alluded to in original considerations on cognitive evaluation theory by Deci (Deci, 1975) where it is proposed that factors which enhance the experience of competence (e.g. receiving positive feedback, acquiring new skills or abilities, or being *optimally challenged*) also enhance intrinsic motivation. As Ryan et al. note, perceived competence would thus be enhanced in games with *intuitive controls* that are *readily mastered*, and when tasks in the game provide *ongoing optimal challenges*, as well as opportunities for *positive feedback* (Ryan et al., 2006). Notably, all of these factors are also described by Csikszentmihalyi as enabling factors for flow experiences (Csikszentmihalyi, 1990). In summary, while SDT provides helpful contributing factors that allow for directed reasoning on game experiences and the impact of adaptability and adaptivity while providing a basis for situations in which intrinsic motivation can manifest, flow theory, despite being challenging to measure due to the immersed nature of the state, provides a closer perspective on the temporal aspects of episodes of strong intrinsically motivated immersed activity and on the aspects that are relevant to maintain such involvement. This is a frequent concern of adaptivity, for example with DDA.

## 3.7   Beyond Adjusting Challenge: Human-centered Adaptive MGH

Although DDA is a frequent focus, the applicability of adaptability and adaptivity in MGH is not limited to adjusting the level of challenge. Also, measuring human performance via indirect measures on one or multiple game variables provides a very limited picture of the objective effects that interacting with a system exerts on any given user. Furthermore, while adjustments to difficulty parameters, game mechanics, the game interface, or the audiovisual appearance may have a negative short-term effect on player performance regarding the intended serious outcomes or on player experience, thereby disregarding a local optimum, they can yet arrive at a better state in the medium or long term that might not have been reached if the original local optimum had been pursued. This calls for an approach to the design, implementation, and evaluation of MGH that considers such possibilities and requires research methods that comprise temporal aspects.

As such, adaptive systems require an approach that goes beyond regular iterative user-centered design. While taking users into account as soon as possible for conceptualization and testing remains an important foundation, adaptable and adaptive systems must be considered in their functioning for individual users over time. This poses challenges within the basic approach of user-centered design. In a typical setting of a user-centered application or game development, different test run participants are preferred for development iterations, in order to avoid the possibility that test users adjust to system shortcomings before these can be detected and adequately resolved. Therefore, user-group-centered iterative design might be a more adequate term. It becomes clear how that approach covers developments where a working solution for a fictitious average user of a larger user group is sought and how it has limitations for developments where that best average solution may not be acceptable, which typically drives the decision to include adaptability or adaptivity in the first place. This, of course, implies the insight that GFH, at the beginning of an interaction with a new player, are always optimized for a certain sample group mean, indicating that there is still room (and often the need) to optimize the game further for individual users. Thus, while user-centered design and participatory methods are important, they often leave room for further improvements. Furthermore, due to the feedback and adjustment cycle that is present in adaptive systems they can be seen to resonate with users and environment in more volatile ways than traditional, more rigid interactive systems, which necessarily produces increases in bottom-up changes that are more difficult to predict and control than classic, more top-down design decisions. This motivates considerations on manual control that allow users to retain (at least partial) control of an adaptive system.

These limitations of established approaches are discussed together with considerations that can support a conscious reflection of the underlying causes. This in turn explains the need to expand upon the commonly established methods of user-centered iterative design, which is discussed in the following subsections along six angles that have proven helpful during the developments and studies that are detailed in chapter 4.

### 3.7.1  Three Temporal Classes of Influence on Level of Abilities and Needs

Human users themselves are highly adaptive systems and subject to ongoing change. Hence, with GFH, player abilities and needs of an individual player are not fixed over time, but instead fluctuate constantly. Considering these changes along three general temporal classes can help designers with structuring their approaches (Smeddinck et al., 2014). *Long-term developments*, such as age-related limitations (Gerling et al., 2012), state of fitness, chronic disease, etc., form an underlying base influence on the abilities and needs of an individual. However, they are also influenced by *medium-term trends* such as learning effects, a temporary sickness, environmental factors such as season and weather, and more. Lastly, *short-term influences*, such as the current mood, or potentially forgotten medication, can also play a considerable role. If plotted over time, as illustrated in Figure 16, the three different temporal classes combine to form a complex, non-linear function that suggests an individual channel of acceptable challenges that is much more complex than the simplistic (non-temporal) illustrations of the flow channel that forms under an adequate balance between challenge and skill (see Figure 3) seem to suggest.

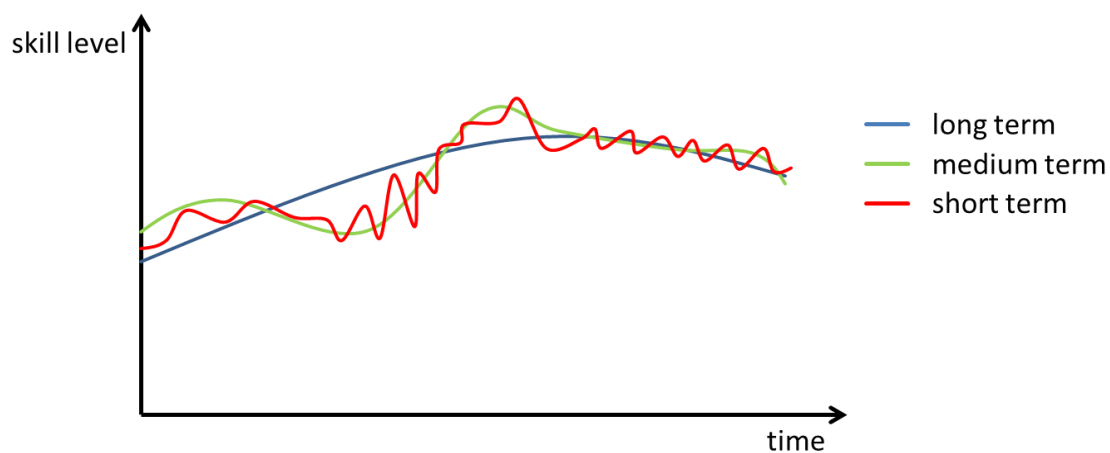

**Figure 16: The change in abilities of a fictitious player over time separated into three temporal classes.**

While it may be adequate for some SG or MGH developments to focus only on long term and/or medium-term trends, in many cases further improvements could result from responding to short term changes as well. This is both a reason to implement adaptability and especially adaptivity, and at the same time poses a challenge by requiring said systems to handle complex developmental patterns.

### 3.7.2  The Source of Change, Explicitness, Autonomy, and Automation

Due to the aforementioned challenges and difficulties that adaptive systems are facing, it is important to consider all sources of information that might benefit the process of determining the optimal system configuration. The original model for dimensions of adaptive systems by Andersson et al. (2009) contains a dimension for the *source* that causes a change and which can be either *internal*, which means inherent to the system, e.g. information resulting from a certain change in application state, or *external*, which means that the cause for change is system exherent, e.g. due to changes in the environment. In the context of MGH, this could mean, for example, that performance analysis can rely on sensing units that are elementary parts of the system, such as a motion-tracking device (representing an instance of a system inherent source for change), or that performance analysis can rely on sensing units that can optionally be linked to the system to provide additional contextual information, such as a wearable fitness tracker, or a thermometer, (representing instances of system exherent sources of information that can cause automatic adjustments). Such sensing can be achieved through a large number of tools, ranging from bio-feedback, over eye tracking, to voice recognition, electroencephalography, interaction logs, and more. An extensive summary of tools and methods for capturing game user experience is provided by Nacke (2009). However, since important aspects that influence the given state of abilities, needs, and last but not least the mood of a given user, cannot always be adequately determined by a fully automated system, the dimension of *explicitness* was added to the extended model presented in section 3.4.2. This accounts for the fact that information which causes a change in the adaptive system can not only take *implicit* forms, such as automatically determined user heart rate, but can also come in *explicit* ways, provided by a user, such as e.g. feedback regarding recent adjustments to difficulty based on a smiley scale. The tools and methods that can be employed in this context thus include a similar set as in the context of the *source* dimension, notably encompassing psychometrics and natural language processing. The *explicitness* dimension is related to the further dimensions of *autonomy* and *automation*, which are both part of the *mechanisms* section of the model (cf. Table 5). However, *explicitness* differs through the focus on information that causes change as opposed to the level of involvement of users in either acknowledging suggested system adjustments (cf. the *autonomy* dimension in the model), or in enacting manually determined adjustments to the system (cf. the *automation* dimension in  the model). Since all three dimensions are related to the level of agency granted to the user which must always be considered in relation to the level of automation, as indicated by Sheridan (2001), and since they form basic considerations regarding human-computer interaction with adaptable and adaptive systems, they are subject to a number of studies that aim to determine the impact that different choices on these modeling dimensions can have on experience and performance, as detailed in section 4.6.1. Besides the relevance due to the entanglement with user agency, the dimensions of explicitness, autonomy, and automation can also be relevant for considerations relating to cold-start problems. Since more explicit systems with more control and adjustment determination being put in the realm of user responsibilities can function well without requiring large amounts of contextual information, such configurations can arguably be employed early on, allowing a system to collect sufficient information, observing manual adjustments, until the levels of explicitness, autonomy, and automation can themselves be adjusted to more optimal configurations, in a form of meta-adaptivity. In a coarse manner, this approach has been used in a number of projects discussed in later sections of this work (cf. sections 4.8 and 4.9), where manual configurations or manually triggered calibration guarantee acceptably adequate settings even before first player interactions take place and before further adjustments are performed via an adaptive system. Notably in the case of these MGH projects, initially used more manual adaptations were executed not by the players but by therapists, as pre-studies had indicated that the exact abilities and needs are not always apparent to the patients, which points at further limitations and important considerations around the aspects of explicitness, autonomy, and automation. The effect of players or users not being fully aware of their objective abilities, needs, and mood, is not exclusive to the application area of MGH, but it is arguably a result of prenoetic (Gallagher, 2006) and imperceptible aspects that are discussed more broadly in their relation to HCI under the umbrella of embodied interaction (Dourish, 2001).

### 3.7.3   The Role of Embodiment in Adaptability and Adaptivity for MGH

Due to the aforementioned dual-task nature of MGH with the simultaneous explicit involvement of body and mind, it is immediately apparent how aspects beyond a purely cognition-focused lens are relevant in models for MGH. Additionally, it was also highlighted above how aspects beyond the immediate psychophysiology of an individual user can influence considerations on adaptability and adaptivity in MGH, GFH, and SG in general. An approach taking such a holistic view, including physiology, cognition, as well as situatedness and the larger usage context, has been discussed by authors such as Paul Dourish (2001), advocating the application of principles form embodied cognition to human-computer interaction. The underlying philosophical method of phenomenology as a general method of inquiry of everyday experience highlights how actions and meaning emerge in specific settings, of a specific physical, social, and cultural nature, and that actors understand actions and meaning through a highly individual lens shaped by prior experiences, activities, and accomplishments (Dourish, 2001). In this light it is also apparent how any static system design is unlikely to optimally accord with the given condition of an embodied user (Bannon, 1990), especially over time, since the condition, following the phenomenological approach, can never be understood as a fixed absolute (Hegel, 1832), as discussed in more practical terms in section 3.7.1 concerning temporal classes of influence on abilities and needs. The broader embodied lens can thus be interpreted to further support the arguments for flexibility through adaptability and adaptivity which can be employed to facilitate that systems correspond better to dynamically changing conditions. Furthermore, it can be argued that the dual-flow perspective, which was discussed and generalized to SG in section 2.5, could be further extended to encompass the context of situated use in addition to cognitive and physical aspects. However, the perspective of embodiment also leads to potentially cautioning considerations regarding adaptability and adaptivity. The theory acknowledges the limited capacity for objective perception, awareness, and conscious processing of a given situation by any individual, which limits the applicability of adaptability and ties in well with the considerations on the different levels of involvement of different parties of stakeholders in the process of adjusting MGH that are discussed in this thesis and will be detailed in the following section. Looking at the suitability of adaptivity for (partially or fully) automating adjustments in GFH, which become part of the immediate surroundings of a given user, for achieving changes that are factually beneficial to the user, the questions in how far the users perceive those changes and are - or become aware - of them, as well as the question in how far they experience their potential to control or influence such adjustments, plays a crucial role, which was discussed in practical terms along the dimensions of autonomy, automation, and explicitness in the prior sections. This is of particular importance in the context of games, since perceived personal causation and control (Burger, 2013; DeCharms, 1968) are important elements of motivation that have notably been adopted by self-determination theory through the needs dimensions of autonomy and competence. Arguably, these considerations also - at least implicitly - appear as a central motivation to Sheridan's considerations on the scale of automation (Sheridan, 2001), who cautions that automation, such as adaptive behavior often interferes directly with the user's current goals and delicate preventions must be made to prevent undesirable exploration through changing behaviors or configurations in situations where these adjustments would interfere with user goals in a negative manner. The practical relevance of considerations regarding player perceptions in relation to difficulty adjustments in games has been discussed in related work (Missura & Gärtner, 2009) and marks a potential challenge to adaptive systems in MGH, since adjustments can make it more difficult to anticipate further expectations of difficulty that exist in players, which can manifest as a form of breaking consistency that is not only cognitively challenging (Hawkins & Blakeslee, 2004), but also endangers usability / playability and the further experience and potential serious outcomes that depend on it (cf. section 3.5). This gave reason to the decision to largely focus on parameter tuning as opposed to larger manipulations, such as dynamic feature introduction or removal, in the approaches to adaptivity that were employed in the practical works that are detailed in chapter 4.

In order to reflect these concerns during system design or analysis, it can be helpful to consider player abilities as they are shaped by the individual embodiment as dynamic and potentially overstrainable resources in game design. As discussed in further detail in the respective underlying publication **[see publication C.3],** players may have limited attention spans, abilities, or special requirements that can be seen as a depletable resource in game design (Gerling et al., 2012). This lens is often helpful when planning interactions around the serious target of the game. Figure 17 presents an according extension of a popular game design model.

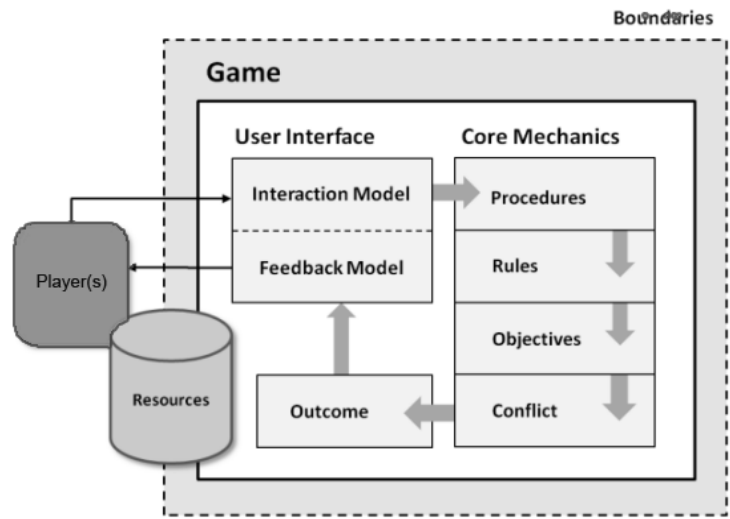

**Figure 17: An extended model of digital games after Fullerton et al. (2004) and Adams (2010); from (Gerling et al., 2012).**

In the context of MGH, it can be similarly helpful to consider the extent and limitations of abilities and capacities of the remaining involved parties, such as therapists, as they function in complex situated settings (e.g. consider social interactions in a therapy practice) with environmental constraints (e.g. limited time and space for treatments; cf. sections 4.5 and 4.9). The game design model illustrated in Figure 17 can thus be extended with a surrounding interaction loop where players and the connected interactive game application additionally interact with third parties whose resources also stand on the boundary of game design aspects and should thus be considered.

Next to providing an underlying framework which ties together the theoretical foundations that were discussed in the prior sections, further practical results of these considerations rooted in embodied interaction are: (a) the approach of human-centric, "embodied" parametrization that puts a focus on adjustments based on human abilities and needs instead of a focus on particular game system variables and is discussed in section 3.7.6, and (b) the more explicit considerations on social context and situatedness that are discussed in the following section.

### 3.7.4   Social Context and Situatedness

While video games are at times played in comparable isolation, the larger societal context still influences even the most isolated interactions. Modern game use with social and multiplayer capabilities in general, and MGH use given the complex constellation of interested parties in particular, occur in constant interplay with the broader situated context, including sociocultural elements. This aspect is discussed in further detail in the related underlying publication [**see publication B.11, section 4.2**], including the notion that adaptive functions may be employed with regard to the interests of the involved heterogeneous parties as well (Smeddinck, Herrlich et al., 2012). Personalized reports based of GFH performance, for example, may be tailored for parents of a child with cerebral palsy who regularly uses a GFH. GFH or their subsystems should also be designed with secondary developers and researchers in mind. As discussed in section 4.10.1, additional challenges arise when GFH feature a multiplayer mode, since adaptivity can have a strong impact on the balancing and the perception of the balancing by the players (Gerling et al., 2014). It is also important to keep in mind that non-competitive multiplayer settings will likely result in different patterns of technique acceptance and resulting game experience (Vicencio-Moreira et al., 2014) and adaptive techniques may, for example, be less intrusive when the player roles are not symmetrical (cf. section 4.10.1).

Considerations on such aspects of social context and situated use can lead to helpful insights regarding practical approaches to MGH design, as the case of multiple stakeholders and their potential involvement in the design and development process of MGH illustrates. Additional stakeholder groups, such as therapists in the case of MGH, cannot only be seen as target users of MGH, which is an important perspective that has been discussed in related work (Ines & Gouaïch, 2010), but they can also be actively involved in the design process of MGH [**see publication C.4**]. Following the application use case of MGH for older adults, non-gamer experts, such as doctors, therapists, and nurses, can provide valuable formative contributions with their insights into common age-related changes and impairments, giving recommendations that help designers implement beneficial movement-based game input, and they can also support game developers throughout evaluation processes. During the development, gaming experts, such as interaction and game designers, can make valuable contributions by providing insights regarding enjoyable game mechanics, appropriate interaction schemes, etc. Based on these theoretical and practical considerations, it is clear that approaches to adaptability and adaptivity in MGH can benefit from a human-centered perspective that understands humans not as isolated individuals, but as complex embodied social beings.

### 3.7.5 Lenses for Tackling Multidisciplinary Complexity

The complexity and extent of the considerations that have been discussed so far indicate that it may be helpful to employ methods for breaking down the tasks around designing, implementing, or evaluating adaptable and adaptive MGH. As noted before, the field of GFH is highly interdisciplinary. Contributions and interest stem from many fields of work with numerous respective sub-fields, including *research* (human-computer interaction, game user research, game studies, psychology, medicine, health sciences, etc.), *engineering* (graphics engines, networking, multi-sensor devices, etc.), *design* (game design, game asset design, user interface and UX design, sound design, etc.), *health* (therapy, nursing, pharmacology, etc.), and potentially others. Hence, if professionals from such a wide variety of disciplines are involved in a GFH project, their interests and needs should be considered in addition to the aforementioned heterogeneous target groups, etc. (cf. section 2.1.5). Furthermore, even co-players or bystanders who are not directly involved with using the GFH for any serious purpose may play a role (cf. section 4.5). In this light, assuming different lenses of reflection related to different interdisciplinary vantage points can help to avoid neglecting important angles (Kessel et al., 2008).

Employing a set of lenses to methodologically consider a complex subject matter from different angles in order to improve overall insight has been suggested in related work, e.g. by Schell (2008), as a viable option for tackling the complexities around game design or more generally processes that require lateral thinking (Bono, 1999). There are many ways to employ the concept of lenses. While some concepts for lenses, such as the thinking hats model (Bono, 1999) employ very general categories that can be applied to any application area, others, such as Schell's lenses for game design (Schell, 2008), are specifically tailored towards aspects of an application scenario. However, in either case, the basic motivation is to avoid getting caught up in a narrow view, which can be helpful with any project but gains importance with growing project complexity. This has also been noted by Deterding (2016) with regard to facing the challenges around designing complex gamification while assuring that the desired ends are met. Often, systems must be "tuned" as they function within a complex setting. This can greatly benefit from the design concept of lenses (for apt analysis of the status-quo). Accordingly, the application of lenses can also support the improvement of subsystems for adaptability and adaptivity as they are observed in action. Since the nature of modern game systems, especially for complex use-cases such as GFH, is arguably typically more software-as-a-service than single-point release products, monitoring and improvements can continue beyond initial release. This illustrates a secondary practical relevance of adaptivity, since it can be seen as a continuous process of iterative user-centered adaptation, and it shows the potential benefit of employing lenses to gain a well-rounded view on the impacts of a functioning system in order to allow for potential larger adjustments, such as adding or removing game features, as well as for meta-adjustments to any systems for adaptability and adaptivity, or mixtures of both.

In the studies and developments that will be discussed in chapter 4 and sub-sections, the lenses of (game user) *researcher* vs. (game) *designer* vs. (game) *engineer* were taken consciously during design, implementation and evaluation phases of the different individual projects, helping with highlighting different challenges and aspects, and leading to insights on what might otherwise have been missed. Each lens typically contributes both towards extending possibilities for further steps in the development process of an MGH, as well as limitations. Other lenses, such as those of the patients or the therapists (i.e. *health*) in the case of MGH are equally important to assume. However, while certain methods, such as working with fictitious persona, as well as the involvement of domain experts can help gain some insight regarding the requirements and project perception of such groups, there often is no way around involving the aforementioned groups directly in order to gain an adequate view through their lens. Applying the concept of lenses can support projects either way with identifying the required points of view and with taking them into account throughout the project development. Arguably, acclaimed game design systems, such as the MDA framework by Hunicke et al. (2004), with its focus on breaking considerations of games into three elements, namely *mechanics*, *dynamics*, and *aesthetics*, can also be thought of as a system of lenses, with each component offering a separate but casually linked lens (or view) of the game (Hunicke et al., 2004). In the context of MGH, it was found in the process of the project detailed in section 4.5 that MDA supports the thinking about a specific game design process very well, offering a form of more detailed modularization of the (game) *designer* lens suggested above, while the overall space of MGH in situated use did call for further and more wide-reaching points of view (thus the addition of the self-employed lenses of game user *researcher* and game *engineer*, as well as the lens fulfilled through third parties regarding *health* aspects, opportunities, and requirements). Furthermore, the application of lenses also shaped many elements of these theory sections, as for example the following considerations on human-centric parameterization are a direct consequence of including the therapist perspective.

### 3.7.6   Human-Centric Parameterization

In some cases of GFH, the mechanics and dynamics of the game suggest a straightforward selection of parameters to be subject to adjustments through adaptability and adaptivity. In the case of a voice treatment game (see Figure 18) that was developed to support therapy for people with Parkinson's disease (Krause et al., 2013) and is included as a related element in the contributions of this thesis **[see publication C.8]**, for example, patients are simply tasked to sustain a certain level of voice loudness over a given period of time. The according parameters of adjustment either through manual settings, or through adaptivity would thus be loudness and duration, which make for a small parameter space that can easily be controlled and tested with thresholds, and that can easily be understood by users or therapists. They simultaneously represent technical game variables and aspects of user embodiment (i.e. the loudness they can achieve and the duration they can sustain that level of loudness).

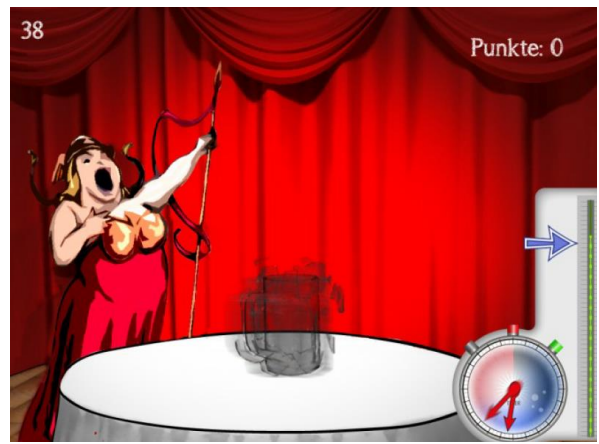

**Figure 18: A screenshot from a game developed to support voice treatment for people with PD. The character on the left changes her facial expression and the glass in the center begins to resonate if a certain loudness threshold (captured via a microphone) is reached. It the loudness level is sustained over a predefined amount of time, the glass will break.**

However, this is not always the case. As the description of the project *Spiel Dich fit* (SDF) (cf. section 4.5) and the related studies (cf. section 4.9) will show in further detail, the technical parameters of a MGH are often not easily understood by related parties that are not immediately involved in game design and development. In the case of an apple picking game, for example, actual in-game parameters such as "apples per minute", or "spread of apples on screen", would not be telling to therapists who are tasked with adjusting settings for their patients. In the context of the development of dynamic difficulty adjustment for the game *Star Money* (Smeddinck, Siegel, et al., 2013) parameters for adjustment were thus expressed through elements of the players' physicality (or embodiment), namely as *reach*, *accuracy*, and *range of motion*, which were easily understood and meaningful to the involved therapists. This concept was later expanded during the SDF project which also included a manual settings interface (cf. section 4.5). In this case, explorations on the evaluative criteria and the characteristics that therapists relate to when discussing patient abilities and needs with regards to their traditional work formed the basis for isolating an extended set of patient-centric parameters. Notably, such a parameter mapping also allows developers and designers of MGH to hide the complexity of a potentially much larger number of in-game settings through mapping them to a reduced and more manageable number of exposed settings parameters without abstracting to an overly generalized single difficulty parameter (e.g. "easy" to "medium" and "hard"), which is also not expressive of the players' embodiment. This human-centric (as opposed to system-centric) approach to adaptability and adaptivity was thus a result of participatory involvement of therapists in the aforementioned projects (including their view, or lens on these MGH) and is in line with the human-centered approach towards design that was famously advocated by Norman and others (Bannon, 1991; Norman & Draper, 1986) and has been discussed in the parent section. In cases where such "don't make me think" (Krug, 2005) parameters are employed in order to improve acceptance and practical use amongst therapists and patients, although they are not directly representative of actual in-game variables, it falls to automation to bridge the gap (Sheridan, 2001) by offering translations between the human-centric "embodied" settings parameters and the actual in-game parameters. Here, the modeling dimensions, especially from the *effects* category (cf. Table 5) can serve to guide design, e.g. by emphasizing considerations on the *breadth* of the parameter mappings (one to one, or one to many, etc.), the *intrusiveness*, or the *saliency*. In a general sense, the mapping can be either *game-specific*, in which case different sets of equally named parameter settings are provided for different MGH, or *player-specific*, in which case the parameters are provided only once per player, describing their embodied abilities and needs directly, while also requiring an additional layer of translation to that generalized parameter representation. For the case of the initial exploration in the direction of adaptive MGH for people with PD (Smeddinck, Siegel, et al., 2013), as well as the later MGH from the SDF suite (Smeddinck, Herrlich, et al., 2015) the human-centric parameters were provided per game in order to avoid the complexity of translating to a generic level of abstraction, which is now being approached in the follow-up project *Adaptify*[16]. In both prior projects, the mapping functions as well as variable boundaries were determined by iterative testing with settings parameters being normalized. Notably, both the mapping functions and the in-game variable boundaries can be subject to adaptivity. Employing a fully generic once-per-user parameter set can arguably allow DDA systems to scale beyond specific MGH in order to facilitate cross use case re-usability of centralized player models and systems for adaptability and adaptivity, once an adequate user model and parameter mapping has been established.

## 3.8   Needs and Abilities Based Human-Centered Design for Adaptive Systems

In bringing together aspects of embodiment with the related approaches from human-centric design to suggest human-centric parameterization that is informed through a combined view of different lenses and draws on the modeling dimensions for adaptive systems to inform implementations, the prior section indicates how the aforementioned aspects come together to form an overarching approach. This section further integrates the prior elements discussed in sections 3.3 to 3.7.6 under the umbrella term of *Needs and Abilities Based Human-Centered Design for Adaptive Systems*, forming a general process for realizing adaptability and adaptivity for serious games. The process model is described in further detail in the underlying publication [**see publication B.11, section 3.5**]. In brief, the approach assumes that making a system, such as an MGH, adaptive is a specific form of automation. It thus builds on the process model for designing and implementing automation systems that was presented by Parasuraman, Sheridan, et al. (2000), introducing additional considerations, as well as design and development goals, that are relevant for adaptable and adaptive serious games. Figure 19 shows the adjusted version, with original parts of the model being shown in blue, and additions or replacements being shown in orange color.

---

[16] https://www.adaptify.de/, last viewed 2016-10-09

In summary, the extended model begins with the question of *"Which aspects of the serious game should be adaptive?"*, which in the case of serious games can range from general difficulty settings, over specific challenging aspects, such as the required range of motion in a game for therapy, up to very complex aspects, such assemblies of learning materials. The following step *"Identify adjustable parameters at steps of automation"* is split into four types in this state of the model, all of which are frequently present in parallel in adaptive serious games, highlighting the complexity of these types of adaptive systems. They relate to the categorical structure of the modeling dimensions for adaptable and adaptive SG as discussed in section 3.4.2 and follow the basic classes of modules for performance evaluation and an adjustment mechanism introduced by Fullerton et al. (2004). Accordingly, the following step suggests conscious decisions on the level of automation, and all other modeling dimensions to determine the exact implementation of the adaptability and adaptivity. See Table 5 for a summary of the levels of automation as suggested for the context of adaptivity and adaptability for SG, GFH, or MGH. Notably, some aspects of these steps are usually fixed or bounded with thresholds in the design phase of a game while some flexibility remains which then gives room to adaptive change during use. Subsequently, the model suggests an evaluation based on criteria as discussed in section 3.5, potentially including UB, UX, PB, PX and measures regarding the intended serious outcomes. Depending on the evaluation results, an initial selection of types and levels of automation is made which is then iterated and eventually evaluated for secondary criteria, such as reliability, outcome-related costs, and ecological validity, leading to a final selection of types and levels of automation. In this light, it also becomes clear how this model is a detailed realization of a general iterative design model, as for instance presented by Hartson & Pyla (2012).

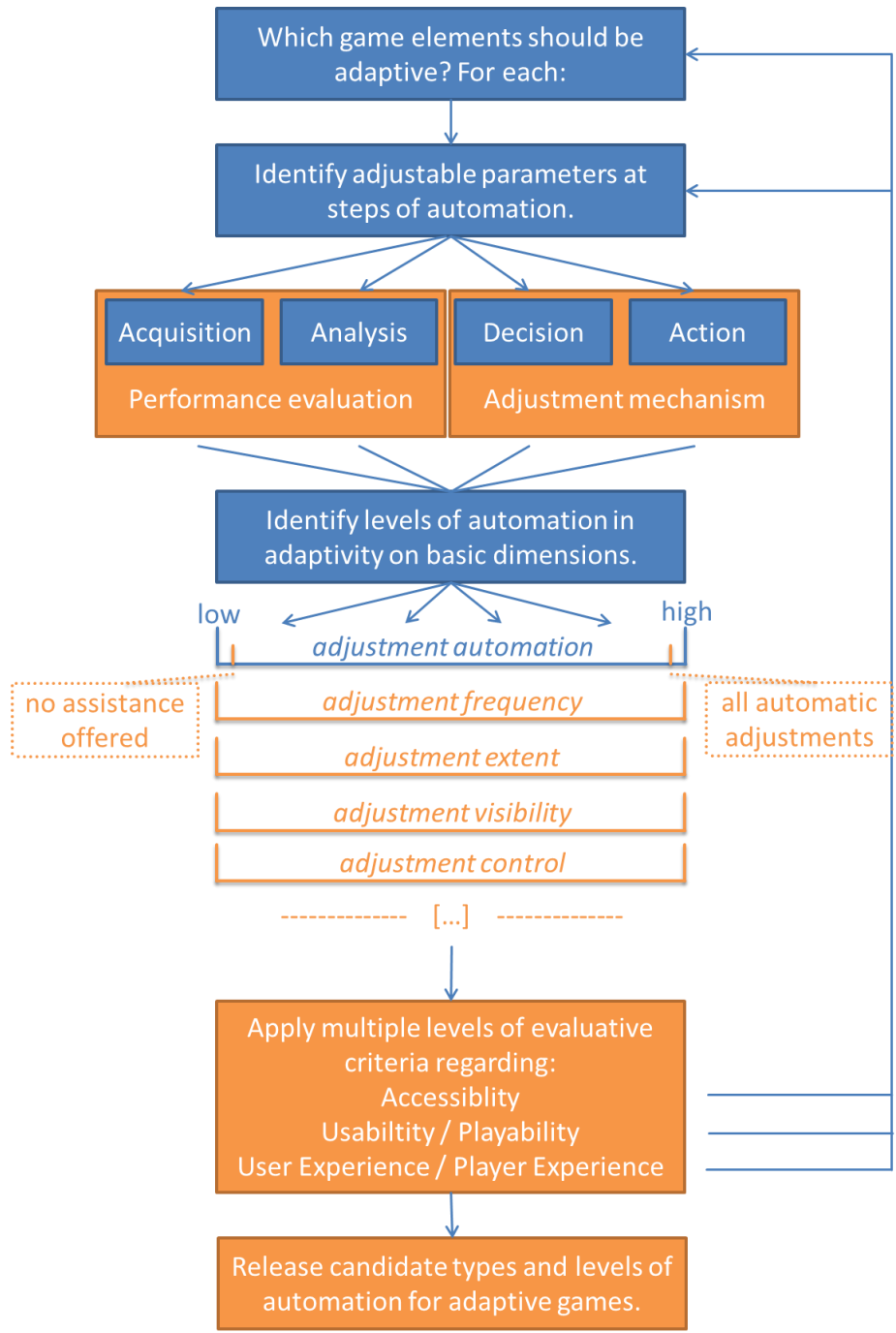

**Figure 19: Levels of adaptation as well as a design and development workflow for adaptive systems in the context of serious games.**

This process, together with the modeling dimensions (cf. section 3.4.2), shows that any adaptive system must take a carefully designed position or subspace in a very complex, high-dimensional design space. Additional complications are added with complex application areas such as serious games. This has implications not only for the design, implementation, and testing of adaptive systems, but also for any research on them, since these aspects can influence the outcomes of studies on the functioning or the acceptance of adaptive systems.

The process described above runs orthogonally to the modularization suggested in section 3.3 and typically interacts with considerations regarding all modules, but is meant to be employed when planning manual adaptability and especially automatic adaptivity. Taken together, considerations on the *modeling dimensions* (cf. section 3.4.2), as well as on *human-centric parametrization* (cf. section 3.7.6) are represented through the elements of *performance evaluation* and *adjustment mechanism* together with the following step of *identifying levels of automation*. *Accessibility, playability, player experience, and effectiveness* (cf. section 3.5) guide the *evaluative criteria,* and *motivational theories* (cf. section 3.6) shape the specific *evaluative endpoints,* together with *performance aspects,* or with the *impact* with respect to the *desired serious outcome.* The sub-sections of section 3.7 again apply orthogonally to each step of the process, with the *temporal classes of influence on levels of abilities and needs* (cf. section 3.7.1), for example, driving considerations on which aspects to subject to adaptivity, as well as how to approach the implementation and parameterization of adaptivity and how to evaluate the system functioning. Considerations on the *source of information selected to cause change* (cf. section 3.7.2), on the *embodiment of target users and the involved third parties* (cf. section 3.7.3), including *social context and situated use* (cf. section 3.7.4) similarly are of relevance to all steps of the process, while the technique of *applying lenses* (cf. section 3.7.5) can help tackling the complexity of keeping the aforementioned aspects in check. Next to the central aspect of modeling (player) *abilities,* which can be expressed through *human-centric parameterization,* the focus on third party and (especially) player *needs,* including the application of SDT as a needs satisfaction theory, is rooted in the unique applicability of SDT to the use-case of MGH. Based on an initial design by principle and constraints, iterative evaluations assure adequate outcomes regarding *emotional* and *physical well-being* (cf. Figure 19). These two factors also inform endpoints both for evaluations before a release candidate is produced, and after a first public release (not pictured), potentially leading to an additional feedback loop, driving meta-adjustments to the choice of adjustable elements and the realization of adaptability and adaptivity for these elements. SDT functions well for simultaneous considerations of emotional and physical well-being, as evidenced by successful applications in related work (see section 2.5). Arguably, *competence, autonomy,* and *relatedness* are needs where the needs satisfaction benefits from a healthy interplay between *cognition* and the *physical body,* showing a firm link not only to the concept of *dual-flow,* as discussed in section 3.6, but also to the foundations of *embodiment* (cf. section 3.7.3). *Autonomy* and *competence* as factors of cognitive evaluation theory appear to be especially closely linked to the design and situated functioning of adaptive systems. Elements of manual involvement can increase autonomy, while increased automation threatens to decrease autonomy (cf. section 3.7.2). Similarly, competence can be influenced by automation, but arguably not in a similarly clearly directed manner. More automation can facilitate greater self-perceived achievements, although this arguably requires a "feeling in control" of the user, whereas a high degree of automation may also hinder a feeling of competence, if the user does not feel "in control" and any achievements that are made are accredited to the automation (or the machine). This can relate not only to the design and engineering of a system (whether feedback is taken / given and whether the user is granted direct or indirect control) but also to its presentation (framing, appearance / disguise either as a powerful tool or as an uncontrollable independent entity), which is also the direction from which a system is initially perceived and understood by its users, as discussed by Hunicke et al. (2004) under their term *aesthetics*. In this way, the basic needs of SDT can be related to central considerations on human-computer interaction with adaptable and adaptive MGH, informing research questions and hypotheses, as well as providing evaluative criteria, as the discussions of the practical projects in sections 4.6.1 and 4.9 show. In summary, the process model suggests to realize adaptability and adaptivity in a way that respects user and third party needs and abilities as they fluctuate over time, embracing situated use with environmental and social factors, and assuring rather holistic assessments of the projected and factual outcomes both during initial design and development, as well as during studies and in public projects.

## 3.9   Adapting to Further Contexts and Devices

It is important to note that even the broad angle that was provided in these theory sections is subject to a number of deliberate limitations resulting from the practical developments and studies discussed in the following chapter that led to the exclusion of potentially important elements that are left to future work. For example, the need to adequately serve potential multiplayer situations, or the context of gameplay, i.e., where it is played, how much space is needed, etc. In order to control for such impacts the possible target environments can be modeled and simulated beforehand. However, this is still an area in need of further research (cf. section 6). The same applies for additional sensor or control devices, which again hints at the complexity of adapting adequately in real gaming situations, even when comparatively simple heuristics are used. These challenges support the need for medium- to long-term studies and for studies that make ecological validity a primary target. They also further underline why commercial motion-based games cannot simply be used for most serious health applications.

## 3.10 Summary of Contributions in Structuring Theoretical Considerations

Building on the background and related work, as well as on outcomes and findings of developments and studies discussed later in this thesis, this chapter introduced and discussed a number of structural theoretical considerations that augment existing models and processes for adaptability and adaptivity in MGH. The elements discussed in this chapter primarily relate to the guiding research question Q1, since they focus on design aspects of adaptable and adaptive MGH. However, the impact on - or relationship to - acceptance and effectiveness have also been discussed. Table 6 presents a summary of the contributions in relation to the referenced related work as well as in relation to the studies and developments discussed in chapter 4.

**Table 6: A summary of the contributions across the different topics of this chapter in relation to specifically related work (e.g. underlying work that was augmented), and to the studies and developments.**

| Topic | Specifically related work | Contributions | Relation to studies and dev. |
|---|---|---|---|
| *Three central promises* | Was found to be related to: (Oinas-Kukkonen & Harjumaa, 2009); other related work had highlighted the promises separately. | Isolation of the three areas (motivation, guidance, analysis) and illustration with examples. | Formed based on experiences during dev. of the SDF project; informed more recent developments with more focus on guidance and analysis. |
| *Approaches to Adaptability* | No direct predecessor, underlying, or related work for the context of GFH / MGH. ( - ) | Concept of game centric vs. player centric parameteriziation; limitations of manual adaptability | Informed by the acceptance study; developed and employed for interface developments in SDF. |
| *Approaches to Adaptivity* | Concept of perf. eval. + adjust. mech. (Adams, 2010); rubberbanding (Pagulayan et al., 2012) | Relation and summary of existing concepts. | Employed in development of adaptivity for acceptance study, SDF, and effectiveness study. |
| *Cold-Start Problem* | General work from ML / AI / automation; no direct references from GFH context. | Suggested approaches for the challenge in the context of MGH. | Informed the approach to manual involvement with semi-automatic adjustments in SDF. |
| *Challenge of Co-Adaptation* | General work from ML / AI / automation; no direct references from GFH context. | Notion of the challenge with regard to adapt. GFH and relation to rubberbanding. | Informed the approach to manual involvement with semi-automatic adjustments in SDF. |
| *Modular Development* | General related work on adaptive games as discussed in section 2.8 [e.g. (Göbel et al., 2010)]. | Summary of common components for adaptable and adaptive GFH; structure: areas / modules. | Informed by the acceptance study; augmented during SDF development; employed for planning of Adaptify. |

| Topic | Specifically related work | Contributions | Relation to studies and dev. |
|---|---|---|---|
| *Dimensions of Adaptivity* | Levels of automation (Parasuraman et al., 2000; Sheridan, 2001); components of DDA (Baldwin et al., 2013); modeling dimensions for adapt. Systems (Andersson et al., 2009) | Relation of existing models to application scenario of adaptable and adaptive GFH / MGH; augmentation of modeling dimensions with further elements. | Informed by SDF; employed in 4.6.1 to clarify differences between two similar adaptive systems; in 4.10.1 to clarify differences between three balancing modes. |
| *Accessibility, Playability, Player Experience, and Effectiveness* | Dual-flow (Sinclair et al., 2009); considerations beyond UB in GFH (Alankus et al. 2010) | Summary and relation of prior work on evaluation criteria for GFH / MGH; fusion to model with UB, UX, PB, and PX. | Employed for planning of human-centered iterative design strategy in SDF. |
| *Adaptability, Adaptivity and the Motivational Pull of Video Games* | Flow in games (Csikszentmihalyi, 1990; Chen 2007); flow as balance between challenges and skills (Denis & Jouvelot, 2005); SDT in games (Ryan et al., 2006) | Relation of the existing theories of flow, dual-flow via the bridge of balance between skills and challenges in the context of adaptable and adaptive GFH and MGH. | Informed by multiple studies on motivation in MGH; informed study on presenting game difficulty choices. |
| *Three Temporal Classes of Influence on Level of Abilities and Needs* | General work from health, therapy, and sports science; no direct references from GFH context. | Introduction of three different temporal classes and discussion of the emerging complex non-linear progressions. | Informed by acceptance study and SDF; motivated milestone definitions in SDF, and remarks on future work. |
| *Source of Information Causing Change, Explicitness, Autonomy, and Automation* | No direct predecessor, underlying, or related work for the context of GFH / MGH. ( - ) | A discussion of the modeling dimensions for adapt. GFH that relate most directly to aspects of HCI. | Informed by modeling dimensions; informed approaches to studies exploring HCI related dims. (4.6.1, 4.10.1). |
| *Role of Embodiment in Adaptability and Adaptivity for MGH* | Interaction design and embodiment (Dourish, 2001); gamedesign (Adams, 2010; Fullerton et al., 2004) | Appropriation of concepts from embodiment for MGH, tied to dual-flow serious targets; augmented game design model with according player resources. | Informed by prior theory; motivated approaches to development and studies in the SDF project. |
| *Social Context and Situatedness* | No direct predecessor, underlying, or related work for the context of GFH / MGH. ( - ) | Consideration of different points for contribution of different invested groups. | Informed by prior theory and early studies; motivated approaches to development and studies in the SDF project. |
| *Lenses for Tackling Multidisciplinary Complexity* | Game design lenses (Schell, 2008); lateral thinking (Bono, 1999); MDA framework (Hunicke et al., 2004) | Appropriation of the concept of lenses for the design of GFH and MGH; relation to the MDA framework. | Formed during development of the SDF project; informed the development of all later projects. |
| *Human-Centric Parameterization* | General related work on human-centered iterative design [e.g. (Bannon, 1991; Norman & Draper, 1986)] | Introduced the concept of human-centered parameterization for relatable and efficient adjustments in MGH. | Formed during development of the SDF project; informed the study on movement cap. configurations (4.6.2). |
| *Needs and Abilities Based Human-Centered Design for Adaptive Systems* | General process model for automation (Parasuraman et al., 2000) | Application of the prior elements to human-centered iterative design of adaptable and adaptive SG / GFH / MGH. | Formed during development of the SDF project; informed the development of all later projects, esp. Adaptify. |

While the relation of the structuring theoretical considerations to the developments and studies has been indicated in brief in the rightmost column in Table 6, the human-centered iterative development of the SDF project serves as an *illustrative case study* of the aforementioned aspects in application with a larger MGH project, as summarized in section 4.5.

# 4  Studies and Developments

The following sections summarize the publications that form the basis of this thesis work and those which closely support it. For each of the projects, a research rationale is provided together with a discussion, while the full details can be found in the respective publications (cf. Publications A - C). The publications are connected to the rationale of the larger theoretical and empirical context of this thesis and linked to the underlying research agenda.

## 4.1  Research Agenda

Based on the finding of a strong need for personalization and flexible and efficient adjustments in MGH that was one of the main outcomes of the *WuppDi* project (see Figure 20) at the University of Bremen (Assad et al., 2011), but had also been concluded in related work (Alankus et al., 2010; Göbel et al., 2010), this thesis set out to study human-computer interaction with respect to systems for adaptability and adaptivity in MGH (cf. chapter 1).

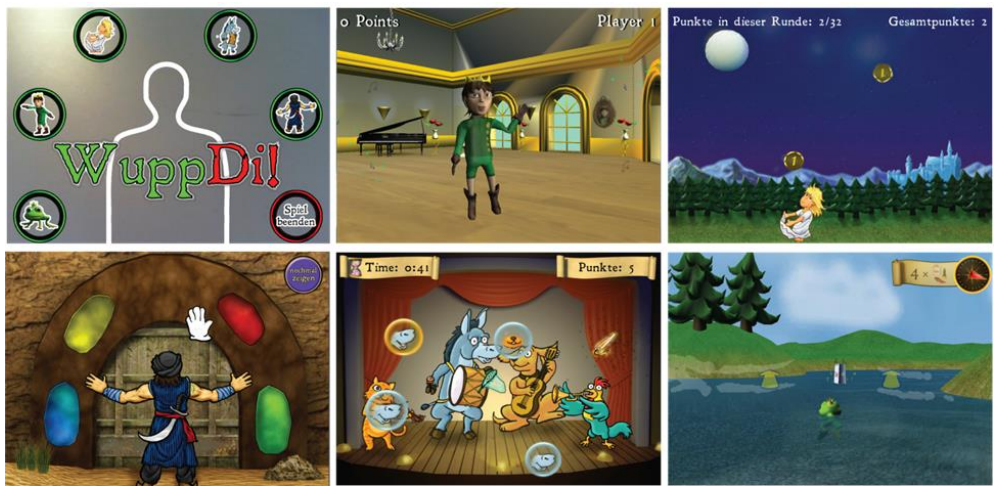

**Figure 20: A series of screenshots summarizing the WuppDi suite of games for the support of physiotherapy for people with Parkinson's disease, showing a game selection screen (upper left) and screenshots from a range of motion-based games set in the theme of a fairy tale world.**

In line with the tenets of playability and player experience, the research agenda evolved around the following central general research question: *How can adaptability and adaptivity in MGH be realized in an efficient, effective, and enjoyable manner?*

Due to the large number of challenges involved with implementing and studying MGH, this general question was approached along three sub-questions (Q1 - Q3), as described in the Introduction, which in turn motivated specific research hypotheses that are described in the individual projects that are summarized in the following sub-sections. Table 7 provides an overview of the contributing area and of the secondary contributions of each study project following the structure of the three sub-questions. The contributions with regard to the larger context of this thesis will also be summarized in the discussion section 5.1.

**Table 7: An overview of the areas of primary and secondary contributions of the study projects presented in the following sections to the guiding research questions motivated by the general research focus on human-computer interaction with adaptable and adaptive motion-based games for health.**

| Research Project | Q1: Design | Q2: Acceptance | Q3: Effectiveness | Approach / Rationale |
|---|---|---|---|---|
| *Perception of Game Difficulty ... [4.3]* | **primary** (formative) | *secondary (formative)* | - | Questionnaires regarding the general stance on (dynamic) difficulty in games. |
| *Motivation with Gamification Interf. [4.4]* | **primary** | - | - | Study of a gamified training information interface. |
| *Human-Centered Iterative MGH ... [4.5]* | **primary** (illustrative cs) | *secondary (illustrative cs)* | - | Illustrative case study (cs) MGH RnD project. |
| ***Presenting Game Difficulty Choices [4.6.1]*** | **primary** | *secondary* | - | Main study of impact of diff. choices on game experience. |
| *Movement Capability Configurations ... [4.6.2]* | **primary** | *secondary* | *secondary* | Research on configuration interfaces for therapists. |
| *Visual Complexity [4.7.1]* | **primary** | *secondary* | *secondary* | Impact of visual detail on PX; potential target for adapt.. |
| *Exercise Instruction Modalities [4.7.2]* | **primary** | *secondary* | *secondary* | Performance and PX with virtual instructor figures. |
| ***Acceptance of Adaptive MGH [4.8]*** | *secondary (expl. case study)* | **primary** *(expl. case study)* | *secondary (expl. case study)* | Case study exploration with three people with PD. |
| ***Prolonged Use and Functional Impact [4.9]*** | *secondary* | *secondary* | **primary** | Explore acceptance and effectiveness over mult. sess.. |

A selection of additional studies and developments that have begun exploring aspects beyond comparatively simple adaptability and heuristic adaptivity are discussed in section 4.10.

In line with the general research question and the three guiding research questions, the work in this thesis explicitly focuses the aspects of human-computer interaction around the topic of adaptability and adaptivity in MGH, as opposed to the development of specific methods for adjustments that is a common focus in the majority of related work.

## 4.2   Research Methods

The nature of the research questions, which are explicitly human-centered, requires the involvement of human subjects in most stages of the research. User studies in the context of adaptability and adaptivity for GFH present a challenging matter that can benefit from a careful consideration of the available research methods. Reliable studies on GFH are difficult to set up, since the environments are either hard to control or artificial, since acquiring participants can be costly, especially adaptability and adaptivity in GFH or MGH can rarely be studied in single sessions and face regulatory challenges. Validated research tools are also often not readily available and the tools that exist are frequently challenging to employ due to potentially conflicting subscales and tradeoffs, for example regarding the importance of efficiency and effectiveness. Overly verbose research methods may feel intrusive to the players, interfering with positive game experiences and also with health-related outcomes. While, some validated game user research questionnaires have been successfully used in recent GFH studies, such as the player experience of needs satisfaction (PENS) questionnaire (Rigby & Ryan, 2007), personality or player types also play an important role (Orji et al., 2013), and health related endpoints (typically dependent variables in studies) are usually most important in clinical terms. Identifying adequate endpoints requires domain knowledge for the application use case. In this light, the careful piloting of studies, expert advice in selecting measurements and target endpoints, as well as a limitation in the implemented features of a GFH and in the number of study conditions, can be strategies to approach the aforementioned challenges. These aspects were discussed in the underlying summative publication [**see publication B.3, section 5.7**].

In practice, for most of the projects that will be discussed in the following sections, mixed research methods were employed to allow for a triangulation of any hypothesized effects. This approach was motivated by the challenging interdisciplinarity of MGH and by the intent to support considerations from different lenses (cf. section 3.7.5). Next to validated psychometric questionnaires, additional custom questionnaire items were sometimes included in order to address more specific research questions, either because available validated questionnaires did not match the application scenario, or for reasons of brevity. Especially in within-subjects comparative study setups, the time required per trial increases considerably with each additional measure that requires manual involvement of the participants. Pre- and post-treatment questionnaires were frequently accompanied by demographic questionnaires and by performance measurements that were either based on in-game metrics, or captured by additional sensor hardware. In order to avoid overlooking unanticipated aspects, the studies were often informed and closed by interviews that featured both close-ended and open-ended questions. While some studies employed convenient subjects (cf. e.g. section 4.6.1 and 4.10.2), in many cases the challenging process of finding study participants from the respective target groups was undertaken, since the research focus explicitly evolves around the goal of considering their specific sets of abilities and needs. While the original publications that are included for the research projects which are summarized in the following sections discuss the respective methods in further detail, a number of common adjustments that were made with regard to research methods in order to accommodate for the specific requirements of the challenging research on MGH are discussed in section 5.3.

## 4.3   Perception of Game Difficulty and Difficulty Adjustments

Aside from controlled studies and more specific explorations that will be discussed in the following sections, the general perception of players regarding game difficulty, as well as manual or automatic difficulty adjustments was explored in a study with 58 participants in Germany and Canada, using a custom questionnaire in order to gain initial leads regarding *whether players tend to appreciate the manual difficulty settings* that they typically encounter in games that they play, *whether they would like to have more or less control*, *how they typically engage with manual difficulty choices*, and *whether they feel that they need or want automatically adaptive systems* for game difficulty adjustments. Custom items were employed since the research was too specific to allow for the application of validated questionnaires and the inclusion of such additional measures was discarded in order to keep the time required to complete the questionnaire in check. The questions were framed as statements explicitly along the lines of difficulty adjustments, since most players could be expected to be familiar with the presence of such choices in the games they play. Agreement to the statements was indicated on *5 pt. Likert scale* items with a scale ranging from *disagree* to *agree*. The analysis employs descriptive results of the whole participant sample, as well as contrasting between subgroup splits by *gender* and *study country*. Inference statistics are performed with *unpaired t-tests* at a significance level of $\alpha = .05$, reporting *Cohen's d* for *effect sizes* in case of significant differences or trends ($\alpha < .1$) when the assumption of normality was not rejected due to *Shapiro-Wilk tests*. In case of a violation of the assumption of normality, the results of *Mann-Whitney's U* tests are reported, with according effect sizes for nonparametric data (Fritz et al., 2012). Correlations are included between the questionnaire items and the *participant age*, as well as the self-reported *number of years that participants have been gaming* if *Pearson* (assuming normality assumption was not rejected) or *Spearman* p values crossed the .05 significance threshold. The questionnaire was completed by *58 convenient subjects* who participated in the studies on interaction with different game difficulty choice modalities that will be discussed in further detail in section 4.6.1. The questionnaire was not related to the experimental manipulation, as it was collected via general question items accompanying the generic demographic questionnaire that was part of the protocol of these studies. Most of the participants were students and the mean (M) *age* was *26.84 years*, with a standard deviation (SD) of 5.71 years and a min-max range from 18 to 45 years of age. 40 subjects participated in the initial study that was performed in *Canada* (CAN), and the remaining 18 subjects participated in the repeat study that was performed in *Germany* (GER), which allows for some contrasting regarding the potential existence of differences in preferences regarding the adaptability and adaptivity of game difficulty in these two countries. The age distributions were comparable for both groups of participants (CAN: M = **26.88**, SD = 5.79; GER: M = **26.78**, SD = 5.71) and the gender splits were almost equal with a slight overrepresentation of self-identifying males (ALL: f = **26**, m = **32**; CAN: f = 19, m = 21; GER: f = 7, m = 11). The participants were rather experienced gamers, with a *mean experience of 11.55 years of gaming* (SD = 6.84) and slightly longer experience in the participant population from the study in Germany (CAN: M = **10.78**, SD = 6.79; GER: M = **13.28**, SD = 6.83), while male and female participants reported comparable experience on average (f: M = **11.15**, SD = 7.97; m: M = **11.88**, SD = 5.88). A later discussion of the *indicated potential influence of gender* on the question item outcomes compared to the amount of experience is hence supported, although the study was not controlled for either factor and the relationships can thus not be seen to establish causation. The individual *proportion of years of gaming experience* (relative to years of age) showed a significant medium size negative Spearman correlation with years of age ($r_s$ = **-0.3**, p < .05), indicating that older participants had comparatively less proportional years of gaming experience. It does not appear unreasonable to expect older potential player-patients to have spent a smaller proportion of their lifetime as active gamers. Most participants indicated that they spend *up to three hours* gaming in a typical week (or *"infrequent"*; **36.7%**), with some playing *three to seven* (or *"about an hour a day"*; **17.2%**) or *eight to fourteen hours* (*"several hours a day"*; **17.2%**). Accordingly, when asked whether they consider themselves to be *"non-video game players"* (N = **11**), *"novice game players"* (N = **8**), *"occasional video game players"* (N = **22**), *"experienced game players"* (N = **15**), or *"expert video game players"* (N = **2**), most considered themselves *occasional players*, followed by *experienced players*. Here, a notable difference existed between *female* participants (only 1 self-reported *experienced* video game player) and *male* participants (constituting the remaining 14 self-described experienced video game players). This was despite comparable mean years of gaming experience, as noted above.

For reasons of brevity, but since this analysis has not been published, the detailed analysis is included as an attachment, stating exact measures of central tendency and variance for the overall group of study participants, as well as for splits by gender and study country for each item, together with inferential statistics in case of significant differences or trends [**see attachment: Analysis for Perception of Game Difficulty and Difficulty Adjustments**].

Based on the negative correlation between proportional gaming experience and years of age, it does thus not appear unreasonable to expect older potential player-patients to have spent a smaller proportion of their lifetime as active gamers, although it must be kept in mind that this study was performed with a range of participants that did not include older adults. Summarizing the average responses (the questionnaire items are rephrased and highlighted in *italics*), the participants appeared slightly interested in *adjusting difficulty settings*, be it *on their own volition*, or *if prompted*. This does seem in line with the findings in later studies that indicate little differences in impact between different difficulty adjustment modalities (cf. section 4.6.1). However, further study still appeared warranted due to the rather notable *awareness* participants expressed with regard to the *presence of difficulty settings*. As could be expected, based on the principle goal of game balancing during development and the presence of difficulty tiers, most agreement was expressed with the statement that the participants *usually play on medium difficulty settings*. A slight agreement to the statement with regard to *high difficulty settings* and the comparatively less agreement to the statement regarding *easy settings* appear in line with the comparatively long years of gaming experience. Interestingly, the participants expressed rather strong agreement with the statement that they *like it when games are challenging*, indicating how the perceived individual level of challenge is not necessarily tied to choices of levels of difficulty. On average, the participants rather agree that they have a *sufficient amount of control over difficulty settings* and that they *just want to get started when playing a new game*, although most seem to be willing to take the time to *initially adjust the settings to their liking*. While this can be interpreted to underline the strength of the traditional low-complexity difficulty choices in menus, the drive to get started and the understandably limited interest in interacting with game difficulty choices can also be seen to support avoiding the introduction of too much complexity to the process of difficulty selections. With regard to *automatic adjustments of difficulty settings*, participants expressed only slight *interest in playing games with such features*, but they strongly agreed with *wanting to be able to influence the resulting settings*, if automatic adjustments were present. This can be seen to motivate approaches with mixed models of automatic difficulty adjustments while still offering manual control on a general level.

The analysis showed a number of effects and correlations based on *study country*, *gender*, *age*, and *years of gaming experience*. A visual summary is provided in Figure 21. Players from the group of *participants in the study conducted in Canada* expressed slightly more *awareness of difficulty settings* and agreement with *adjusting settings when they play a new game for the first time*, while *participants in the study conducted in Germany* expressed more agreement with the statement that they *just want to get started with playing the game*. However, participants from Canada expressed slightly more agreement to *playing on low difficulty settings*.

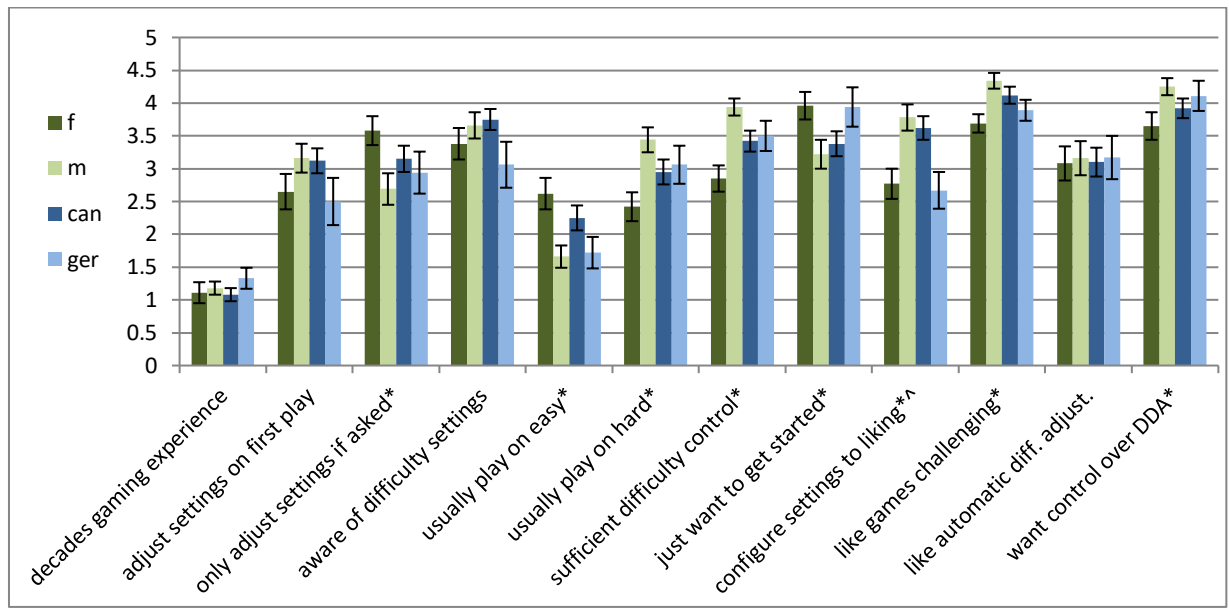

**Figure 21: A selection of items with telling results. Significant differences based on gender are indicated with a '\*' marker, whereas sig. diffs based on country are indicated with '^'. Error bars show standard error.**

Unsurprisingly, increasing *years of gaming experience* appeared weakly linked to increasingly *frequent readjustments of game difficulty settings*, to increased agreement with *liking games to be challenging*, and to less agreement with *selecting lower difficulty settings*. Hence, with increasing gaming experience, there appears to be more interest in technical details and in control of difficulty settings. However, over all measures, gaming experience appears to play a less pronounced role than what might have been expected, especially with regard to the presence or control of automatic difficulty adjustments.

The *participant age* was positively correlated to *playing on low difficulty settings*, but also to *playing on high difficulty settings*, which may be interpreted to suggest that older players may have a clearer view regarding the level of challenge they would like to face. At the same time, a trend pointed at a weak negative correlation to *appreciating challenging games*, with *having a sufficient amount of control over difficulty settings*, as well as with *wanting games with automatic difficulty adjustments*. Accordingly, increasing age was also positively correlated with *wanting to be able to influence outcomes of potential automatic adjustments*. This might be summarized as slightly increased interest in control over aspects of game difficulty. However, regarding the influence of *study country*, *years of gaming experience*, and *participant age*, the results did not appear to be unequivocally pointing towards strong directed conclusions.

This does appear to be the case when the data are split based on *participant gender*. Female participants stated less agreement with *adjusting game difficulty settings when playing for the first time*, while agreeing significantly more to the statement that they *only adjust difficulty settings if prompted*, and to the statement that they *just want to get started*. Male participants were found to express significantly less agreement with *playing on low difficulty settings* and significantly more agreement with *playing on high difficulty settings*, as well as with *liking games to be challenging*, compared to female participants. Male participants were also found to express significantly more agreement with having *sufficient amounts of control over difficulty settings*, and with *wanting to be able to influence settings resulting from automatic difficulty adjustments*. While this study is limited due to taking a narrow perspective on gender and it was not controlled for in the sampling strategy, these effects appear to be consistently pointing towards more interest and involvement with technical details of difficulty adjustments on average in the male participant population. Alternatively, it can also be argued that there may be a tendency to be more trusting of the adequacy of predetermined settings or of automatic adjustments in the female participant population.

Limitations of the study due to being merely based on opinions regarding statements, and potential sources of bias in snowball sampling, as well as gender differences in responding to the wording of questionnaire items in statements that may be subject to differences in interpretation between subjects, etc. have to be considered and the results cannot be generalized with certainty in the absence of studies that observe actual play. A detailed consideration of scatterplots showed that the results with observed effects still display considerable variance, indicating notable heterogeneity between individuals. However, contrary to what might have been expected, years of gaming experience did not appear to play a similarly pronounced role as gender did. Furthermore, female participants did not report less game experience on average, which further solidifies a difference in preferences based on gender. Most participants are rather satisfied with the difficulty choices they are usually met with in the games that they play, although notably this was not related to serious games or games for health. On the other hand, the participants, on average were rather positively inclined towards the idea of automatic adjustments, as long as an adequate amount of manual control is offered.

Regarding the modeling dimensions for adaptable and adaptive MGH these results can be interpreted to suggest that most players are neither for, nor against the introduction of adaptivity, while many are also not very interested in interacting with game difficulty adjustments, lending room for large degrees of *automation* in determining adjustments that are based largely on *implicit* feedback. However, if the adjustment systems are *salient*, their design should aim at avoiding negative *intrusiveness*, by offering a satisfying amount of *control* that allows users to provide *explicit* feedback, avoiding fully *autonomous* adjustments. The semi-automatic adjustment systems presented in sections 4.5, 4.8, and 4.9 correspond to these indications.

## 4.4 Motivation with Gamification in Interfaces for Electrical Muscle Stimulation Training

The concepts of GFH and MGH are usually understood to focus on games with game mechanics that are interwoven with activities relating to the desired serious outcomes. Gamification is often discussed in a more restricted fashion, limited to considerations on rewards that can be added to any activity without deeper integration, such as adding badges, or score systems for predefined blocks of activity. However, such elements are typically also part of the mix of design elements that are considered for GFH or MGH. Concerning the question how successful MGH can be designed, it is thus reasonable to explore the motivational potential of common gamification reward types in the context of physical activities that are augmented with digital companion applications *[see publication B.7]*. Arguably, there is no clear line that separates playful digital health applications from GFH. Any development can rather be located on a continuous *scale of gamification* that ranges from fully serious applications to fully entertaining games. Depending on the point of origin, game elements or serious application elements are added or removed. The understanding of gamification is often limited to reward schemes that are not closely related to the core activities in an application. However, game mechanics, design aesthetics, and storytelling, are viable game elements for GFH that should be considered during conceptualization and design. And yet, even the addition of a simple reward layer, such as the one illustrated in Figure 22 in playful applications has shown to provide measurable benefits (Smeddinck, Herrlich, et al., 2014). It can be expected that the introduction of a broader repertoire of game elements, if done in a harmonious fashion, can lead to more far-reaching positive impacts on game experience, performance, and health outcomes.

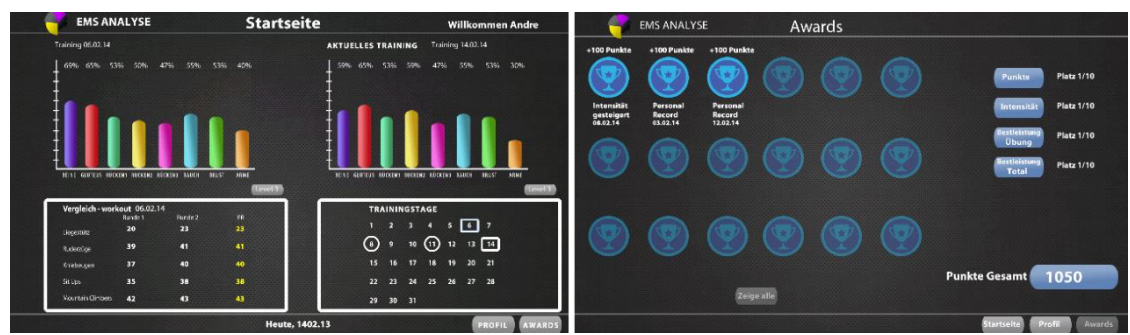

**Figure 22: The figure shows two screenshots of a gamified information application for users of electro muscle stimulation training (Smeddinck, Herrlich, et al., 2014). Augmenting trainer discussions with small interaction sessions with such performance-oriented applications was shown to lead to improvements in some motivational measures.**

More specifically, the accompanying study (Smeddinck, Herrlich, et al., 2014) indicates that simple gamified interfaces that provide an overview of the performance and loosely integrated badge rewards can improve user motivation amongst comparatively experienced athletes who participated in electrical muscle stimulation workouts to improve their existing training. While the self-determination related subscale of *tension pressure* as surveyed via the use of the *intrinsic motivation inventory* (McAuley et al., 1989) was not found to be affected for an experimental group that used a slightly gamified training interface as compared to a group that did not use the interface, the remaining IMI measures of *interest enjoyment*, *perceived competence*, and *effort importance* were all found to be notably increased, as indicated in Figure 23 (cf. to the paper for further details).

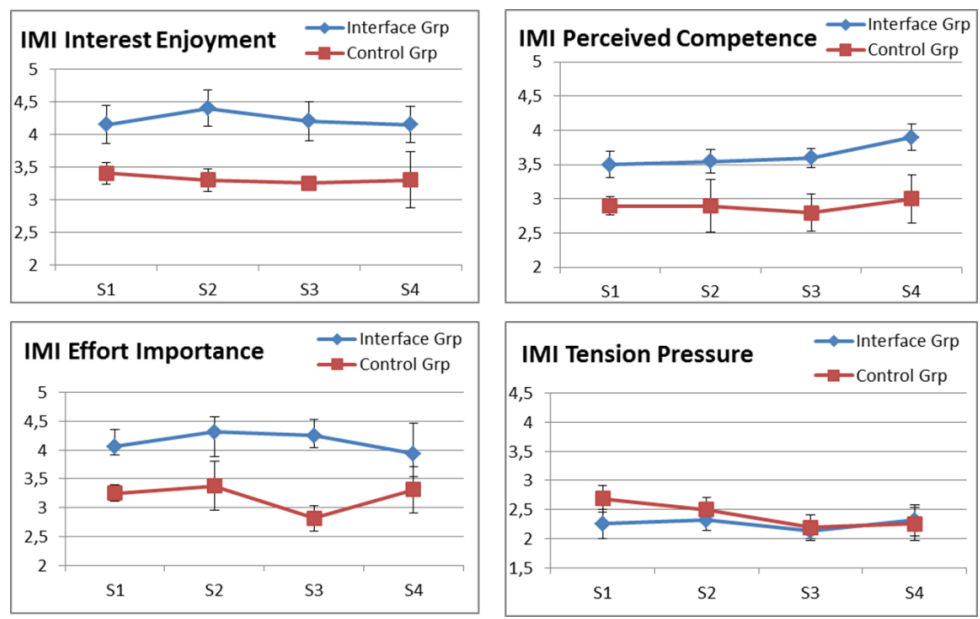

**Figure 23: EMS study IMI results over consecutive sessions. Error bars show standard error.**

It can thus be argued that even comparatively simple interfaces with gamification elements for exercise applications can improve motivation, which has, in turn, been related to improved adherence in the context of exercise motivation as measured via SDT methods (Ryan et al., 1997). Such elements should thus be considered viable candidates for inclusion in any MGH system. This does not render considerations regarding the specific target groups unnecessary, since older adults, for example, have been found to express lack of preference of competition and achievement oriented gamification elements compared to other elements, such as appealing game environments or social aspects (cf. section 4.5). Yet, such gamification elements also make for promising targets for adaptive adjustments or for adaptability, e.g. though manually determined custom reward tiers, or through automatic adjustments to the presentation or presence of such elements based on player types.

## 4.5   Human-Centered Iterative MGH Design and Difficulty Settings Parameterization

Relating mostly to the guiding research question QI (*How to design...?*), this section reports on the development process of the project *Spiel Dich fit* (SDF). As opposed to the hypothesis driven empirical studies that form the main scientific contribution of this thesis, this report summarizes a use case that illustrates the theoretical considerations on the design, implementation, and study of MGH discussed in the sections on *Structuring Theoretical Considerations* (chapter 3 and subsections) along the lines of the planning and execution of a practical project. In turn, the execution of the project contributed to the further forming of the body of theory described above, which was based on related work and supported by a number of specific explorations as the execution of SDF progressed, which will be detailed in the following sections.

**[see publication B.3, section 4.2]**

*On the basis of the lessons from the WuppDi project [(Assad et al., 2011)], and given the strong positive response expressed not only by PD patients, their relatives, and therapists, but also through notable public interest and requests by therapists for similar programs for other target groups, the project* **Spiel Dich fit und gesund** *(SDF), which translates roughly to "play to become fit and healthy", was set up to focus on exploring the use of MGH for older adults. Gerontologists and social care workers had suggested that games similar to those that they had seen as part of the WuppDi suite of games for people with PD could work well for general movement motivation with older adults. Thus, a suite of games was envisioned around the cornerstones of supporting motivation to improve upper body movement, flexibility, and balance, as well as exploring general movement motivation, cognitive training, and MGH with a strong musical, or rhythm and timing component. The possibility to personalize the games to individual users was also an integral part of the concept. [...]*

*Since related work and the prior project had underlined the importance of user-centered iterative design in the context of MGH, SDF was designed around that approach from the start. The project aimed for continuous iterative testing alongside the project development to start as soon as interactive prototypes were available. The continuous brief iterative testing was flanked by selective, more quantitative evaluations and comparative studies around specific questions that arose during the development.*

*Since SDF targeted the implementation of prototypes around three topic areas, namely* **movement activation**, **cognitive training**, *and* **music, rhythm, and timing guided movement dual-tasks**, *the first year of the project started with early brainstorming and conceptualization sessions together with experts from social support services for older adults and* *experts from a game development studio. These initial planning steps involved the creation of persona to facilitate a guided discussion and pre-evaluation of various design approaches and concepts, as well as discussions around topics such as the movement patterns that were to be implemented, the settings or scenarios in which the players should be placed, and potential game concepts which could serve to connect both of the prior aspects. Early on, group interviews with attendees of various older adult meeting centers, including the center staff, were conducted to gather more information regarding target group preferences regarding suggestions for game world scenarios, their musical taste, favored types of games, exercising, and preferences regarding potential visual styles for the game prototypes. The outcomes were largely in agreement with related work on similar sets of preferences (Ijsselsteijn et al., 2007; Nap et al., 2009).*

*In these early explorations and throughout the iterative process of the SDF project it was regularly challenging to find groups with balanced numbers of male and female participants. While this largely represents the gender proportions present in older adult populations, differences in interests and motivation of the gender groups should still be considered if the target group definition does not explicitly exclude either of them. In summary, and not separating by gender, the results hinted at "familiar scenarios", such as a garden or a shopping mall being frequently selected over more exotic ones. The importance of popular songs stood out regarding musical tastes, although later on, music that was not well-known also turned out to be accepted. Members of the target group mostly favored classic board or card games, and those that played computer or video games also most often reported to be playing digital versions of such classics or games of a very similar nature. Participants were split regarding exercising with some being regularly active and a larger portion not being active, although many said that they had been regularly active in the past. After showing members of the target group pictures of different levels of visual complexity to determine target design styles the responses were mixed.*

*Excerpt from: (Smeddinck, 2016)*

Due to these mixed responses, a study regarding the preferences of the target group with respect to visual complexity, as well as the impact of various grades of visual complexity in MGH on game experience and performance was set up. As one example of a number of empirical studies that were inspired by research questions becoming evident during the development process of the SDF project, it is summarized in further detail in section 4.7.1, referencing the included publication (Smeddinck, Gerling, & Tiemkeo, 2013). The outcomes led the SDF project to adopt a reduced complexity visual style as a cost-efficient approach.

Since it occurred rather early in the process of the overall project, the study on visual complexity was carried out with modified versions of non-SDF prototypes. The first SDF prototypes of games with "activating movements" set in a garden scenario were tested early on with older adults in social meetup facilities. This allowed for detailed observations, as well as pre- and post-play discussions. Further early trials were conducted in public spaces, such as malls, which provided a broad exposure, testing the reactions of heterogeneous "walk-in subjects". As a side effect of this public exposure, requests by a number of physiotherapists who showed interest in the approach of MGH led to the more formal definition of the support of their application scenarios in physiotherapy, rehabilitation, and prevention (PRP) with the target group of older adults as a development target for SDF.

Remarks by therapists, by caregivers, and by social meetup center workers who were leading regular physical activity sessions for older adults, showed a prevalence of wide and soft activating exercises, targeting the support of elementary functional movements such as *bending down*, *rotating*, and *keeping one's balance* that were usually instructed via metaphorical relation to real-world or fictional activities that are easily relatable to for members of the target audience (e.g. standing up and reaching out with one's arms as if picking apples).

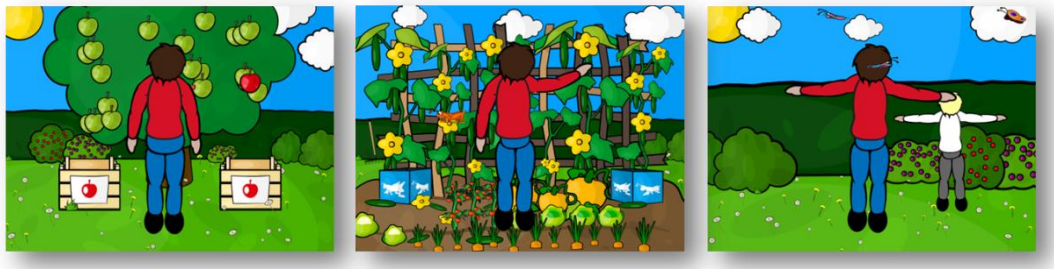

**Figure 24: Screenshots from the three games in the SDF garden scenario: apple picking (players move to changing targets; threshold based minimum speed), catching locusts (players move their hands to changing targets; threshold based maximum speed), and attracting butterflies (players strike and hold poses).**

The initial garden scenario (see Figure 24) was thus set up to implement such metaphorical activities and it was well received, with most participants in early public testing being immediately able to employ known behavioral schemata. Such an "easy accessibility" (cf. section 3.5) arguably provides a foundation of *competence* (Anderson, 1984) that is much needed when aiming to convince members of a largely non-technical and non-video-gamer audience to openly approach the concept of MGH. Outdoor scenarios, with their usually positive attributions and potential to increase feelings of vitality and even the likelihood of pro-social goals (Weinstein et al., 2009), were frequently highlighted as desirable, which led SDF to consider further outdoor scenarios for the remaining game settings. The physical exercises observed in therapy and group activity sessions were often executed in a purposefully slow and controlled manner, which informed the use of speed-oriented thresholds or timed poses in the SDF games. *Clear instructions* were noted to be important while at the same time *being active and participating* was seen as more important by the involved professionals than perfectly accurate executions. For a broader accessibility, the regular physical exercises were frequently adjusted to accommodate participation while remaining seated, which was reflected through a focus on upper body movements in the further development of the SDF games. While *"fun"* was seen as an important factor, especially in group sessions, the involved professionals were cautious of complex humor (e.g. sarcasm or irony). Similarly, music was made out to be an important measure for *motivation* and *timing*, but employing *popular harmonic music* with *clear melodies and beats* was encouraged. At the same time, *short blocks of repetitions* for each exercise were advised, leading to initial play durations of approx. two minutes per exercise with target durations for full sessions of less than 20 minutes. *Periods of rest* were frequently noted as important explicit elements of movement activity sessions, leading SDF to adopt *activity breaks* that were bridged by entertaining but not exertion-based quizzes. *Physical contact* was often highlighted as an important element of the light intensity group exercising sessions for older adults, but due to the technical challenge and the delicate human-relations aspects, the SDF project focused on a single-player setup.

However, it was noted early on that the games were frequently *appropriated to the liking of the professionals* or the users in ways that were not expected during development (cf. section 3.7.4). This can be interpreted as a mechanism of increasing task interest (Sansone et al., 1992), or in terms of technical appropriation in order to increase the range of contexts in which the given MGH are applicable. The early public exposure tests, for example, highlighted the importance of considering the highly social use case of employing the games, which were designed as single-player games, in a sort of *ad-hoc turn-taking multiplayer with onlookers* who may or may not participate in taking turns playing. Other forms of such user generated context (Dourish, 2004) were the use of the games with players who were either *seated*, or *using a walking frame*, or the *use of analog balance boards* for increased difficulty.

Finally, therapists remarked that patients should be enabled to start the MGH and work with them without assistance being necessarily required, which induces not only the needs for technical accessibility, good playability, and player experience (cf. section 3.5), but also for precautions for preventing overexertion, as participants had been observed to get deeply involved in the gameplay up to the point where *some participants moved into very strenuous positions* or positions that were considerably close to being off balance. In this regard, SDF was designed to primarily focus on *augmenting in-practice treatment* with some professional oversight. Overall, professionals were found to play a critical role (cf. sections 3.7.4 and 3.7.5), as *contributors* to the potentially beneficial use of MGH, but also as *gatekeepers* who must be enabled to work efficiently and with a good user experience with the MGH and the accompanying systems, if they are to be successfully installed for regular use in therapy practices or in exercise activity group facilities.

*After initial user feedback and observations (both collected following loosely structured protocols) were integrated to improve the initial prototypes, regular bi-weekly testing sessions with roughly two to five participants per test run were started in cooperation with a large physiotherapy practice to accompany the further development and fine-tuning. Player feedback was collected with single-sheet questionnaire featuring smiley scales (regarding the game experience and the experience of performing the exercises) for brevity and clarity. Therapist feedback on their impression of interacting with the system, as well as their evaluation of the way that the respective patients interacted with the system, was also collected with a questionnaire accompanying each test session. The first major development cycle was rounded up by interviews and discussion rounds with therapists regarding their requests for settings and parameters to make the games adaptable to individual users, leading to the user-centric parameters **range of motion**, **speed**, **accuracy**, **endurance**, **cognitive complexity**, and **resilience** being targeted for settings interfaces with a threshold-based mapping onto game variables.*

*In addition to the continued iterative testing, the games were also employed in a first exploratory evaluation in a nursing home. While most patients were aware of the connection between their body movements and actions by a player character on screen after a thorough introduction, they still required close guidance or direct physical support and in many cases tracking was complicated by obstructive devices, such as wheelchairs or walking aids, or by limitations in the tracking of less pronounced movements. These findings are in line with the limitations encountered by other researchers when attempting to employ MGH with frail older adults (Gerling, 2014a). Due to the challenges encountered in this exploratory trial, the target group for SDF was limited to not include frail older adults.*

*Excerpt from: (Smeddinck, 2016)*

While such extremely differing target groups were thus not in the focus of further developments in the SDF project, the target group of non-frail older adults nevertheless displays considerable heterogeneity. Demographic and background questionnaires in a number of studies that were carried out in the larger context of SDF (see publications B.5 and A.2) showed that some individuals were still practicing ambitious sports routines, while others were already notably limited with regard to their ability to perform crucial physical day-to-day activities. As discussed in section 3.7.1 (following indications from publication A.1), this can also vary within a person over time, or depending on the type of activity. Accordingly, adaptability and adaptivity remained important aspects of the SDF project. In order to allow the project to start working with a system without extensive prior data collection, the approaches in SDF were based on heuristics that were subject to iterative testing and that encompassed manual monitoring and control by professionals. A configuration interface for therapists was designed to allow these manual functions to be executed efficiently (see Figure 25 and Figure 26). During prototyping for the configuration interface it became evident that professionals would have to be very familiar with any given MGH if they were to enact settings based on actual in-game variables, such has how many apples appear per minute in an apple picking game. Since this would endanger initial adoption and entail considerable learning periods, the concept of human-centric parametrization (cf. section 3.7.6) was developed, allowing therapists to enact their settings on a set of parameters that expresses the level of abilities of each player in terms that they can immediately relate to (see section 4.9 for a more detailed description).

*With first prototypes of a settings and configuration interface for therapists at hand, usability focused testing helped with resolving general design challenges. A small study with therapists confirmed the adequate breadth, understandability, and flexibility of the chosen user-centered parameter set. Since movement capability configurations were deemed important by the therapists, a special grid-based settings interface was implemented that allowed therapists to configure intensities of activation [...] of zones for the placement of interactive targets in the game (see Figure 25).*

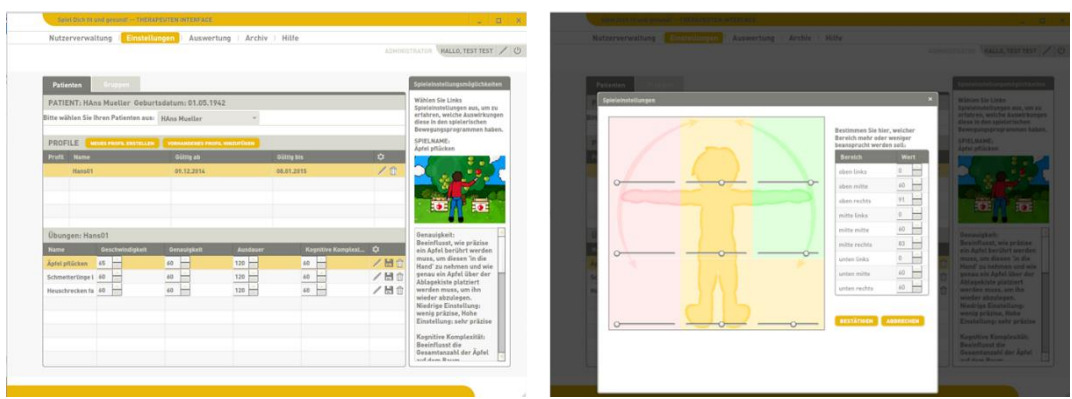

**Figure 25: Screenshots from the configuration interface for therapists designed in the context of the SDF project. Left: Settings are performed per player / per group on separate controls for speed, accuracy, etc. Right: A custom grid-based component that allows for configuring range of motion via active or inactive motion-target zones.**

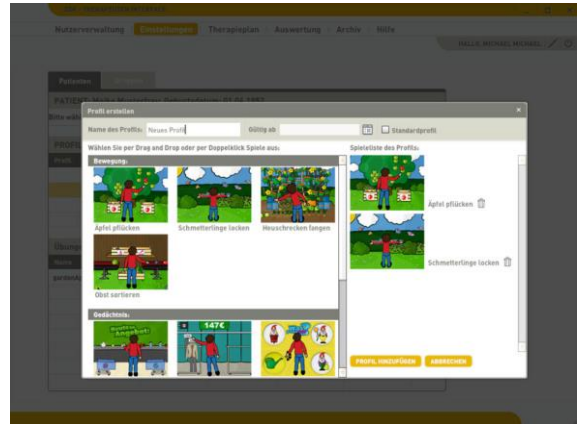

**Figure 26: The configuration interface for therapists in SDF also features a sequence editor that can be used to determine in which order the individual games of the suite should be played and whether or how often they will occur in a session for either an individual player, or a specific group of players.**

Next to the subsequent introduction of more game prototypes from the second and third development target areas, which were tested in social meetup facilities for older adults and in public spaces, further dedicated [small scale] studies investigated the **impact of modality and delay of audio instructions**, especially in the games focused on rhythm and timing, the **optimal activation time for motion-based interactions with hover activated buttons** [leading to longer activation times than common guidelines for NUIs recommend] for the target group, and the **preference and impact on player experience and performance of different modalities for instructing dance movements** in a rhythm and timing game. The latter included a comparison of instructions between dance moves shown by (a) an instructor character that was similar in visual style to the player character (the player character is not shown in the figure), (b) instructions shown by icons approaching an "action zone" akin to popular dance games in the tradition of Dance Dance Revolution, and (c) a mixture of both approaches (see Figure 27).

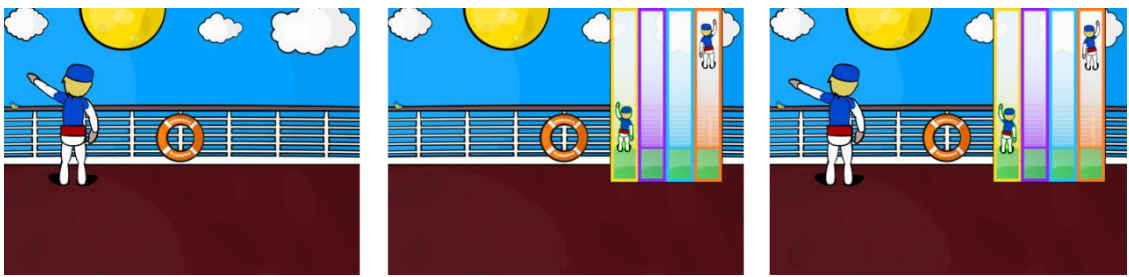

**Figure 27: Screenshots of the dancing game from the rhythm and timing development target in SDF showing the different experiment conditions (left to right: a, b, c) for motion instructions.**

While quantitative results on game experience and performance showed no consistent differences, observations and player comments indicated that the participants were able to play with all three modalities. However, the overlay version (b) required more introduction and *training than the instructor character version (a) and while the overlay version was eventually used by some participants to foresee the coming actions, many participants had troubles with understanding the instructions indicated by the icons, thinking, for example, that poses should be held indefinitely. The version with both displays (c) was confusing to some participants, which led to SDF adopting the instructor character for the continuing development. [...]*

*The last major development and evaluation cycle included further implementations of the interaction work-flow surrounding the core games (game menus, pausing the game, etc.), validating the acceptance of intermittent non motion-based quizzes that were implemented to ensure breaks between sessions of more physically intense gameplay in order to avoid overstraining, as well as prototype finalizations and polishing (including, for example, visual notifications and guides if players left the trackable space in front of the sensor device).*

*Excerpt from: (Smeddinck, 2016)*

A number of additional studies that were carried out in the context of the development of SDF have not yet been published (cf. section 5.2) and still had an impact on game design decisions, especially with regard to adaptability and adaptivity. This includes an investigation of the usability, user experience, and performance of varying interaction modalities for enacting range of motion configurations and other movement related capability adjustments, leading to the selection of a grid-based method rather than more classic windows-icons-mouse-pointer components or motion-based input as modalities for configuring MGH. Further studies explored the impact of being *informed* or *uninformed* about automated difficulty adjustments, relating to the modeling dimension of *saliency* (cf. section 3.4.2), and of having a *choice* or *no choice* on the activation of such systems, relating to the modeling dimension of *automation* (cf. section 3.4.2). The study outcomes led to the decision to spare investing into attempts at avoiding any mentioning or saliency of the adjustments, since the results did not suggest the presence of significant negative effects on player experience or performance.

The impact of different modalities for presenting movement instructions, comparing a virtual character to video based and real human instructor based instructions, was also studied (cf. section 4.7.2), with outcomes underpinning the adoption of a virtual instructor for SDF, since the player performance was found to be significantly improved compared to the performance achieved with a video instructor. Lastly, the medium-term impact on motivation and functional therapeutic measures of working with the SDF games under either semi-automatic or fully manual adjustment conditions was compared to traditional therapy sessions in a situated study over the course of multiple weeks (cf. section 4.9).

Relating back to the structuring theoretical considerations discussed in chapter 3, this case study of a MGH development highlights the importance of considering the perspectives of different involved parties, indicates the importance of human-centered adjustments that can be adopted to function across applications, and underlines the importance of human-centered iterative design that does not only involve members of the target group and further involved parties, but also considers situated use and user generated context. In addition to reviewing related literature, explorative methods, such as target group surveys, interviews, expert involvement, paper prototyping, or participatory design, play an important role with GFH design, since these methods make requirements and challenges evident that designers and developers might otherwise overlook. Overall the development of this project inspired the studies that are discussed in the following sections with specific research questions and informed the theoretical considerations that are discussed in section 3.1 to 3.9.

## 4.6 Interacting with Difficulty Adjustments

As mentioned above, manual involvement in making adjustments to MGH in order for them to better respect player abilities and needs is not only helpful for technical reasons, especially when first introducing adaptivity to MGH, but manual involvement can also be expected to interact with player experience, motivation to engage in activities, performance, sense of self, program adherence, and more. Within the larger context of the guiding research questions Q1 (design) and Q2 (acceptance), the following sections summarize and contextualize studies that focus on the role of manual involvement of players and third parties, such as motion-based therapy professionals, in the process of personalizing MGH experiences through adaptability or semi-automated adaptivity. This includes studies on the reaction to different levels of *explicitness* in *changes* that cause adjustments to the system, as well as different levels of *autonomy* and *automation* with respect to the functioning of the adaptive system. These studies thus focus on the central aspects when considering adaptability and adaptivity in MGH through the lens of human-computer interaction, which are discussed in relation to the modeling dimensions for such systems in section 3.7.2.

### 4.6.1 Presenting Game Difficulty Choices

As indicated above, the question *whether and how to present game difficulty choices to players* is relevant to both the design of motion-based games and video games in general. While many aspects count into the overall player experience, the balance of skills and challenges (Chen, 2007; Crawford, 1984) is arguably key to enabling rewarding player experiences (see sections 2.5 and 3.6). This is crucial in SG and MGH which aim for a serious outcome, and where sustained motivation is a prerequisite to achieving that outcome. While user-centered iterative design and prudent testing can produce good average settings and difficulty progressions, no two players are alike (see section 2.1.5) and abilities and needs fluctuate over time (see section 3.7.1). Figure 28 shows a selection of different players of motion-based games to illustrate this argument. However, while providing means to adjust the challenge a game presents is clearly an important design consideration, there is little research on how to present game difficulty choices. The details to an according study on the impact of different modalities for game difficulty choices on player experience and performance are provided in [**publication A.3**] (Smeddinck et al., 2016).

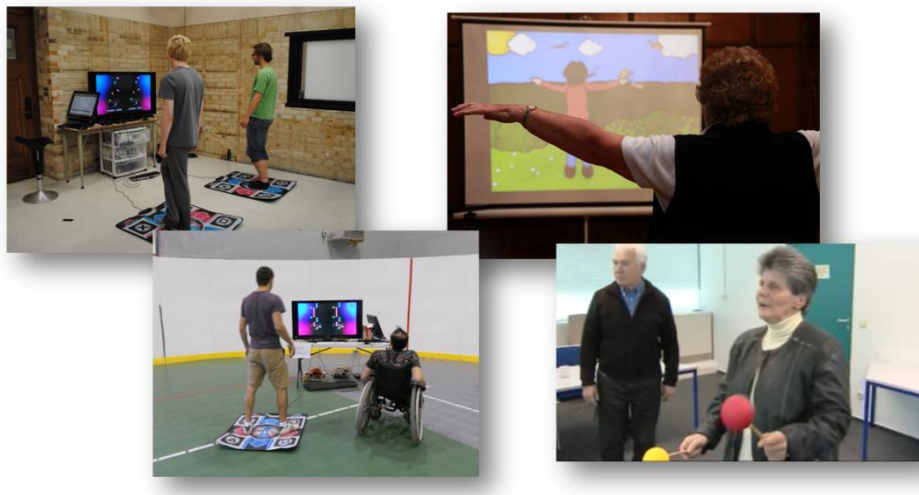

**Figure 28: This figure illustrates how different players of MGH can be, both between use cases and even within the same use case. The top left shows two sportive young adult players playing a competitive dance game. The bottom left shows a pair with one player being a young adult wheelchair user. The top right shows an older adult playing the butterfly game from the SDF suite of games, while the bottom right shows two older adults with Parkinson's disease taking part in early tests of the WuppDi project.**

In summary, this study compared *traditional menus for adjusting game difficulty* with *automatic adaptivity* and the concept of *embedded difficulty choices* (or player-oriented DDA), which was described by Chen in his work on flow in games (Chen, 2007), as alternative modes of presenting difficulty choices (see Figure 29). It aimed to determine whether the differences in difficulty choice modality would lead to differences in game experience or performance.

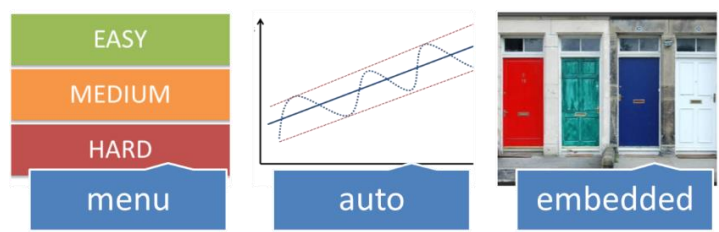

**Figure 29: The different game difficulty choice interaction modalities that were compared in the study. Left to right: traditional game difficulty menu, automatic adjustments via DDA, and manual embedded difficulty choices.**

As discussed in further detail in the paper, Chen argues that *menus can interrupt the flow of the game experience*, while *DDA can strip players of control and autonomy*. The concept of *embedded difficulty choices* that are presented to the players *as elements of the game world* as opposed to being presented in separate menus can arguably provide *control* while avoiding breaking *presence* and *immersion* or the flow of the game experience. In the paper these arguments based on flow theory are connected to *competence*, *autonomy*, and *presence/immersion* as they are discussed in SDT (Smeddinck et al., 2016), hypothesizing that difficulty choice modalities that allow for explicit difficulty choices (here: *embedded* and *menu*) should lead to better *autonomy* need satisfaction compared to difficulty choice modalities that do not allow for explicit difficulty choices (here: *auto*). Furthermore, *presence* and *immersion* were hypothesized to be increased in modalities that allow for uninterrupted game experiences (here: *embedded* and *auto*), compared to difficulty choice modalities that lead to interrupted game experiences (here: *menu*). Notably, *embedded* is the only modality that is on the positive side in both cases. Figure 30 shows a visual summary of these directed hypotheses that formed the basis for the comparative study summarized below.

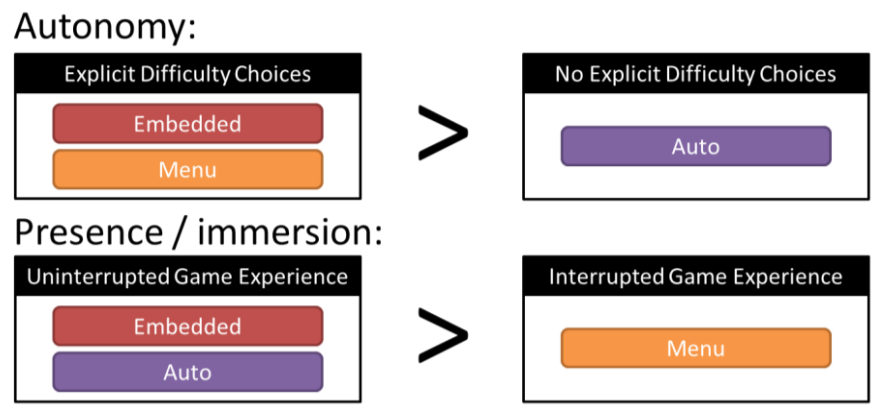

**Figure 30: A visual summary of the hypotheses for the study showing the expected relation of the conditions with regard to the SDT dimensions of autonomy and presence / immersion.**

In order to provide the conditions, a game called THYFTHYF (see Figure 31) was developed with the goal to allow for salient changes to the game experience depending on changes in difficulty, providing dense in-game and post-game audio-visual feedback. Details on the functioning, implementation, and the design rationale, are provided in the paper.

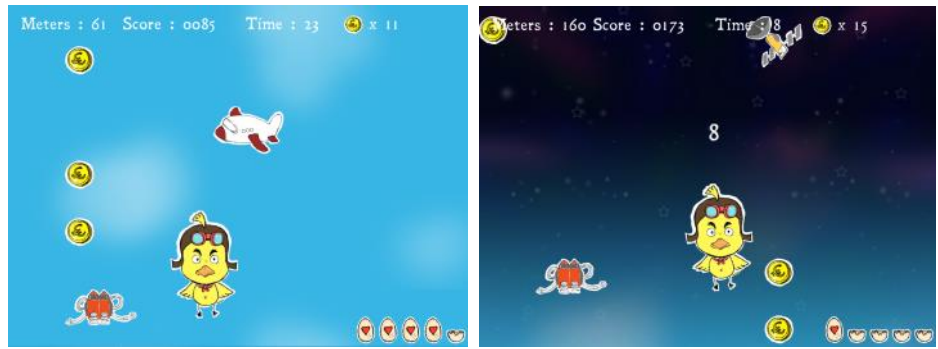

**Figure 31: Two screenshots from two different points of advancement in a level of THYFTHYF a casual collection and aversion game with motion-based or alternative sedentary controls.**

Following pilot study runs the game balancing was deliberately adjusted to include the possibilities of feeling overstrained, or of losing, even within a short time. While the automatic adjustments were performed between rounds and did not have a distinct visual component other than changes to the game dynamics, Figure 32 provides a visual impression of the *embedded* and the *manual* condition.

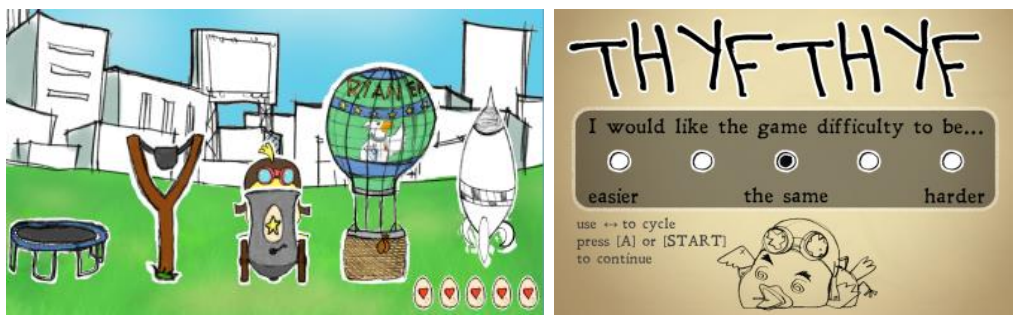

**Figure 32: Screenshots of the difficulty selection in THYFTHYF for both the *embedded* (l) and the *manual* (r) condition.**

Figure 33 provides a summary of the study procedure with is described in detail in the paper. Notably, the results that will be described below appeared somewhat surprising, which led to a re-run of the same setup with modified versions of the game *fl0w* which Chen had originally based his arguments upon, in order to determine whether design choices in the game had an influence on the results.

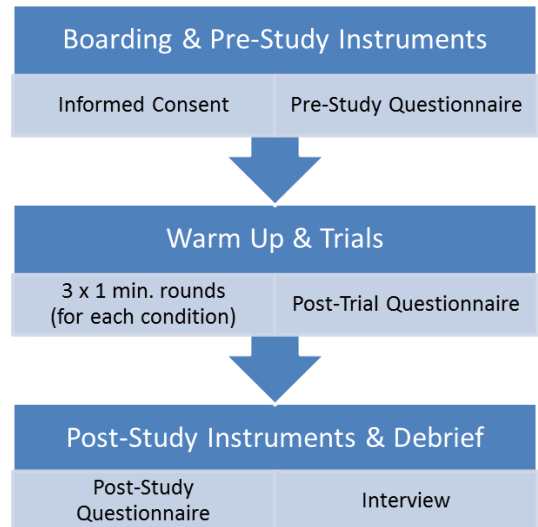

**Figure 33: The study procedure for the study on game difficulty choice modalities.**

An overview of the resulting full set of conditions of both studies is provided in Figure 34. While the *menu* and *embedded* conditions are described in further detail in the paper, their functioning is also rather evident. The exact nature of the *automatic adaptivity* in both games was only described briefly in the paper. In the context of this thesis, a more specific classification can serve to clarify the role of this study with regard to the modeling dimensions of adaptive and adaptable systems (as discussed in section 3.4).

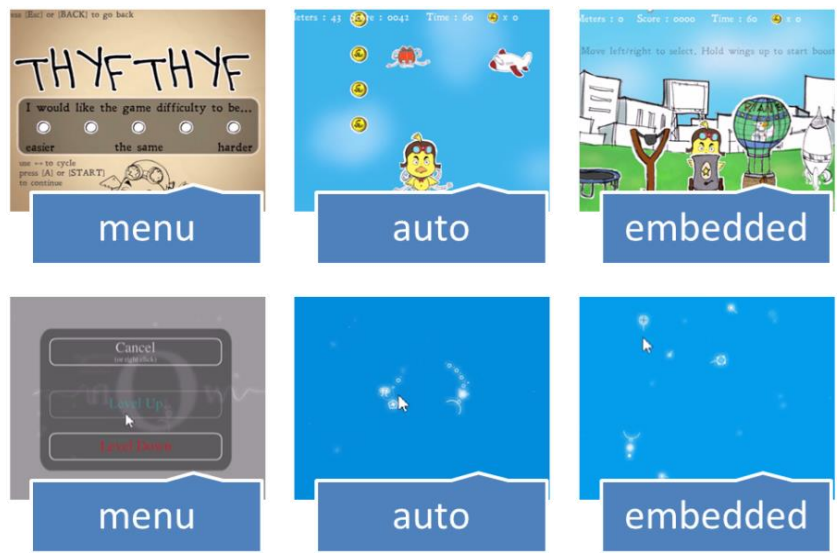

**Figure 34: Full set of conditions in both studies with the games THYFTHYF (top) and fl0w (bottom row).**

The adaptive systems for both games were deliberately kept comparatively simple, since the focus was set on the presence or absence of automatic adjustments rather than on comparing between multiple candidate systems for automatic adjustments. In short, player performance is compared to a pre-defined target range and adjusted if upper or lower thresholds are violated (cf. section 3.2.2). Regarding the *goals* (cf. section 3.4.2), the adaptive systems were *static* (not subject to *evolution* of goals), *rigid* (goals were not *flexible*), with the *persistent* and *single goal* of adjusting the game difficulty, which was thus also *independent* of other goals. Both systems provided clear and notable adjustments to game mechanics related to difficulty, so that achieving an improved outcome based on more relaxed difficulty was only marginally *skill-dependent*. Regarding the cause for *change* both systems reacted to *external* changes in player performance that are *non-functional*, occur *frequently*, especially when players are first interacting with a new game, and are on a *foreseeable* trajectory but still difficult to *predict* accurately. The games employed *implicit* performance measures as triggers for changes. Due to the nature of the underlying games, a small difference existed in the *determination* of changes. Since *THYFTHYF* is played in multiple short rounds, change was considered *after* each round. In *fl0w*, change was *continuously* observed during play. The functioning with respect to *mechanisms* was also mostly similar with both systems, featuring *parametric* adjustments that were enacted completely *autonomously*, organized through a *centralized* module with comparatively *local* scope and changes lasting *indefinitely* until further adjustments were made. The *timeliness* of the adjustments was *guaranteed* due to single thread local execution and rather *exploitative* due to following predefined heuristics although they included some rubberbanding around the estimated personal optima. *Step sizes* for adjustments were kept in predefined thresholds but purposefully *large* enough to be notable. While the determination of adjustments was fully *automated* for both systems, again due to the nature of the structure of the games, the adjustments were *triggered* based on *events* in *fl0w*, while a *time-based* trigger (after each round) was employed for *THYFTHYF*. Regarding *effects*, the adjustments were of *low criticality*, with *non-deterministic* consequences due to the unpredictable reaction of the affected human players. The systems produced insignificant performance *overhead* due to the simple heuristic nature of the adaptivity, functioned in a very *resilient* manner, affected a single player as the *recipient*, affected *control* and mechanics rather than *feedback* (such as score multipliers; cf. section 4.10.1 for an example of the study on these modeling dimensions), featured a single parameter to multiple-variables *breadth* in mapping, were designed to be *salient* in order to allow for comparing to the manual modalities for game difficulty adjustment which are also clearly salient, and their adaptivity was designed to be positively *intrusive* concerning the player goals, aiming to provide distinct positive impact.

Strictly speaking, both the *embedded* and the *menu* modality for difficulty adjustments can be interpreted as adaptable software systems with a fully *explicit* source for *change* (the player deciding to make adjustments), which makes the *mechanism* fully *assisted* and *manual* with regard to *autonomy* and *automation*. The major difference between the *embedded* condition and the *menu* condition in the studies then lies in the *saliency* of the adjustments and potentially in how far they are perceived as *intrusive*, which can clearly be related to the expectations with regard to the aspects of *autonomy* and *presence / immersion* respectively as they are discussed based on SDT in the setup of the studies.

The first study included 40 participants and a between groups variable of *controller type* with the conditions *sedentary* and *motion-based* in order to allow for more likely transfer of these studies with convenient subjects and entertainment games to the research interest of MGH. The primary research interest (as outlined above) relates to the within-subjects variable of *difficulty selection interaction modality*. Conditions were ordered via Latin square randomization, and the results did not show consistent interaction effects between the group variable and the within-subjects variable for the initial study run with the game THYFTHYF. For a full report on the results, please refer to the paper. The results show a significant difference between conditions on the SDT related sub-scale of *autonomy*, as indicated in Figure 35, with the *embedded* condition showing higher autonomy needs satisfaction than the *auto* condition, indicating an effect in the direction that was assumed by the first hypothesis described above.

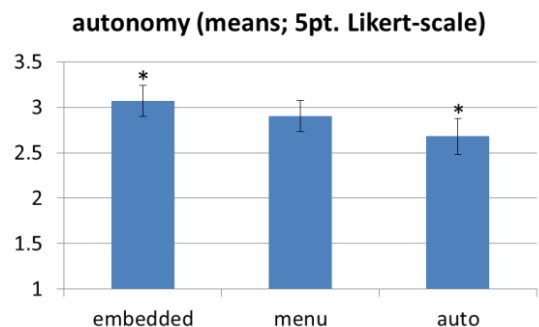

**Figure 35: Results from the initial study run with the game THYFTHYF on the SDT sub-scale autonomy. Error bars indicate standard error.**

While some significant differences exist on the between groups variable, they can be explained with expectable differences between the *motion-based* version and the *sedentary* version of the game (e.g. increased self-perceived *physical demand* and *intuitive control* in the motion-based version, since the game was primarily designed as a motion-based game). These differences show that the measures were in principle sensitive to differences in game design under the conditions of this study. However, the remaining experiential sub-scales of the employed validated questionnaires did not only show no further significant differences on the within-subjects variable. To the contrary, the means appeared extremely consistent across conditions (cf. Figure 36). While this result does not prove the absence of effects due to differences in difficulty interaction modality beyond the reported presence of an effect on autonomy, the consistency of almost identical means across all remaining measures indicates that large effects are unlikely to exist.

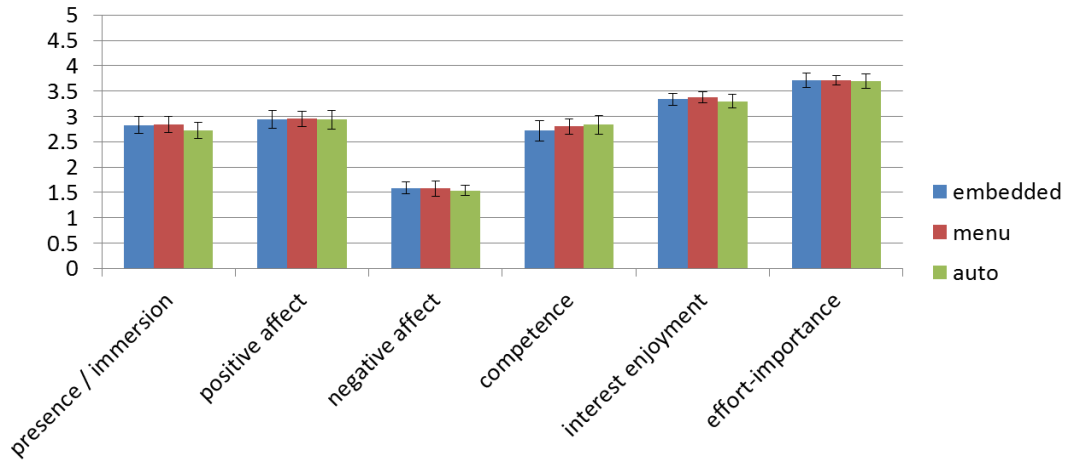

**Figure 36: A summary of experiential measures from the initial study with the game THYFTHYF, showing the absence of significant effects and the extreme similarity in means and variance across the within-subject conditions. Error bars indicate standard error.**

As indicated above, this partially unexpected result led to the implementation of a replication study using the game *fl0w* that Chen had employed as his primary example when discussing the concept of (player-oriented) embedded difficulty adjustments, since the lack of further effects, especially considering the specific hypothesis with regard to *presence / immersion* partially contradicted the expectations that were constructed based on the original arguments by Chen. The study was designed to be identical in setup and procedure with the exception of a reduction to 20 participants and leaving out the between groups variable of *controller type* (focusing exclusively on sedentary gameplay in order to avoid radical deviations from the original game *fl0w*). Again, the detailed setup and results are reported in the paper. In summary, an effect on the *autonomy* sub-scale could be confirmed while no further effects on the game experience measures were found, with results showing similar almost identical means across conditions. Notably there the second study showed somewhat deviant results with regard to the *task load index* (TLX) (Hart & Staveland, 1988) that are summarized in Figure 37.

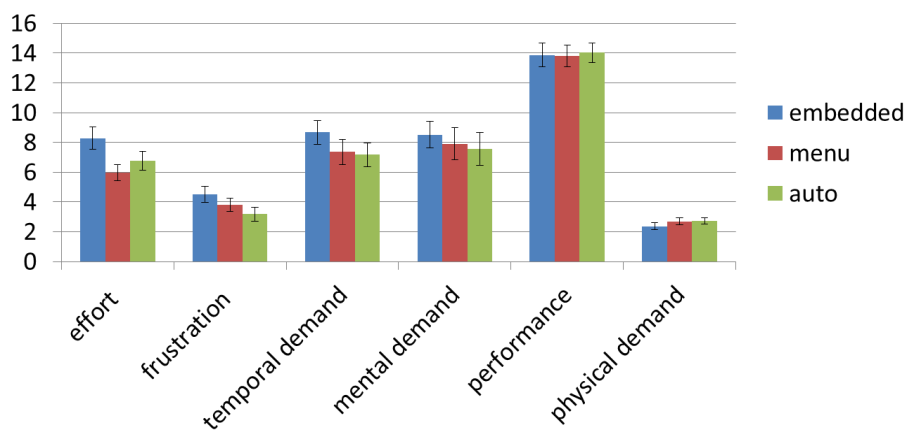

**Figure 37: The TLX results from the repeat study with the game fl0w. Error bars show standard error.**

Participants felt they had to put in significantly more *effort* when playing the *embedded* difficulty choice mode compared to the *menu* mode, with the mean of the *auto* mode being located between the other two conditions. Participants also felt significantly more *frustration* in the *embedded* condition than in the *auto* condition. This can perhaps be attributed to the game design of the embedded difficulty choice modality in *fl0w* that required players to seek out specific food items in order to move up or down in difficulty levels, making the process arguably more demanding than a menu based or automatic selection and also more complicated than the selection of a different start boost in the *embedded difficulty choice* modality in the game THYFTHYF where no such differences had been found in the initial study. This argument for significantly increased *effort* and *frustration* aligns well with the visibly increased *temporal* and *mental demand* in the embedded condition.

Nevertheless, in this second study, comparative ranking after the completion of all trials most often put the *embedded* condition in the first rank, indicating how high task loads are not necessary undesirable in game settings, where challenge matters. This argument is further supported by the player experience results, since the explicitly game-related experiential measures showed no significant differences beyond the difference on *autonomy*. In other words: while challenges are undesirable from the perspective of usability in non-game applications, they are – to a certain extent – central to game design (Deterding, 2016). However, game designers will arguably still benefit from being aware of potential causes of frustration or overall task load with their games.

The studies are, of course, limited in a number of ways. They tackle only short term fluctuations and immediate reactions to adjustments and interaction with setting modalities in casual games. The ecological validity still needs to be determined due to the limited context of lab study setups with short chunks of forced play and due to leaving out more complex social contexts such as multiplayer or onlooker scenarios. The parametrization of the automatic adjustment modalities was not human-centric for practical reasons, since both games were not MGH, but consisted of game-specific parameters bundling multiple variables. Since most of the potentially interesting independent variables (e.g. the remaining modeling dimensions of adaptable and adaptive SG) had to be fixed for experimental control, such as the *point of intervention* for adjustments, many opportunities for further studies remain.

However, this work contributes empirical insights relating to the impact of different *modalities* for *interacting with game difficulty choices* on *player experience* and *preferences*. In summary, with respect to the question of *how to present game difficulty choices*, the studies do not suggest a clear-cut response. While *embedded* difficulty choices can lead to *increased autonomy* needs satisfaction, they can also lead to *increased effort / frustration* and do not appear to be a magic bullet. The *strong similarity in game experience measures* across conditions suggests that difficulty choice interaction mode in casual games might play a minor role compared to other design choices and that the effect on *autonomy* does not carry over to other game experiential measures, which might have been expected based on theory (Deci et al., 2001). In the case of complex MGH, for example, where automation is desirable, well-implemented and tested automatic adaptivity can be expected to lead to similar overall game experience results compared to well-implemented and tested manual settings. This argument appears in line with findings from related work that showed that visible difficulty adjustments for balancing in multiplayer scenarios were better accepted than what might have been expected based on conventional wisdom (Depping et al., 2016).

### 4.6.2   Movement Capability Configurations on Mobile Devices

*[see publication B.9]*

As argued throughout this thesis, strong adaptability is a major requirement, but also a challenge, for advancing the therapeutic use of MGH. Explorative research and studies involving physiotherapists (cf. sections 4.5 and 4.9) in the context of the projects *Spiel Dich Fit* and *Adaptify* have indicated that configuration interfaces for MGH must function efficiently in situated use in practices. A limit of around two minutes per player / patient was repeatedly noted as the maximum acceptable amount of time required to perform an initial boarding process when first setting up the system for a new user in the busy day to day operations of a physiotherapy practice. Interactions for later checkups or adjustments should arguably work even faster. While earlier developments (such as the SDF project) have focused on making configuration options available that are understood by therapists and players alike, and on making them accessible through standalone or web applications in the first place, situated use and device form factors must be considered, if efficient and effective use is to be guaranteed. For adaptation tool development, tablets are a promising platform due to their similarity in affordance compared to traditional clipboards, which still are the dominant tools for managing patient treatments and practice workflows in most therapy practices. With the increasing prevalence of affordable, rugged, performant, secure and reliable tablet devices, and given the continuing growth and pervasion of cross-platform electronic health records, tablets are likely to gradually replace clipboards, and thus are an especially interesting development target for next generation configuration and management applications for MGH.

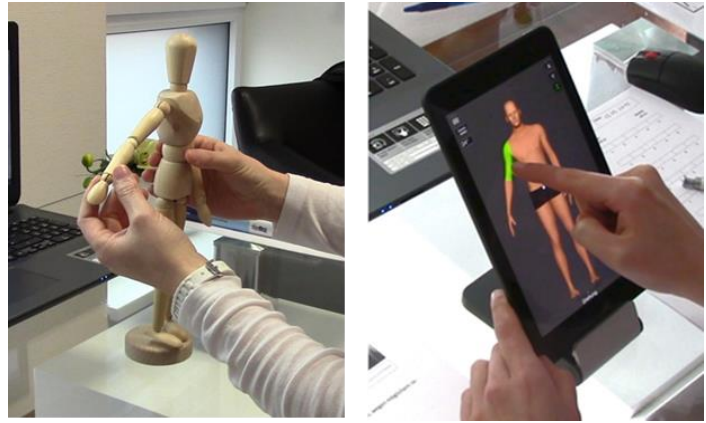

**Figure 38: A participant enacting movement capability adjustments either on a physical wooden (left) or digital manikin (right). The study explored approaches to realizing efficient and well-integrated patient movement capability configurations for physiotherapists (Smeddinck, Hey, et al., 2015).**

Accordingly, a study with therapist participants was set up to explore more efficient interaction modalities for mobile devices (cf. Figure 38 on the right). In a comparative setup, three different input modalities in use on tablet devices were examined that allow for configuring joint angles. The underlying application was designed to facilitate patient capability configurations for milestone-based semi-automatic adaptivity (akin to the system that was employed in the context of the SDF project and the according studies; cf. sections 4.5 and 4.9). The goal to facilitate movement capability configurations by employing a virtual manikin that illustrates the physical body of a patient was motivated by prior findings on the importance of human-centric parameterization (cf. sections 3.7.6 and 4.5). The input modalities that were compared in the study are (a) *direct-touch* (direct manipulation of the virtual figure via touch), (b) *classic interface* components (i.e. buttons and sliders), and (c) a *combination* of both (see Figure 39). While *direct-touch* emerged as the least preferable modality, the results highlight the benefits of a combination of direct-touch and classic interface components as the most accessible modality for configuring joint angle ranges.

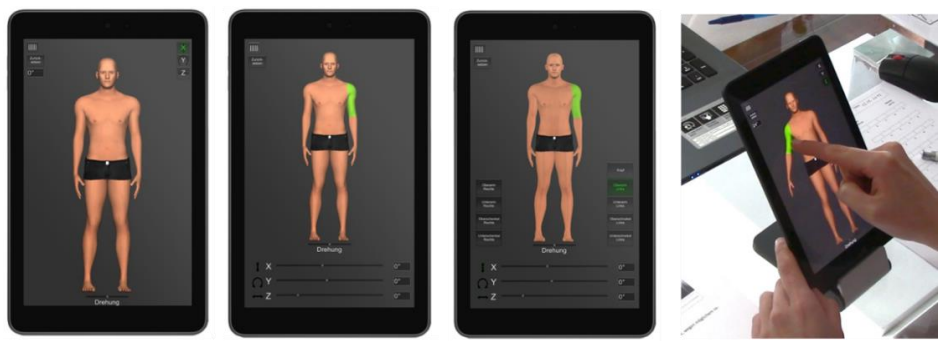

**Figure 39: The study conditions (left to right) *direct-touch* (a), *combined*; with body part selection via direct-touch with joint angle configurations via classic interface components (c), and *classic interface components* for selection as well as angle configurations (b). The rightmost image shows an example of a therapist using the interface on a tablet device.**

Furthermore, based on a pre-study that included asking therapists to perform movement capability configurations for a fictitious persona on a wooden manikin (see Figure 38, left), in order to observe their intuitive approach to bimanual interaction and to gather clues on which aspects of the movement capability configurations are important to them through a thinking out loud technique, the importance of configuring joint angles along three distinct axes relative to the body orientation of the patient representation became clear. An unexpected additional potential benefit of employing such movement configuration interfaces also emerged, since therapists repeatedly highlighted their interest in using the study application as communication support when discussing patients with colleagues, or when discussing progress or other aspects of their therapy with the patients themselves.

While this study with an early interactive prototype found that a mixture of direct manipulation of a 3D avatar with traditional input elements and clearly separated movement axes worked best for the therapists, it was the early involvement of the therapists that led to the unexpected finding that the therapists highlighted the usefulness of such a tool for communicating with other therapists and with patients about movement capabilities and the development of individual patients. These findings are practical examples of the importance of including interested parties beyond the players in the development of MGH (cf. section 3.5), of considering situated use (cf. 3.7.4), and of assuming the lenses of explorative design and of health professionals (cf. section 3.7.5). With regard to the guiding research questions Q1 – Q3, these outcomes indicate support for the aforementioned design approaches. Furthermore, a combined design with direct-touch and classic interface components was found to be most accessible, showing promising acceptance, while system usability scale scores showed room for improvements in efficiency even with the superior methods.

## 4.7   Communicating Feedback for Motion-based Health Applications

Due to the extremely versatile nature of games in applications for serious contexts, many aspects ranging from their appearance, over game mechanics, and sound design, to menus, and surrounding ecosystem applications or devices are of interest in design, development, and for game user research, as they potentially influence accessibility, playability, player experience, and thus influence the resulting motivation, as well as the desired positive outcomes (cf. section 3.5). While most work on adaptability and adaptivity in games focuses on the adjustment of parameters that influence the game difficulty, in the larger context of aiming to adjust games and especially MGH to the individual abilities and needs of their players, other aspects beyond game difficulty also make for candidates for adjustments through adaptability or adaptivity. This section reports on a number of studies on aspects that were considered as potential targets for adaptivity next to DDA. By regarding the studies in the light of this question, the contributions in the context of this thesis fall largely into the area of the guiding research question Q1 (*How to design...?*).

### 4.7.1 Visual Complexity

Since older adults were identified as the primary target group for the SDF project, the question arose, which level of fidelity in visual design would best accommodate their preferences and needs. Considerations based on the limitations or impairments in vision that grow more prevalent with increasing age motivated a comparative study on the impact of different levels of visual complexity on player experience, performance, and physical exertion in motion-based games for older adults. A detailed description is available in the attached paper (Smeddinck, Gerling, et al., 2013) [**see publication B.5**]. In the context of adaptability and adaptivity for MGH, visual complexity was also considered as a candidate aspect of game design that could be opened to adjustments for personalization. This would, however, come at the cost of investing into the design of different levels of complexity, or of methods to scale between different levels of complexity of a given MGH.

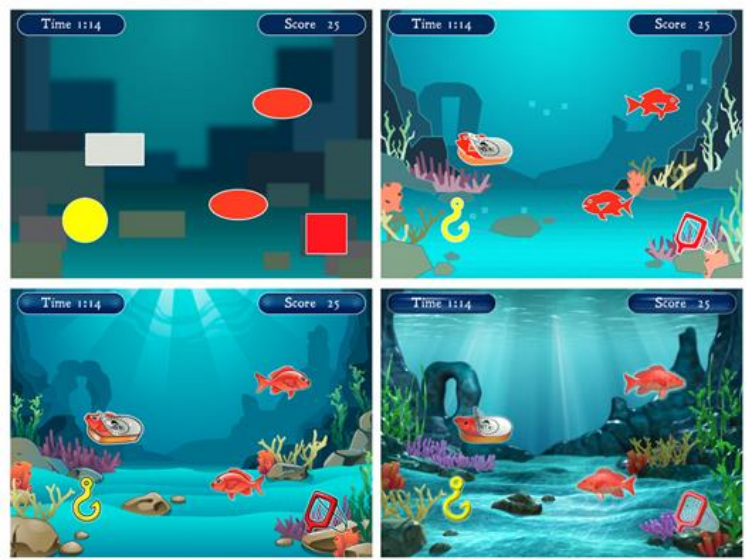

**Figure 40: Four levels of visual complexity of the same game. The abstract version (top left) was shown to have inferior effects on player experience compared to the other versions.**

Hence, the study decomposes the player experience of older adults engaging with MGH, focusing on the effects of manipulations of the game representation through the visual channel (visual complexity); the primary feedback modality for interaction in most games. The results from the within-subjects setup with fifteen participants show that the degree of visual complexity affected the way the game was perceived in two ways: (1) while the older adult participants expressed different preferences in terms of visual complexity of video games, notable effects were only measurable following drastic variations, separating a version with simple shapes as game objects from all remaining versions, while no significant differences were observed between the remaining versions of increasing visual complexity (see Figure 40). (2) Perceived exertion shifts depending on the degree of visual complexity, with higher perceived exertion being reported in the abstract condition.

It was noted in the section on motivation with gamification (section 4.4) that it can be expected that the introduction of a broader repertoire of game elements, if done harmoniously, can lead to more far-reaching positive impacts on game experience and both game and health related performance. Arguably, the outcomes of the study support this hypothesis, since notable differences only resulted when the visual fidelity was reduced to a level that removed the micro-story that was inherent in the fishing scenario of the game. For the SDF project these results led to the preference of a reduced complexity visual style as a cost-efficient approach, as well as to the conclusion that the potential benefits of implementing visual complexity as a dynamically adjustable aspect of the game would likely not warrant the required investment.

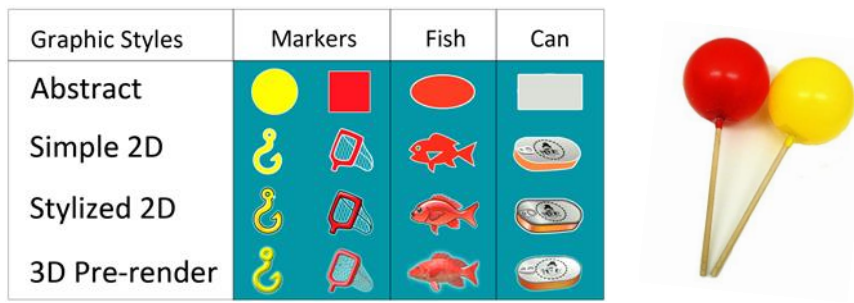

**Figure 41: Classes of visual complexity employed for the study (left). Handheld markers (right) for color-blob tracking were used to map the location of the players hand to a hook and a net that were used in the game to catch and collect.**

While the study was limited due to comparatively short interaction durations, it contributed to a classification schema for visual complexity in games (cf. Figure 41) and resulted in findings on the impact on game experience and perceived exertion due to playing a MGH for older adults. The study also informed the development of further projects with regard to acceptable levels of visual complexity, as well as by suggesting that visual complexity does not appear to be a highly promising target for adaptability or adaptivity. In the context of this thesis, the work contributes primarily to the guiding research question Q1. The work for this study with older adults in the setting of a community meetup center also helped inform the theory sections on the inclusion of third parties and members of the target group in the design of MGH, as well as the section on temporal classes of influence on the level of abilities and needs and following (sections 3.7.1 to 3.7.5).

### *4.7.2   Exercise Instruction Modalities*

Another design aspect that emerged for consideration in the implementation phase of the SDF project was the question how the exercise instructions should be realized in the games regarding the visual feedback provided to the user. It was not clear, how different modalities for conveying exercise instructions would compare to instructions provided by a live human instructor with regard to user acceptance, experience, and performance. The related publication that is appended to this thesis (Smeddinck, Voges et al., 2014) presents an experimental comparison of three instruction modalities for kinesiatric exercises: (a) a live human instructor (*human*), (b) recorded video (*video*) and (c) a virtual figure displayed next to the representation of the users' approximate skeleton (*interactive*). These conditions correspond to typical exercise instruction modalities that players of MGH might already be accustomed to (live or recorded human instructor) and compare them to the case of a virtual instructor figure (cf. Figure 42) that appeared desirable in the context of adaptable and adaptive MGH since it allows for flexibility, e.g. in personalizing exercise instructions to accurately reflect movement speed or range that would be difficult to achieve, or prohibitively costly, if prior recordings of each different version would have to be produced with a human instructor.

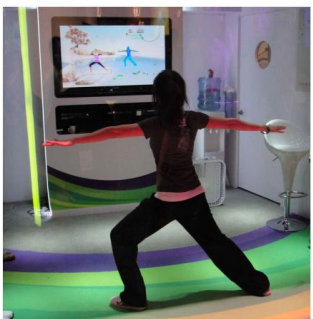 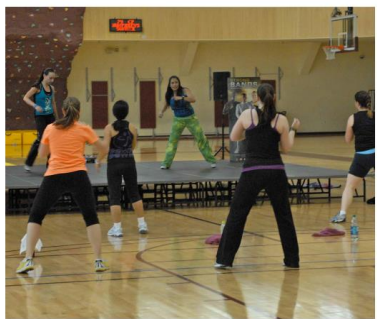 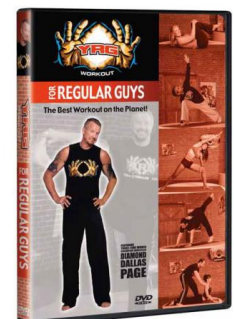

Photo by Doug Kline (CC-BY-2.0). See attribution below large version (above).

Photo by USAG-Humphreys (CC-BY-2.0). Available online at: https://secure.flickr.com/photos/usaghumphreys/7220583710

Image via Wikimedia Commons (by ladiedee28; CC-BY-SA-2.5). Attribution: Diamond Dallas Page.

**Figure 42: Common exercise instruction modalities that were compared in the study (left to right): virtual instructor figure, a live human instructor, a recorded video instructor.**

Furthermore, although Uzor et al. (2012) had already found that instruction booklets and videos as modalities for exercise instruction that do not require the presence of an instructor can lead to a number of problems (e.g. skipped repetitions / exercises, movements executed faster than intended, passive nature, no progress display), an experimental comparison regarding experience, acceptance, and performance had not yet been reported. The study thus contributes to all three guiding research questions (Q1–3). While instruction sheets are the de-facto standard for therapy exercise instructions for exercising at home, human instructors are arguably the "gold standard". They can provide motivation, feedback, and potentially also relatedness needs satisfaction. However, therapists are not always available or affordable, and human instructors can also not be expected to be completely consistent. Since Uzor et al. had already reported conclusively on the drawbacks of exercise instruction sheets, instructions by a recorded human instructor were included as a technically viable alternative baseline. It was not clear, how exercise instructions with a virtual instructor figure would compare, because they might have drawbacks due to not exhibiting fully natural movements and having less textural detail, or because humans are not as accustomed to "reading" virtual avatars, as they are to looking at – and understanding – other people. The study conditions are summarized in Figure 43, while the details are discussed in the accompanying publication (Smeddinck, Voges, et al., 2014) [**see publication B.10**].

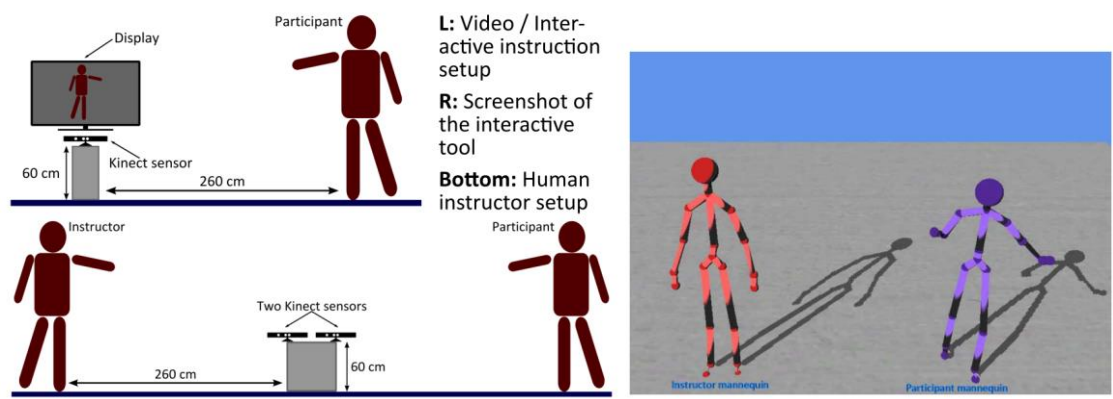

**Figure 43: A summary of the study conditions with a *video instructor* (top left), a *human instructor* (bottom left), and a *virtual instructor figure* (right; only a sample screenshot is shown; setup was identical to the top left).**

The results of the study regarding user experience, preferences, and exercise accuracy indicate a preference for the *human instructor* across measures (see Figure 44). A disparity in results exists between *exercise accuracy* and the *perceived ease of understanding* when comparing the *video* with the *interactive* modality. While perception measures indicate a slight preference for the *video* modality, the performance data analysis shows a significantly higher accuracy in the *interactive* condition (see Figure 45). This may be due to the presence of a player avatar on screen, which was not the case in the *video* condition and which may allow for a better comparability and proprioception (cf. section 3.7.3). This result informed the design of the *"butterflies"* pose-striking game in the SDF suite of games (see Figure 24).

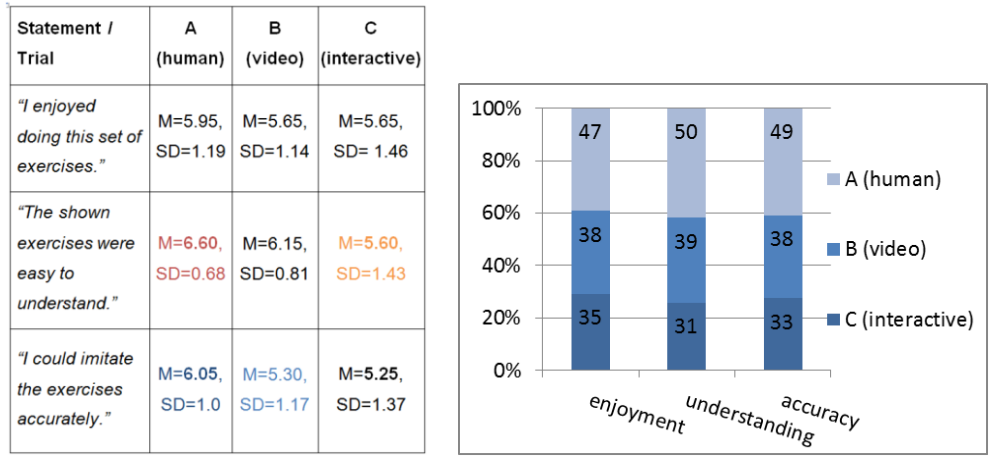

| Statement / Trial | A (human) | B (video) | C (interactive) |
|---|---|---|---|
| *"I enjoyed doing this set of exercises."* | M=5.95, SD=1.19 | M=5.65, SD=1.14 | M=5.65, SD= 1.46 |
| *"The shown exercises were easy to understand."* | M=6.60, SD=0.68 | M=6.15, SD=0.81 | M=5.60, SD=1.43 |
| *"I could imitate the exercises accurately."* | M=6.05, SD=1.0 | M=5.30, SD=1.17 | M=**5.25**, SD=1.37 |

**Figure 44: Left: The use case specific Likert-type scale items presented after each condition together with the response means and standard deviations. Significant differences highlighted through colors / contrasts. Right: Results of the final user preference ranking for each condition, regarding how enjoyable working with the instruction modality appeared to them, how easy they found it to understand the instructions, and how accurate they thought they were when working with the respective instruction modality. Numbers on the bars represent sums of rank scores.**

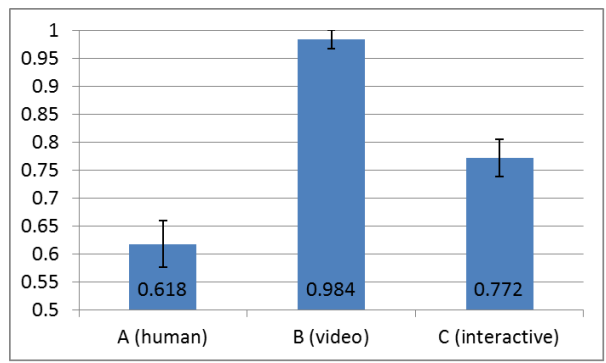

**Figure 45: The mean NMSRD (normalized mean squared rotation distance) across all participants for each condition (error bars indicate standard error); lower distance indicates higher accuracy.**

In order to acquire a measure of the exercise execution performance, participant performances in the different conditions were recorded with a Kinect (VI) device and the resulting skeleton data was compared to a gold-standard recording that was taken while producing the video instruction material and the skeleton recordings for the virtual instructor figure. Since the accompanying publication did not leave room for a technical explanation of the approach of the normalized mean squared rotation distance (NMSRD) it is included here:

Determining a distance between two posture data sets (which can be interpreted to represent a simplified skeleton) is not a trivial task. Every person has different proportions between individual body parts, which means that accuracy becomes a somewhat subjective measure (e.g. when a recipient who is taller than the instructor reaches for the exact same [mirrored and translated] position in space with her hand as indicated by the instructor, if her arms were longer than those of the instructor, her elbow would be displaced in comparison to the elbow position of the instructor). Related research indicates that in questions of exercise performance quality and accuracy, even trained professionals, like physiotherapists, do not always agree in their judgments (Pomeroy et al., 2003). The problem of unequal proportions and the resulting unpredictable displacements indicates that a comparison based on plain Euclidian distance between joint positions may not be adequate. In order to avoid strong comparison bias, the NMSRD instead relies on comparisons between hierarchical rotations of the joints, which also cannot be expected to be identical between subjects, but are a closer metric when aiming to reproduce the same relative posture setup in two skeleton representations with varying proportions. The distance $d$ between two joint rotations which are represented in Matrix form can be calculated with the *Frobenius norm* (Moakher, 2002), which takes the following form for two rotation matrices $R_1$ and $R_2$:

$$d_F(R_1, R_2) = \|R_1 - R_2\|_F \qquad (1)$$

Where the Frobenius norm for a (*m* x *n)* Matrix $R$ can be calculated as:

$$\|R\|_F = \sqrt{\sum_{i=1}^{m} \sum_{j=1}^{n} |r_{ij}|^2} \qquad (2)$$

Before applying the distance measure, the following steps for preparation and adjustment were applied:

*Relevant posture data time segments are pre-determined based on the audio cues. All timestamps for the posture data recordings (at 30 frames per second) are corrected to be relative to the individual starting frames.*

1. *If the condition is human: mirror the skeleton of the recipient.*
2. *Translate both skeletons so that the central hip joint resides in the origin (overlay A and B on top of each other).*
3. *Uniformly scale the recipient's skeleton so that the central hip bone to central shoulder bone distance matches that of the instructor's skeleton.*
4. *Calculate the local error based on sum-of-squares distances between the gold standard hierarchical joint orientations of the instructor and those of the recipient, for a preselected subset of joints that are relevant to interpreting each exercise.*
5. *Provide a grand mean of the sum of all offsets over all frames for each exercise and report the result as an accuracy performance difference metric.*

Due to the interpersonal differences in body proportions, the distance metric results following such an analysis procedure are not fully comparable in absolute values. Therefore, the sums of mean rotation differences for each exercise in every condition were normalized per participant against the condition in which that participant performed with the largest difference in comparison to the instructor posture data, resulting in a comparable measure which can be summarized as *normalized mean squared (relevant hierarchical joint) rotation distance* (NMSRD). The resulting relative differences between conditions of all participants were then subjected to a one-way repeated-measure analysis of variance. In order to minimize synchronization errors at the beginning and the end of each exercise block, only the repetitions 3 to 8 (out of 10) of each exercise were considered.

Since the NMSRD results are difficult to contextualize, because they do not correspond directly to an absolute measure but rather represent a measure of relative offset, a comparison between the non-normalized mean differences across trials was also computed, including the mean performance of the instructor (as captured during his performances in the *human* instructor condition).

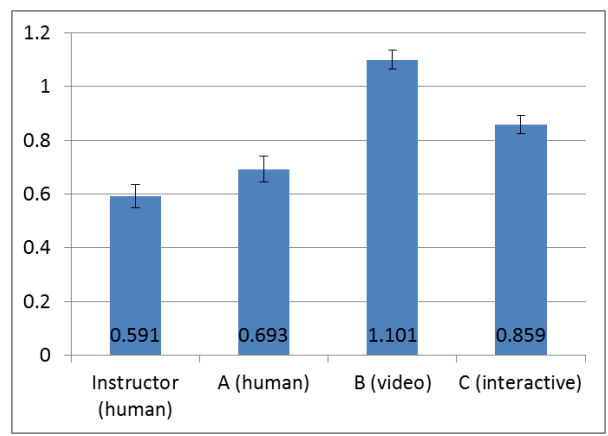

**Figure 46: Mean squared rotation distances in all conditions, including the mean performance of the trained instructor (error bars indicate standard error).**

This analysis (see Figure 46) shows that the performance of the trained instructor already displays a considerable mean difference from the original "gold standard" performance and illustrates that the mean performance difference in the *human instructor* condition is rather close to the best performance that a trained instructor was capable of achieving with the given exercise sets and under the conditions of the study. Furthermore, the differences resulting from the worst condition for the participants of this study (*video*) were less than twice as distant from the gold standard as the differences displayed by the instructor.

In summary, the *human instructor* was preferred and resulted in the highest performance accuracy, as expected on a naïve theoretical basis. While the *interactive* condition at times appeared difficult to understand, and led to varying perception, participants showed significantly better performance accuracy than in the *video* condition. This was somewhat surprising given the comparatively low quality of the employed virtual instructor figure and runs contrary to more recent findings from a study on the judgment of quality of motion of exercise executions [**see publication C.9**] by convenient subjects (Sarma et al., 2015). However, it is not unreasonable to assume that there may be a difference with regard to the abilities of judging the quality of motion executions compared to physically imitating movement executions based on a video or a virtual instructor modality. Improvements to the visual appearance and detail of the instructor figure could arguably be expected to improve the perception. The results informed the SDF project in proceeding with a virtual instructor figure in those games of the suite that required specific movement instructions.

While future work is needed to separate the impact of interpretability of a photorealistic / real human, consistent reproduction of instructions, and the role of constant feedback in more detail, the findings support the further investigation of digital interfaces to support physical therapy and rehabilitation as a cost-effective and potentially more efficiently customizable addition to traditional exercise instruction forms. The work also resulted in a measure for exercise execution distance from a gold standard performance called NMSRD. Since live and present human instructors are not an option for MGH and the results did not conclusively indicate major drawbacks when employing virtual instructor figures, the findings can be seen to support their use in the context of adaptable and adaptive MGH. However, explorations on the potential effects of personalizing adjustments to the avatar appearance or to the exercise instruction animations are open targets for further study.

## 4.8   Acceptance of Adaptive MGH

As the following paragraphs will further underline, the early explorations on MGH for the support of physiotherapy for people with PD clearly pointed towards a need for flexible adjustments that could not be met by employing traditional difficulty choices via a single difficulty parameter with a few coarse discrete steps. In addition to exploring a combination of initial calibration and following adaptivity based on predefined heuristics that processed player performance, a first study served primarily to gather insights regarding the question whether adaptivity in MGH would be accepted (thus contributing in the direction of the guiding research question Q2). The study also served to start gathering insights regarding the effectiveness of such an approach (cf. guiding research question Q3).

> *Although the game prototypes [from the original WuppDi suite] were repeatedly adjusted to be easier to use and understand, and despite the fact that difficulty choices were available (as classic options or changes in levels), early evaluations suggested that the games were not able to fully cover for the very heterogeneous abilities and needs of individual users from the target group. As a result the game Sterntaler (Figure 47) was augmented with an adaptive system based on heuristic DDA (Smeddinck, Siegel, et al., 2013).*

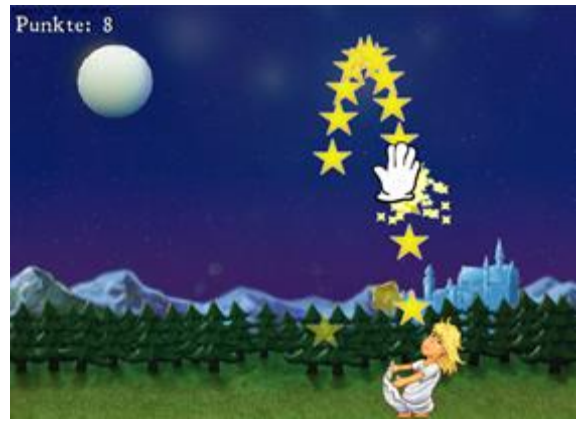

**Figure 47: The game** *"Sterntaler"*. **The hand cursor is controlled by a person standing in front of a screen or display. Players have to collect streaks of stars that mirror therapeutically helpful motion.**

*Details regarding the calibration of **range of motion** (ROM), the adaptivity module and the results of a first evaluation can be found in Smeddinck, Siegel et al. (2013) **[see publication A.1]**. In brief, while the calibration focuses on ROM, the game was changed to be adaptive with regard to three aspects: ROM, accuracy, and speed (Figure 48).*

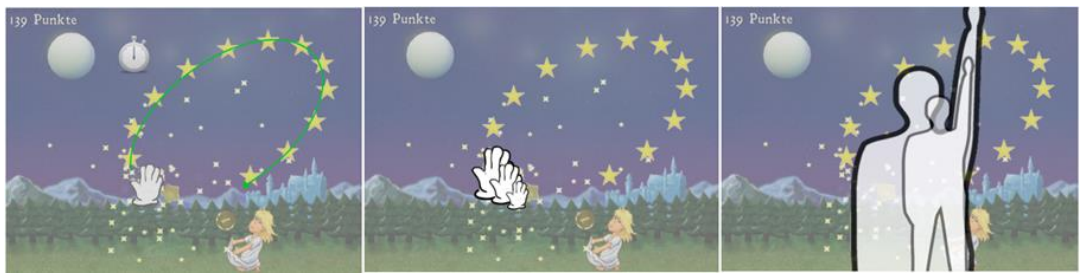

**Figure 48: Three adjustment mechanisms employed in the adaptive version of Sterntaler: speed, accuracy, and range of motion (left to right).**

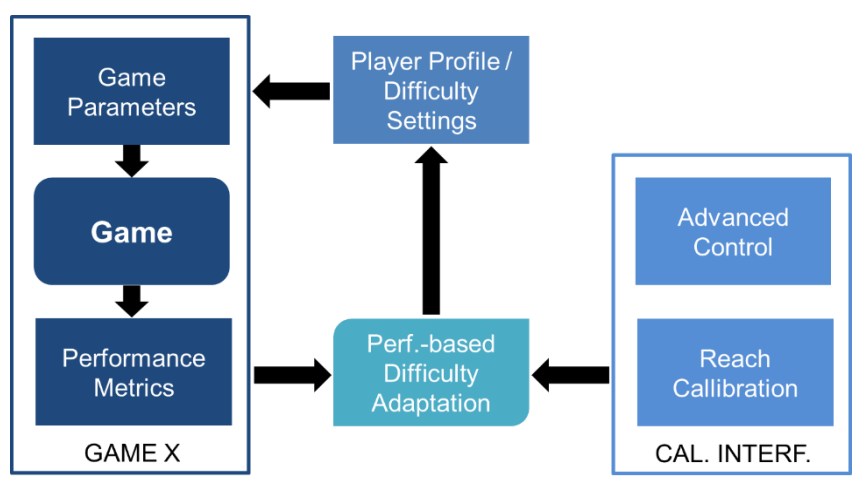

**Figure 49: Schematic overview of a game architecture with mixed adaptability and adaptivity. Elements for advanced control allow for manual adjustments, while reach calibration and performance-based difficulty adaptation implement automated adjustments.**

*Figure 49 shows an extension of the general model with a performance evaluation and a difficulty adjustment mechanism component mentioned above, by introducing system boundaries, configuration, and calibration. A study of the adaptive components was performed with three participants over a course of five sessions (Smeddinck, Siegel, et al., 2013).*

*The results were promising in that the participants managed to reach the calibrated target ROM thresholds almost without exception during the study period. They also accepted the games and were not irritated by the updated adjustment mechanisms. However, some challenges became clearer and additional challenges also became evident. The visibility of adjustments, for example, proved to be potentially problematic, even though this was a single player situation […]. Some questions also arose around the scoring. Remarks by the participants in different directions indicated that the question whether scoring should be adjusted to reflect the current internal difficulty level requires further study. Lastly, even with only three participants, the detailed analysis underlined the existence of extreme interpersonal differences and even the same person could perform very differently with one arm compared to the another, or perform differently from day to day, for example due to variations in medication (Smeddinck, Siegel, et al., 2013).*

*Excerpt from: (Streicher & Smeddinck, 2016)*

The iterative human-centered development process for this project was carried out with the support of PD patients and therapists. It included extending the underlying game from the *WuppDi* suite of MGH for PD patients with more robust Kinect body tracking for game control, and an integration of the calibration tool and modules for adaptivity produced the foundations for the modular development approach for MGH that was refined during development process for the SDF project and discussed in section 3.3. The study also makes for an early example of a situated study, since the participants with PD completed the trial play sessions in meeting room facilities of the *German Parkinson's Association*, which was a partner for the project. Due to the explorative nature of the work, a close case study observation of a few patients over multiple sessions was preferred over the possibility of a shorter observation period with more patients. This allowed the gathering of a detailed picture of the abilities, needs, and motivation of each individual participant as they were exposed to and kept on interacting with an adaptive MGH. These detailed observations clearly showed substantial differences between just three participants that were sampled from a specific target group, further underlining the need for adaptability and adaptivity in MGH (cf. section 2.1.6). The system was found to function well and to be well received. However, some observations warrant cautious consideration. During development, the initial adjustment method for the *range of motion* parameter was based on scaling the shapes that the stars would form on the screen. This was clearly noticeable and irritated some test participants (leading them to question their abilities, especially if the stars were brought closer in order to make the game easier to play). A solution was found in scaling the input that was mapped on an invisible player skeleton representation. This observation regarding the potential impact of adjustments being unmistakably visible and potentially perceived as intrusive, or, in contrast, being indiscernible and unobtrusive, triggered the approach to a later study on the impact of these factors on the perception of balancing for physical abilities (cf. section 4.10.1) and also motivated the inclusion of the *saliency* and the *intrusiveness* modeling dimensions for adaptable and adaptive MGH (cf. section 3.4.2). Furthermore, while *perceived difficulty* correlated with *personal performance* for the study participants, it did not notably correlate with the actual absolute difficulty, which may be one reason explaining instances of participants questioning the composition and meaning of the final scores that they managed to achieve, since they wondered about not necessarily achieving greater results in later play sessions although they felt that they had shown an increased performance. This later led to the design of reward systems that would include a score multiplier to reflect increases in difficulty. Lastly, multiple statements by the participants indicated that they were well aware of the targeted serious outcome and they expected to be challenged and would be willing to accept (or even expect) limited success given their condition. Arguably, such a stance can allow for an internalization of the extrinsic motivation of playing games for health, delivering an approach for explaining the positive impact on aspects of intrinsic motivation that was observed in a later study (cf. section 4.9). Related work has since explored and discussed the role of framing motion-based game use either as exercising or as game play (Zaczynski & Whitehead, 2014). These observations also informed the discussion of the double role of users of MGH as both players and patients in this thesis.

The decision to work with pre-designed heuristics with parameter optimization in controlled boundaries as opposed to more explorative approaches to adaptivity was carried over into the later SDF project, since the outcomes of this study underlined the considerable heterogeneity of MGH players, which presents a challenge for any methods that are based on group similarities or supervised learning. This, in turn, informed the according discussions in the prior sections of this thesis (cf. sections 2.1.5, 3.2.3, 3.2.4, and 3.7.1).

While the extremely heterogeneous participants underlined the need for personalization via adaptability or adaptivity, the tools developed for these early explorations were found to be unfit for third party use outside of the context of the study, and the parameter set that was designed for the specific application interest of extending the range of motion was found to be too limited for more general use with MGH for the support of PRP. Furthermore, the system employed in this study allowed for only one pre-calibrated target to be defined for each player and any adjustments beyond reaching that first goal were considered beyond the scope of the study. These limitations, together with the low number of participants, motivated the development of a more capable adjustment and configuration interface for the project SDF that included a system of unlimited milestones for each player (or group of players), which were defined on a broader set of parameters (see the discussion in section 3.7.6 and the description of the interface in sections 4.5 and 4.9), as well as the implementation of a larger and comparative medium-term situated study that will be discussed in the following section.

## 4.9   Prolonged Use and Functional Impact of MGH

Following a range of studies on specific design questions and early explorations on the acceptance and potential effectiveness both for improving motivational factors, and with facilitating functional improvements, as described in the previous sections, a study that aimed for a medium-term situated, controlled comparison between non-MGH physiotherapy, therapy with the support of MGH, as well as therapy with the support of adaptive MGH, was still outstanding in order to gain further insights into the acceptance and effectiveness of MGH in a realistic setting. The following sections summarize the according study that corresponds to the guiding research questions Q2 and especially Q3.

> *[As described in the prior sections, f]ollowing up on the WuppDi project, the project Spiel Dich Fit (SDF) aimed at integrating a broader number of options for adaptability and more advanced semi-automatic adaptivity. A suite of MGH was developed implementing activating movement games for older adults and for explorative use in physiotherapy, prevention, and rehabilitation (Smeddinck, Herrlich, et al., 2015) **[see publication A.2]**. Accordingly, a configuration tool for therapists was developed alongside the games (Figure 50). Both elements were developed with a user-centered iterative design process featuring multiple formative exploratory studies and continuous evaluations. [...]*

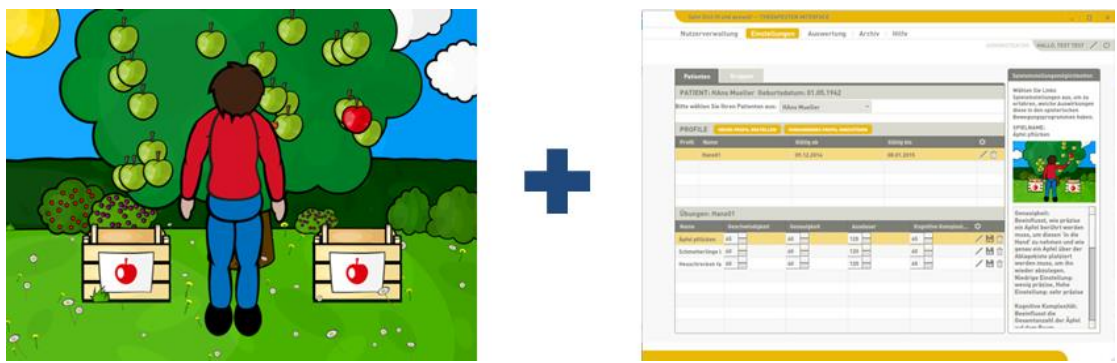

**Figure 50: Settings tool that allows therapists to perform detailed manual adaptations.**

*The project was followed by a medium-term study of the situated use of the games and settings interface in a physiotherapy practice over the course of five weeks (Smeddinck, Herrlich, et al., 2015). The player performance, functional development, and experience have been compared, as well as the therapist experience between a group working with activating movement games in a garden setting with a manual settings interface. The study included a group of therapists and participants working with the same games and interface, but with added semi-automatic adaptivity, and a control group that performed traditional physiotherapy exercises without MGH. Initial results showed mixed impacts on experiential measures while the physiological measure of functional reach increased significantly more in both games groups than it did in the traditional therapy group over the course of five weeks (Smeddinck, Herrlich, et al., 2015).*

*With regard to adaptability, the settings interface allowed the therapists to perform personalized settings for range of motion (by active game screen zones; Figure 51), required speed, motion accuracy, endurance (level duration) and complexity (amount of active objects) for each level. These settings were subject to a pre-study with ten therapists and were evaluated as being "easily understandable", "useful for making meaningful adjustments", and "allowing for efficient configurations". Since therapists are likely to be able to make informed decisions about the physical abilities much more so than about specific game performance abilities of their patients, the decision was made to express options for manual adaptation as parameters that relate to the player abilities (such as at what speed a person can or should move) and not to specific ingame variables (e.g., how many apples per minute should appear on a tree to be picked by the player). However, this does result in the need for a translation of the difficulty settings that were performed by therapists into difficulty adjustments of actual ingame variables. In SDF, these were achieved via the linear mapping of thresholded variable ranges to normalized difficulty parameters from the interface. Since the project did not encompass extended user models or a heavy focus on translatable settings, the therapists supplied settings per player per game. An additional layer of translation would be necessary to achieve the most efficient way of providing settings; only once per player; which could then be transformed into normalized parameters for a potentially broad number of games. This method was found to be preferred by the therapists, but it requires extensive balancing and would likely benefit from complex adaptive techniques that exceeded the scope of the project.*

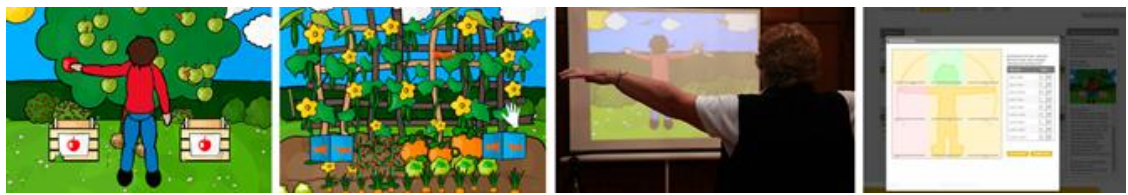

Figure 51: (Left to right) Screenshots of the games from Spiel Dich Fit employed in the medium-term study; apple picking and catching locusts; balancing with butterflies (with a player from the target group); the settings interface for controlling adaptations and adaptivity.

*The approach to adaptivity in SDF is illustrated in Figure 52. It featured a mixed model for semi-automatic adaptivity. In that approach, manual settings provide keyframes [also called "milestones"] between which optimal, but manually determined candidate settings are interpolated. Player performance is then allowed to fluctuate around the target level of performance within a certain threshold (Figure 52). Dynamic difficulty adjustments are performed if the thresholds are violated in order to assure an adequate game experience while still driving the player towards the targeted serious goals. This model was found to be well-accepted by the therapists in the study and allowed for steady performance increases without notable frustration among the patients (Smeddinck, Herrlich, et al., 2015).*

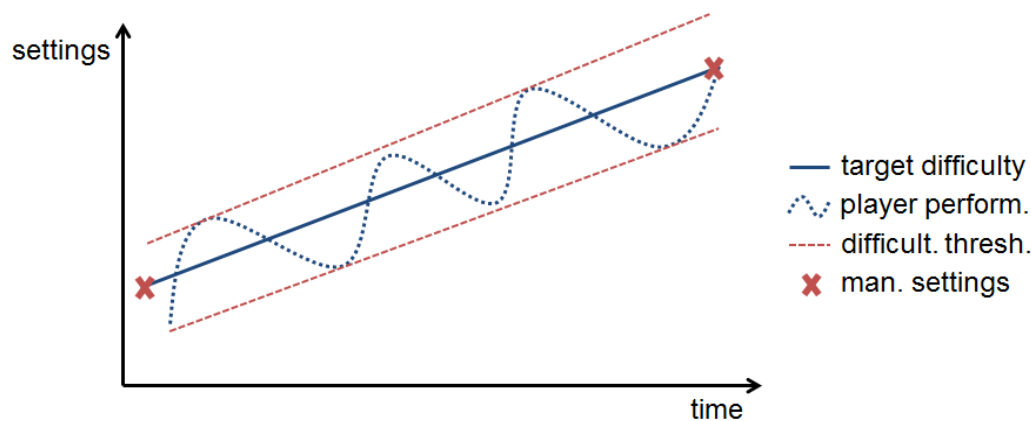

Figure 52: Adaptivity approach in the Spiel Dich Fit project by dynamic difficulty adjustment.

*Excerpt from: (Streicher & Smeddinck, 2016)*

As indicated above, the study was carried out in the facilities of a large physiotherapy practice in northern Germany, with the support of 12 participating therapists who were allotted to the three separate treatment groups. A therapist was present as the contact person for the patients in each of the five weekly sessions that the participants attended and that lasted 20 minutes. Patients were sampled aiming for an average age of 65+, and sampling targeted patients with chronic upper-body back and neck afflictions, leading to 29 subjects being allotted across the three treatment groups. The study employed a large range of measures in order to facilitate the triangulation of outcomes, since the results were expected to be noisy, given the complex situated setting. Further details are discussed in the accompanying publication. The results discussed in this section are based on contrasting questionnaire responses drawn before the first and after the last trial session of each participant. The findings indicated that using the games were beneficial to the perceived *autonomy* and *presence* expressed as need satisfaction components, while the classic therapy sessions showed higher *tension-pressure* and *effort-importance*. Each of the measures is positively related to intrinsic motivation. While no significant differences were found on these measures when comparing the two versions of the game suite, automatic adjustments were preferred by the therapists, and the game groups showed significantly increased *functional reach* (FRT) compared to classic treatment (Smeddinck, Herrlich, et al., 2015). In the context of the SDF project, these findings were interpreted to support the integration of semi-automatic adaptivity, since it was not found to have notable negative impacts on the player experience while therapists indicated a preference for automated support in adjusting the games to the individual patients.

Thus, the study produced a number of interesting observations, significant outcomes, and served to inform ongoing and future projects. However, despite the comparatively complex medium-term situated setup with quantitatively telling group sizes, the study still leaves room for improvements. It showed promising effects with non-frail older adults, and the options for adaptability were found to be telling and easy to use for therapists. However, the semi-automatic adaptivity did not lead to large differences between the two game groups (i.e., compared to purely manual settings). The mixed model with human-centered difficulty parameters which where configured on a per-player and per-game basis requires at least semi-regular active involvement by therapists. Furthermore, automatic adjustments in some cases quickly reached thresholds that did not respect the abilities and needs of all members of the target population, so that notably suboptimal situations did occur. However, while the lack of challenge for some participants towards the end of the study period may have led to periods of light workout, relating to the model presented in Figure 15 in section 3.6, the improvements in the FRT results compared to the regular therapy sessions indicate that the settings did not lead the participants to cross into deterioration. Longer-term studies with larger group sizes and controlled conditions that hold up to the requirements for clinical trials are still outstanding for this branch of MGH. To such ends, the project did inform the planning of future clinical studies with grounded effect size estimates for the motivational and functional outcomes.

Regarding the guiding research question Q2 and Q3, the outcomes indicate that the games and the surrounding application workflow were well received, with *acceptance* lasting over the medium-term duration of the study. The MGH showed to be clearly *effective* with regard to the significant improvements on the functional reach test, which can serve as an indicator for balance and thus hints at a positive impact in fall prevention (Sherrington et al., 2008). Fall prevention plays a major role in preventing complications for older adults (Kannus et al., 2005). These functional indications are in line with approaches in more recent related work (Marston et al., 2015). The differences in motivational measures were not unequivocally in favor of the MGH compared to therapy without MGH, which underlines the need to carefully consider the motivational impact of MGH. The interviews and observations obtained as parts of the study procedure hinted at potential reasons for these results, e.g. since the therapists were observed to be interacting more closely and gave more directions to the patients in the traditional therapy condition. This underlines the importance of taking a wide perspective into account, and of potentially triangulating measures (cf. sections 3.7.4 and 3.7.5). It also points towards considering virtual therapists, or patient-to-therapist communication features as a potential source for broader motivational needs satisfaction in future developments. All participants were able to immediately start working with the games, using custom settings that were enacted by their therapists. The therapist involvement in this form of semi-automated assisted mechanism that relied on implicit performance measures for the heuristic adaptivity that was employed between milestones, yet allowed for explicit manual control by therapists and for manual adaptability at any time, facilitated personalized experiences without any cold-start problems and without the risk of a fully automated system becoming locked in a local optimum (cf. section 3.2.3). It was noted, however, that the predefined maximum parameter thresholds were reached for a number of participants towards the end of the study. This resulted in further adjustments of the thresholds for the ongoing SDF project and highlights the importance of an iterative consideration of the impact of automated adjustments (cf. section 3.8). Furthermore, the therapists only rarely made use of the opportunity to adjust treatment plans that were already active. This may be due to the medium-term duration of the study, which was planned to cover the minimum duration for engaging in physiotherapy that typically leads to first notable improvements, as indicated by therapists for the given conditions when planning the study. Nevertheless, therapists clearly preferred the mode with automatic adjustments when considering the different conditions in a final interview session, indicating that they would not be able to find the time to frequently enact manual settings, or a manually triggered calibration, for each individual patient. The range of motion configuration option via a special interface with zones was well received, indicating that the conclusions from the prior exploration on configuration modalities (cf. section 4.5) appear validated in situated use. Lastly, therapists expressed the intention to experiment with employing the games from the study in contexts that exceed the originally intended target group (e.g. with younger audiences, in occupational therapy, or in a modification where patients would be asked to stand on a non-digital balance board while playing the games in order to achieve further increases in difficulty and secondary training effects). These observations further informed the prior remarks on user generated context (cf. section 2.1.6 and 4.5).

While further studies are needed, as indicated above, in order to generalize these findings, the indicated partial improvements can be considered a successful demonstration of *acceptance* (Q2) and *effectiveness* (Q3) of adaptable and adaptive MGH in a realistic setting in the targeted place of use. Arguably, due to the potential cost benefit of MGH, because of the range of possibilities they offer for guided training without requiring a physiotherapist to be present, even results that are on par - or almost on par - with traditional techniques, would warrant further consideration of MGH as a promising augmentation for use cases in PRP.

## 4.10 Beyond Adaptability and Heuristic Adaptivity for Casual Style MGH

The studies and developments summarized above have delivered insights of practical relevance to ongoing, future, and related projects, and contributed to the surrounding theory on adaptable and adaptive MGH regarding design aspects, acceptance, and effectiveness. Next to the individual limitations and avenues for future work that are discussed in the respective publications, the projects share a number of common limitations. The focus on aspects of human-computer interaction with interfaces for adaptability and heuristic adaptivity in MGH for the support of PRP for older adults or people with PD using straightforward casual types of games results in limited generalizability and leaves many avenues for future work. While the limitations and the remaining directions for future work will be discussed in further detail in section 5.2, a number of additional developments and studies have already been carried out in the broader context of the thesis.

Akin to the prior explorations regarding design aspects of MGH (see e.g. section 4.7.1), further aspects can be scrutinized with regard to their pertinence as targets for flexible adjustments to personalize a MGH, such as the impact of rhythm and timing on player experience and performance (Lilla et al., 2012). Additionally, different types of targeted usage contexts are likely to interact with the acceptance and effectiveness, as well as with the player experience and the resulting motivation to engage in playing and beneficial activities, of interacting with MGH. The following sections discuss multiplayer and social aspects in relation to adaptability and adaptivity, as well as expanding use cases. This includes explorations on further application areas for GFH, on encompassing further interaction modalities, on approaches to improving medium to long-term motivation and adherence, as well as on more loosely coupled configurations of exercising and gaming.

### 4.10.1 Multiplayer and Social Aspects

Techniques from adaptivity cannot only be applied to single-player games, but also to multiplayer games. In this context, dynamic difficulty adjustments are often referred to as balancing, and typically aim to help players with different skill levels to be met with equally rewarding game experiences. Little is known, however, regarding the impact of different types of *adjustment mechanisms* on *game experience* and *performance*. Furthermore, impacts on self-esteem can be expected due to the complex social dynamics of joint play, especially when differences in player proficiency result from given abilities, rather than learned skill, which can become especially evident in motion-based games when players with a notably different physique engage in joint play. The attached publication (Gerling et al., 2014) **[see publication B.1]** presents two studies comparing three balancing approaches in a dance game for two players that implemented the basic game mechanics of *Dance Dance Revolution* (see Figure 53).

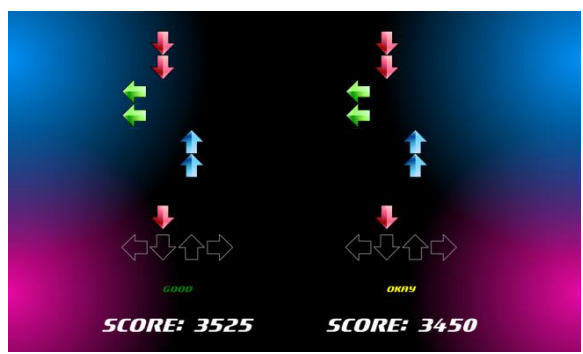

**Figure 53: A screenshot from the dance game that was implemented for the study. Players use a dance mat with four active sectors to step on, aiming to hit the right switch at the moment where an according falling arrow enters the "hit zone" (grey arrow outlines).**

The different balancing approaches were (a) *input balancing*, which presented adjusted step charts for each player, (b) *time balancing*, which notably increased the time window for scoring good hits by invisibly scaling up the "hit zone", and (c) *score balancing*, which simply set different internal score multipliers for each player. Further details of the implementations are described in the paper. In terms of the modeling dimensions for adaptive games, these conditions thus represent different levels of *saliency* (with input balancing being more salient in principle than the remaining two approaches, due to visible changes in arrow patterns), and also potentially according *intrusiveness* (with score balancing being less intrusive in principle, since it did not interfere with the gameplay itself), while the *positioning* of the adjustments also differed between conditions (score balancing represents an adjustment to the feedback, whereas the other two approaches reflect adjustments to the player control and input processing). The first study with pairs of regular and generally healthy individuals indicated that *input balancing* best balanced the achieved step count between players, yet as the most visible (or salient) approach, it made the difference in ability between the participants explicit. This led to negative effects on the weaker participants' feelings of self-esteem and of their own performance compared to that of their partner. Stronger players experienced reduced feelings of relatedness. *Time balancing* allowed for the most balanced score differential, but did not enable the weaker player to win more games. This approach turned out to be the least visible and least perceived by stronger players. It also had positive effects on self-esteem for both the weaker and the stronger players. *Score balancing* allowed the weaker players to win more games, but it also led to weaker players outscoring the stronger players in the score differential, arguably overbalancing the games. Despite being a rather crude approach, it was the least perceived approach for weaker players and led to higher self-esteem for them. Regarding the *saliency* of adaptivity for motion-based multiplayer games, the results indicate that openly visible adjustments, which are often employed in commercial games (cf. section 2.8.1) may lead to reduced self-esteem and feelings of relatedness in pairs of players. This runs contrary to the insight that salient adjustments can be surprisingly well accepted in single-player games that was discussed in the prior sections. Furthermore, the results of this study indicate that more hidden balancing methods could improve self-esteem and, in the case of *time balancing*, reduced score differentials without affecting the game outcome. It is notable that the adjustments in *time balancing* occurred through invisible adjustments to the control mapping, in ways that are comparable to the *input scaling* that was discussed above (cf. section 4.8), suggesting that inconspicuous adjustments to user control mapping (cf. to the *positioning* dimension) to game mechanics can support adaptivity with good results in single-player and multiplayer settings alike.

While the dyads of players in the first study showed somewhat different physical abilities and game skills, the follow-up study set out to examine the impact of employing an improved combination of the rather indiscernible methods of *time balancing* and *score balancing* on players with more radical differences, by focusing on dyads where one player had a mobility disability and used a wheelchair (see Figure 2). The game was found to be accessible and enjoyable to people using wheelchairs following adjustments and a control scheme based on prior work on motion-based games for wheelchair users (Gerling et al., 2015) **[see publication C.2]**. The experience ratings and player feedback showed that people using wheelchairs enjoyed the integration of their assistive device into the game. However, performance differences within dyads showed that balancing may have to be increased to bridge gaps in scores. While players using wheelchairs did note that they did not mind frequently losing to the able-bodied players, some able-bodied persons expressed discomfort beating a person with a disability. This can be seen as further motivation to improve approaches to dynamic difficulty balancing in multiplayer games to simultaneously empower weaker players and improve the experience of (much) stronger players.

While social aspects and individual perceptions of fairness and respective abilities certainly play a role in competitive multiplayer games, different outcomes can be expected with regard to cooperative multiplayer exergaming. A study was setup to compare the impact on player experience and performance of either *strong cooperation*, in which players take different roles and can only reach the goals of a game if they interlace their activities in adequate patterns, or *loose cooperation*, in which players take equal roles and do not depend on complex interactions in order to reach the goals of a game **[see publication C.5]**. Employing a "window washer" game that was developed as a MGH for people with PD. In the game, players had to clean a window in order to answer the question what is hidden behind it (Figure 54). It supported both a strong and a loose cooperation mode (see Figure 55), and the study led to the finding that strong cooperation can entail benefits in increased coordination and communication between players, resulting in higher overall scores. However, these strengths did not notably impact the player preference, since 50% of the participants preferred the loose cooperation mode compared to only 27% who clearly favored the strong cooperation mode.

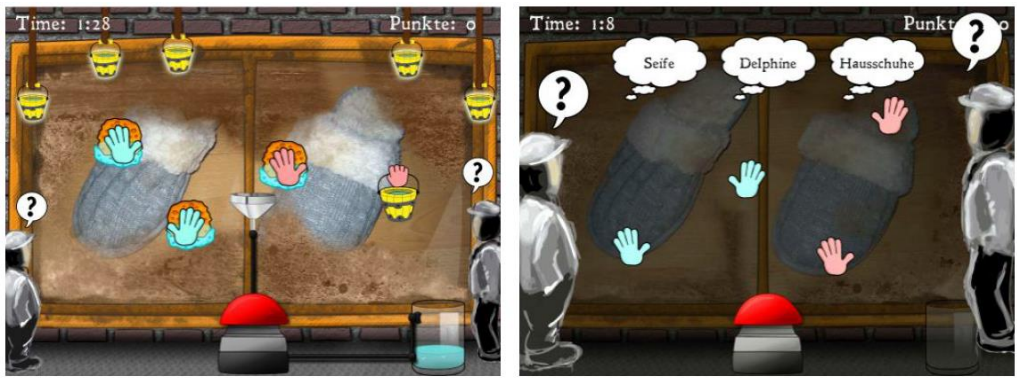

**Figure 54: Main game screen with partially cleaned windows (left) and answer screen.**

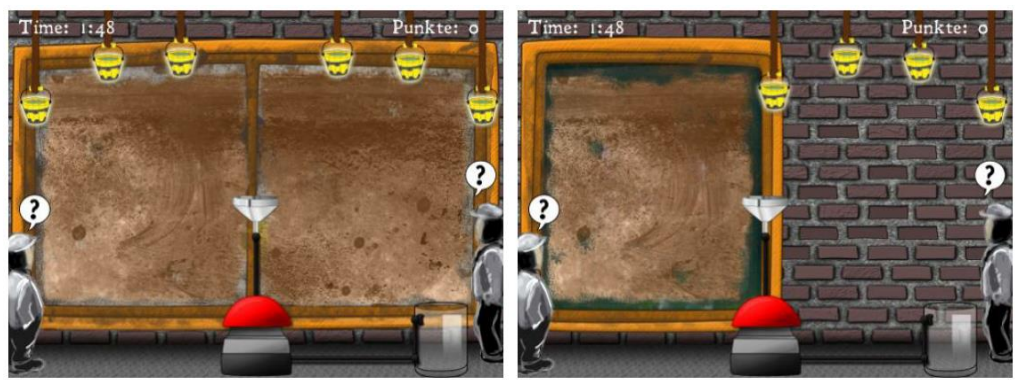

**Figure 55: Loose cooperation mode (left) and strong cooperation mode (right).**

Regarding adaptability and adaptivity, these results may appear beneficial, since balancing symmetric roles is arguably less complex than balancing asymmetric roles. However, future work is needed to further discern potential effects of anxiety and relatedness in the context of cooperative multiplayer MGH. Drawn together, the considerations from these studies suggest that adding social aspects, such as multiplayer gaming, can lead to good player motivation in motion-based games both with comparatively similar and with extremely different players, potentially increasing the motivation to play and thereby exercise via increased relatedness needs satisfaction. However, adding multiplayer induces social aspects that can lead to additional challenges for adaptability and adaptivity (cf. section 3.7.4).

### 4.10.2 Expanding Use Cases

Expanding use cases for SG where adaptability and adaptivity may play a considerable role can include both the exploration of alternative target groups for MGH, and the exploration of additional GFH application areas, such as games that are physical but not motion-based. One example for the latter group is the game designed for supporting voice therapy for PD patients that was mentioned in section 3.7.6 (see also Figure 18) that is controlled using microphone input and was shown to be well received. This game led to considerable increases in loudness over time in a first explorative evaluation (Krause et al., 2013), and offers a much more simple parameter space regarding game controls than full-body motion-based games [**see publication C.8**]. It can thus be considered as an interesting testbed for controlled comparative studies on adaptability and adaptivity for GFH in future work.

As an alternative to employing non-motion based games to other health purposes, motion-based games can also be used for other serious purposes outside of health. A modified tower-defense game that was designed to generate samples for *robot imitation learning* using motion-tracking of the players' hands (Walther-Franks et al., 2015) makes for a good example in this regard (see Figure 56). The according study compares between using a game for action executions and demonstrating actions in a plainer virtual environment. It showed that such a form of human computation can lead to increased interaction times and better user experiences when playing the game. Although aspects of adaptability or adaptivity were not yet considered as study variables in that application area they are of interest, since a large number of motion executions are required and motivating players to provide a large number of sample executions can arguably benefit from optimized levels of challenge [**see publication C.11**].

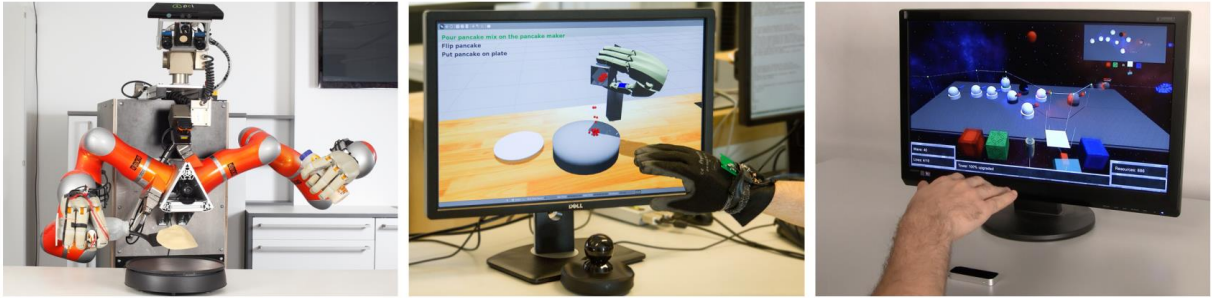

**Figure 56: Many robot manipulation tasks (left) are learnt best by observing humans. Human motion can be demonstrated using virtual environments (center). Data acquisition with motion-based games makes demonstrating a fun activity (right).**

Another area that is of interest for expanding use cases of MGH is the aim to overcome purely heuristic adaptivity with parameter definitions that are configured separately for each game, by offering fully human-centered adaptability and adaptivity that evolve around one single detailed patient-player model per user, which is then employed to map adjustments to a number of different MGH. Accordingly, this is a central aim of the project *Adaptify* that is currently underway as a third generation MGH project developed at the *TZI Digital Media Lab* at the *University of Bremen*. The same project also explores the development and integration of a pressure sensitive training mat as an input device that can overcome limitations due to limited view angles of optical tracking devices such as the Kinect, aiming to facilitate a more reliable tracking of exercises that are executed while kneeling or laying down, as these play an important role in PRP for chronic lower back afflictions, which is the primary health application target in that project. Such alternative input devices (see Figure 57) can play an important role for adaptability and adaptivity, since prior projects have shown that largely hidden adjustments in input interpretation and mapping to game actions can provide a promising approach for unobtrusive yet beneficial adjustments.

Additional input devices have been explored in the context of locomotion in virtual worlds that can arguably support other forms of PRP. In an early qualitative comparison study, physical locomotion with a suspended torso harness that can be used for specific forms of supported walking rehabilitation and potentially increases the accessibility of motion-based games to user groups who have so far been excluded, has been explored for walking motion control in FPS games, comparing the user experience to walking-in-place without a suspension harness and investigating techniques for implementing common in-game actions, such as turning, pushing buttons, etc. The early outcomes indicated that *suspended walking* has potential as a natural walking controller, although correct harness adjustment was challenging and walking-in-place without a harness was perceived as more comfortable (Walther-Franks et al., 2013b) **[see publication C.13]**.

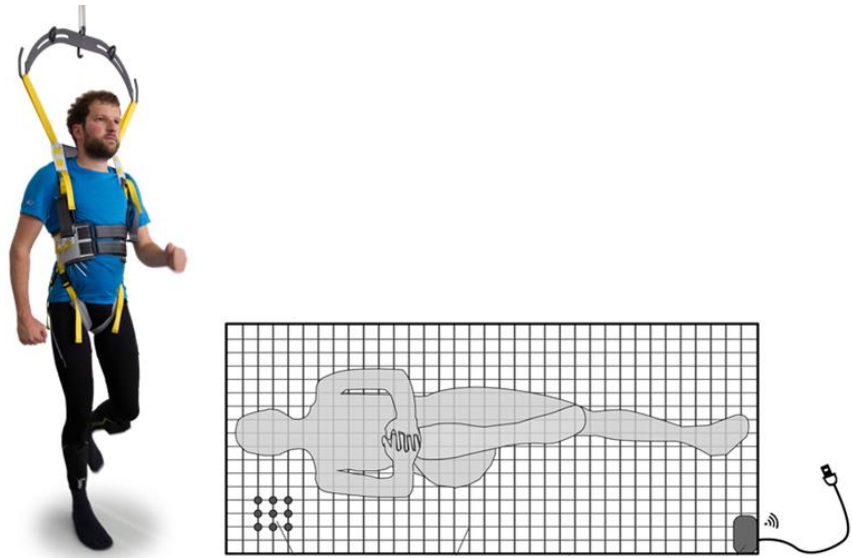

**Figure 57: The figure shows alternative input devices that are being explored as input technologies for MGH. Left: A suspension harness that allows for mapping comparatively natural walking movements to navigating virtual spaces. Right: A functional schema for a pressure sensitive mat that can be employed to provide an additional view of the body configuration for MGH players, facilitating more accurate tracking of near-ground exercises or exercises, as well as more accurate body balance estimates.**

A further important avenue towards expanding use cases for MGH that requires considerations on adaptability and adaptivity concerns *means to achieve sustained motivation* (especially medium- to long-term). Interesting and entertaining content can be one of the best sources of motivation in the context of video games. However, well-crafted manually designed game worlds, mechanics, and storytelling are challenging to produce and often beyond the budget of MGH productions. Therefore, the possibility to convert off-the-shelf video games with high production quality into exergames or MGH has been explored. In early work in this direction, the *Exercise My Game* (XMG) design framework was developed for this purpose, including approaches to the challenges of finding adequate mappings from control input to game-actions, and of blending active input feedback with the interface of a given game [**see publication C.12**]. XMG was illustrated along the example of a conversion of the popular first-person action game *Portal 2* into an exergame with promising results in an early informal evaluation. Figure 58 shows a screenshot from the conversion where an overlay is used to convey feedback regarding the motion-tracking and recognized control, while exergame-related movement instructions are embedded into the game world, which features custom levels that were designed for exergame purposes using a modular level editor (Walther-Franks et al., 2013a). Related work has since continued explorations in this direction, finding that such game conversions can support anti-sedentary levels of exertion while not affecting players' enjoyment (Ketcheson et al., 2016). However, a successful tight integration can be difficult to achieve and the pacing of the regular video games is often not adequate for engaging motion-based control.

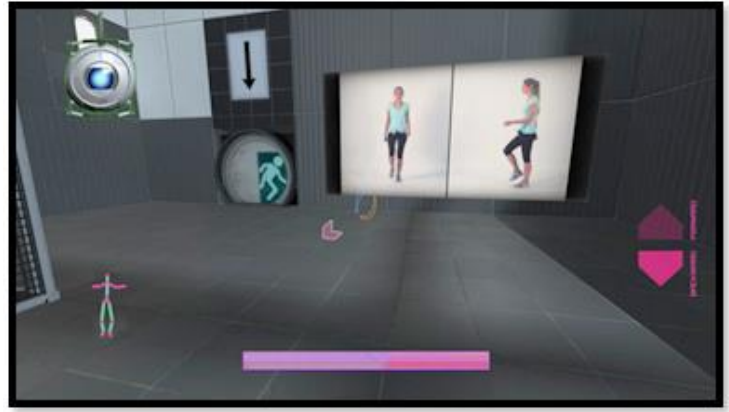

**Figure 58: A screenshot from a conversion of the game Portal 2 into an experimental exergame, featuring a full-screen overlay to provide feedback about the motion-tracking and the recognized motion actions, as well as motion-game specific instructions that were embedded into levels that were custom designed to support the exercising.**

Next to more loose mappings of exercising activities to in-game actions, concepts for harnessing the motivational effects of video games to support exercising by providing specific rewards for prior exercising in a later game session have also been explored. Building on the concept of so-called *pervasive accumulated context exergames* (PACE) (Stanley et al., 2011), which have less special requirements for mapping and motion-control feedback, and which may appeal to a broader audience, a study was setup to compare the impact of playing games after exercising that present explicitly linked rewards with the effects of playing games after exercising that do not contain explicitly linked rewards on motivation and exercise performance (Smeddinck et al., 2018) **[see publication B.8].**

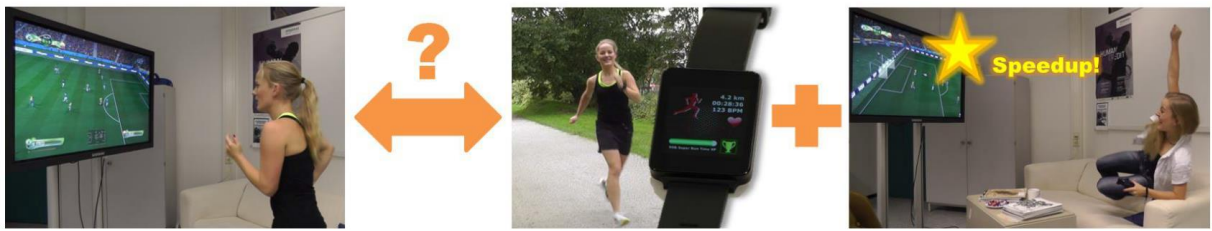

**Figure 59: Left: Exercising is input control for regular (synchronous) exergames; immediate feedback. Right: Data from tracked exercises provide the basis for feedback in PACE (or asynchronous exergames); these are controlled by standard input devices.**

Figure 59 illustrates the contrast between regular exergames and PACE. The results suggest that a setup with explicitly linked rewards can lead to motivational benefits and to increased physical activity, delivering quantitative support for the concept of PACE. With this concept, games and exercises can be combined quite arbitrarily. However, the exact amount and type of reward to be supplied for different forms of exercising requires further study and should also be considered as an adjustable parameter for personalization via adaptability or adaptivity. Since the exercising must be tracked, using e.g. wearable fitness devices, this concept can also be thought of as an augmentation of the available spectrum of MGH, with additional data sources for patient-player models in human-centered adaptive systems. Furthermore, the potential of PACE to alleviate social anxiety and to facilitate a broader accessibility of such asynchronous exergames due to the spatio-temporal uncoupling of the exercising activity from the gaming is promising but requires further study.

# 5   Discussion and Future Work

Although a focused discussion and contextualization was already included with most sub-topics of this thesis, the following paragraphs contain a condensed discussion in order to provide a better overview and to indicate further relations between the different sub-topics that were covered by the various projects and papers that provide the foundation of this thesis.

As noted in the introduction, this thesis explores the following general research question: *"How can adaptability and adaptivity in MGH be realized in an efficient, effective, and enjoyable manner?"* This question was further motivated, and existing approaches were introduced, by providing a summary of the background and related work on games for health in general and on motion-based games for health in particular. The discussion contained the argument that there is *a considerable need and market for GFH and MGH* (section 2.1.1) with especially *promising applications in PRP* (section 2.1.2), which arise - amongst other reasons - from *the dual-task nature of motion-based games* (section 2.1.3). Given the application areas in health, rather *uncommon target groups* must be considered (section 2.1.4) as patient-players in design, development, and evaluations. Even if specific target group requirements are explicitly considered, their individual members bring along *heterogeneous sets of abilities and needs*. Further varying requirements, needs, and abilities are brought in by *other interested parties*, such as involved professionals, or family members (section 2.1.5). Hence, there exists *a need for GFH to be adjustable, or to offer personalization* according to these requirements, either through manual adaptations, through automated adaptivity (section 2.1.6), or through a combination of the two. *Advances in sensor devices* have not only augmented the range of devices for MGH control that can support a growing range of application scenarios, but they can also support data collection to better inform adaptivity (section 2.4). Next to performance outcomes, endpoints from *motivational psychology* and research on *behavioral change* (section 2.5) should be considered both for estimating appropriate adjustments, and for evaluating the achieved outcomes. Based on *developments in mass-market consumer motion-based games* (section 2.6), a growing number of projects in research and development are exploring the field of MGH. *Seminal work* from this area was summarized (section 2.7), highlighting a number of projects that have shown considerable technical achievements accompanied by convincing research. Lastly, *common approaches to adaptability and adaptivity* in general video games (section 2.8.1), GFH, and MGH (section 2.8.2) were discussed, including methods for advanced automation with *learning or patient-player modeling* (section 2.8.3), together with considerations on *privacy, data protection, security, and ethics* (section 2.9) in the light of the data collection and processing that the aforementioned approaches entail.

Building on this background and related work, as well as on outcomes and findings of developments and studies discussed later in this thesis, a number of structural theoretical considerations were introduced and discussed that augment existing models and processes for adaptability and adaptivity in MGH (chapter 3). The *potential to offer motivation, guidance, and analysis* was discussed, highlighting three central promises of MGH (section 3.1). Following a reminder of the need for adaptability and adaptivity, both approaches to facilitating adjustments were addressed. The *general approach to adaptability* in games through iterative testing and difficulty menus was described, together with potential challenges entailed by manual adaptability, such as breaking immersion, and the separation between game-centric and player-centric parameterization that was motivated by observations during the MGH project SDF (section 3.2.1). A *general approach to adaptivity* was discussed, separating performance evaluation from adjustment mechanisms as two general classes of elements of such systems (section 3.2.2). Additionally, an overview of dynamic difficulty adjustments was provided as a common form of adaptivity in games that can be targeted with approaches based either mainly in heuristics, or mainly in learning methods. The common challenges of *cold-start problems* (section 3.2.3) and *co-adaptation* (section 3.2.4) were briefly touched upon, together with approaches to overcoming the obstacles they entail. Based on these foundational considerations, a *modular architecture for adaptable and adaptive MGH* was introduced (section 3.3), featuring a classification of modules into the areas of *game*, *manual adaptability*, *feedback*, or *automatic adaptivity*. The *modeling dimensions* of adaptivity and adaptability in GFH were subsequently discussed based on prior work on the *scale of automation*, on *components for DDA* design and evaluation, as well as on detailed *modeling dimensions for adaptive systems* in general (section 3.4.1), adding further dimensions that were inspired by the developments and studies that were discussed in later sections of this thesis. The presentation of the modeling dimensions included an illustration of the role of the four classes of dimensions of *goals*, causes for *change*, *mechanisms* for adjustment, and resulting *effects*, as elements of a process model for the development or analysis of adaptive SG that can provide a bird's eye view on the complex matter (section 3.4.2). *Accessibility, playability, player experience, and effectiveness* were then discussed as important considerations when planning and evaluating the functioning of adaptable and adaptive systems for MGH, pointing towards iterative design, usability, and user experience as related concepts, and including a discussion on the balance between challenges and ease of use in games (section 3.5). This discussion was inspired by the seemingly contradictory results of simultaneously increased task load *effort* and *frustration* in the absence of negative effects on player experience that were outcomes of a study on modalities for game difficulty choices which was discussed later on (section 4.6.1). Approaches to explaining the *motivational pull of video games* in the context of adaptability and adaptivity were then discussed, bringing together different existing theories, including *self-determination theory*, *flow*, and *dual-flow* (section 3.6). A number of aspects were then considered that combine to motivate – and show approaches to – *human-centered adaptive MGH*. The *fluctuations of abilities and needs* in individuals were discussed, including a general classification into *short-*, *medium-*, and *long-term* changes that combine to form a complex non-linear progression (section 3.7.1). The following section (3.7.2) elaborated a detailed explanation of the meaning and relation of those modeling dimensions of adaptive MGH that are arguably of central relevance when considering human-computer interaction with such systems, namely the *source* of information causing change, the *explicitness* that states how directly feedback that causes *change* points towards that change, the level of *autonomy* of a system in triggering adjustments, and the degree of *automation* that is present for determining candidates for adjustments, connecting back to prior discussions on the involvement of users in automation. This discussion also indicated a relation to the concept of embodiment, which can arguably support reasoning on the question *which aspects of an adaptive system are of interest for manual involvement and control*, and which parts can be fully automated. Accordingly, the concept of *embodiment* was discussed in further detail (section 3.7.3), establishing further links to game design in suggesting the consideration of the attention spans, abilities, or special requirements of players as depletable resources. *Social context and situatedness*, as important aspects of embodied interaction, were then discussed with regard to how they further complicate adaptivity and adaptability for MGH, and it was shown how these considerations can inform decisions on when to include game design experts, patients, or health professionals in different phases of MGH projects (section 3.7.4). This thorough discussion of challenges was then extended by a discussion of the *concept of lenses* and how it can be applied to MGH (section 3.7.5), including the lenses of *design*, *engineering*, *research*, and *health* that were employed during the development of the SDF project. Furthermore, the concept of *human-centric parameterization for adaptability and adaptivity* that was developed during the studies and developments in the context of the SDF project in order to support better usability and user experience of interfaces for manual adjustments, as well as *adaptivity that scales beyond individual games* and their specific spaces of game variables that are targets for adjustments, were introduced (section 3.7.6). Finally, the prior considerations and approaches were combined into a process model for the development and analysis of GFH called *Needs and Abilities Based Human-Centered Design for Adaptive Systems* (section 3.8). This model provides an extension to a prior model of types and levels of automation that describes the general process of designing and implementing automation systems by Parasuraman et al. (2000) and it was discussed in the terms of the prior theory.

## 5.1  Contributions Regarding the Guiding Research Questions

The three research questions that were introduced in the research agenda for the studies and developments guided the specific directions of explorations, developments, and comparative studies that were summarized in this thesis and which were the subject of the technical papers that are included as attachments. The contributions of the projects and developments regarding the three guiding research questions will be discussed in the following sections, with a focus on the main elements of (a) the development of the project SDF as a case study of a development process for an MGH (cf. section 4.5), (b) the comparative study on modalities for interacting with game difficulty choices (cf. section 4.6.1), (c) the early work on adaptive MGH for people with PD (cf. section 4.8), as well as (d) the medium-term situated study comparing adaptive MGH with the use of MGH with exclusively manual adjustment options, and with traditional therapy for chronic back afflictions (cf. section 4.9).

### 5.1.1  Design and Implementation: How Can Flexible MGH Be Realized?

In detail, this first guiding research question was worded as follows: *What are the requirements of adaptable and adaptive systems for MGH and how can such systems be designed and implemented to respect the requirements of their specific application use cases (such as games for the support of physiotherapy for people with Parkinson's disease, games for the support of people with chronic unspecific lower back afflictions, or the requirements of more general groups, such as older adults)?*

As discussed in the theory sections, there are several requirements that are common to most adaptable and adaptive MGH. The according approaches to modular development and modeling dimensions were presented together with aspects of human-centered iterative design, as well as evaluations with regard to accessibility, playability, player experience, and effectiveness, indicating means to assure that the requirements of the specific application use cases are respected. The developments and evaluations in the context of the SDF project form a compelling use case for the application of these principles (cf. section 4.5). Furthermore, a number of insights regarding design aspects were included in the discussion, such as the instruction modality for movement instructions in a dance game, an analysis of gaming preferences within the older adult target group, and the specific human-centered parameterization that was developed for SDF. Additional larger studies were motivated by the SDF project that have been discussed in dedicated sections and make additional contributions to all three guiding research questions. The series of studies on the impact of different modalities for presenting difficulty choices in games (cf. section 4.6.1) found that while game difficulty choices had an impact on autonomy needs satisfaction, they had surprisingly little influence on overall game experience. This was interpreted to support the use of adaptivity if it is needed for other reasons. Furthermore, this research was discussed as a use case for employing the modeling dimensions for adaptive and adaptable SG in order to provide a detailed differentiation between the two adaptive systems employed in the initial and the repeat study.

Additional contributions with respect to this guiding research question were drawn from a number of the remaining studies and developments. The early study on motivation with gamification in interfaces for EMS training (cf. section 4.4) indicated that incorporating even comparatively simple gamification elements in interfaces for exercise applications can improve motivation compared to working without such gamified training information interfaces. Tools for an embodied configuration of movement capabilities for semi-automatic adjustments via therapy plans with milestones were shown to be comparatively efficient and easily understood (cf. section 4.6.2), highlighting secondary usage contexts in therapist-to-therapist and therapist-to-patient exchange about patient abilities, needs, or progress that were not expected. While the study on visual complexity, as a design aspect regarding feedback that is presented to players, was interpreted to indicate that visual complexity did not appear to make for a worthwhile candidate for adaptability or adaptivity, the importance of (micro-)stories and telling visual scenarios was underlined (cf. section 4.7.1). Furthermore, the outcomes were interpreted to support rather flexible design decisions regarding adaptability and adaptivity, based on other considerations for the ongoing projects. The outcomes of the study comparing different exercise instruction modalities did not indicate major drawbacks when employing virtual instructor figures (cf. section 4.7.2). Given the impracticality of instructions by a live human instructor, the findings can be seen as a support for virtual instructor figures for use in the context of adaptable and adaptive MGH.

While the major focus of contributions for the initial study of adaptivity in MGH for people with PD (c; see the prior section) was on questions about acceptance and effectiveness, it also contributed design insights with regard to the functioning of calibration procedures and the benefits of hiding overly salient adjustments (in the case of range of motion adjustments), as well as a need to adjust scoring mechanisms to reflect increased or decreased difficulty (cf. section 4.8). This can be seen to suggest that the individual adjustment mechanics should not be obtrusive, while it can be important to saliently communicate the presence of automatic adaptivity, as well as the adjustments that have occurred. Lastly, the medium-term situated study employing a subset of the SDF games indicated that manual adjustment interfaces for use in adaptive MGH for PRP are required to function very efficiently (cf. section 4.9), and that employing human-centric parameter sets with automatic adjustments seems advisable. This conclusion was drawn as the opportunity to enact manual adjustments after an initial adjustment session was almost never made use of, while the automatic adaptivity was appreciated by therapists and did not appear to have any notable negative effects on player experience, performance, or on functional measures, given the absence of significant effects between the MGH with and without adaptivity.

### 5.1.2 Acceptance and Experience: How are Flexible MGH Perceived?

In detail, this second guiding research question was worded as follows: *Will manual adaptation options and automatic adaptivity in MGH systems be accepted, and can they offer adequate user experience for both patients and professionals?*

While initial feedback was collected in early testing of the games and the configuration interface of the SDF project, the main contributing study on acceptance of MGH (c) provided a first formal exploration regarding this guiding research question (cf. section 4.8). The outcomes indicate that adaptivity in MGH for special target groups, such as people with PD, can lead to good acceptance and player experience, given a prudent development process to detect problems such as the negative perception of overly salient adjustments, as indicated in chapter 3. Support for the acceptance and for a positive impact of interfaces with simple gamification elements on intrinsic motivation was shown by a study with people in EMS training (cf. section 4.4). Furthermore, movement capability configurations using embodied representations of patients were found to be well accepted by therapists (cf. section 4.6.2). In addition to the contribution of the medium-term situated study on a subset of the SDF games in regular practical use in a physiotherapy practice to the guiding research question on effectiveness discussed in the following section, the study also served to show good sustained acceptance and experience over multiple sessions of weekly game use (cf. section 4.9). However, the experience and acceptance were found to be neither significantly improved nor negatively impacted by the presence or absence of automatic adaptivity. Remarks by therapists in follow-up interviews did indicate a preference towards the presence of automatic adaptivity, suggesting that MGH with adaptivity, which can potentially decrease the required workload for therapists, might be better accepted in fully unrestricted situated use outside of a study situation. The related study on balancing in multiplayer games indicated that adaptivity in motion-based games can be accepted and lead to positive player experience and an improvement of self-esteem for this potentially more challenging scenario, both for dyads of players with comparable sets of abilities and skills, and for dyads with extremely different sets of abilities, if the adjustment method is not overly salient (cf. section 4.10.1).

### 5.1.3 Function and Benefits: Do Flexible MGH Work?

In detail, the third guiding research question was worded as follows: *Will adaptation and adaptivity in MGH work effectively and efficiently in practice with different specific use-cases and both for patients and professionals?*

Building on earlier investigations that had already shown good indications for effective and efficient functioning of both MGH and accompanying configuration interfaces for manual adaptation through therapists (cf. section 4.5), the detailed case study of acceptance of MGH in use over multiple sessions delivered first indications for positive effects with regard to the desired serious outcome, in this case the participating people with PD largely reached the preconfigured development targets for range of motion (cf. section 4.8). The later, and more extensive, medium-term situated study on MGH for lower back afflictions indicated significantly improved functional reach when using MGH over a period of five weeks, compared to regular physiotherapy (cf. section 4.9). The addition of adaptivity in one MGH condition, however, did not show further significant benefits or drawbacks compared to the version of the MGH with manual adaptability only. Lastly, more recent work on pervasive accumulated context exergames indicates that providing rewards in gaming sessions after prior exercising can lead to positive impacts on motivation and exercise performance, if the rewards are explicitly linked to prior exercise achievements (cf. section 4.10.2). This indicates potential avenues for more flexible combinations of different types of exercising with later gaming that can also be sedentary, and thus more easily employ high quality games that have the potential to keep players highly motivated over extended periods of time, providing interesting potential for future work.

## 5.2 Limitations and Future Work

Despite notable contributions in theory and to the body of empirical evidence that is available on human-computer interaction with adaptable and adaptive MGH, a considerable number of limitations apply to this work. While the limitations of the individual studies and developments are discussed in the publications, a shared set of limitations will be discussed here.

Most apparently, the bulk of work in this thesis focused on single-player and casual-style MGH with optical body tracking for PD patients or older adults that employed interfaces for manual adaptation by therapists, or used semi-automatic adaptivity between sessions, based on simple threshold heuristics. Generally speaking, future work is thus required concerning multiplayer and social aspects (cf. section 4.10.1), more complex types of games (cf. section 4.10.2), for health or for other purposes, with carefully balanced information about adjustments being provided to all relevant stakeholders, offering them efficient, yet not overstraining interfaces for exerting control over adjustments. Additionally, more fully automated adaptivity that can be activated even during ongoing play sessions, and which includes components that make use of learning over time, or of player modeling to facilitate cross use case adaptivity, should be examined. These limitations can be further augmented along the example of the SDF project. Along with simultaneously developing related work, it had demonstrated positive indications of the applicability of MGH in the context of PRP. However, the games were largely tailored to motivate movements of broad characteristic classes (e.g. the metaphor of picking apples from a tree led to broad movements at medium or high self-controlled speed, catching locusts required slow movements at higher accuracy, striking balancing poses with stretched out arms involves very slow, highly controlled, endurance poses). Many PRP programs can incorporate such elements, but many exercises that therapists regularly work with were not explicitly present in those games. Furthermore, the SDF suite of mini games were very accessible but did not aim to provide a lot of alternation over time and all games relied on a direct player-body to player-character mapping as the core mechanism. While semi-automatic adaptivity was present, it required the configuration of milestones and no explicit relation to the content, duration, or progression of ongoing traditional physiotherapy was given. Lastly, the optical tracking limited the number of exercises that could be reliably and accurately detected.

Future and currently ongoing work, such as the project *Adaptify,* is hence targeting these potential points for improvements. *Adaptify,* for example, includes the design and implementation of a new series of prototypes for MGH that focus on the *modular extensibility* to multiple exercises, the implementation of more complex game mechanics and elements of storytelling, as well as on generative content for improved alternation in game play experiences respectively. At the same time, the games include methods for patient *guidance* and the software infrastructure with accompanying manager applications also explicitly embraces the potential to provide *analysis* (cf. section 3.1). *Adaptify* targets adaptivity both with regard to the personalization of therapeutic exercise programs, expressed in series of exercising sessions over time, as well as in user capability and needs modeling, which can facilitate the automatic generation of personalized exercise series, guidance, and feedback.

However, even compared to such more complex approaches to adaptability and adaptivity combined with elements of generative content, more radical automation can be considered that takes over parts of the game design process as a whole and may be employed to provide not only parameter tuning and content adjustments but to provide unique personalized games. First explorations of such computational creativity in this regard have already been reported (Apken et al., 2014; Cook & Colton, 2011).

Further limitations that provide leads for future work that were found in the remaining projects are, for example, the fixed nature of badge and score reward tiers of loosely integrated gamification elements (cf. section 4.4) that could be personalized to better correspond to personal preferences and to the impact of any ongoing adaptivity on the level of difficulty. This - in turn - would likely have an impact on the game performance that these aforementioned reward schemes (cf. section 4.8) usually reflect. Based on the study comparing exercise instruction modalities, it was argued that virtual instructor figures offer an opportunity for personalized instruction executions, depending on the state of the player, which also presents an opportunity for future work in the direction of personalization, since this potential has not yet been explored.

Regarding interfaces for adaptability and the interaction with them, this work suggests a human-centric parameterization for a better usability and user experience, as well as for potentially improved cross use case applicability. As indicated above, the feasibility of these promises requires further study. Also, while some explorations regarding more "embodied" interface components for such applications have been documented (cf. section 4.6.2), the focus was on the range of motion, and the findings indicated that usability should be further improved. Furthermore, such interfaces have yet to be extended to the remaining parameters that could also be expressed along the lines of a virtual figure, or with similar metaphors that can arguably support a more natural interaction. For example, maximum weight-loading could be modeled by attaching visual representations of weights to joints.

The current systems that employ a semi-automatic form of adaptivity relying on manual guidance regarding start points, end points, and offering opportunities for a manual control of upper and lower performance thresholds, indicate a need of explorations on further automation. More comfortable or flexible manual adjustment methods, such as a curve editor for non-linear interpolations of adjustment progressions, also offer interesting opportunities for future work. Lastly, replacing metaphorically embodied interaction through common digital interaction interfaces with truly embodied control for adaptability and adaptivity arguably makes an interesting case for future work, since this would facilitate a closed-loop of physical control. In such a case, therapists might adjust existing motion patterns, or record and add new ones [e.g. using a layered compositing editor, akin to video editing tools (Walther-Franks et al., 2012)], using their bodies with full-body tracking that produces "gold standard" and potentially highly customized exercise execution sequences for individual players or groups of players. These motion sequences could then be employed to generate personalized game experiences that require patient-players to imitate the targeted motion patterns in an adequate quality in order to reach the game goals.

The theory discussed in this work was partially based on existing approaches and the augmentations and additional arguments regarding the models and processes were supported by findings and observations from the practical developments and studies that were discussed in later sections of this work. However, future work is required to aggregate additional evidence in support of most assumptions and arguments, especially if a generalization beyond the limited scope (see above) of this work is to be achieved. This would include more structured explorations of the modeling dimensions for adaptive and adaptable MGH both in isolation and in interaction, as well as explicit studies of the application of the proposed design process for needs and abilities based human-centered design for adaptive systems in the context of MGH and beyond.

Next to further studies of the potentials and benefits of MGH, possible negative aspects also require the attention of future work. This includes considerations on the topic of addiction (Grüsser et al., 2007), which is apparently relevant not only because of the large SDT needs satisfaction potential of modern games (Przybylski et al., 2009), but also since exercising and sports can lead to their own addiction patterns, arguably aggregating to a combined risk that has not yet been considered in research. Furthermore, depression can be masked by gaming which is a considerable risk factor with target groups that suffer from lasting health afflictions. Lastly, it was noted that the situated and prolonged studies that have been carried out so far each contributed a set of unexpected insights regarding practical challenges of shifts in usage patterns that warrant further research. In addition to the findings summarized above it was noted in related work that a reduced comparability between different player's performances results from adaptability and adaptivity that can make it even more challenging to interpret their performance, and that misleading data can result from different players using one account (Charles et al., 2005). As discussed above, some therapists point out that therapeutic need is not always clear to patients, and that they should sometimes be forced to accept a certain level of physical strain, pointing to limitations with regard to adaptation based on explicit patient feedback. The practical effects of co-adaptation are a challenging subject for empirical study, yet also require future work. Lastly, with adaptable and adaptive MGH, the heterogeneity of the players is driving the need for adjustments, while the usage context is typically assumed fixed from the game-development point of view. Observations of spontaneous re-appropriation and user generated context (cf. section 4.5) have shown that this assumption does not always hold. Adapting to different usage contexts was out of scope for this work but would make for interesting future work.

Lastly, a number of projects that have been carried out in the context of this thesis but have not yet been published were excluded for brevity although they suggest future contributions especially with regards to sampling different levels of the modeling dimensions for adaptive games. This includes a comparative study on different levels of the automation in games with DDA, on the choice of – and/or presence of - adaptivity, on the interaction of therapists with further configuration interfaces, on different modalities of audio-feedback for providing cues on exercise timing and counting, as well as on the applicability of the Kinect as a measuring device alternative to gonimeters.

## 5.3 Challenges in Evaluation and Relation to General Game User Research

It was indicated in the prior section on research methods that a number of challenges arise for the application of established research procedures and methods in the area of adaptable and adaptive MGH due to the complex nature of the subject matter, and especially due to the special target groups that can be involved. The attached publication on challenges in studying exer-games with older adults [**see publication B.6**] contains an overview of recurring challenges in evaluations with that target group, which also largely apply to other potential GFH target groups. These challenges can be summarized as (a) *general challenges* due to age-related changes and low levels of computer literacy, (b) *anxiety* related to technology necessary for evaluations, (c) *negative feelings* related to *personal performance*, (d) *social factors* that can introduce biases such as good participant effects, (e) *overstraining* participants, which is connected to (f) *health risks*, such as the risk of falls, that are specific to (motion-based) games for health and pose a particular problem in the older adult target group. Furthermore, since measures are often *noisy*, triangulating methods may have to be employed, and considerations of functional outcomes require *long-term* investigations (Smeddinck et al., 2012). In addition, further challenges in the general design and evaluation of games for health have been discussed in the appended publication B.3, including *interdisciplinary and multiple party interests*, *truly user-centered iterative design*, *heterogeneity* and a *broad range of serious goals*, *the role of machine learning and data analysis*, *sensing and tracking*, the *practical integration* of GFH, *safety and clinical validation*, *evaluation methods and long term use*, as well as *ethics, data privacy and regulations*, which have mostly been discussed separately in this thesis.

Specific approaches to some of these challenges with studies and evaluations that occurred in the SDF project and during the surrounding investigations included the use of smiley scales instead of regular Likert-type sales, reading questionnaires out loud as a form of structured interviews, simplified questionnaires (e.g. only including items for necessary subscales), an emphasis on observations to augment self-report measures, the inclusion of health professionals as observers, frequent re-evaluations, and the use of paper-based questionnaires instead of PC- or tablet-based questionnaires for reduced technological burden. These approaches for coping with the aforementioned challenges can, however, introduce additional biases (e.g. when reading questionnaire items out loud, the study conductor may create biases by putting an emphasis on certain words), and can push the limits of studies (e.g. with a large amount of paper-based questionnaires leading to survey fatigue). In addition to controlled studies and small-scale evaluations that are conducted "in the wild", larger scale observations from unconstrained real-world use play an important role in commercial game user research and can also inform research studies (Smeddinck, Krause, & Lubitz, 2013) **[see publication C.10]**, especially for ongoing adjustments to adaptable and adaptive systems after release in MGH as a service. In some cases, such as the study on game difficulty modalities (cf. section 4.6.1), it was found to be beneficial to tightly integrate the experiment procedure into the flow of game menus in order to remove burden from the experiment conductors in complex study setups. Lastly, while early studies on the subject of MGH often attempted to employ the game experience questionnaire (GEQ) by Ijsselstejn et al. (2007), it was found to rarely return significant results even in the light of directed hypotheses that were well supported by related work. Instead of this specific questionnaire, the PENS and IMI questionnaires were employed in later studies (cf. section 4.2).

A number of challenges result from the dynamic nature of adaptable and adaptive MGH, such as the fact that, with a personalizing adaptive system on a per-user basis, every user becomes a case study and regular statistical inference based on group means is difficult to apply, suggesting the application of methods that allow for statistical inference based on individual participants and then aggregate the results. Furthermore, evaluating affect is challenging, especially in laboratory settings (Kappas, 2010) and the potential presence of systematic biases has rarely been considered in research on MGH or adaptive games (Denisova & Cairns, 2015).

Lastly, in the context of more complex GFH developments, the reporting of evaluation and study results as a feedback for development teams in a timely and meaningful way brings additional challenges (Charles et al., 2005).

# 6   Outlook

Based on the work that was summarized in this thesis, and based on the surrounding related, ongoing, and developing work, as well as on looming technological trends, the following sections provide an outlook into five directions that represent subjectively promising avenues for future work on adaptable and adaptive MGH.

## 6.1   Improving Devices for Sensing and Feedback

Arguably, each generation of motion-based games has been driven by advances in hardware for sensing and feedback. In a transitive sense, this also applies to facilitating use cases in health, as well as to facilitating advances in adaptability and adaptivity. While the generation of MGH discussed in this thesis focused largely on full-body motion tracking with optical sensors, a broadening of the scope can be expected with the inclusion of further advanced, yet affordable, sensors, actuators, and integrated devices. With regard to adaptivity, employing advanced ubiquitous and unobtrusive sensors can allow devices to capture data to better inform systems about aspects of game or player state, such as player emotions (Sykes & Brown, 2003), or frustration levels (Gilleade & Dix, 2004). Furthermore, additional devices will become available for input, or for supporting input (cf. suspended walking) and feedback (e.g. exercise mats), as many fitness devices already have digital tracking and information features that can be transferred to more niche health or assistive devices. Lastly, the currently growing fields of virtual reality (VR) and augmented reality (AR) will shape approaches to future MGH. AR in particular promises methods to deliver feedback in the direct reference frame of the body of a patient-player. Furthermore, room-scale body, face, eye, and hand-tracking are being pushed forward as enabling techniques for VR and AR that can each play important roles in both MGH and adaptability and adaptivity for MGH.

## 6.2   Improving Performance Analysis and Adjustment Mechanisms

Together with the progress in hardware for human-computer interaction, software methods will be developed to facilitate the application of this hardware both for serious and for entertainment purposes. The according methods, prominently including *machine learning* and *human computation*, will likely influence the approaches to performance analysis and adjustment mechanisms for adaptability and adaptivity in MGH.

As indicated above (cf. section 2.8.3), methods from machine learning are likely to play an increasingly important role, especially with regard to player modeling and profiling (Fawcett & Provost, 1999) that can be useful for ensuring more appropriate adaptivity (Charles et al., 2005).

Recent work on the automatic detection of *player roles* (Eggert et al., 2015) [**see publication C.1**] shows additional directions that can facilitate more targeted adjustments based on player grouping with methods from collective intelligence, similar to suggestions from existing work on player types (Charles & Livingstone, 2004). Furthermore, games with online functions and account management, and games with large player numbers can allow for a continuous adaptation to individual players whilst also adapting the parameters of the underlying models themselves (Lomas et al., 2013). Overall, methods from machine learning, or modern statistics and artificial intelligence can be expected to see further use with games in (a) supporting manual adaptations, in (b) automatic adaptivity, or in (c) iterative player-centered game design. Arguably, (a) can benefit from classifiers, clustering or recommendations and search to provide personalized adjustment recommendations, while (b) requires recommendations (search) for general solutions but might also depend on clustering and classifiers in pre-processing. With (c) the pool of available methods would be applied not only for online adjustments with players after release, but for partially automated balancing and parameter tuning during development, or with test groups after release for ongoing service patches. Next to the generation of suggestions for manual adjustments that are produced with the support of methods from ML and AI, initial player grouping can also be supported without the need for collecting first observations. Related work has employed questionnaires that were modeled after decision trees to facilitate such groupings (Rosenfeld et al., 2012) to boost adaptive systems with collective intelligence and allow for better classification while data on the new individual user is still sparse to non-existent.

Although considerable progress can be expected with novel applications of fully automated methods, the complexity of the involvement of human abilities and needs in dynamic situated use will remain challenging. Arguably, methods from *human computation* can play a role in areas that are particularly challenging to tackle based on machine computation only, such as *intuitive decision making, aesthetic judgment, contextual reasoning,* or *embodiment issues* (Krause & Smeddinck, 2012) [**see publication C.7**]. For MGH, all of these aspects play a role in the extended human-centered iterative design and balancing cycle. With human contextual reasoning, less conservative approaches to the extent of adjustments can likely be supported, moving beyond the tuning of pre-defined parameters within fixed boundaries towards more explorative and even generative aspects that may include introducing or removing game features. Notably, the involvement of members of online crowds can itself be realized by employing serious games for motivation in the form of so-called human computation games (Krause & Smeddinck, 2011) or games with a purpose (von Ahn, 2006). In a recent study, for example, different candidate modalities for presenting human movement executions for quality of motion judgments have been compared following a human computation approach. In this comparison between RGB video, a virtual rendered character, a virtual rendered skeleton, and greyscale depth video (based in infrared), RGB video led to the best inter-rater agreement and was also most preferred (Sarma et al., 2015). Such quality of motion judgments with human computation can arguably inform performance evaluation in adaptive MGH and supply sample data for a future automation of such processes.

Altogether, human and machine computation can be expected to support more adequate adjustments regarding the individual abilities and needs of heterogeneous player groups. However, next to the established paradigm of adaptivity being supported by *performance evaluation* and *adjustment mechanisms* (cf. section 3.2.2) that relies on observing the reaction and performance of players under a certain given system configuration and which is therefore fundamentally based on "reacting" to sub-optimal configuration states, more predictive systems are promising targets for future work. Such systems would aim at anticipating patient-player developments in order to facilitate "acting ahead of sub-optimal states" that are foreseen but do not have to come to be, avoiding temporary disservice to users. The underlying included publication provides a more detailed discussion *[see publication B.2]*. In principle, by integrating user and world models, the impact of considered adjustments can be estimated without requiring prior execution, allowing for a smaller number of suboptimal situations. Figure 60 (left) outlines a typical cycle of adaptive gameplay with non-anticipatory adjustments. Such a process can be augmented with a predictive anticipatory loop before the actual adjustments are performed (as illustrated by the box with red outline in Figure 60). Depending on the player reactions following the execution of adjustments, world and user models can be updated.

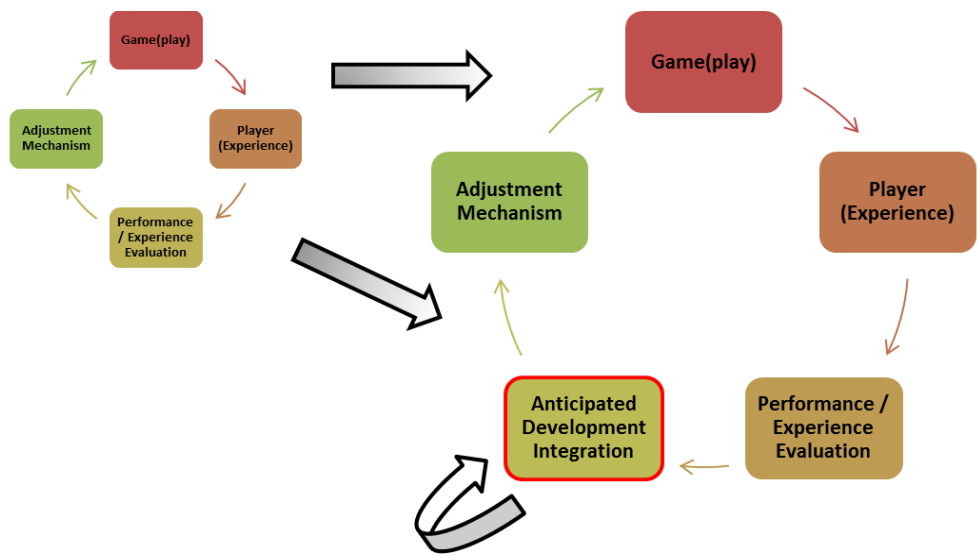

**Figure 60: Reactive adaptive systems can potentially be extended to prevent maladaptations by introducing a predictive anticipatory loop that simulates various settings candidates and the likely effects.**

While more anticipatory approaches are promising, they are also very challenging and require complex user and world models for enacting simulations of candidate futures, akin to cognitive architectures, forming second order intentional systems that carry beliefs about other systems (Bannon, 1991). However, next to avoiding maladaptations, such an anticipatory perspective could also be employed to perform meta-adjustments on the adaptive system itself, effectively forming an ongoing outer control loop around the design process for needs and abilities based human-centered design for adaptive systems presented in this thesis (cf. section 3.8) that has the potential to facilitate cross use case applicability of adaptive systems not only in the sense that existing user models are carried over to another application (e.g. a new GFH), but even in the sense that a given adaptive system can self-adjust to support a novel application with different parameters and contextual requirements.

## 6.3   Analysis for Objective Health

The projected rise in cross use case applicability of adaptive systems for MGH, be they reactive or anticipatory, also plays an important role when considering the potential of MGH to support healthcare with objective information. As argued above (cf. section 3.1), the potential to support the objective analysis and diagnosis of aspects of the individual state of health of the patient-players, is a powerful promise of a more widespread adoption of MGH, as well as of GFH in general. If GFH can be designed in a manner that allows them to be truly embedded in practical day to day healthcare (cf. section 3.7), large amounts of player performance and sensor data become available for secondary processing that can not only inform the adaptability and adaptivity of existing and the design of new GFH, but that can also play a role in supplying more structured and dense information on the effectiveness of treatments or aspects of treatments in comparison to currently available standards in most areas where treatments are not fully supervised and documented. While the acceptance of games as serious elements of current and future treatments amongst professionals still requires notable progress, steps in this direction are already being taken, for example in the public debate about seals of quality that foster credibility without the requiring the full extent of strict testing and proofs that is involved with the traditional process of national and international medical device and application certification and which would be prohibitively expensive for most developers of GFH (Beeger, 2016). At the same time, the requirement to engage with stationary or wearable sensor and actuator devices that form the basis of the data collection, and which are required to support the described trend for analyses concerning objective health measures, can arguably more willingly be accepted and appear less obtrusive when the devices occur as parts of games or playful applications from the point of view of the targeted patient-players.

## 6.4   Sustained Motivation and Effectiveness

If sensor-actuator-devices are seamlessly embedded as controllers, or in the larger technological setup of GFH, and if the offered interaction patterns appear natural or are easy to learn, the patient-players can truly engage with the game mechanics, the game world, and the game story or scenario; i.e. they can enter the "magic circle" (Huizinga, 2008) that is a prerequisite for the exceptional motivation that games can provide. Ways to sustain that motivation, and to study and reliably achieve lasting effectiveness of MGH and GFH are much needed and have not yet been thoroughly researched. First explorations, as described in section 4.10.2, towards integrating GFH with high quality and longer-lasting game experiences are gaining more attention in current work (Ketcheson et al., 2016) and are likely to gain additional attention in future work. Arguably, the development of MGH and GFH can be expected to follow the development of regular video games in an accelerated fashion. If the view described by Rigby and Ryan (Rigby & Ryan, 2011) is assumed, following the SDT needs dimensions of *competence* (e.g. skill-based casual-style gaming), and *autonomy* (more choice in game worlds, or between games; e.g. MGH suites), the factor of *relatedness* needs satisfaction is a promising candidate for fostering further improvements in motivation and longer longer-term adherence (e.g. via multiplayer, surrounding social platforms, or captivating – potentially AI-driven – NPCs). The results regarding needs satisfaction that were attained in the studies presented in this thesis support this argument (cf. section 4.9) and the increasing focus on surrounding social architectures (as shown by some of the seminal related work projects discussed in section 2.7.1), as well as on targeting the integration with existing gaming (such as *Steam*) and health platforms for overarching achievement and progress tracking, as well as for their embedded social features, deliver further evidence of a trend towards supporting more complete relatedness need satisfaction.

## 6.5   User Generated Context and Self-Empowering Health Care

Notably, if the competence, autonomy, and relatedness needs satisfaction is considered not only in the context of the motivation of engaging with an individual GFH, but in a context spanning across multiple GFH and potential ecosystem applications, the considerations approach overall self-determination based need satisfaction as an important facilitator of general well-being (Deci, 1975). In this light, GFH can be seen as potentially facilitating elements for more self-empowering health care, since patient-players are not passive recipients of treatments, but actively take part in working towards health goals. This argument is in line with the development of health as a promising aspect of quantified-self and the increasing availability of both sensing and tracking devices, as well as of data aggregation and analysis platforms (cf. section 2.4), since the underlying technologies are – in turn – central drivers of MGH and GFH. While current generation MGH are becoming ready for use in augmenting established approaches in many areas of health, such as PRP (cf. section 4.9), future generations of GFH and MGH that are designed for patient-to-therapist, patient-to-patient, and self-reflexive interaction from the start can conceivably not only improve on the SDT factors as described above, but also facilitate more self-directed, or community-driven approaches to health. At the same time, developments such as PACE show paths towards a more pervasive reach of GFH (cf. section 4.10.2). Hints pointing at self-directed re-appropriation and emerging user generated context have already been observed in the context of the studies and developments that were discussed in this thesis, and they indicate a great potential of considering self-direction, empowerment, or even the support of citizen-science projects with MGH or GFH from conceptualization, through to their design, development, and use. Since such approaches could easily be endangered if the transparency and trust of underlying technological platforms appear vulnerable, considerations in this regard, and on the broader topics of privacy, security, and data protection, as well as ethics (cf. section 2.9), would continue to play an increasingly critical role. Hence, current generation research and development projects, such as *Adaptify*, are already partnering with institutes for legal research, in order to start exploring the feasibility and limitations of such approaches, which include the same technological foundations as systems for adaptability and adaptivity that scale across multiple MGH. Further dangers in the context of self-empowerment with GFH include the potential of self-overpowering given the considerable motivational potential of games that has to be considered with respect to the eventuality of crossing into patterns of addiction (Rigby & Ryan, 2011). If future GFH and MGH developments aim at embracing user generated context and self-empowering health care as desired outcomes, the creative process will arguably have to shift even further from the established foci on engineering for playability, and design for player experience, towards more radically constituent-driven (and less top-down) methods. This transition towards a gradual superseding of a primarily instrumental perspective on GFH and MGH is not likely to come easy. However, if taken seriously, and implemented in a conscious manner, such approaches promise to allow developments in GFH and MGH to support a broader shift of attention in human-computer interaction from the traditional foci of usability and user experience, towards more holistic and persistent considerations on user well-being, adding an explicitly eudemonic perspective (Ryff & Singer, 2006) to the classic mixture of the utilitarian (~UB) and hedonistic (~UX) lenses.

# 7  Conclusion

Given the pressures on health care systems due to increasing life expectancy and the prevalence of lifestyle diseases across all age groups, and in the light of the promises of games for health and motion-based games to play a role in tackling the resulting challenges if they can meet the complex requirements and abilities of potential patient-players, this thesis considered the question how adaptability and adaptivity can be realized in an efficient, effective, and enjoyable manner in MGH. Based on a discussion of the background and related work of human-computer interaction with adaptable and adaptive motion-based games for health, and a discussion of structuring theoretical considerations that encompass the central promises of MGH, approaches to adaptability and adaptivity, modular development, a consolidation of existing work on modeling dimensions for adaptivity in games and on user agency in automation, targeted outcomes, foundations of motivation in MGH, as well as needs and abilities based human-centered design for adaptive systems in general, and GFH and MGH in particular, this thesis presented a number of practical developments and research projects that are documented in detail in the accompanying publications. The projects were structured along the lines of three guiding research questions that contribute to the general research question noted above, focusing on *design approaches* for adaptable and adaptive MGH, on their *acceptance*, and on their *effectiveness*. Next to a number of specific collateral contributions that were provided by the summative and supportive publications, which have been discussed in the according subsections of the studies and developments part of this thesis, and which are summarized in section 5.1, the three foundational publications included in this work have contributed to each guiding research question. Concerning *design aspects*, a study comparing different difficulty interaction modalities, suggests that the compared methods of manual, embedded, or automatic difficulty settings each carry their own advantages and disadvantages but do not differ in overall player experience as much as the underlying theory might be interpreted to suggest. Hence, if adaptivity is desired due to the design requirements of a particular MGH, it can be expected that automatic adaptivity can be implemented in an enjoyable manner that is on par with the viable manual adaptability alternatives. Regarding *acceptance*, an initial explorative case study with three patients using an adaptive MGH for the support of physiotherapy for people with PD indicated positive resonance both with patient-players and therapists, which was later underlined by the outcomes of a medium-term situated study that evaluated the usage of MGH for the support of PRP mainly for older adults with chronic lower back afflictions. Both studies also indicated that the involved systems for adaptability and adaptivity worked efficiently enough as to not interfere with the complex requirements of their situated use with real therapists and patients. Lastly, regarding *effectiveness*, first indications from the case study with a MGH for people with PD were strongly supported by the outcomes of the situated medium-term study that showed significant functional benefits after using the MGH with either purely manual adaptability, or with a mixture of semi-automatic adaptivity, compared to conventional therapy sessions in the setting of the study. Further design aspects were drawn from the extensive iterative human-centered design processes that facilitated the latter two studies and are documented in this thesis. Considering the overall outcomes, this thesis suggests that adaptable and adaptive MGH can be designed to be accepted and efficient, and a number of approaches to developing efficient, effective, and enjoyable adaptability and adaptivity in MGH were presented. Next to a discussion of specific limitations and challenges that apply to these outcomes (cf. section 5.2), the studies included in this thesis repeatedly indicated that the impact on patient-player performance and motivation under conditions of either manual adaptability or automatic adaptivity does not appear to follow clearly predictable patterns, suggesting the application of iterative methods and the consideration of goals and outcomes through different lenses in order to adequately gauge the results achieved for a given state of development of a given MGH or GFH system. While future work may lead to more specific models, at least for constrained scenarios given specific target groups, underlying technologies, or application areas, this thesis has shown how partial improvements with adaptable and adaptive MGH can be achieved that already present a contribution towards facilitating MGH that are accepted by professionals and patients alike, which can lead to measurable effects, and which are enjoyable to use. Such MGH can motivate patient-players to perform exercises that might otherwise be perceived as dull and repetitive, and they allow therapists to work towards personalized augmenting treatments for their patients that do not require extensive manual configuration or ongoing adjustments.

# PART II: Publications

The following list of publications contains works that have benefitted substantially from the participation of students, or from the involvement of the listed co-authors, who have contributed to the underlying projects, as well as during the writing. Each reference is accompanied by a brief summary of the contribution of the publication with regard to the context of this work, as well as by a brief statement on the personal contribution that provide a general impression regarding the scope of the involvement of the author of this thesis in the respective works. The rough percentage estimates are concerned with the contribution specifically regarding the write-up and publication. The original publications or manuscripts follow in order.

## A. Foundational publications:

*[A.1] Smeddinck, J., Siegel, S., & Herrlich, M. (2013). **Adaptive Difficulty in Exergames for Parkinson's disease Patients**. In Proceedings of Graphics Interface 2013 (pp. 141–148). Regina, SK, Canada: Canadian Human-Computer Communications Society. https://doi.org/10.20380/GI2013.19.*

**Contribution of the work in the context of this thesis:** An early implementation of dynamic difficulty adjustments for a game for the support of physiotherapy for people with Parkinson's disease, focusing on the difficulty parameters speed, accuracy, and range of motion, which were increased based on a calibration over the course of a three week study with three Parkinson's disease patients, indicating that the system was well accepted and facilitated a challenging yet suitable game experience for all three patients who were found to display very heterogeneous abilities and needs. The study revealed insights about the efficacy and limitations of the parameter set, as well as insights into the acceptance of prolonged use of the adaptive game and informed the design of follow-up work.

**Personal contribution to the work:** Close support in the supervision of Sandra Siegel, including the survey of related work, study planning, game and calibration tool implementations, study analysis, and interpretation; led the write-up for publication (~75%).

*[A.2] Smeddinck, J. D., Herrlich, M., & Malaka, R. (2015). **Exergames for Physiotherapy and Rehabilitation: A Medium-term Situated Study of Motivational Aspects and Impact on Functional Reach**. In Proceedings of the 33rd Annual ACM Conference on Human Factors in Computing Systems (CHI 2015) (S. 4143–4146). New York, NY, USA: ACM. https://doi.org/10.1145/2702123.2702598.*

**Contribution of the work in the context of this thesis:** A report on initial results from a medium-term study of a suite of motion-based games for health with patients with chronic unspecific back pain afflictions, comparing manually adaptable games with games with additional semi-automatic adaptivity and with regular physiotherapy sessions. Results indicate significantly improved functional reach test results, as well as improved self-reported patient autonomy, presence and reduced tension-pressure as well as effort-importance in the games conditions compared to treatment with regular physiotherapy sessions, together with a generally positive and sustained acceptance and quality of user experience in patients and therapists alike when using the games. The similarities between both types of game treatments and the differences between game and non-game treatments indicate a good applicability and acceptance of motion-based games in physiotherapy, rehabilitation, and prevention with a focus on upper body exercises, while therapist interviews and observations indicate a preference for the adaptive version of the employed suite of games.

**Personal contribution to the work:** Close support in the design and development of the suite of games in the context of the *Spiel Dich fit* project; led the research on related work, the planning, piloting, implementation, execution, analysis, and interpretation of the study, as well as the write-up for publication (~85%).

*[A.3] Smeddinck, J. D., Mandryk, R. L., Birk, M. V., Gerling, K. M., Barsilowski, D., & Malaka, R. (2016).* **How to Present Game Difficulty Choices? Exploring the Impact on Player Experience.** *In Proceedings of the 2016 CHI Conference on Human Factors in Computing Systems (S. 5595–5607). New York, NY, USA: ACM. http://doi.org/10.1145/2858036.2858574.*

**Contribution of the work in the context of this thesis:** The work presents a discussion of the theoretical foundations of the impact of user-involvement in game difficulty choices as a game design aspect that relates to the players' perceived autonomy and might conflict with immersion, employing a basis in flow theory and self-determination theory. An according study and a repeat study are discussed with two separate games that offer three different levels of user-involvement in game difficulty adjustments, comparing *manual settings via a difficulty selection menu* with a *manual game difficulty selection that is seamlessly embedded* in the game world and *automatic adaptivity*. Autonomy needs satisfaction was found to be lower under the fully automatic difficulty selection performed in the adaptive condition, although other player experience factors were not found to be affected as expected. The study suggests that the overall game experience was not notably impacted by the choice of game difficulty adjustment, which supports the prioritization of other game design decisions over the selection of difficulty interaction modality with regard to design dimensions around questions of user-involvement. This work can therefore be interpreted as a specific evaluation of discrete positions on the modeling dimensions of adaptivity (as discussed in this thesis).

**Personal contribution to the work:** Led the adjustments and further development of the game for the first study, supported the development and adjustments of the game for the second study; close support in the supervision of Dietrich Barsilowski; led the study design, implementation, testing, execution, full dataset analysis, interpretation, as well as the write-up for publication (~95%).

## B.  Summative and supportive publications:

*[B.1] Gerling, K. M., Miller, M., Mandryk, R. L., Birk, M. V., & Smeddinck, J. D. (2014). **Effects of Balancing for Physical Abilities on Player Performance, Experience and Self-esteem in Exergames**. In Proceedings of the 32Nd Annual ACM Conference on Human Factors in Computing Systems (S. 2201–2210). New York, NY, USA: ACM. http://doi.org/10.1145/2556288.2556963.*

**Contribution of the work in the context of this thesis:** A discussion on balancing in sports and games with a transfer to potential effects on player experience in multiplayer games, accompanied by a study comparing different modes for accounting for inter-player differences in skill in a dance game based on a mat for step-detection. The modes *input balancing* (asking players to perform step sets with differing density), *score balancing* (providing more points for a lesser performance to the weaker player), and *time balancing* (allowing more or less time to accurately perform a step) differ in their saliency and potentially in intrusiveness, which allows for this work to be interpreted as a specific evaluation of discrete positions on the dimensions of adaptivity with a focus on the saliency dimension. Results indicate that input balancing worked best towards a more equal performance while it impacted the stronger player negatively in terms of a reduction of self-esteem, and it reduced relatedness for both players; all likely due to the high saliency of the approach. Findings from a second exploratory study with dyads of wheelchair users and non-wheelchair users found a need for stronger balancing, as well as evidence for a more accepting stance of the players towards balancing, regardless of the level of saliency. Some able-bodied players felt uncomfortable beating a person with a disability, arguably underlining the need for balancing, although some wheelchair-users also expressed that they did not mind achieving comparatively weaker scores.

**Personal contribution to the work:** Support of the study planning, piloting, execution, and interpretation; minor support of the write-up for publication (5~%).

*[B.2] Malaka, R., Herrlich, M., & Smeddinck, J. (2017). **Anticipation in Motion-Based Games for Health**. In M. Nadin (Ed.), Anticipation and Medicine (pp. 351–363). Springer International Publishing, Berlin / Heidelberg. https://doi.org/10.1007/978-3-319-45142-8_22.*

**Contribution of the work in the context of this thesis:** A summary of practical and theoretical approaches to adaptability and adaptivity in motion-based games for health that were pursued in the project *Spiel Dich fit*, including a summary of system modules, sensor technologies, and an early discussion of the dimensions of adaptivity that are discussed in this thesis. The chapter closes with a discussion on the potential of anticipatory techniques, e.g. with predictive user- and context-modeling, in the context of games for health.

**Personal contribution to the work:** Close support and involvement in the development of the underlying works and concepts that are summarized in this publication; support of the write-up for publication (~40%).

*[B.3] Smeddinck, J. D. (2016). **Games for Health**. In R. Dörner, S. Göbel, M. Kickmeier-Rust, M. Masuch, & K. Zweig (Eds.), LNCS Entertainment Computing and Serious Games (Vol. 9970, pp. 212–264). Springer International Publishing, Berlin / Heidelberg. https://doi.org/10.1007/978-3-319-46152-6_10*

**Contribution of the work in the context of this thesis:** This publication is a chapter of a lecture / tutorial educational book on serious games. The chapter provides a summary of a number of background aspects and discusses challenges and strategies in the design, implementation, and evaluation of games for health. General examples and theory are accompanied by practical examples along the lines of three motion-based games for health projects that are also featured in this thesis. Many sections appear as excerpts throughout this thesis. The publication was the first to draw together aspects of flow and self-determination theory in a way that is expanded upon in this thesis.

**Personal contribution to the work:** Close support and involvement in the development of the underlying works and concepts that are summarized in this publication; fully responsible as sole author of this publication (100%).

*[B.4] Smeddinck, J. D., Gerling, K. M., & Malaka, R. (2014). **Anpassbare Computerspiele für Senioren**. Informatik-Spektrum, 37(6), 575–579. doi:10.1007/s00287-014-0835-z.*

**Contribution of the work in the context of this thesis:** A summary of promises and challenges of serious games for older adults with a focus on adaptability and adaptivity, discussing the prerequisites of accessibility, playability, and player experience for the targeted dual-purpose of motivation with a positive experience and efficiency in achieving (health) benefits.

**Personal contribution to the work:** Led the conceptual development and writing (~70%).

**Contribution of the work in the context of this thesis:** A within-subjects study with older adult participants of four versions of increasing visual complexity of the same motion-based game for the support of physiotherapy for people with Parkinson's disease. Results indicated that significant differences in player experience and preference occur with drastic reductions in visual complexity to simple geometry where the inherent micro-story of the game (here: fishing) is no longer clearly perceivable, whereas the remaining three levels of visual complexity, ranging from plain 2D cartoonish appearance to 3D realistic pre-renderings did not lead to notable differences in player experience or physical exertion. The outcomes of this study were employed to inform game design decisions in the *Spiel Dich fit* project and can be seen to support the prioritization of other game design decisions (e.g. production cost) over visual fidelity, as long as the perception of the narrative is not endangered.

**Personal contribution to the work:** Close support and supervision of Saranat Tiemkeo in the preparation of the foundations of the study, in piloting, game implementation, study execution, data analysis, and interpretation; led the write-up for publication (~90%).

**Contribution of the work in the context of this thesis:** This workshop contribution summarizes a number of common challenges that occur with evaluations and studies with games for older adults, focusing on motion-based games. The paper also discusses strategies to cope with a subset of these challenges and points towards directions for future work that aim at developing strategies and methods that improve means to cope with the challenges. The list of challenges is discussed in further detail and contextualized along the example of specific games for health projects in this thesis, building on the foundation provided in this paper, as well as in (Jan D. Smeddinck, 2016).

**Personal contribution to the work:** Led the collection, discussion, and consolidation of common challenges that had occurred in projects that were spearheaded by the co-authors; coordination of the writing and submission (~40%).

*[B.7] Smeddinck, J. D., Herrlich, M., Roll, M., & Malaka, R. (2014). **Motivational Effects of a Gamified Training Analysis Interface**. In A. Butz, M. Koch, & J. Schlichter (Hrsg.), Mensch & Computer 2014 - Workshopband (pp. 397–404). Berlin: De Gruyter Oldenbourg.*

**Contribution of the work in the context of this thesis:** A study comparing the impact on motivation and emotional valence resulting from using a training information system for professional electronic muscle stimulation training featuring quantified-self aspects and simple gamification elements with regular feedback by a human coach only. The results indicate a notable positive impact on intrinsic motivation as well as valence, hinting at the potential role and benefits of similar information / secondary feedback and analysis interfaces for private use or use together with a therapist in the context of motion-based games for health.

**Personal contribution to the work:** Close support and supervision of Max Roll in the preparation of the foundations of the study, in piloting, data analysis and interpretation; led the write-up for publication (~90%).

*[B.8] Smeddinck, J. D., Herrlich, M., Wang, X., Zhang, G., & Malaka, R. (2018). **Work Hard, Play Hard: How Linking Rewards in Games to Prior Exercise Performance Improves Motivation and Exercise Intensity**. Entertainment Computing (Vol. 29, pp. 20–30). Elsevier, Amsterdam, Netherlands.*

**Contribution of the work in the context of this thesis:** The work presents a discussion of a design space for asynchronous exergames where a performance context is accumulated using sensing devices during physical activities over some period of time and rewards according to the performance are presented during regular sedentary gaming sessions, which are the preferred style of video game play for many gamers. A comparative study reveals significantly increased motivation when games played after an exercise session offer asynchronously linked rewards, compared to not offering linked rewards. Such pervasive accumulated context exergames can broaden the appeal of games for sustained motion-based treatments and behavior change towards increased physical activity, since they do not have the share the explicit requirements - and limits in appeal - of synchronous exergames.

**Personal contribution to the work:** Close support and supervision of Xiaoyi Wang and Guangtao Zhang in planning, testing, conducting, and analyzing the underlying studies; spearheaded the research on related work, the contextualization, formal discussion of the design framework, and the write-up for publication (~85%).

*[B.9] Smeddinck, J. D., Hey, J., Runge, N., Herrlich, M., Jacobsen, C., Wolters, J., & Malaka, R. (2015). **MoviTouch: Mobile Movement Capability Configurations**. In Proceedings of the*

**Contribution of the work in the context of this thesis:** The development and experimental comparison based on usage by physiotherapists of three different body movement capability configuration modalities (direct touch, buttons and sliders, direct touch and sliders), finding a preference for the mixed model which allows for the quick selection of body parts via direct touch (on a humanoid 3D figure) and fine-grained angular movement capability configurations via classic slider components. The implementation prepared configuration components for tablets based on the Unity game engine while the study hints at a primary direction for manual adaptability implementations and unearthed the unexpected usage pattern of configuration for demonstration of patient abilities in therapist-to-therapist and therapist-to-patient communication.

**Personal contribution to the work:** Close support and supervision of Jorge Hey in the preparation of the foundations of the study, the tool implementation, the piloting of two studies, as well as in the data analysis and interpretation; led the write-up for publication (~95%).

**Contribution of the work in the context of this thesis:** A study of the preference of modality, ease of understanding, and exercise performance accuracy, when following either a live human instructor, an instructor recorded on video, or a virtual figure derived from motion-tracking recordings of an instructor. While the live human instructor was most preferred, and the virtual figure lead to comparatively worst ease of understanding, the virtual instructor mode facilitated better exercise execution accuracy than the video recording. The study informs the design of instruction modalities in the absence of a human instructor (for cost saving and automation), highlighting the importance of visual feedback of the user performance, potential pitfalls when employing virtual character visualizations, and provides a novel measure of exercise execution accuracy (normalized mean squared rotation distance), which are all elements of interest in the design and implementation of motion-based games for the support of physiotherapy, rehabilitation, and prevention.

**Personal contribution to the work:** Supervision of Jens Voges and support with the theoretical foundations, related work, study planning, piloting, analysis, and interpretation; led the secondary analysis and write-up for publication (~85%).

*[B.11] Streicher, A., & Smeddinck, J. D. (2016). **Personalized and Adaptive Serious Games**. In R. Dörner, S. Göbel, M. Kickmeier-Rust, M. Masuch, & K. Zweig (Eds.), Entertainment Computing and Serious Games (Vol. 9970, pp. 332–377), LNCS. Cham: Springer International Publishing. https://doi.org/10.1007/978-3-319-46152-6_14.*

**Contribution of the work in the context of this thesis:** This publication is a chapter of a lecture / tutorial educational book on serious games. The chapter provides a summary of a number of background aspects and discusses challenges and strategies in the design, implementation, and evaluation of adaptive and adaptable serious games. General examples and theory are accompanied by practical examples along the lines of two application areas, namely learning games and games for health. Many sections appear as excerpts throughout this thesis. The publication was the first to draw together the considerations on the scale and dimensions of automation for the use-case of adaptable and adaptive serious games in a way that is expanded upon in this thesis.

**Personal contribution to the work:** Contributed to the writing of all sections; led the definition of the basic terminology, the sections on the scale of automation, modeling dimensions, as well as the background and practical examples from the area of games for health (~55%).

## C.  Additional related publications:

*[C.1] Eggert, C., Herrlich, M., Smeddinck, J., & Malaka, R. (2015). **Classification of Player Roles in the Team-Based Multi-player Game Dota 2**. In K. Chorianopoulos, M. Divitini, J. B. Hauge, L. Jaccheri, & R. Malaka (Eds.), Entertainment Computing - ICEC 2015 (Vol. 9353, pp. 112–125), LNCS. Cham: Springer International Publishing. https://doi.org/10.1007/978-3-319-24589-8_9.*

**Contribution of the work in the context of this thesis:** An exploration of the applicability of a range of common machine learning methods for the automatic classification of player roles that arise out of gaming communities for specific titles, such as multi-player online battle arena games, including the construction and discussion of features, finding logistic regression as the most successful of a range of well-functioning methods. Player role classification augments established classification approaches in game user research based on average performance tiers or player types and has application use-cases in game testing and evaluation, online analysis for broadcasting, as well as in informing adaptivity.

**Personal contribution to the work:** Minor participation in study planning, support of analysis / interpretation, and in the final write-up (~15%).

*[C.2] Gerling, K. M., Mandryk, R. L., Miller, M., Kalyn, M. R., Birk, M., & Smeddinck, J. D. (2015).* **Designing Wheelchair-Based Movement Games**. *ACM Trans. Access. Comput., 6(2), 6:1–6:23. https://doi.org/10.1145/2724729.*

**Contribution of the work in the context of this thesis:** The paper presents a toolkit for integrating wheelchair movements into motion-based games, as well as a discussion of the the-oretical foundations of technical challenges, the accessibility and potential empowerment with motion-based for people using wheelchairs. It also contains reports on the development of a wheelchair-tracking game for older adults and a wheelchair-based implementation of a popular dance game which was employed in a study on approaches to balancing between non-wheelchair users and mixed dyads of wheel-chair users and non-wheelchair users, finding that wheelchair users are appreciative of the notably positive play experiences enabled by these developments and do not express concerns about their performance compared to a non-wheelchair using co-player, regardless of the visibility of the present balancing mechanism. The work summarizes insights regarding the impact of game adaptation as an element of multi-player motion-based game experiences that are inclusive to people using wheelchairs.

**Personal contribution to the work:** Support of the study planning, piloting, execution, and interpretation for the project, using the game Wheelchair Revolution; minor support of the write-up (~5%).

*[C.3] Gerling, K. M., Schulte, F. P., Smeddinck, J., & Masuch, M. (2012).* **Game Design for Older Adults: Effects of Age-Related Changes on Structural Elements of Digital Games**. *In M. Herrlich, R. Malaka, & M. Masuch (Eds.), Entertainment Computing - ICEC 2012 (pp. 235–242), LNCS. Springer Berlin Heidelberg. https://doi.org/10.1007/978-3-642-33542-6_20.*

**Contribution of the work in the context of this thesis:** Following a basic discussion of the promises and challenges of implementing motion-based games for health for the target group of older adults, this paper presents an extension of a popular model for game design and evaluation with a factor that represents human abilities as a resource with limits. This model contributes to the theoretical considerations presented in this thesis, especially with regards to aspects of embodiment, situated use, and heterogeneous players and interest groups.

**Personal contribution to the work:** Contribution to the model extension and all elements of the write-up for publication (~15%).

*[C.4] Gerling, K. M., & Smeddinck, J. (2013).* **Involving Users and Experts in Motion-Based Game Design for Older Adults.** *In Proceedings of the CHI Game User Research Workshop. Presented at CHI 2013, Game User Research Workshop, Paris, France.*

**Contribution of the work in the context of this thesis:** This work discusses the application of a user-centered iterative design process to motion-based games for health for older adults, pointing at additional considerations that are required and highlighting when experts or members of the target audience can each respectively be expected to provide the most valuable feedback, depending on the state of the project development and the type of feedback they are typically able to provide. The line of arguments is also referenced in this thesis.

**Personal contribution to the work:** Contributed to the underlying discussion and model construction, as well as to the write-up for publication (~40%).

*[C.5] Hermann, R., Herrlich, M., Wenig, D., Smeddinck, J. & Malaka, R. (2013).* ***Strong and Loose Cooperation in Exergames for Older Adults with Parkinson's Disease****. In Workshop Entertainment Computing: Mensch & Computer 2013 - Workshopband: Interaktive Vielfalt, Oldenbourg Verlag, München.*

**Contribution of the work in the context of this thesis:** A study comparing strong (mutually dependent) and loose (independent parallel play) cooperation as potential motivating factors in multiplayer motion-based games for older adults with Parkinson's disease. Strong cooperation was found to entail benefits such as increased communication and coordination between players with higher overall game scores, while 50% of players still expressed a preference towards the loose cooperation mode. Insights in multiplayer motion-based games for health can inform game design, and strategies for balancing, as well as adaptability and adaptivity.

**Personal contribution to the work:** Minor participation in study planning, support of interpretation, and in the final write-up (~5%).

*[C.6] Herrlich, M., Wenig, D., Walther-Franks, B., Smeddinck, J. D., & Malaka, R. (2014). „****Raus aus dem Sessel****" – Computerspiele für mehr Gesundheit: Eine Übersicht und aktuelle Beispiele****. Informatik-Spektrum, 37(6), 558–566. https://doi.org/10.1007/s00287-014-0825-1.*

**Contribution of the work in the context of this thesis:** This publication summarizes the concept as well as the basic promises and the history of exergames and motion-based games, highlighting games to support physiotherapy for older adults, and the augmentation of existing commercial games with components for natural user interface control and feedback as contemporary trends.

**Personal contribution to the work:** Support of most projects referenced as underlying work and of the write-up for publication (~10%).

*[C.7] Krause, M., & Smeddinck, J. (2011).* ***Human Computation Games: A Survey****. In 19th European Signal Processing Conference, 2011 (pp. 754–758). Barcelona, Spain: EURASIP.*

**Contribution of the work in the context of this thesis:** The paper discusses motivation as a central challenge in human computation that can be tackled with games and discusses a number of viable application use-cases for such games in application areas which are difficult to approach through fully automated computation, since they require intuitive decisions, aesthetic judgment, contextual reasoning, or embodiment. This discussion of potential application areas informs the applicability of methods from human computation in the context of adaptable and adaptive motion-based games for health.

**Personal contribution to the work:** Participation in the identification of the viable application use-cases, the facilitation of the technical implementation of exemplary human computation services, and support of the writing of the paper (~15%).

*[C.8] Krause, M., Smeddinck, J., & Meyer, R. (2013).* **A Digital Game to Support Voice Treatment for Parkinson's Disease**. *In CHI '13 Extended Abstracts on Human Factors in Computing Systems (pp. 445–450). New York, NY, USA: ACM. https://doi.org/10.1145/2468356.2468435.*

**Contribution of the work in the context of this thesis:** This publication explores the applicability of games for health for the support of voice therapy for people with Parkinson's disease. Based on a game that was specifically built for this purpose using a user-centered iterative design process that included both Parkinson's patients and therapists, the paper presents results from an early study with eight participants that show increases in peak loudness, as an important key indicator for vocal functional ability, following test sessions with the game. In the context of this thesis, this work presents an example of application areas beyond motion-based games for health, provided further insights regarding the applicability of human-centered design in the context of games for health with complex target groups such as people with Parkinson's disease, and also represents a potential target for the future addition of adaptivity with a comparatively simple parameter space.

**Personal contribution to the work:** Support in the supervision of Ronald Meyer, the study analysis, results interpretation, and in the write-up for publication (~20%).

*[C.9] Sarma, H., Porzel, R., Smeddinck, J., & Malaka, R. (2015).* **Towards Generating Virtual Movement from Textual Instructions A Case Study in Quality Assessment**. *In Proceedings of the 3rd AAAI Conference on Human Computation and Crowdsourcing (HCOMP '15), Works-in-Progress. San Diego: AAAI.*

**Contribution of the work in the context of this thesis:** A brief introduction of the applicability of methods from human computation in the context of games for health (performance analysis and content generation, as well as the respective quality control) and a pre-study on the applicability of alternative types of movement visualizations for crowd-based ratings of exercise execution quality.

**Personal contribution to the work:** Support of the supervision of Himangshu Sarma, of the study planning, execution, and results interpretation, as well as of the write-up for publication (~45%).

*[C.10] Smeddinck, J., Krause, M., & Lubitz, K. (2013).* **Mobile Game User Research: The World as Your Lab?** *Presented at the CHI 2013, Game User Research Workshop, Paris, France.*

**Contribution of the work in the context of this thesis:** A comparison between the results and insights of a study on game controller modalities in a small-scale partially controlled "in-the-wild" setting with the results of a large-scale study of game controller modality choice in a released product underlines a discussion of the importance of evaluating games in conditions of real-world situated use.

**Personal contribution to the work:** Support in the setup and analysis of both studies, as well as spearheading the write-up (~80%).

*[C.11] Walther-Franks, B., Smeddinck, J., Szmidt, P., Haidu, A., Beetz, M., & Malaka, R. (2015).* **Robots, Pancakes, and Computer Games: Designing Serious Games for Robot Imitation Learning**. *In Proceedings of the 33rd Annual ACM Conference on Human Factors in Computing Systems (CHI 2015) (S. 3623–3632). New York, NY, USA: ACM. https://doi.org/10.1145/2702123.2702552*

**Contribution of the work in the context of this thesis:** Employing the motivational pull of video games in order to improve the collection of human-made samples for programming by demonstration presents a novel use-case for serious games and is shown to lead to increased interaction time and experience compared to demonstrating actions in a virtual environment that is not a game. Hand-movement tracking presents a different type of body-movement tracking that is typically not included in contemporary motion-based games and can facilitate a broader range of games for health.

**Personal contribution to the work:** Support of the game design process with feedback and recommendations, support of the write-up with a focus on game-related aspects (~15%).

*[C.12] Walther-Franks, B., Wenig, D., Smeddinck, J., & Malaka, R. (2013).* **Exercise My Game: Turning Off-The-Shelf Games into Exergames**. *In J. C. Anacleto, E. W. G. Clua, F. S. C. da Silva, S. Fels, & H. S. Yang (Eds.), Entertainment Computing – ICEC 2013 (pp. 126–131), LNCS. Springer Berlin Heidelberg. https://doi.org/10.1007/978-3-642-41106-9_15.*

**Contribution of the work in the context of this thesis:** The paper introduces *Exercise My Game*, a framework consisting of the design aspects *game choice*, *control overlay*, *feedback overlay*, and *workout adaptations* for turning off-the-shelf-games into exergames. The approach can be applied for the affordable improvement of strong medium- to long-term motivation to actively engage with exergames or games for health, by piggybacking on existing - potentially high quality - game content that would otherwise be prohibitively expensive to produce.

**Personal contribution to the work:** Support of the student team behind the underlying student project *sPortal* through semi-regular advising, support of the write-up for publication (~25%).

*[C.13] Walther-Franks, B., Wenig, D., Smeddinck, J., & Malaka, R. (2013).* **Suspended Walking: A Physical Locomotion Interface for Virtual Reality**. *In J. C. Anacleto, E. W. G. Clua, F. S. C. da Silva, S. Fels, & H. S. Yang (Eds.), Entertainment Computing – ICEC 2013 (pp. 185–188), LNCS. Springer Berlin Heidelberg. https://doi.org/10.1007/978-3-642-41106-9_27.*

**Contribution of the work in the context of this thesis:** The development, discussion, and comparison of suspended walking as a technique for locomotion in virtual reality and game applications that is supported by a ceiling-mounted harness and results in interaction which is similar to walking-in-place. A locomotion technique based on a harness can be accessible to people who could otherwise not participate in motion-based games that require free locomotion. The according motion-detection techniques for walking, as well as direction of sight, and pointing can augment the commonly used camera-based approaches.

**Personal contribution to the work:** Support of the student team behind the underlying student project *sPortal* (semi-regular advising), support of the write-up for publication (~25%).

# List of Figures

# List of Tables

## Analysis for Perception of Game Difficulty and Difficulty Adjustments

*This attachment presents the full analysis of the study summarized in section 4.3.*

### Procedure:

Agreement to the statements was indicated on *5 pt. Likert scale* items with a scale ranging from *disagree* to *agree*. The analysis employs descriptive results of the whole participant sample, as well as contrasting between subgroup splits by *gender* and *study country*. Inference statistics are performed with *unpaired t-tests* at a significance level of $\alpha = .05$, reporting *Cohen's d* for *effect sizes* in case of significant differences or trends ($\alpha < .1$) when the assumption of normality was not rejected due to *Shapiro-Wilk tests*. In case of a violation of the assumption of normality, the results of *Mann-Whitney's U* tests are reported, with according effect sizes for nonparametric data (Fritz et al., 2012). Correlations are included between the questionnaire items and the *participant age* and the self-reported *number of years that participants have been gaming* if notable *Pearson* (assuming normality assumption was not rejected) or *Spearman* p values cross the .05 significance threshold.

### Results:

A number of statements were concerned with the interaction of players with manual difficulty choices. When asked whether *[they] always adjust the difficulty settings, when [they] play a game for the first time*, the participants on average expressed only very slight agreement (M = **2.93**, SD = 1.34), appearing rather ambivalent. While there were some differences in means based on the country and self-identified gender (CAN: M = **3.12**, SD = 1.2; GER: M = **2.5**, SD = 1.54; f: M = **2.65**, SD = 1.38; m: M = **3.16**, SD = 1.27), suggesting more agreement by the participants in the study that was conducted in Canada and by male participants, exact Mann-Whitney's U tests for contrasting did not indicate statistical significance. In a similar vein, there was some agreement on average to the statement *"When I play a game for the first time, I only adjust the difficulty settings if I am prompted to make a choice."* (M = **3.09**, SD = 1.3). In this case, a significant difference was indicated between male and female participants that runs inversely to the difference in means on the prior item (f: M = **3.58**, SD = 1.1, Mdn = 4; m: M = **2.69**, SD = 1.33, Mdn = 2.5; U = 579, p < .01, r = .34), showing a medium effect size. The finding persists in the participants in Germany as a significant difference (GER_f: M = **3.71**, SD = 1.11; GER_m: M = **2.45**, SD = 1.29; t(14) = 2.2, p < .05, d = 1.03) with a large effect size, and in the participants in Canada as a trend (CAN_f: M = **3.53**, SD = 1.12, Mdn = 4; CAN_m: M = **2.81**, SD = 1.36, Mdn = 3; U = 262, p < .1, r = .28).

On average, the participants noted that they are rather *aware of the difficulty settings when they play a game* (M = **3.53**, SD = 1.19). The data contain a trend towards a difference based on study country (CAN: M = **3.75**, SD = 0.98, Mdn = 4; GER: M = **3.06**, SD = 1.47, Mdn = 3; U = 456, p < .1, r = .22), suggesting a small effect size. This trend does not appear to originate from the male subsamples (CAN_m: M = **3.76**, SD = 1, Mdn = 4; GER_m: M = **3.45**, SD = 1.44, Mdn = 4), but rather from a difference in female subsamples (CAN_f: M = **3.74**, SD = .99, Mdn = 4; GER_f: M = **2.43**, SD = 1.4, Mdn = 3; U = 103, p < .05, r = .43), which amounts to a significant difference with medium effect size. In this case, there was no apparent difference due to gender (f: M = **3.38**, SD = 1.24, Mdn = 4; m: M = **3.66**, SD = 1.15, Mdn = 4).

Although being rather aware of the difficulty settings, the participants, on average neither disagreed nor agreed with the statement *"I frequently readjust game difficulty settings"* (M = **2.59**, SD = 1.3), and there were no notable differences based on study country (CAN: M = **2.67**, SD = 1.42; GER: M = **2.39**, SD = .98) or gender regarding this statement (f: M = **2.65**, SD = 1.29; m: M = **2.53**, SD = 1.32). A trend was found for a weak correlation between the item and the participant age (*Spearman's r* ($r_s$) = **.24**, p < .1), indicating slightly larger agreement with increasing participant age. The quasi-inverse statement *"I never readjust game difficulty settings"* led to comparable results.

A separate series of three items asked participants to indicate their agreement to the statements *whether they usually play on low / normal / hard difficulty settings* and led to a number of differences. While overall average agreement to playing on *low difficulty settings* was rather low (M = **2.09**, SD = 1.17), the *sample from Canada* on average tended to agree more to this statement than the *sample from Germany* (CAN: M = **2.25**, SD = 1.21; GER: M = **1.72**, SD = 1.02), although the difference was not pronounced enough to trigger a trend or significance. A separation based on *gender* shows a significant difference (f: M = **2.62**, SD = 1.24, Mdn = 2; m: M = **1.66**, SD = .94, Mdn = 1; U = 609, p < .01, r = .42) with a medium effect size with more agreement by females. The effect persisted in the country subsets. A correlation with the *participant age* revealed a significant medium positive relationship ($r_s$ = **.37**, p < .05), while a correlation with *years of gaming experience* showed a small negative relationship ($r_s$ = **-.28**, p < .05). Regarding the *normal (or medium) difficulty* choice, the participants on average rather agreed (M = **3.59**, SD = 1.14), and the item did not lead to pronounced or significant differences based on *country* (CAN: M = **3.58**, SD = 1.13; GER: M = **3.61**, SD = 1.2) or *gender* (f: M = **3.81**, SD = .98; m: M = **3.41**, SD = 1.24), although female participants showed slightly higher agreement on average. Compared to this item, the participants showed less, yet still slightly positive agreement to the statement regarding *high difficulty settings* (M = **2.98**, SD = 1.21). In this case, a difference based on country could not be observed (CAN: M = **2.95**, SD = 1.22; GER: M = **3.06**, SD = 1.21), however, in reversed agreement with the item on low (or easy) difficulty settings, *male participants* were found to be significantly more likely to respond in agreement (f: M = **2.42**, SD = 1.14, Mdn = 2; m: M = **3.44**,

SD = 1.08, Mdn = 4; U = 218.5, p < .01, r = .42), with a medium effect size. The according correlation to *participant age* revealed a trend for a *weak relationship* ($r_s$ = **.23**, p < .1). Regarding the more general statement *"I like it when games are challenging"*, the participants on average expressed notable agreement (M = **4.05**, SD = .78), and the effect based on *gender* that could be expected based on the prior items on difficulty was confirmed (f: M = **3.69**, SD = .74, Mdn = 4; m: M = **4.34**, SD = .7, Mdn = 4; U = 228, p < .01, r = .42) through a significant difference with medium effect size, while no significant difference was found based on country (CAN: M = **4.12**, SD = .82; GER: M = **3.98**, SD = .68). A trend for a *weak negative relationship* with *participant age* was found ($r_s$ = **-.24**, p < .1), as well as a significant yet also *weak positive relationship* with *years of gaming experience* ($r_s$ = **.28**, p < .05).

When asked whether they agree with the statement *"I have a sufficient amount of control over difficulty settings in most games I play"*, the participants on average tended to agree (M = **3.45**, SD = 1.01), with no notable difference based on country (CAN: M = **3.42**, SD = 1.03; GER: M = **3.5**, SD = .99). However, *male participants* expressed significantly larger agreement on average (f: M = **2.85**, SD = 1.01, Mdn = 3; m: M = **3.94**, SD = .72, Mdn = 4; U = 176.5, p < .001, r = .52) with a large effect size. A trend for *weak negative relationship* was indicated in correlation with *participant age* ($r_s$ = **-.25**, p < .1). Regarding the statement *"When first playing a new game, I just want to get started"*, the participants on average express agreement (M = **3.55**, SD = 1.23), with only a slight difference between *study countries* that amounts to a trend (CAN: M = **3.38**, SD = 1.19, Mdn = 3.5; GER: M = **3.94**, SD = 1.26, Mdn = 4; U = 253, p < .1, r = .24), yet again, a significant difference based on *gender* (f: M = **3.96**, SD = 1.08, Mdn = 4; m: M = **3.22**, SD = 1.26, Mdn = 3; U = 559.5, p < .05, r = .3) with a medium effect size. For the related item *"When first playing a new game I configure the settings to my liking"*, the participants on average also expressed agreement (M = **3.33**, SD = 1.25), and there was again a sig. difference with a medium effect size based on *gender* (f: M = **2.77**, SD = 1.18, Mdn = 3; m: M = **3.78**, SD = 1.13, Mdn = 4; U = 219, p < .01, r = .42), accompanied by a significant effect with medium effect size based on *country* (CAN: M = **3.62**, SD = 1.17, Mdn = 4; GER: M = **2.67**, SD = 1.19, Mdn = 3; U = 514, p < .01, r = .35).

A further set of statements was concerned with the participants' perspective explicitly on *automatic difficulty adjustments*. The statement *"I would like to play games which adjust the difficulty settings automatically"* was met with slight agreement on average (M = **3.12**, SD = 1.39), with no notable difference based on country (CAN: M = **3.1**, SD = 1.41; GER: M = **3.17**, SD = 1.38), or based on gender (f: M = **3.08**, SD = 1.32; m: M = **3.16**, SD = 1.46). A weak negative correlation was indicated in relation to *years of gaming experience* ($r_s$ = **-.26**, p < .05). However, when confronted with the statement *"If a game I played would adjust the difficulty settings automatically* *depending on my performance, I would like to be able to influence the resulting settings"*, the participants on average expressed agreement (M = **3.98**, SD = .95), and while there was no pronounced difference based on country (CAN: M = **3.92**, SD = .94; GER: M = **4.11**, SD = .96), a significant difference with medium effect size was found based on *gender* (f: M = **3.65**, SD = 1.06, Mdn = 4; m: M = **4.25**, SD .76, Mdn = 4; U = 277.5, p < .05, r = .3). The latter difference was also found in subsets by country. In this case, a trend for a weak positive correlation to years of gaming experience was found ($r_s$ = **.24**, p < .1).